\documentclass[thesis,a4paper,12pt,oneside]{ucdThesis} 

\usepackage{acronym}
\usepackage{amsfonts}
\usepackage{amsmath,amssymb}
\usepackage{array}
\usepackage[greek,english]{babel} 
\usepackage[font={small}]{caption}
\usepackage{cite}
\usepackage{color}
\usepackage{comment}
\usepackage{enumerate}
\usepackage{enumitem}
\usepackage{eurosym}
\usepackage{float}
\usepackage{footnote}
\usepackage{graphics}
\usepackage{graphicx}
\usepackage{scrextend}
\usepackage{multicol}
\usepackage{multirow}
\usepackage{pdflscape}
\usepackage{pifont}
\usepackage{ragged2e}
\usepackage{xpatch}

\usepackage{rotating}

\usepackage{subcaption}
\usepackage{upgreek}

\usepackage[noprefix]{nomencl}

\usepackage{epsf}
\usepackage{wasysym}
\usepackage{tikz}

\usepackage{esvect}
\usepackage{shortvrb}
\usepackage{empheq}
\usepackage{flushend}
\usepackage{lastpage}

\newcommand{\bfg}[1]{\boldsymbol{#1}}

\newcommand{\bfi}[1]{\tilde{\boldsymbol{#1}}}
\newcommand{\bfb}[1]{\boldsymbol{\rm #1}}

\newcommand{\jj}{\jmath}
\newcommand{\T}{^{\scriptscriptstyle \rm T}}
\newcommand{\CT}{^{\scriptscriptstyle \rm H}}

\newcommand{\eop}{\ensuremath{\mbox{$\blacksquare$}}} 

\newcommand{\sss}{\scriptscriptstyle}

\newcommand{\dGi}[1]{\delta_{{\rm r},#1}}
\newcommand{\wGi}[1]{\omega_{{ \rm r},#1}}
\newcommand{\dG}{\delta_{{\rm r}}}
\newcommand{\wG }{\omega_{ \rm r}}

\newcommand{\torque}{\uptau}

\newcommand{\Inert}{M}
\newcommand{\Gain}{K}
\newcommand{\point}{\Sigma}
\newcommand{\TFUN}{H}

\newcommand{\dax}{{\text d}}
\newcommand{\qax}{{\text q}}
\newcommand{\dqax}{{\text dq}}

\newcommand{\Ohm}{$\Omega$}
\newcommand{\Mho}{$\Omega^{-1}$}

\newcommand{\x}{\ensuremath{\mbox{$ \text{x}$}}}
\newcommand{\xs}{\ensuremath{\mbox{$ \textbf{x}$}}}
\newcommand{\Xs}{\ensuremath{\mbox{$ \textbf{X}$}}}
\newcommand{\xy}{\ensuremath{\mbox{$ \text{z}$}}}
\newcommand{\xys}{\ensuremath{\mbox{$ \textbf{z}$}}}

\newcommand{\bfbt}[1]{\tilde{\boldsymbol{\rm #1}}}

\newcommand{\nx}{n}
\newcommand{\ny}{m}
\newcommand{\nxy}{l}
\newcommand{\nin}{p}
\newcommand{\nout}{q}

\newcommand{\nd}{r}

\newcommand{\nf}{\nu}
\newcommand{\nfi}[1]{\nu_{#1}}
\newcommand{\ninf}{\mu}
\newcommand{\njb}{\alpha}
\newcommand{\rjb}{\beta}

\newcommand{\Dt}[1]{\dot {#1} }

\newcommand{\jacs}[2]{{\bfg #1}_{\bfg #2}}

\newcommand{\jac}[2]{{ #1}_{ #2}}
\newcommand{\jacsinv}[2]{{\bfg #1}_{\bfg #2}^{-1}}

\newcommand{\poly}{\ensuremath{\mbox{$\varphi$}}}
\newcommand{\reigv}{\ensuremath{\mbox{$ \textbf{v}$}}}
\newcommand{\reigvel}{\ensuremath{\mbox{$ \text{v}$}}}
\newcommand{\reigvmat}{\ensuremath{\mbox{$ \textbf{V}$}}}
\newcommand{\leigv}{\ensuremath{\mbox{$ \textbf{w}$}}}
\newcommand{\leigvel}{\ensuremath{\mbox{$ \text{w}$}}}
\newcommand{\leigvmat}{\ensuremath{\mbox{$ \textbf{W}$}}}
\newcommand{\PF}{{\uppi}}
\newcommand{\PFmat}{\ensuremath{\mbox{$ \bfb{\Pi}_{\scriptscriptstyle \rm PF}$}}}
\newcommand{\PFmatw}[1]{{\bfb \Pi}_{{\scriptscriptstyle \rm PF}, #1}}
\newcommand{\hPFmatw}[2]{{#1 {\bfb \Pi}}_{{\scriptscriptstyle \rm PF}, #2}}

\newcommand{\res}{\ensuremath{\mbox{$\text{R}$}}}

\newcommand{\dome}{{Dome} }
\newcommand{\newengland}{IEEE 39-bus }
\newcommand{\threebus}{3-bus }
\newcommand{\Threebus}{3-bus }

\newcommand{\Efr}{ {\bfi E} }
\newcommand{\Afr}{ {\bfi A} }

\newcommand{\Esng}{ {\bfb E} }
\newcommand{\Asng}{ {\bfb A} }
\newcommand{\Bsng}{ {\bfb B} }
\newcommand{\Csng}{ {\bfb C} }
\newcommand{\Dsng}{ {\bfb D} }

\newcommand{\Edae}{ {\bfg E_a} }
\newcommand{\Adae}{ {\bfg A_a} }
\newcommand{\Bdae}{ {\bfg B_a} }
\newcommand{\Cdae}{ {\bfg C_a} }
\newcommand{\Ddae}{ {\bfg D_a} }

\newcommand{\AS}{ {\bfg A_s} }
\newcommand{\BS}{ {\bfg B_s} }

\newcommand{\cmark}{\ding{51}}%
\newcommand{\xmark}{\ding{55}}%

\title{Small-Signal Stability Techniques for Power System Modal Analysis, Control, and Numerical Integration}
\author{Georgios Tzounas}  
\headSchool{Prof. Peter Kennedy}
\supervisor{Prof. Federico Milano}	
\graduationMonth{February}
\copyrightYear{2021} 
\major{Electrical Engineering}	
\degree{Doctor of Philosophy}	   
\college{Engineering and Architecture}     
\dept{Electrical \& Electronic Engineering}

\DeclareGraphicsExtensions{.pdf}

\makenomenclature

\makeindex

\begin{document}

\pagenumbering{alph} 
\addToPDFBookmarks{0}{Front Matter}{rootNode} 
\addToPDFBookmarks{1}{Title}{a} 

\makeTitlePage 

\newpage
\pagenumbering{roman}

 
\pagestyle{plain}

I hereby certify that the submitted work is my own work, was completed while registered as a candidate for the degree stated on the Title Page, and I have not obtained a degree elsewhere on the basis of the research presented in this submitted work.
`
\vfill
\begin{center}
\copyright ~Georgios Tzounas, 2020.
All Rights Reserved. 
\end{center}
\vspace{30mm}

\begin{spacing}{1.5}

\newpage
\chapter*{Acknowledgements}

The content of this thesis is the result of my three-year long research
in Dublin, Ireland, 
working under the supervision of Prof.~Federico Milano and within the AMPSAS (Advanced Modelling for Power System Analysis and Simulation)
project. 
%
%
Therefore, on completion of writing this thesis,
I wish to thank my mentor, Federico,
for the constant support, motivation, and feedback he provided throughout my PhD. 
Working with him has been a great experience, mainly for two reasons: first, 
his immense knowledge and passion about power systems worked as a source of inspiration and made me 
always 
try to give my best;
and second, he always encouraged me to explore different directions and implement my ideas, 
which 
made the PhD an enjoyable adventure.
%

In addition, I would like to thank all the 
members of the group 
for their invaluable help, our interesting discussions, and mostly, for sharing 
everyday life 
on and off campus. 
Special thanks to Dr.~Ioannis Dassios, first for his friendship and then for the
continuous assistance and 
collaboration, which have definitely influenced my path toward the PhD and, in turn, the content of this thesis.
Special thanks also to my pals and colleagues Dr.~Mohammed Ahsan Adib Murad and Dr.~Muyang Liu, with whom I had extended collaboration within the group. 

Many thanks to Prof.~Rifat Sipahi, for his guidance and pleasant collaboration during my 
stay at Northeastern University in Boston, MA. 

Finally, I am very grateful to some people who 
have supported me along 
this way in my PhD but most importantly, in my life in general. 
These are my parents and siblings, 
my friends in Greece, and Avgi. To all of you, thanks for your love and for putting up with me being far from home for such a long period. 

\begin{flushright}
\textit{Georgios Tzounas}

\textit{Dublin, 
October 2020}
\end{flushright}

\newpage

\cleardoublepage
\thispagestyle{empty}
\vspace*{\stretch{1}}
\begin{center}
  \large\itshape
  \textgreek{Στους γονείς μου, Σοφία και Γιώργο.} 
  \\\phantom{xx}
\end{center}

\vfill
\newpage
\section*{Abstract}

This thesis proposes novel Small-Signal Stability Analysis (SSSA)-based techniques that contribute to electric power system modal analysis, automatic control, and numerical integration.

Modal analysis is a fundamental tool for power system stability analysis and control. The thesis proposes a SSSA approach to determine the Participation Factors (PFs) of algebraic variables in power system dynamic modes. The approach is based on a new interpretation of the classical modal PFs as eigen-sensitivities, as well as on the definition of adequate inputs and outputs of the system's state-space representation.  Both linear and generalized eigenvalue problems are considered for the calculation of PFs and a theorem to cope with eigenvalue multiplicities is presented. 

SSSA is also ubiquitous in the synthesis of controllers for power systems. The thesis explores SSSA techniques for the design of power system controllers. The contributions on this topic are twofold, as follows: 

(i) Investigate a promising control approach, that is to synthesize automatic regulators for power systems based on the theory of fractional calculus. In particular, using eigenvalue analysis, a comprehensive theory on the stability of power systems with inclusion of Fractional Order Controllers (FOCs) is provided. Moreover, the software implementation of FOCs based on Oustaloup's Recursive Approximation (ORA) method is discussed. A variety of FOC applications are illustrated, namely, automatic generation control of synchronous machines; frequency control of a converter-interfaced energy storage system; and voltage control through a static synchronous compensator. 

(ii) Propose a novel perspective on the potential impact of time delays on power system stability. In general, measurement and communication of control signals in electric energy networks introduces significant time delays that are known to be a threat for the dynamic performance of power systems. However, research in control theory has shown that, by nature, delays are neutral and, if properly introduced, can also stabilize a dynamical system. Through SSSA, the thesis systematically identifies the control parameter settings for which delays in Power System Stabilizers (PSSs) improve the damping of a power system. Both analytical and simulation-based results are presented.

Finally, SSSA is utilized in the thesis to systematically propose a delay-based method to reduce the coupling of the equations of power system models for transient stability analysis.  The method consists in identifying the variables that, when subjected to a delay equal to the time step of the numerical integration, leave practically unchanged the system trajectories.  Automatic selection of the variables and estimation of the maximum admissible delay are carried out by SSSA-based techniques.  Such an one-step-delay approximation increases the sparsity of the system Jacobian matrices and can be used in conjunction with state-of-the-art techniques for the integration of Differential-Algebraic Equations (DAEs).  The proposed approach is evaluated in terms of accuracy, convergence and computational burden.

Throughout the thesis, the proposed techniques are duly validated through numerical tests based on real-world network models.



\newpage
%

\tableofcontents

\cleardoublepage
\phantomsection
\addcontentsline{toc}{chapter}{\listfigurename}
\listoffigures

\cleardoublepage
\phantomsection
\addcontentsline{toc}{chapter}{\listtablename}

\listoftables

\makenomenclature
{\RaggedRight \printnomenclature[2cm]}

\cleardoublepage
\phantomsection
\addcontentsline{toc}{chapter}{Acronyms}
\chapter*{List of Acronyms and Abbreviations}

\begin{acronym}[STATCOM]

\acro{agc}[\textsc{AGC}]{Automatic Generation Control}
\acro{aiits}[\textsc{AIITS}]{All-Island Irish Transmission System}
\acro{avr}[\textsc{AVR}]{Automatic Voltage Regulator}
\acro{bbd}[\textsc{BBD}]{Bordered-Block Diagonal}
\acro{cih}[\textsc{CIH}]{Contour Integration with Hankel matrices}
\acro{cirr}[\textsc{CIRR}]{Contour Integration with Rayleigh-Ritz}
\acro{coi}[\textsc{CoI}]{Center of Inertia}
\acro{csr}[\textsc{CSR}]{Compressed Sparse Row}
\acro{dae}[\textsc{DAE}]{Differential-Algebraic Equation}
\acro{ddae}[\textsc{DDAE}]{Delay Differential-Algebraic Equation}
\acro{dde}[\textsc{DDE}]{Delay Differential Equation}
\acro{emt}[\textsc{EMT}]{Electro-Magnetic Transient}
\acro{entsoe}[\textsc{ENTSO-E}]{European Network of Transmission System Operators for Electricity}
\acro{erd}[\textsc{ERD}]{Explicitly Restarted and
  Deflated}  
\acro{ess}[\textsc{ESS}]{Energy Storage System}
\acro{ewic}[\textsc{EWIC}]{East-West
Inter-Connector}
\acro{facts}[\textsc{FACTS}]{Flexible AC Transmission System}
\acro{FO}[\textsc{FO}]{Fractional Order}
\acro{foc}[\textsc{FOC}]{Fractional-Order Controller}
\acro{foi}[\textsc{FOI}]{Fractional-Order Integral}
\acro{fopi}[\textsc{FOPI}]{Fractional-Order Proportional Integral}
\acro{FOPID}[\textsc{FOPID}]{Fractional-Order Proportional Integral Derivative}
\acro{fopss}[\textsc{FOPSS}]{Fractional-Order Power System Stabilizer}
\acro{gd}[\textsc{GD}]{Generalized Davidson}
\acro{gep}[\textsc{GEP}]{Generalized Eigenvalue Problem}
\acro{grpr}[\textsc{GRPR}]{Generator Reactive Power Regulator}
\acro{hvdc}[\textsc{HVDC}]{High Voltage Direct Current}
\acro{IO}[\textsc{IO}]{Integer Order}
\acro{itm}[\textsc{ITM}]{Implicit Trapezoidal Method}
\acro{jd}[\textsc{JD}]{Jacobi-Davidson}
\acro{lep}[\textsc{LEP}]{Linear Eigenvalue Problem}
\acro{LM}[\textsc{LM}]{Largest Magnitude}
\acro{LRP}[\textsc{LRP}]{Largest Real Part}
\acro{lti}[\textsc{LTI}]{Linear Time-Invariant}
\acro{mpi}[\textsc{MPI}]{Message
Passing Interface}
\acro{nnz}[\textsc{NNZ}]{Number of Non-Zero}
\acro{ode}[\textsc{ODE}]{Ordinary Differential Equation}
\acro{omib}[\textsc{OMIB}]{One-Machine Infinite-Bus}
\acro{ora}[\textsc{ORA}]{Oustaloup's Recursive Approximation}
\acro{pde}[\textsc{PDE}]{Partial Differential Equation}
\acro{pf}[\textsc{PF}]{Participation Factor}
\acro{PI}[\textsc{PI}]{Proportional Integral}
\acro{pid}[\textsc{PID}]{Proportional Integral Derivative}
\acro{pod}[\textsc{POD}]{Power Oscillation Damper}
\acro{pr}[\textsc{PR}]{Proportional Retarded}
\acro{pss}[\textsc{PSS}]{Power System Stabilizer}
\acro{rci}[\textsc{RCI}]{Reverse
Communication Interface}
\acro{RL}[\textsc{RL}]{Riemann-Liouville}
\acro{rocof}[\textsc{RoCoF}]{Rate of Change of Frequency}
\acro{SM}[\textsc{SM}]{Smallest Magnitude}
\acro{sssa}[\textsc{SSSA}]{Small-Signal Stability Analysis}
\acro{statcom}[\textsc{STATCOM}]{STATic synchronous COMpensator}
\acro{svc}[\textsc{SVC}]{Static Var Compensator}
\acro{svr}[\textsc{SVR}]{Secondary Voltage Regulation}
\acro{tdi}[\textsc{TDI}]{Time Domain Integration}
\acro{tg}[\textsc{TG}]{Turbine Governor}
\acro{vsc}[\textsc{VSC}]{Voltage Source Converter}
\acro{wadc}[\textsc{WADC}]{Wide-Area Damping Controller}
\acro{wams}[\textsc{WAMS}]{Wide-Area Measurement System}
\acro{wscc}[\textsc{WSCC}]{Western Systems Coordinating Council}


\end{acronym}

\chapter*{Notation}

This section states the notation adopted throughout the thesis.

\xpatchcmd{\labeling}
  {\settowidth}
  {\itemsep=-1.5pt\parsep=0pt\topsep=0pt\partopsep=0pt\settowidth}
  {}{}

\subsubsection*{Vectors and Matrices}

\begin{labeling}{xxxxxxx} 
\item[$a$, $A$] scalar
\item[$\bfg a$, $\bfb a$] vector
\item [$\bfg A$, $\bfb A$] matrix 
\item [$\bfg A\T$] matrix transpose
\item [$\bfg A\CT$] matrix conjugate transpose (Hermitian)
\item [$\bfg I_\nd$] identity matrix of dimensions $\nd \times \nd$
\item [$\bfg 0_{\nd,m}$] zero matrix of dimensions $\nd \times m$
\end{labeling}

\subsubsection*{Sets and Units}

\begin{labeling}{xxxxxxx} 
\item[$\mathcal{C}^n$] $n$ time continuously differentiable functions
\item[$\mathbb{C}$] complex numbers
\item[$\jj$] unit imaginary number 
\item[$\mathbb{I}$] imaginary numbers
\item[$\mathbb{N}$] natural numbers
\item[$\mathbb{R}$] real numbers
\item[$\mathbb{Z}$] integer numbers
\end{labeling}

\subsubsection*{Time and Frequency Domain}

\begin{labeling}{xxxxxxx} 
\item[$a(t)$] time domain quantity
\item[$\Dt a(t)$] first order derivative
\item[$\ddot a(t)$] second order derivative
\item[$a^{(n)}(t)$] $n$-th order derivative (of fractional order or of integer order $\geq 3$)
\item[$a(s)$] frequency domain quantity
\item[$\mathcal{L}$] Laplace transform 
\item[$s$] complex Laplace variable
\item[$z$] spectral transform
\end{labeling}



\subsubsection*{Eigenvalues and Eigenvectors}

\begin{labeling}{xxxxxxx} 
\item[${\rm f}_{\rm n}$] natural frequency of eigenvalue
\item[$\reigvel$] element of right eigenvector
\item[$\reigv$] right eigenvector
\item[$\reigvmat$] right modal matrix
\item[$\leigvel$] element of left eigenvector
\item[$\leigv$] left eigenvector
\item[$\leigvmat$] left modal matrix
\item[$\zeta$] damping ratio of eigenvalue
\item[$\lambda$] eigenvalue
\item[${\rm Im}(\lambda)$] imaginary part of eigenvalue $\lambda$
\item[${\rm Re}(\lambda)$] real part of eigenvalue 
$\lambda$
\item[$\ninf$] multiplicity of infinite eigenvalue
\item[$\nf$] number of finite eigenvalues
\item[$\PF$] participation factor
\item[$\PFmat$] participation matrix
\end{labeling}



\subsubsection*{Parameters}

\begin{labeling}{xxxxxxx}  
\item[$B$] susceptance
\item[$c$] constant
\item[$D$] damping coefficient
\item[$G$] conductance
\item[$h$] time step size
\item[$K$] control gain
\item[$M$] mechanical starting time
\item[$R$] resistance  
\item[$\mathcal{R}$] droop constant of primary frequency control
\item[$T$] time constant 
\item[$X$] reactance
\item[$\tau$] time delay
\end{labeling}

\subsubsection*{Variables and Functions}

\begin{labeling}{xxxxxxx}  
\item[$e$] electromotive force
\item[$P$] active power
\item[$Q$] reactive power
\item[$t$] time 
\item[$u$] input signal  
\item[$v$] voltage magnitude
\item[$\mathcal{V}$] Lyapunov function
\item[$w$] output signal   
\item[$x$] state variable 
\item[$y$] algebraic variable 
\item[$\delta$] angular position
\item[$\theta$] voltage phase angle
\item[$\uptau$] torque
\item[$\omega$] angular speed
\end{labeling}


\subsubsection*{Superscripts and Subscripts}
\begin{labeling}{xxxxxxx}   
\item[$\rm d$] direct axis of the $\rm dqo$ transform
\item[$d$] delayed quantity
\item[$\rm m$] mechanical
\item[$\rm min$] minimum
\item[$\rm max$] maximum
\item[$o$] initial condition
\item[$\rm q$] quadrature axis of the $\rm dqo$ transform
\item[$\rm r$] rotor
\item[$\rm ref$] reference
\item[$\rm s$] stator
\end{labeling}


\newpage
\pagenumbering{arabic}

\chapter{Introduction}

\section{Research Motivation}

Electric power systems around the world are currently undergoing a deep structural transformation. Arguably the most important change 
is the gradual replacement of conventional synchronous generator-based fossil fuel power plants -- that have been dominating the dynamic response of power systems
for more than a century now -- 
by 
converter-based, intermittent renewable energy resources, such as wind and solar photovoltaic generation. 
Other significant changes are
the increasing flexibility of energy consumption --
partially due to the electrification of
transportation and heating systems --,
as well as the integration 
of power networks along with digital technologies and communication systems. Meanwhile, there is a continuous expansion of national distribution and transmission networks, as well as of international 
interconnections and power exchanges, especially by means of \ac{hvdc} connections.

As dynamical systems, power systems are large-scale, highly non-linear systems which include continuous, discrete and stochastic variables. 
 Moreover, following from the aforementioned changes, 
the size, uncertainty and dynamic complexity of power systems are further increasing.
Therefore, stability assessment, optimal control synthesis, and accurate and efficient computer-based simulation of power systems are challenging tasks which, in order to be adequately addressed, require the development of ad hoc analytical and numerical tools. 

Undoubtedly, the most successful method for assessing the dynamic behavior of a power system model after a disturbance is to carry out a numerical time domain simulation. On the other hand, assessing the overall performance of a power system by means of time domain simulations requires considering a large number of disturbances and scenarios. Even so, and despite some efforts that have been made, see e.g.~\cite{ajjarapu:11}, time domain analysis typically does not answer crucial quantitative  questions such as: 
What is the stability margin of the system?
What are the properties, e.g.~natural frequency and damping, of the most critical for the stability dynamic modes?
What are the couplings between the critical modes 
and the variables of the power system? 
These questions are typically addressed by means of 
stability analysis.

Power system stability is defined as \textit{the ability of an electric power system, for a given initial operating condition, to regain a state of operating equilibrium after being subjected to a physical disturbance, with most system variables bounded so that practically the entire system remains intact} \cite{stability:04}. 
There are various mechanisms that may lead a power system to instability. For this reason, power system stability has been classified in categories, which helps identify the causes of instability and simplify the analysis by using appropriate models and tools. 
Following from \cite{stability:04}, 
the ability of a power system, following a disturbance, to maintain 
{(i)} synchronism, defines \textit{rotor angle stability}; 
{(ii)} steady voltages at all buses, 
defines
\textit{voltage stability}; 
and {(iii)} steady frequency, i.e.~balanced generation and load, defines \textit{frequency stability}. 
Due to the
the increasing penetration of 
power electronic converter interfaced technologies, 
this classification has been recently revisited in \cite{stability:20}, 
to include two 
new types of stability, namely, 
(iv)
\textit{resonance stability}, which is concerned with resonances of electromechanical and electrical nature;
and (v) \textit{converter-driven stability}, which is concerned with fast and slow interactions caused by the operation of power electronic converters.
Finally,
from a system-theoretic point of view,
the conditions that may lead a dynamical system and hence also a power system to collapse after a disturbance are, ultimately, three, as follows: (i) a
post-disturbance operating equilibrium does not exist;
(ii) a post-disturbance equilibrium exists but it can not be reached, because the trajectory that the system follows is unstable; and (iii) a post-disturbance equilibrium exists but it is unstable.

Among the various stability analysis techniques available for power systems, this thesis focuses
in particular on
\ac{sssa}, which
studies the third condition, i.e.~the stability of equilibria.
\ac{sssa} 
has been mostly used for the analysis of rotor angle stability of small-disturbances. However, the definition and most importantly the tools of \ac{sssa} also apply to other types of 
power system stability. In particular, 
resonance stability, both torsional and
electrical, as well as the 
fast-interaction converter-driven stability, can be studied
and addressed using \ac{sssa}.
The main advantage of \ac{sssa} is that,
provided that a stationary condition exists, it can be always applied to a system, as opposed to other
stability analysis techniques, such as Lyapunov's energy function, which cannot be defined for all systems. On the other hand,
\ac{sssa}
is valid only in the neighborhood of an examined stationary point and thus it is not suitable for assessing the stability following a large disturbance. Despite this limitation, \ac{sssa} provides valuable insights on a power system model by capturing its structural characteristics and hence, it is a fundamental component of power system dynamic analysis. 
Among other applications, 
\ac{sssa} has been employed in power systems as a tool for
modal analysis, see e.g. \cite{arriaga:82_1, hamdan:86, arriaga:89}, control design, see e.g. \cite{chow:00, heydt:04, gurrala:10}, and numerical analysis, see e.g. \cite{ajjarapu:06, milano:psa:16}.
The objective of this thesis is to explore all three above directions.
In particular, the thesis 
employs \ac{sssa} techniques
to feature novel aspects in
modal participation analysis,
automatic control design, and
time domain integration.


\section{Thesis Overview}

\subsection{Contributions}

The main goal of this thesis is to contribute to the stability analysis and control of power systems by developing a handful of novel analytical and computational tools, based on \ac{sssa}. 
In particular, the main contributions of the thesis are in three directions, namely, modal participation analysis,
automatic control design, and
time domain integration.

\subsubsection*{Modal Participation Analysis}

Modal analysis studies the properties of a dynamical system in the frequency domain. An important component of power system modal analysis is \textit{participation analysis}, which is an approach to efficiently
determine the sensitivities of the 
dynamic modes of a power system model 
to variations of its variables.
In its classical formulation, modal participation analysis quantifies the coupling of the states of a linear system of \acp{ode}
with its dynamic modes (eigenvalues of the state matrix), and is considered a standard tool of power system \ac{sssa} \cite{arriaga:82_1}. 
The non-linear power system model for 
angle and voltage stability
analysis, however, is typically formulated as a set of \acp{dae}. That is, it also includes a variety of algebraic equations and
variables, e.g.~power and/or current flows in network branches, that constrain the system and define its dynamics. To study the impact of algebraic
variables on the system dynamic response through participation factors is one of the scopes of this thesis. Such a study is relevant since, very
often, the measurements taken on the transmission network and used by local and wide area controllers are modeled as algebraic variables.


This thesis proposes 
a measure for the modal participation of the algebraic variables of a power system model in its dynamic modes, through the definition of appropriate input-output vectors of the
system's state-space representation.
To this aim, an alternative interpretation of the classical participation factors as
eigen-sensitivities is also proposed.
The new interpretation removes the basic assumptions of classical participation analysis, since it assumes that the power system is modeled as singular system of differential equations with eigenvalue multiplicities. 

\subsubsection*{Control Design}

Proper control design is crucial
to ensure a stable operation of a power system.
%
This thesis employs \ac{sssa} techniques for the purpose of control design in two ways: (i) by
studying an extension of classical control theory which stems from the theory of fractional order differential equations; (ii)
by exploring the impact of delay-based controllers on power system stability.

\begin{enumerate}
\item \textit{Fractional Order Control:}

The selection of a proper control scheme is a critical decision during a control design.
With this regard, despite the 
recent developments in the theory of 
robust and advanced control, see 
\cite{zhou:99, camacho:07},
the vast majority of controllers
employed in industrial applications 
are still based on classical schemes, such as the \ac{pid} controller. 
This is mostly due to the fact that classical control schemes combine simple structure, easy tuning, and overall good performance. With this in mind, the thesis investigates a promising extension of classical controllers, which stems from 
the theory of fractional calculus.
Fractional calculus is the
mathematical analysis that studies
differentials and integrals
of non-integer order.   
Control schemes based on fractional calculus have gained momentum in
power system applications due to their ability to enhance performance
and increase the stability margin, under the presence of topological
changes, parameter uncertainty and noise. 

In this thesis, 
a theory on how to carry out \ac{sssa} of
power systems with \textit{exact} fractional dynamics is developed.
In addition, a step-by-step analytical study on the modeling and parameters
selection of \ac{ora}-based \acp{foc}
is provided. 
Finally, the thesis carries out a systematic analysis of \ac{foc} applications to power system
controllers.  These include 
automatic generation
control of synchronous machines; frequency control of a
converter-interfaced energy storage system; and voltage control
through a static synchronous compensator. 

\item \textit{Delay-based Control:} 

Measurement and communication delays in local \acp{pss} and \acp{wadc} are known to be a potential
threat for the overall dynamic performance of power systems \cite{heydt:2002, heydt_stochastic:08, delay1, muyang:wams}.
How to properly study the impact of
delays through accurate yet robust numerical techniques
is still an open and active field of research.
In spite of their bad reputation, delays
are not always detrimental, but can also have unexpectedly
beneficial effects on the stability of dynamical systems
\cite{Sipahi2011_CSM, Pyragas:1992-PLA, Sipahi:2005-AUTOMATICA,
Stepan:1989-BOOK, Abdallah:1993}.  
It has been shown, for example,
that intentionally inserting a certain amount of delay in a feedback
control system can enhance disturbance rejection capabilities, improve
response time, and add the required damping to avoid undesired
oscillations in a closed-loop system, see, e.g.~\cite{Olgac:2005}.
More recently, analytical tuning techniques have been proposed to adjust
time delays and controller gains to achieve fast response
\cite{Ramirez:2015-ISA, Ramirez:2016-TAC, Ramirez:2017-SICON}.  These
new results motivate the use of intentional time delays as part of
controllers, e.g. to effectively suppress poorly damped synchronous machine electromechanical oscillations.

In this thesis,
the structure of stability crossing boundaries and the damping characteristics in the
delay-controller gain parameter space of power systems with delay-based control are featured using two complementary approaches. First, through an
analytical proof-of concept, by using the one-machine infinite-bus system.
Second, through a numerical analysis on a larger, more
realistic system. The main novel result of these studies is that proper design of a
two-channel \ac{pss} allows unifying disconnected
stability regions.

\end{enumerate}

\subsubsection*{Time Domain Integration}

The power system model for rotor-angle and voltage stability analysis is conventionally formulated as a set of non-linear \acp{dae}. These equations are mutually dependent due to the meshed topology of transmission networks and the action of secondary controllers.  

This thesis proposes
a technique to decouple the power system \acp{dae} by introducing a delay that is equal to the time step of the numerical integration. 
Such delay, while not altering the overall dynamic response of the system, allows reducing the coupling of the \acp{dae} by removing off-diagonal elements of the system Jacobian matrix.
The impact of the proposed \emph{one-step-delay} approximation on
the accuracy, convergence and computational burden of the time domain integration routine are rigorously and systematically discussed.
In addition, a 
method to identify the elements of a power system \ac{dae} model
that can be delayed by one time step, as well as a technique to
estimate the maximum admissible delay, so that the approximation is
within a given tolerance are provided.
\vspace{-0.5cm}
\begin{center}
* * *
\end{center}

Simulations in this thesis 
are carried out using the Python-based power system analysis software tool Dome \cite{vancouver}.
These include solution of 
power flow problems, \ac{sssa}, and time domain simulations. In addition,
the models and techniques developed in the course of this thesis are implemented and included in Dome. 

\subsection{Organization}

The remainder of the thesis is organized as follows.

Chapter~\ref{ch:sssa} provides the fundamentals of
\ac{sssa}, which are 
then utilized throughout the thesis.
The formulation, eigenvalue analysis and stability condition of linearized power systems are presented.
In addition, the chapter provides an overview of existing algorithms, and a comprehensive comparison of 
available open-source libraries,
that are suitable for the solution of non-Hermitian eigenvalue problems.

Chapter~\ref{ch:pf} focuses on modal participation
analysis as a measure of the coupling between the variables and dynamic modes of a power system. Classical definitions of 
participation factors are provided first. 
The modal participation analysis of a power system modeled as a singular system of differential equations with eigenvalue multiplicities, as well as a new interpretation of participation factors as eigensensitivities, are presented next. 
Then, the chapter proposes
an approach to determine the participation of
algebraic variables in power system dynamic modes, 
which considers adequate input/output variables of the system's state-space representation. Both the linear and generalized eigenvalue problems are considered for the calculation of the participation factors.
An illustrative example on the two-area system, as
well as a study on the 1,479-bus \ac{aiits} model
are carried out to support the theory and
illustrate the features of the proposed approach.

Chapter~\ref{ch:foc} presents the theoretical foundation and practical
implementation aspects of \acp{foc} for
power system applications. First, essential definitions and concepts from fractional calculus are described.
Second, mathematical theory on the stability analysis of
power systems with inclusion of \acp{foc} is presented. Next, Chapter~\ref{ch:foc}
discusses the software implementation of \acp{foc} based on the \ac{ora} method. A
variety of examples of \ac{ora}-based \acp{foc} are illustrated, namely, integral \ac{foc} for \ac{agc}; lead-lag \ac{foc} for frequency regulation of an \ac{ess}; and multiple \ac{PI} \acp{foc} for voltage regulation provided by a \ac{statcom}. The \ac{wscc}
9-bus test system and the 1,479-bus \ac{aiits} model are employed to test and compare the examined \acp{foc} with their integer-order versions.

Chapter~\ref{ch:tdc} discusses how to utilize intentional time delays as part
of controllers to improve the damping characteristics in electromechanical oscillations of power system synchronous machines. First, stability theory on
the spectral analysis of small and large 
time-delay systems is provided. The control parameter settings for which time delays in \acp{pss} improve the small-signal stability of a power system are systematically identified. Analytical results are presented by applying a \ac{pr} control scheme to the \ac{omib} electromechanical power system model. Finally, to demonstrate the opportunities in more realistic models, the obtained
results are tested via numerical analysis on the IEEE 14-bus system.

Chapter~\ref{ch:osda}
proposes the inclusion 
of a delay -- equal to the time step of the numerical integration -- to reduce the coupling of the equations of the non-linear \ac{dae} power system model. 
At first, the conventional implicit numerical integration of power systems is described. 
Subsequently, the proposed one-step delay approximation is presented. 
The selection of the variables that when subjected to one-step delay leave practically unchanged the system trajectories, as well as estimation of the maximum admissible delay, are discussed using small-signal stability analysis. 
Finally, the proposed approach is evaluated in terms of accuracy, convergence and computational burden, by means of {(i)} the IEEE 39-bus system; 
{(ii)} the 21,177-bus model of the \ac{entsoe}.

Finally, Chapter~\ref{ch:conclusions} summarizes the most relevant conclusions of the thesis and suggests directions for future work.

\subsection{Publications}

This section provides the list of publications that gave rise to the work presented in this thesis.



\subsection*{Journal papers}

\subsubsection*{(Closely related to the content of the thesis)}

\begin{enumerate}
%
\leftskip-0.13in 
\item \textbf{G. Tzounas}, R. Sipahi, and F. Milano, Damping power system electromechanical oscillations using time delays, IEEE Transactions on Circuits and Systems I: Regular Papers, accepted in Feb.~2021, in press.

\item \textbf{G. Tzounas}, I. Dassios, M. Liu, and F. Milano, Comparison of numerical methods and open-source libraries for eigenvalue analysis of large-scale power systems,
Applied Sciences, MDPI, \textit{Special Issue: Methods in Dynamical Systems, Mathematics of Networks, and Optimization for Modelling in Engineering}, 
vol.~10, no.~21, 7592, Oct.~2020. 
DOI:~\href{https://doi.org/10.3390/app10217592}{10.3390/app10217592}.

\item \textbf{G. Tzounas}, F. Milano,
Delay-based decoupling of power system models for transient stability analysis,
IEEE Transactions on Power Systems, 
vol.~36, no~1, 
pp.~464-473, Jan.~2021.
DOI:~\href{https://doi.org/10.1109/TPWRS.2020.3009172}{10.1109/TPWRS.2020.3009172}.
 
\item I. Dassios, \textbf{G. Tzounas}, F. Milano,
Generalized fractional controller for singular systems of differential equations,
Journal of Computational and Applied Mathematics, Elsevier, vol. 378,
Nov. 2020. 
DOI: \href{https://doi.org/10.1016/j.cam.2020.112919}{10.1016/j.cam.2020.112919}.

\item \textbf{G. Tzounas}, I. Dassios, M. A. A. Murad, F. Milano,
Theory and implementation of fractional order
controllers for power system applications,
IEEE Transactions on Power Systems, vol.~35, no.~6, pp.~4622-4631, Nov.~2020.
DOI: \href{https://doi.org/10.1109/TPWRS.2020.2999415}{10.1109/TPWRS.2020.2999415}.
  
\item \textbf{G. Tzounas}, I. Dassios, F. Milano, Modal participation factors of algebraic variables, IEEE Transactions on Power Systems, 
vol. 35, no. 1, pp. 742-750, Jan. 2020.
DOI: \href{https://doi.org/10.1109/tpwrs.2019.2931965}{10.1109/TPWRS.2019.2931965}.

\item I. Dassios, \textbf{G. Tzounas}, F. Milano, Participation factors for singular systems of differential equations, Circuits, Systems, and Signal Processing, Springer, 
vol. 39, no. 1, pp. 83-110, Jan. 2020.
DOI: \href{https://doi.org/10.1007/s00034-019-01183-1}{10.1007/s00034-019-01183-1}.
\end{enumerate}

\subsubsection*{(Other)}

\begin{enumerate}
\setcounter{enumi}{7}
  \leftskip-0.13in 

\item I. Dassios, \textbf{G. Tzounas}, F. Milano,
Robust stability criterion for perturbed singular systems of linearized differential equations, Journal of Computational and Applied Mathematics, Elsevier, vol. 381, Jan. 2021.
DOI: \href{https://doi.org/10.1016/j.cam.2020.113032 }{10.1016/j.cam.2020.113032}.  

\item M. Liu, I. Dassios, \textbf{G. Tzounas}, F. Milano,  Model-independent derivative control delay compensation methods for power systems, Energies, MDPI,
\textit{Special Issue: Advanced Solutions for Monitoring, Protection and Control of Modern Power Transmission System},
vol. 13, no. 2, Jan. 2020. DOI: \href{https://doi.org/10.3390/en13020342}{10.3390/en13020342}.
  
\item I. Dassios, \textbf{G. Tzounas}, and F. Milano, The M{\"o}bius transform effect in singular systems of
differential equations, Applied Mathematics and Computation, Elsevier, vol. 361, pp. 338-353, Nov. 2019.
DOI: \href{https://doi.org/10.1016/j.amc.2019.05.047}{10.1016/j.amc.2019.05.047}.  

\item M. Liu, I. Dassios, \textbf{G. Tzounas}, and F. Milano, Stability analysis of power systems with inclusion
of realistic-modeling WAMS delays, IEEE Transactions on Power Systems, vol. 34, no. 1, pp. 627-636, Jan. 2019.
DOI: \href{https://doi.org/10.1109/tpwrs.2018.2865559}{10.1109/tpwrs.2018.2865559}.
\end{enumerate}

\subsection*{Books}

\begin{enumerate}
\setcounter{enumi}{11}
  \leftskip-0.13in 
   \item F. Milano, I. Dassios, M. Liu, and \textbf{G. Tzounas}, \textit{Eigenvalue Problems in Power Systems}, CRC Press, Taylor \& Francis Group, 2020.
   ISBN:~\href{https://www.routledge.com/Eigenvalue-Problems-in-Power-Systems/Milano-Dassios-Liu-Tzounas/p/book/9780367343675}{9780367343675}.
\end{enumerate}

\subsection*{Conference Papers}

\begin{enumerate}
\setcounter{enumi}{12}
  \leftskip-0.13in 
  \item M. A. A. Murad, \textbf{G. Tzounas} and F. Milano, Modeling and simulation of fractional order PI control limiters for power systems,  21st IFAC World Congress (IFAC 2020), Berlin, Germany, 12-17 Jul. 2020.

  \item \textbf{G. Tzounas} and F. Milano, Impact of the estimation of synchronous machine rotor speeds on wide area
damping controllers,  IEEE PES General Meeting, Atlanta, GA, 4-8 Aug. 2019. 
DOI: \href{https://doi.org/10.1109/pesgm40551.2019.8973611}{10.1109/pesgm40551.2019.8973611}.
  
  \item M. A. A. Murad, \textbf{G. Tzounas}, M. Liu, and F. Milano, Frequency control through voltage regulation
of power system using SVC devices,  IEEE PES General Meeting, Atlanta, GA, 4-8 Aug. 2019. DOI: \href{http://dx.doi.org/10.1109/pesgm40551.2019.8973807}{10.1109/pesgm40551.2019.8973807}.

  \item \textbf{G. Tzounas}, M. Liu, M. A. A. Murad, and F. Milano, Impact of realistic bus frequency measurements on wide-area power system stabilizers, IEEE PowerTech, Milano, Italy, 23-27 Jun. 2019. DOI: \href{http://dx.doi.org/10.1109/ptc.2019.8810695}{10.1109/ptc.2019.8810695}.

  \item M. Liu, \textbf{G. Tzounas}, and F. Milano, A model-independent delay compensation method for power
systems, IEEE PowerTech, Milano, Italy, 23-27 Jun. 2019. DOI: \href{http://dx.doi.org/10.1109/ptc.2019.8810975}{10.1109/ptc.2019.8810975}.

\item \textbf{G. Tzounas}, M. Liu, M. A. A. Murad, and F. Milano, Stability analysis of wide area damping controllers with multiple time delays, 10th Symposium on Control of Power and Energy Systems (IFAC CPES), Tokyo, Japan, 4-6 Sep. 2018. DOI: \href{http://dx.doi.org/10.1016/j.ifacol.2018.11.753}{10.1016/j.ifacol.2018.11.753}.

\end{enumerate}

\newpage
\chapter{Small-Signal Stability Analysis}
\label{ch:sssa}

\section{Introduction}

The objective of this chapter is to provide
definitions, formulations, theorems, and 
software tools related to
\ac{sssa} and linearized power systems. 
The chapter is organized as follows. Section~\ref{sssa:sec:nlmodel} describes the formulation of power system models for transient stability analysis.
Section~\ref{sssa:sec:lmodel} discusses the power system model linearized around an equilibrium. Section~\ref{sssa:sec:linsys}
provides the outlines of linear systems of differential equations.
In particular, this section discusses the definition of matrix pencils, the solution, 
the formulation and properties of the \ac{lep} and \ac{gep} and defines asymptotic stability.
The most relevant numerical methods and open-source libraries for the solution of the eigenvalue problems that arise in power systems are discussed in
Section~\ref{sssa:sec:numerical}.
A comprehensive comparison of these libraries
is carried out through two real-world power system models in the case studies discussed in 
Section \ref{sssa:sec:case}.
Finally, conclusions are drawn in Section~\ref{sssa:sec:conclusions}.

\section{Power System Model}
\label{sssa:sec:nlmodel}

The power system model for transient stability analysis can be formulated as a set of non-linear, \textit{semi-implicit} \acp{dae}, as follows \cite{semi:2016}:
\begin{align}
\label{sssa:eq:sidae}
  \begin{bmatrix}
    \bfg T & \bfg 0_{\nx,\ny} \\
    \bfg R & \bfg 0_{\ny,\ny}
    \end{bmatrix}
      \begin{bmatrix}
       \Dt{\bfg x} \\
       \Dt{\bfg y}
    \end{bmatrix}
    &=
    \begin{bmatrix}
        \bfg f( \bfg  x, \bfg y , \bfg u) \\
      \bfg   g( \bfg  x, \bfg y,  \bfg u)
    \end{bmatrix} \, ,
\end{align}
%
%
where
$\bfg f : \mathbb{R}^{\nx+\ny+\nin} \rightarrow \mathbb{R}^{\nx}$,
$\bfg g : \mathbb{R}^{\nx+\ny+\nin} \rightarrow \mathbb{R}^{\ny}$; 
$\bfg x = \bfg x(t)$,
$\bfg x \in \mathbb{R}^{\nx}$, are the
state variables,
$\bfg y = \bfg y(t)$, $\bfg y \in \mathbb{R}^{\ny}$,
are the algebraic variables; 
$\bfg u = \bfg u(t) $,
$\bfg u \in \mathbb{R}^{\nin}$, are 
controlled and/or uncontrolled inputs;
$\bfg T \in \mathbb{R}^{\nx \times \nx}$,  
$\bfg R\in \mathbb{R}^{\ny \times \nx}$,
are assumed to be constant matrices;
and $t \in [0,\infty)$ is the simulation time. Finally, $\bfg 0_{\nx,\ny}$ denotes the zero matrix of dimensions $\nx \times \ny$.

In the formulation of \eqref{sssa:eq:sidae}, discrete variables are modeled implicitly,
i.e.~each discontinuous change in the system leads to a new continuous set of equations in the form of \eqref{sssa:eq:sidae}.
In addition, 
dynamic components of
\eqref{sssa:eq:sidae} are modeled
following the \textit{phasor} or \textit{quasi sinusoidal} approximation. That is, the three phases of AC electric networks and machines are assumed symmetric; 
stator transients of electric machines, as well as 
\ac{emt} phenomena of transmission lines are ignored.


A special case of \eqref{sssa:eq:sidae} is when
%
  \begin{align}
  \label{sssa:eq:explcon}
   \bfg T &= \bfg I_\nx \, , \ \ 
   \bfg R = \bfg 0_{\ny,\nx} \, ,
\end{align}  
%
where $\bfg I_\nx$ denotes the $\nx \times \nx$ identity matrix. This case
leads to an \textit{explicit} set of \acp{dae}, that is
the 
formulation most commonly employed for transient stability analysis. 
In this thesis, the more general form of \eqref{sssa:eq:sidae} is used, unless otherwise explicitly stated.

%

%
%


\section{Linearized Power System Model}
\label{sssa:sec:lmodel}

\subsection{DAE Formulation}

The objective of \ac{sssa} is to study the stability properties of a power system model around an equilibrium point. An equilibrium point $(\bfg x_o, \bfg y_o, \bfg u_o)$ of \eqref{sssa:eq:sidae} satisfies:
\begin{align}
\bfg 0_{\nx,1} &=  \bfg f( \bfg  x_o, \bfg y_o , \bfg u_o) \, ,
\nonumber \\
\bfg 0_{\ny,1} &=  \bfg g( \bfg  x_o, \bfg y_o , \bfg u_o) \, .
\nonumber
\end{align}

For sufficiently small disturbances,
\eqref{sssa:eq:sidae} can be linearized
around $(\bfg x_o, \bfg y_o, \bfg u_o)$ for the purpose of analysis, as follows:
\begin{align}
\label{sssa:eq:sidaelin}
  \begin{bmatrix}
    \bfg T & \bfg 0_{\nx,\ny} \\
    \bfg R & \bfg 0_{\ny,\ny}
    \end{bmatrix}
      \begin{bmatrix}
      \Delta \Dt{\bfg x} \\
      \Delta \Dt{\bfg y}
    \end{bmatrix}
    &=
    \begin{bmatrix}
      \jacs{f}{x} \, \Delta \bfg x + \jacs{f}{y} \, \Delta \bfg y +
  \jacs{f}{u} \,  \Delta \bfg u \\
     \jacs{g}{x} \, \Delta \bfg x +\jacs{g}{y} \, \Delta \bfg y  +
  \jacs{g}{u} \, \Delta \bfg u
    \end{bmatrix} \, ,
\end{align}
where $\Delta \bfg x = \bfg x - \bfg x_o$, $\Delta \bfg y = \bfg y -
\bfg y_o$, $\Delta \bfg u = \bfg u -
\bfg u_o$; 
$\jacs{f}{x}$, $\jacs{f}{y}$, $\jacs{f}{u}$, $\jacs{g}{x}$, $\jacs{g}{y}$,
$\jacs{g}{u}$, are the
Jacobian matrices calculated at 
$( \bfg x_o, \bfg y_o, \bfg u_o )$.
%
Obtaining the linear \ac{dae} system \eqref{sssa:eq:sidaelin} from \eqref{sssa:eq:sidae}
is straightforward, by considering
Taylor's expansion and ignoring all
derivative terms of order 
higher than one.
System \eqref{sssa:eq:sidaelin} is an autonomous linear system, i.e.~the elements of the Jacobian matrices
are not functions of time $t$.  
This system can be rewritten in the following form:
\begin{equation}
  \Edae  \, \Dt \xys = 
  \Adae  \, \xys
+ \Bdae  \, \Delta \bfg u \, ,
  \label{sssa:eq:singps}
\end{equation}
where 
\begin{equation}
\xys = 
  \begin{bmatrix}
    \Delta \bfg x  \\
    \Delta \bfg y  \\
  \end{bmatrix}  
  , \
  \Edae = 
  \begin{bmatrix}
    \bfg T & \bfg 0_{\nx,\ny} \\
    \bfg R & \bfg 0_{\ny,\ny} \\
  \end{bmatrix}
  , \
  \Adae = 
  \begin{bmatrix}
   \jacs{f}{x} & \jacs{f}{y} \\
   \jacs{g}{x} & \jacs{g}{y} \\
  \end{bmatrix} \, , \ 
 \Bdae = 
  \begin{bmatrix}
    \jacs{f}{u} \\
    \jacs{g}{u} \\
  \end{bmatrix}  \, . 
  \nonumber
\end{equation}
In the special case of 
\eqref{sssa:eq:explcon}, system 
\eqref{sssa:eq:sidaelin} takes the following explicit \ac{dae} form:
\begin{align}
\label{sssa:eq:exdaelin}
\begin{bmatrix}
\bfg I_{\nx} & \bfg 0_{\nx,\ny} \\
\bfg 0_{\ny,\nx} & \bfg 0_{\ny,\ny} \\
\end{bmatrix}
\begin{bmatrix}
\Delta \Dt{\bfg x} \\
\Delta \Dt{\bfg y}
\end{bmatrix}
    &=
    \begin{bmatrix}
      \jacs{f}{x} \, \Delta \bfg x + \jacs{f}{y} \, \Delta \bfg y +
  \jacs{f}{u} \,  \Delta \bfg u \\
     \jacs{g}{x} \, \Delta \bfg x +\jacs{g}{y} \, \Delta \bfg y  +
  \jacs{g}{u} \, \Delta \bfg u
    \end{bmatrix} \, .
\end{align}
In this case, matrix
$\Edae$ has the diagonal form:
\begin{equation}
\Edae = 
  \begin{bmatrix}
   \bfg I_{\nx} & \bfg 0_{\nx,\ny} \\
    \bfg 0_{\ny,\nx} & \bfg 0_{\ny,\ny} \\
    \end{bmatrix} \, .
    \nonumber
\end{equation}

\subsection{ODE Formulation}

Under the assumption that $\jacs{g}{y}$ is not singular, system
\eqref{sssa:eq:exdaelin}
can be reduced to a system
of \acp{ode}, by eliminating the deviations of the algebraic variables.
Rewrite \eqref{sssa:eq:exdaelin} as:
\begin{align}
  \label{sssa:eq:daelin}
  \Delta \Dt{\bfg x} &=  
  \jacs{f}{x} \, \Delta \bfg x + \jacs{f}{y} \, \Delta \bfg y +
  \jacs{f}{u} \,  \Delta \bfg u \, ,
  \\
  \bfg  0_{m,1} &= \jacs{g}{x} \, \Delta \bfg x +\jacs{g}{y} \, \Delta \bfg y  
  +
  \jacs{g}{u} \, \Delta \bfg u
  \, .
  \label{sssa:eq:daeliny}
\end{align}
Solving \eqref{sssa:eq:daeliny} for
$\Delta \bfg y$ yields:
\begin{equation}
  \label{sssa:eq:algeb}
  \begin{aligned}
    \Delta{\bfg{y}} &= - \jacsinv{g}{y} \, \jacs{g}{x} \, \Delta{\bfg{x}} - \jacsinv{g}{y} \, \jacs{g}{u}
    \, \Delta \bfg{u} \, .
   \end{aligned}    
\end{equation}
Substitution of
\eqref{sssa:eq:algeb} in \eqref{sssa:eq:daelin} leads to the following linear
system:
\begin{equation}
\label{sssa:eq:odelin}
 \Delta \Dt {\bfg x} = \AS \, \Delta \bfg x +  \BS \, \Delta \bfg u \, ,
\end{equation}
where $\AS = \jacs{f}{x}-\jacs{f}{y} \, \jacsinv{g}{y} \, \jacs{g}{x}$, 
$\BS = \jacs{f}{u}-\jacs{f}{y} \, \jacsinv{g}{y} \, \jacs{g}{u}$,
are the state matrix and  
input matrix, respectively, of the \ac{ode} system. System \eqref{sssa:eq:odelin} is the model conventionally used in power system \ac{sssa}.
%

In general, \eqref{sssa:eq:odelin} (or \eqref{sssa:eq:sidaelin}) can be employed to accurately assess the small-signal stability of the non-linear power system model \eqref{sssa:eq:sidae}. An exception 
occurs at points where the system undergoes structural changes, e.g.~at bifurcation points, where linearization as a technique to assess the stability of equilibria is inconclusive and where, thus, the non-linear dynamics of the system must be taken into account.

\section{Linear System of Differential Equations}
\label{sssa:sec:linsys}

\subsection{Formulation}

The properties of a linearized power system model can be systematically studied using theory of linear differential equations. To this aim, this section considers the following system:
\begin{equation}
\label{sssa:eq:sing}
\Esng \, \Dt \xs(t)=
\Asng \, \xs(t)
+ \Bsng \, \bfg{u}(t)  
\, ,
\end{equation}
where  $\Esng, \Asng \in \mathbb{R}^{\nd \times \nd}$, $\xs:[0,+\infty)\rightarrow\mathbb{R}^{\nd}$. 

Matrix $\Esng$ in \eqref{sssa:eq:sing} can be:

\begin{itemize}
\item non-singular, i.e.~${\rm det}(\Esng)\neq 0$. This is 
the case of the \ac{ode} power system model \eqref{sssa:eq:odelin}. In particular, \eqref{sssa:eq:odelin} can be obtained from \eqref{sssa:eq:sing}
for $\nd =\nx$, 
$\xs \equiv \Delta \bfg x$,
$\Asng \equiv \AS$, 
$\Esng \equiv \bfg I_\nx$, 
$\Bsng \equiv \BS$.

\item singular, i.e.~${\rm det}(\Esng) = 0$. This is the case of the \ac{dae} power system
model \eqref{sssa:eq:sidaelin}.
In particular, \eqref{sssa:eq:sidaelin} can be obtained from \eqref{sssa:eq:sing}
for $\nd = \nx + \ny$,
$\xs \equiv \xys$,
$\Asng \equiv \Adae$, 
$\Esng \equiv \Edae$,
$\Bsng \equiv \Bdae$.
\end{itemize}

In case that 
$\Esng$ in \eqref{sssa:eq:sing} is a singular matrix, 
system \eqref{sssa:eq:sing} is called a \textit{singular system} of 
differential equations. 
Singular systems are relevant in many fields of engineering, such as automatic control, circuits, and robotic systems \cite{Dai, Lew, Du}. Note that, the theory 
of singular systems of differential equations involves a class of systems that is more general than \eqref{sssa:eq:sing}, 
e.g.~it includes systems with non-square matrices. However, 
for the needs of this thesis, it is adequate to study systems in the form of \eqref{sssa:eq:sing}.
The reader interested in a comprehensive theory on singular systems of differential equations 
may refer to \cite{gantmacher:59}, while some applications of that theory to power systems can be found in \cite{book:eigenvalue, moebius, dassios:robust}.

This section studies next
some important properties of system 
\eqref{sssa:eq:sing}, and in particular, 
defines its matrix pencil, discusses existence and uniqueness of its solutions, and provides the conditions for its stability.   

\subsection{Matrix Pencil}

Applying the Laplace transform 
to \eqref{sssa:eq:sing}, 
one gets:
\begin{equation}
\label{sssa:eq:singL}
\Esng \, \mathcal{L}
\{ \Dt \xs(t) \} = \Asng
\, \mathcal{L} \{ \xs(t) \} 
+ \Bsng
\mathcal{L} 
\{ \bfg{u}(t) \}
\, ,
\end{equation}
where $s\in\mathbb{C}$ denotes the complex Laplace variable.
Employing Laplace transform properties, 
one has:
\begin{equation}
\label{sssa:eq:singL2}
\Esng \, 
\big ( s \mathcal{L}\{ \xs(t) \} - \xs(0)
\big ) = \Asng \, 
\mathcal{L}\{ \xs(t) \}
+\Bsng \, \mathcal{L} \{ \bfg u(t) \}
\, ,
\end{equation}
or, equivalently,
\begin{equation}
\label{sssa:eq:singL3}
( s \Esng - \Asng ) \,
\mathcal{L}\{ \xs(t) \} = \Esng \, \xs(0) 
+\Bsng \, \mathcal{L} \{ \bfg u(t) \}
\,  .
\end{equation}

The structure of the polynomial matrix $ s \Esng - \Asng $, hereafter referred as
the \textit{matrix pencil}
of system \eqref{sssa:eq:sing},
defines the existence of solutions and the stability properties of \eqref{sssa:eq:sing}.

This thesis considers only matrix pencils that are \textit{regular}, 
i.e.~the associated matrices are square and, in addition,
${\rm det}(s\Esng-\Asng)= \poly(s)\not\equiv 0$, where $\poly(s)$ is a polynomial of $s$ of degree ${\rm deg} ( \poly(s) ) \leq \nd$.
This is, in fact, the 
form of matrix pencils
that most commonly appear
for the purpose of power system \ac{sssa}.

\subsection{Solutions}

{ \remark{ (\textit{Existence of Solutions}).
Consider system \eqref{sssa:eq:sing} with ${\rm det}(s\Esng-\Asng)\not\equiv 0$. Then, $s \Esng-\Asng$ is invertible and  \eqref{sssa:eq:singL3}
can always be solved for 
$\mathcal{L} \{ \xs(t) \}$ as:
\[
\mathcal{L} \{ \xs(t) \}  = (s \Esng-\Asng)^{-1}\, \Esng \, \xs(0)
+ (s \Esng-\Asng)^{-1} \,
\Bsng \, \mathcal{L} \{ \bfg u(t) \}
\, .
\]
Consequently, 
a solution $\xs(t)$ 
always exists and is given by:
\begin{equation}
\label{sssa:eq:existence}
 \xs(t)=\mathcal{L}^{-1} \big \{(s \Esng-\Asng)^{-1}\, \Esng \, \xs(0)
+ (s \Esng-\Asng)^{-1} \, 
\Bsng \, \mathcal{L} \{ \bfg u(t) \}
\big \} \, .   
\end{equation}
}
\label{sssa:remark:existence}
}

{\remark { (\textit{Uniqueness of Solutions}).  
There are two types of initial conditions \xs(0): \textit{consistent}
and \textit{non-consistent}.
If $\Esng$ is non-singular,
then for any given initial conditions the solution is unique. However, 
it is not guaranteed that for given initial conditions a singular
system with a regular pencil has a unique solution. In this case, if the given
initial conditions are consistent, then 
the solution is unique. 
Otherwise, there are infinitely many solutions, 
the general solution \eqref{sssa:eq:existence}
holds for $t>0$ and the system is called \textit{impulsive}.
}
\label{sssa:remark:init}
}

\subsection{Eigenvalue Problem}
\label{sec:eigenvalue}

\subsubsection{Formulation}

The stability of \eqref{sssa:eq:sing} can be assessed by calculating its eigenvalues, which are defined as the roots of the 
\emph{characteristic equation}: 
\begin{align}
\label{eq:sing:char}
{\rm det}( s \Esng- \Asng)=0 \, .
\end{align}
where ${\rm det}( s \Esng-\Asng )$ is called the
\emph{characteristic polynomial}
of system \eqref{sssa:eq:sing}. 
In general, analytical solution of
\eqref{eq:sing:char} is possible only if $\nd \leq 4$. For higher degrees,
general formulas do not exist and only the application of a numerical
method is possible.
In addition, algorithms that explicitly determine the characteristic polynomial 
${\rm det}( s \Esng - \Asng \big )$ 
and then numerically calculate
its roots, may be extremely slow even for small problems.
Alternatively, the eigenvalues of $s\Esng - \Asng$ can be found from the solution of the \ac{gep}:
\begin{equation}
\begin{aligned}
  \label{sssa:eq:gep}
  (s \Esng - \Asng) \, \reigv &= \bfg 0_{\nd,1} \, , \\
  \leigv \, (s \Esng - \Asng)  &= \bfg 0_{1,\nd}\, ,
\end{aligned}
\end{equation}
where $\reigv \in \mathbb{C}^{\nd \times 1}$
and $\leigv \in \mathbb{C}^{1 \times \nd}$.
Every value of $s$ that satisfies \eqref{sssa:eq:gep} is 
an eigenvalue of 
the pencil $s \Esng - \Asng$, with
the vectors $\reigv$, $\leigv$ being the corresponding right and left eigenvectors,
respectively. Thus, the solution 
of the \ac{gep} consists in 
calculating the eigenpairs,
i.e.~eigenvalues and eigenvectors, that satisfy \eqref{sssa:eq:gep}.
Depending on the analysis that needs to be carried out, it may be required that only right (or left)
or both right and left eigenvectors are calculated.
In general, the pencil 
$s \Esng - \Asng$ has 
$\nf = {\rm rank}(s \Esng - \Asng)$
finite eigenvalues and 
the infinite eigenvalue with multiplicity
$\ninf = \nd - {\rm rank}(s \Esng - \Asng)$. 
Note that, unless otherwise stated, when this thesis refers to 
infinite eigenvalues, it implies eigenvalues that are at infinity and
not infinitely many.
Note also that, if $\Esng$ is singular, the pencil will have the 
infinite eigenvalue with multiplicity
at least one.

In the special case that the left-hand side matrix of \eqref{sssa:eq:sing} is the identity matrix (as is the case of system \eqref{sssa:eq:odelin}), \eqref{sssa:eq:gep} is equivalent to the \ac{lep}:
\begin{equation}
\begin{aligned}
  \label{sssa:eq:lep}
  (s \bfg I_\nd - \Asng) \, \reigv &= \bfg 0_{\nd,1} \, , \\
  \leigv \, (s \bfg I_\nd - \Asng)  &= \bfg 0_{1,\nd} \, .
\end{aligned}
\end{equation}
%
%
Every value of $s$ that satisfies \eqref{sssa:eq:lep} is 
an eigenvalue of 
the pencil $s \bfg I_{\nd} - \Asng$, with the vectors $\reigv$, $\leigv$ being the corresponding right and left eigenvectors,
respectively.
Thus, 
the solution of the \ac{lep} consists in calculating the 
eigenvalues and eigenvectors of $s \bfg I_{\nd} - \Asng$. In particular, the pencil 
$s \bfg I_{\nd} - \Asng$, has $\nd$ finite eigenvalues. 



\subsubsection{Properties}

{ \remark{ (\textit{Jordan Decomposition}). 
Consider the pencil $s\Esng-\Asng$,
with ${\rm det}(s\Esng-\Asng)\not\equiv 0$.
There exist non-singular matrices 
$\leigvmat$, $\reigvmat$ $\in \mathbb{C}^{\nd \times \nd}$ such that \cite{gantmacher:59}:
\begin{equation}
\label{sssa:eq:decomp}
\begin{aligned}
\leigvmat \, \Esng \, \reigvmat &=
\bfg I_\nf  \oplus \bfg H_\ninf \, ,  \\
\leigvmat \, \Asng \, \reigvmat &=
\bfg J_\nf  \oplus \bfg I_\ninf \, , 
\end{aligned}
\end{equation}
where $\nf+\ninf=\nd$; $\bfg J_\nf$, $\bfg J_\nf\in \mathbb{C}^{\nf \times \nf}$, is constructed by the $\nf$ finite eigenvalues 
$\lambda_1,\lambda_2, \ldots, \lambda_\nf$,
and their multiplicities, and has the Jordan canonical form \cite{gantmacher:59};
%
$\bfg H_\ninf$,
$\bfg H_\ninf \in \mathbb{C}^{\ninf \times \ninf}$, is a nilpotent matrix with index $\ninf_*$, constructed by
using the algebraic multiplicity $\ninf$ of the infinite eigenvalue.
Alternatively, the matrix $\bfg H_\ninf$ can be perceived as
the Jordan matrix of the zero eigenvalue of the dual pencil $z\Asng-\Esng$, where
$z= 1 \big / s$. 

The following notation is used:
\[
\leigvmat=
\begin{bmatrix}
\leigvmat_\nf \\ 
\leigvmat_\ninf
\end{bmatrix} \, , 
 \ 
\reigvmat=
\begin{bmatrix}
\reigvmat_\nf & 
\reigvmat_\ninf
\end{bmatrix}
\, ,
\]
with 
$\leigvmat_\nf \in \mathbb{C}^{\nf \times \nd}$,
$\leigvmat_\ninf \in \mathbb{C}^{\ninf \times \nd}$ and $\reigvmat_\nf\in \mathbb{C}^{\nd \times \nf}$, $\reigvmat_\ninf\in \mathbb{C}^{\nd \times \ninf}$. 
$\leigvmat_\nf$ is a matrix with rows $\nf$ linear independent left eigenvectors 
(including the generalized) 
of the $\nf$ finite eigenvalues of $s \Esng-\Asng$; 
$\leigvmat_\ninf$ is a matrix with rows $\ninf$ linear independent (including the generalized) left eigenvectors of the infinite eigenvalue of $s \Esng-\Asng$ with algebraic multiplicity $\ninf$;
$\reigvmat_\nf$ is a matrix with columns $\nf$ linear independent (including the generalized) right eigenvectors of the $\nf$ finite eigenvalues of $s \Esng-\Asng$; 
and $\reigvmat_\ninf$ is a matrix with columns $\ninf$ linear independent (including the generalized) right eigenvectors of the infinite eigenvalue of $s\Esng-\Asng$ with algebraic multiplicity $\ninf$. 
By applying the above expressions into \eqref{sssa:eq:sing}, one gets the following eight equalities:
\[
\begin{array}{c}
\leigvmat_\nf\Asng\reigvmat_\nf=\bfg J_\nf \, ,\\
\leigvmat_\nf\Asng\reigvmat_\ninf=\bfg 0_{\nf,\ninf}
\, , \\
\leigvmat_\ninf\Asng\reigvmat_\nf=\bfg 0_{\ninf,\nf}
\, , \\
\leigvmat_\ninf\Asng\reigvmat_\ninf=\bfg I_\ninf \ ,
\end{array}
\quad
\begin{array}{c}
\leigvmat_\nf\Esng \reigvmat_\nf=\bfg I_\nf
\, , \\
\leigvmat_\nf\Esng \reigvmat_\ninf=\bfg 0_{\nf,\ninf}
\, , \\
\leigvmat_\ninf\Esng \reigvmat_\nf=\bfg 0_{\ninf,\nf}
\, , \\
\leigvmat_\ninf \Esng \reigvmat_\ninf=\bfg H_\ninf \ .
\end{array}
\]
}
\label{sssa:remark:jordan}
}


{\theorem
{
Consider system \eqref{sssa:eq:sing} with a regular pencil.
Then the general solution of 
\eqref{sssa:eq:sing} is given by:
\begin{equation}
\label{sssa:eq:sol}
\xs (t)=
\reigvmat_\nf \, e^{\bfg J_\nf t} 
\, \bfb c
 +
\reigvmat_\nf
 \int_0^\infty
  e^{\bfg J_\nf(t-\tau)}
 \, \leigvmat_{\nf} 
  \, \Bsng  
  \, \bfg u(\tau) 
  \, d \tau
 -  \reigvmat_\ninf
 \sum^{\ninf_*-1}_{i=0}
   \bfg H_\ninf^i \leigvmat_{\ninf} 
    \, \Bsng \, 
{\bfg u}^{(i)}(t)
\, ,
\end{equation}
where $\bfb c \in \mathbb{C}^{\nf \times \nf}$ is constant vector and
$e^{\bfg J_\nf t}$ is the matrix exponential of $\bfg J_\nf t$.
}
\label{sssa:theorem:sol}
}

The proof of Theorem~\ref{sssa:theorem:sol} is given in Appendix~\ref{app:proofs}.

\subsection{Stability}

{
\definition (\textit{Asymptotic stability}) 
{ 
Consider an autonomous non-linear system $\Dt {\xs}(t) = \bfg h(\xs(t)) $ 
with 
equilibrium $\xs_o$. 
Then, the 
equilibrium $\xs_o$ is said to be \textit{asymptotically stable} if there exists $\delta>0$ 
such that if $||\xs(0)-\xs_o||< \delta$,
then $\lim_{t \rightarrow \infty}||\xs(t)-\xs_o||=0$.
}
\label{sssa:definition:lyap}
}

Simply put, asymptotic stability implies that solutions starting close enough to the equilibrium will eventually converge to it. 
Asymptotic stability of equilibria is a local property for non-linear systems.
On the other hand, for linear systems,
asymptotic stability is a global property,
which
means that
the solution 
will eventually
converge to the equilibrium 
for any given initial condition. 
In this case, applying the above definition to the general solution \eqref{sssa:eq:sol} of system
\eqref{sssa:eq:sing}, yields that 
stability is guaranteed if and only if no element of the matrix exponential 
$e^{\bfg J_\nf t}$ goes to infinity, when $t \rightarrow \infty$,
which,
in turn, leads to the following
well-known stability criterion.

{
\definition
{ 
System \eqref{sssa:eq:sing} 
is said to be 
\textit{asymptotically stable} if all finite eigenvalues $\lambda^*$ of
its 
matrix pencil $s \Esng - \Asng$ 
satisfy:
\begin{equation}
\label{sssa:eq:stability}
{\rm Re}(\lambda^*) < 0 \, .
\end{equation}
}
\label{sssa:definition:lyaplin}
}
The stability condition \eqref{sssa:eq:stability} can be also obtained using Lyapunov stability theory \cite{TAKABA199549, ishihara:02}. In particular, condition \eqref{sssa:eq:stability} is equivalent to considering the Lyapunov function:
\begin{equation}
  \mathcal{V}(\xs) = \xs\T \, \Esng\T \, \bfb M \, \xs \, ,
\end{equation}
with $\Esng\T \bfb M$ symmetric and positive definite,
%
and
\begin{equation}
  \label{sssa:eq:negative}
  \Asng\T \, \bfb M + \bfb M\T \, \Asng 
\end{equation}
negative definite.
If such a matrix $\bfb M$ 
exists, then
\eqref{sssa:eq:sing} is
asymptotically stable and, hence, also Lyapunov stable.

Finally, calculation of eigenvalues allows
measuring 
the characteristics of the most critical 
or \textit{dominant} for the stability dynamic modes 
on the system. In particular, 
the
\textit{damping ratio} and 
\textit{natural frequency} 
of
a dynamic mode are defined as follows.


{
\definition
{
Let $\lambda = a + \jj \, b$ be a finite, complex eigenvalue of 
$s \Esng - \Asng$.  The damping
ratio and the natural frequency
of $\lambda$ are defined as follows:
\begin{align}
\label{eq:model:zeta}
\zeta &= -
\frac{a}{|\lambda|} = -
\frac{a}{\sqrt{a^2 + b^2}} \, ,
\quad 
\text{(Damping ratio)} 
\\  \nonumber \\ 
\label{eq:model:fn}
{\rm f}_{\rm n} &= \frac{|\lambda|}{2\pi}  = \frac{\sqrt{a^2 + b^2}}{2\pi} \, ,
\quad \quad \ \,
\text{(Natural frequency)} 
\end{align}
}}
 
%
    
The power system is said to be well-damped,
if for all eigenvalues $\lambda^*$, the damping ratio is higher than a threshold, typically
$\zeta^* > 5 \% $.

\section{Numerical Methods}
\label{sssa:sec:numerical}

\subsection{Eigenvalue Algorithms}
\label{sssa:sec:num}

There is a rich literature on numerical algorithms that compute the full or a partial solution of a given \ac{lep} or \ac{gep}.
Relevant monographs on the topic are,
for example, \cite{saad:2011} and \cite{kressner}.  However, not all
available algorithms are suitable for \ac{sssa} of
power systems.  
The vast majority of numerical algorithms, in fact, solve exclusively
symmetric eigenvalue problems.
Such algorithms are, for example, the
ones described in \cite{Sorensen,LOBPCG}.
However, the matrices that describe a linearized power system model
are typically non-symmetric.  Compared to symmetric problems,
non-symmetric eigenvalue problems are more difficult and
computationally demanding to solve.   

The scalability of the numerical solution of
eigenvalue problems is also very important since, 
real-world power networks are large-scale dynamic systems. Unfortunately, 
the most reliable methods to find the full spectrum of an eigenvalue problem are 
dense-matrix methods, and their computational complexity and memory requirements
increase more than quadratically (in some cases even cubically) as
the size of the matrix increases. This is further discussed in Section~\ref{sssa:sec:complexity}.
Even using sparse matrices and limiting the search to a subset of the
spectrum, the solution of large-scale power system eigenvalue problems is challenging.

A coarse taxonomy of existing algorithms for the solution of non-symmetric
eigenvalue problems is as follows: 
vector iteration methods, Schur decomposition methods, Krylov subspace methods, and
contour integration methods.

Vector iteration methods are
in turn separated
to single and simultaneous vector iteration methods. Single vector
  iteration methods include the power method and
  its variants, such as the inverse power and
Rayleigh quotient iteration.   
Simultaneous vector iteration methods include the subspace iteration method \cite{subspace} and its
variants, such as the inverse subspace method.

Schur decomposition methods mainly include the
QR algorithm \cite{Francis}, the QZ
algorithm  \cite{molerqz}, and their variants, such as the QR algorithm with shifts. 
Schur decomposition based methods have been the standard methods employed for the eigenvalue analysis of small to medium size power systems \cite{kundur:1990,dual:17}. 

Krylov subspace methods basically include the
  Arnoldi iteration \cite{Arnoldi} and its variants,
  such as the implicitly restarted Arnoldi
  \cite{Lehoucq} and the Krylov-Schur method  \cite{KrylovSchur}. 
  In this category belong also preconditioned extensions
  of the Lanczos algorithm, such as 
  the non-symmetric versions of 
  the Generalized Davidson and Jacobi-Davidson method.
  
Finally, contour integration methods 
include a
moment-based Hankel method \cite{sakurai1}  and a
Rayleigh-Ritz-based projection
method \cite{sakurai2} proposed by Sakurai; and the FEAST algorithm \cite{polizzifeast}.

\subsection{Open-Source Libraries}
\label{sssa:sec:libs}

Available
free and open-source software libraries that solve non-symmetric
eigenvalue problems are a small subset of all existing eigensolvers.
This section provides an overview of the open-source
solvers that implement state-of-art numerical algorithms for non-symmetric eigenvalue problems \cite{app10217592}.
These are
LAPACK, 
ARPACK, Anasazi, SLEPc, FEAST
and z-PARES. 

LAPACK  \cite{lapack} is a standard library aimed at solving problems of
numerical linear algebra, such as systems of linear equations and eigenvalue problems.
A large part of the computations required
  by the routines of LAPACK are performed by calling the
  BLAS \cite{blas}.
As an eigensolver, LAPACK includes the QR and QZ algorithms.
Although it cannot handle general sparse matrices,
LAPACK is functional with dense matrices and, in fact, is the
standard dense matrix data interface used by all other eigenvalue libraries. A powerful 
GPU-based implementation of LAPACK routines is provided by MAGMA which, for general non-symmetric matrices, supports only the
solution of the \ac{lep}.

ARPACK \cite{arpack} is a
library developed for solving large eigenvalue problems with the IR-Arnoldi method. 
ARPACK depends on a number of
subroutines from LAPACK/BLAS. An important feature 
of ARPACK is the support of a \ac{rci}, which provides to the user the freedom to customize the matrix data format as desired.
An implementation of ARPACK for parallel computers is provided by PARPACK. 
The message parsing layers supported by PARPACK are 
\ac{mpi} \cite{mpi} and BLACS.

Anasazi
\cite{anasazi} is a library that implements block versions of both symmetric and non-symmetric algorithms for the solution 
of large-scale eigenvalue problems.
Regarding non-symmetric problems, it provides a block 
extension of the Krylov-Schur method and the 
\ac{gd} method. Anasazi depends on Trilinos \cite{trilinos} and uses LAPACK as an interface for dense matrix and Epetra as an interface for
sparse \ac{csr} matrix formats. 

SLEPc \cite{slepc} is a library that includes a variety of
 symmetric and non-symmetric methods, for the solution of
large sparse eigenproblems. For
  non-symmetric problems, it provides the following methods:
  power/inverse, power/Rayleigh quotient in a single
  implementation; subspace iteration
  with Rayleigh-Ritz projection and locking; 
  \ac{erd} Arnoldi; Krylov-Schur; GD;
  \ac{jd}; \ac{cih} and \ac{cirr}
  methods. SLEPc depends on PETSc
  \cite{petsc} and employs LAPACK as
  an interface for dense matrix, MUMPS \cite{MUMPS}
   as an interface for
  sparse \ac{csr} matrix formats and
  supports custom data formats, enabled by \ac{rci}.

FEAST \cite{feast_guide} is the eigensolver
  that implements the FEAST algorithm, first proposed in
  \cite{polizzifeast}.  
It depends on LAPACK as
  an interface for dense matrix, on SPIKE 
  as an interface
  for banded matrix and on MKL-PARDISO \cite{pardiso} for sparse \ac{csr}
  matrix formats.  In addition, FEAST includes
  \ac{rci} and thus, data formats can be customized by the user.
  Using the sparse interface requires linking FEAST with
  \index{Intel MKL} Intel MKL. 
Finally, FEAST includes parallel
  implementations which
  support 3-Level \ac{mpi} message parsing layer. 

z-PARES \cite{zpares_guide} is a complex
  moment-based contour integration eigensolver for \acp{gep} that implements the \ac{cih}, and \ac{cirr} methods to find
  the eigenvalues (and corresponding eigenvectors) that lie into a
  contour path defined by the user. The library depends on LAPACK for dense
  matrices, on MUMPS
  for sparse \ac{csr} matrices, while it supports
  custom data formats, enabled by \ac{rci}. Moreover,
  z-PARES includes a parallel version,
  which exploits 2-Level \ac{mpi} layer and employs MUMPS as its sparse solver.

Tables~\ref{tab:libs:num_twosided} and \ref{tab:libs:num_summ} 
provide a synoptic summary of the methods and
relevant features 
of open-source
libraries that solve non-symmetric eigenvalue problems. As it can be seen from Table~\ref{tab:libs:num_summ},
all libraries can handle both real and complex arithmetic types. On the other hand,
not all libraries are 2-sided, i.e~provide algorithms that allow calculating both left and right eigenvectors at once. 

\begin{table}[ht!]
  \centering
  \renewcommand{\arraystretch}{1.3}
  \caption[Methods of open-source libraries]{Methods of open-source
    libraries for non-symmetric eigenvalue problems.}
  \begin{tabular}{l|l}
    \hline 
    Library & Method   \\
    \hline
    LAPACK  & QR, \index{QR algorithm} QZ \index{QZ algorithm}   \\
    ARPACK  & IR-Arnoldi \\
    SLEPc & Power/Inverse Power/Rayleigh Quotient Iteration, Subspace,   \\
            & ERD-Arnoldi, \index{Arnoldi iteration}
        Krylov-Schur, GD, JD, CIH, CIRR \\
    Anasazi & Block Krylov-Schur, GD \\      
    FEAST & FEAST  \\
    z-PARES & CIH,  CIRR  \index{Contour integration methods} \\
    \hline 
  \end{tabular}
  \label{tab:libs:num_twosided}
\end{table}
\begin{table}[ht!]
  \centering
  \renewcommand{\arraystretch}{1.3}
  \setlength{\tabcolsep}{2pt}
  \caption[Relevant features of open-source libraries]{Relevant
    features of open-source libraries for non-symmetric eigenvalue
    problems.}
  \begin{minipage}{\textwidth}
  \centering 
    \begin{tabular}{l|cccc|cc|c|c|ll}
      \hline 
      Library & \multicolumn{4}{c|}{Data formats} &
      \multicolumn{2}{c|}{Computing}  & 2-sided & Real/ & \multicolumn{2}{c}{Releases}\\
      & dense & CSR & band& RCI & GPU  & parallel & & complex & first & latest\\
      \hline
      LAPACK & \cmark & \xmark & \xmark & \xmark &
      \cmark \footnote{With MAGMA.}  &
      \cmark\footnote{Parallel implementations of LAPACK routines are provided by ScaLAPACK \cite{scalapack}.} & \cmark & \cmark &1992 & 2016 \\
      ARPACK  & \xmark & \xmark & \xmark& \cmark  & \xmark & \cmark  & \xmark
      & \cmark & 1995 & 2019\footnote{ARPACK has been forked into ARPACK-NG.}\\
      SLEPc & \cmark  & \cmark  & \xmark & \cmark  & \cmark  & \cmark  & \cmark\footnote{In SLEPc, only the \index{Power method} power and the Krylov-Schur methods are 2-sided. }  & \cmark & 2002 & 2020 \\
      Anasazi & \cmark & \cmark & \xmark & \xmark & \xmark & \xmark
       & \xmark & \cmark & 2008 & 2014  \\
      FEAST & \cmark  & \cmark  & \cmark  & \cmark  & \xmark & \cmark  & \cmark & \cmark & 2009 & 2020\\
      z-PARES & \cmark  & \cmark  & \xmark & \cmark  & \xmark & \cmark  & \xmark & \cmark & 2014 & 2014 \\
      \hline \multicolumn{10}{c}{}
    \end{tabular}
  \end{minipage}
  \label{tab:libs:num_summ}
\end{table}

\subsection{Spectral Transforms}
\label{sssa:sec:moebius}

In general, solving the eigenvalue problem involves finding the full or partial spectrum of 
the pencil $s \Esng - \Asng$. However, depending on the applied numerical method as well as on the structure of the system matrices, it is common that the eigenvalues are not found by using directly $s \Esng - \Asng$, but through the pencil that arises after the application of a proper spectral transform. Spectral transforms are utilized by eigenvalue numerical
methods to find the eigenvalues of interest, address a singularity issue, or
accelerate convergence.

The M{\"o}bius transformation, which is a general variable transformation that 
includes as special cases all
spectral transforms used in practice by eigenvalue algorithms, is discussed here. 
The formulation of the M{\"o}bius transformation is:
  \begin{equation}\label{eq82}
    s:= \frac{az+b}{cz+d}\,,\quad a,b,c,d\in\mathbb{C}\,,\quad ad-bc\neq 0 \, .
  \end{equation}
Applying the transform \eqref{eq82} in 
\eqref{eq:sing:char} one has
\[
  {\rm det} \bigg ( \frac{az+b}{cz+d} \, \Esng-\Asng \bigg )=0 \, ,
\]
or, equivalently, by using determinant properties
\[
 {\rm det} \big((az+b)\, \Esng-(cz+d)\, \Asng \big )=0 \, ,
\]
or, equivalently, 
\[
{\rm det} \big ( (a \Esng -c \Asng)\, z -
  (d \Asng-b \Esng) \big )=0 \, ,
\]
which is the \index{Characteristic equation} characteristic equation
of the linear dynamical system
\begin{equation}\label{eq83}
(a \Esng-c \Asng) \,\Dt{\bfbt x}(t) =
 (d \Asng-b \Esng)\, \bfbt x(t) \, ,
\end{equation}
with pencil $z(a \bfb E-c \Asng) - (d \Asng-b \bfb E)$.
System \eqref{sssa:eq:sing} will be referred as the prime system, and
the family of systems \eqref{eq83} will be defined as the proper
``M-systems''. An important property is that the solutions and stability properties of system \eqref{sssa:eq:sing} can be studied through \eqref{eq83} without resorting to any further computations, see \cite{moebius}. 
The utilities of the family of systems of type \eqref{eq83} have been
further emphasized by the features of some particular special cases.
The most commonly employed 
M{\"o}bius transforms and the corresponding matrix pencils for the
\ac{gep} are 
summarized in
Table \ref{tab:moeb:special}. 
The values of the parameters $a,b,c,d$
that lead to each of these transforms are given in Table \ref{tab:moeb:pars}. 
In case that $\sigma>0$, the Cayley transform is equivalent to the bilinear transform 
$z:={(\frac{T}{2}s+1) }/{(\frac{T}{2}s-1)}$, where $T=\frac{2}{\sigma}$. 
Finally, the selection of the best 
transform for a specific system and eigenvalue problem is a
challenging task to solve, since the selection of shift
values is, ultimately, heuristic.

\begin{table}[ht!]
  \centering
 \renewcommand{\arraystretch}{1.2}
  \caption[Common linear spectral transforms]{Common linear spectral transforms.}
  \begin{minipage}{\textwidth}
  \centering
  \begin{tabular}{l|c|c|c}
    \hline
    Name 
    & $z$ & Pencil & $s$ \\ 
    \hline
    Prime system& $s$ & $s \Esng - \Asng$ & $z$\\
    Invert  & ${1}/{s}$  & $z\Asng - \Esng$ & ${1}/{z}$ \\
    Shift \& invert & ${1}/{s-\sigma}$ & $z( \sigma \bfb E - \Asng) + \Esng$ & ${1}/{z}+\sigma$\\
    Cayley & ${(s+\sigma) }/{(s-\sigma)}$
    & $z(\sigma \Esng - \Asng) - (\Asng + \sigma \bfb E)$   
    & $\sigma {(z-1)}/{(z+1)}$    \\
    Gen. Cayley  & ${(s+\nu)}/{(s-\sigma)}$  & $z(\sigma \Esng - \Asng) - (\Asng + \nu \bfb E)$  & ${(\sigma z-\nu)}/{(z+1)}$   \\
    M\"obius & ${(-d s+ b)}/{(cs-a)}$ & 
    $z(a \Esng - c \Asng) - (d\Asng - b \bfb E)$ & ${(az+b)}/{(cz+d)}$\\
    \hline
  \end{tabular}
\end{minipage}
\label{tab:moeb:special}
\end{table}

\begin{table}[ht!]
  \centering
  \renewcommand{\arraystretch}{1.05}
  \caption[Coefficients
    of special M{\"o}bius transformations]{Coefficients
    of special M{\"o}bius transformations.}
    \vspace{1mm}
  \begin{tabular}{l|rrrr}
 M-system & $a$ & $b$ & $c$ & $d$ \\ 
      \hline
Prime& $-1$ & $0$ & $0$ & $-1$\\
      Dual & $0$  & $1$ & $1$ & $0$ \\
Shift \& invert &$\sigma$& $1$ & $1$ & $0$\\
Cayley  &$\sigma$& $-\sigma$ & $1$ & $1$\\
      Gen. Cayley  &$\sigma$& $-\nu$ & $1$ & $1$\\   
      \hline
    \end{tabular}
  \label{tab:moeb:pars}
\end{table}

\subsection{Computational Complexity}
\label{sssa:sec:complexity}

The computational complexity of an eigenvalue algorithm is in general dependent upon the particular implementation provided by a given software library. However, library manuals typically do not detail their memory and computational requirements, and thus to provide 
a systematic and precise comparison with this regard is not a trivial task. Yet, for an eigenvalue problem with a pencil of size $n$, one may provide a rough summary of the costs associated to a generic algorithm that searches for $k \leq n$ eigenvalues. That is, the algorithm:
\begin{itemize}
\item constructs a subspace 
of order $k$ associated to the eigenvalue problem\footnote{Note this is a rough estimation.  In order to accurately capture $k$ eigenvalues, practical algorithms often work with subspaces of size that is larger than $k$, thus increasing the overall computational cost.}. A basis of the subspace, that is $k$ vectors of size $n$, needs to be stored.
\item carries out computations to guarantee that the basis vectors of the subspace
are orthogonal. The associated cost of such computations is $\mathcal{O}(k^2 n)$.
\item projects the matrix pencil to the subspace which yields a reduced eigenvalue problem with a pencil of size $k$. The dense matrices of dimensions $k \times k$ that comprise this pencil need to be stored.
\item solves the projected
dense eigenvalue problem. This
problem is typically solved using QR factorization, with an associated cost of $\mathcal{O}(k^3)$.
\end{itemize}

The total computational cost sums to $\mathcal{O}(k^2 n+k^3)$. 
%
For a dense matrix method, i.e.~the QR algorithm and its variants, a complete basis of vectors is used, which yields a computational complexity of $\mathcal{O}(n^3)$.
It follows that the cost of the resources required to solve a very large problem using a dense-matrix algorithm is very high.
As a matter of fact, the largest ever eigenvalue
analysis with a dense algorithm to date was the solution of a $10^6 \times 10^6$ problem in about 1~h, and it
was carried out in 2014 by the Japanese K computer in Riken. To be able to obtain this result, the K
computer includes $88,000$ processors that draw a peak power of $12.6$~MW, while its operation costs annually US\$10 million. 


The computational burden associated to the numerical algorithms and open-source libraries described in this thesis is further discussed 
through numerical simulations in the case study of Section~\ref{sssa:sec:case}.

\section{Case Studies}
\label{sssa:sec:case}

In this section simulation results
are presented based on two real-world size power system models. The first system is a detailed model of the \ac{aiits} which includes $1,443$
state variables and $7,197$ algebraic variables. The second system is a dynamic model of the \ac{entsoe} system, which includes 
$49,396$ state variables and $96,770$ algebraic
variables.
The versions and dependencies 
of the open-source
libraries considered in this section are summarized in
Table \ref{tab:libs:libraries3}. Note that
this section considers only the open-source
libraries that were successfully 
compiled and installed on Linux and Mac OS X operating systems and that
worked for relatively ``large'' eigenvalue problems. 

\begin{table}[ht!]
  \centering
  \renewcommand{\arraystretch}{1.3}
  \caption[Versions and dependencies of open-source
  libraries]{Versions and dependencies of open-source libraries for
    non-symmetric eigenvalue problems.}
  \begin{tabular}{l|l}
    \hline 
    Library (Version) & Dependencies (Version) \\
    \hline
    LAPACK (3.8.0) & ATLAS (3.10.3) \\
    MAGMA  (2.2.0) & NVidia CUDA \index{NVidia CUDA} (10.1) \\
    ARPACK-NG (3.5.0) & SuiteSparse KLU (1.3.9) \\      
    z-PARES (0.9.6a) & OpenMPI (3.0.0), MUMPS (5.1.2) \\
    SLEPc (3.8.2) & PETSc (3.8.4), MUMPS (5.1.2) \\
    \hline 
  \end{tabular}
  \label{tab:libs:libraries3}
\end{table}

All simulations
are obtained using Dome \cite{vancouver}.  
The Dome version
utilized for this chapter is based on Fedora Linux 28, Python 3.6.8,
CVXOPT 1.1.9 and KLU 1.3.9.
Regarding the computing times reported in
both examples, two comments are relevant.  
First, all simulations were executed
on a server mounting two quad-core Intel Xeon 3.50 GHz CPUs, 1 GB
NVidia Quadro 2000 GPU, 12 GB of RAM, and running a 64-bit Linux OS.
Second, since, not all method implementations include 2-sided
versions and in order to provide as a fair comparison as possible,
all eigensolvers are called so as to return only the calculated
eigenvalues and not eigenvectors.

\subsection{All-Island Irish Transmission System}
\label{sssa:sec:aiits}

This case study considers a real-world model of the \ac{aiits}.
  The topology and the steady-state operation data of the system have
  been provided by the Irish transmission system operator, EirGrid
  Group. Dynamic data have been determined based on current
knowledge about the technology of the generators and the controllers.  
The system
  consists of 1,479 buses, 796 lines, 1,055 transformers, 245 loads,
  22 synchronous machines, with \acp{avr} and \acp{tg}, 6 \acp{pss}
  and 176 wind generators.
 In total, the dynamic model has $\nx=1,443$
  state variables and $\ny=7,197$ algebraic variables. 
The map of the \ac{aiits} is given in Appendix~\ref{app:data}.
  \begin{table}[ht!]
    \centering
    \renewcommand{\arraystretch}{1.2}
    \caption[AIITS: dimensions
      of the LEP and GEP]{AIITS: dimensions
      of the LEP and GEP.}
    \vspace{1mm}
    \begin{tabular}{lll}
      \hline 
      Problem & Pencil & Size \\
      \hline
      LEP & $s  \bfg I_{\nx} - \AS$ & $1,443 \times 1,443$ \\
      GEP &$s 
      \Edae - \Adae$ & $8,640 \times 8,640$ \\
      \hline 
    \end{tabular}
    \label{tab:librarirish}
  \end{table}

  \index{QR algorithm}
  \index{QZ algorithm}
  %
  

Results of the eigenvalue analysis of the \ac{aiits} are discussed for both
\ac{lep} and \ac{gep}
and 
for a variety of different numerical methods, namely, QR and QZ
algorithms by LAPACK, GPU-based QR algorithm by MAGMA, subspace
iteration, \ac{erd}-Arnoldi and Krylov-Schur
methods by SLEPc, IR-Arnoldi by
ARPACK; and \ac{cirr} by 
z-PARES. 

\begin{figure}[ht!]
    \begin{center}
      \resizebox{0.85\linewidth}{!}{\includegraphics{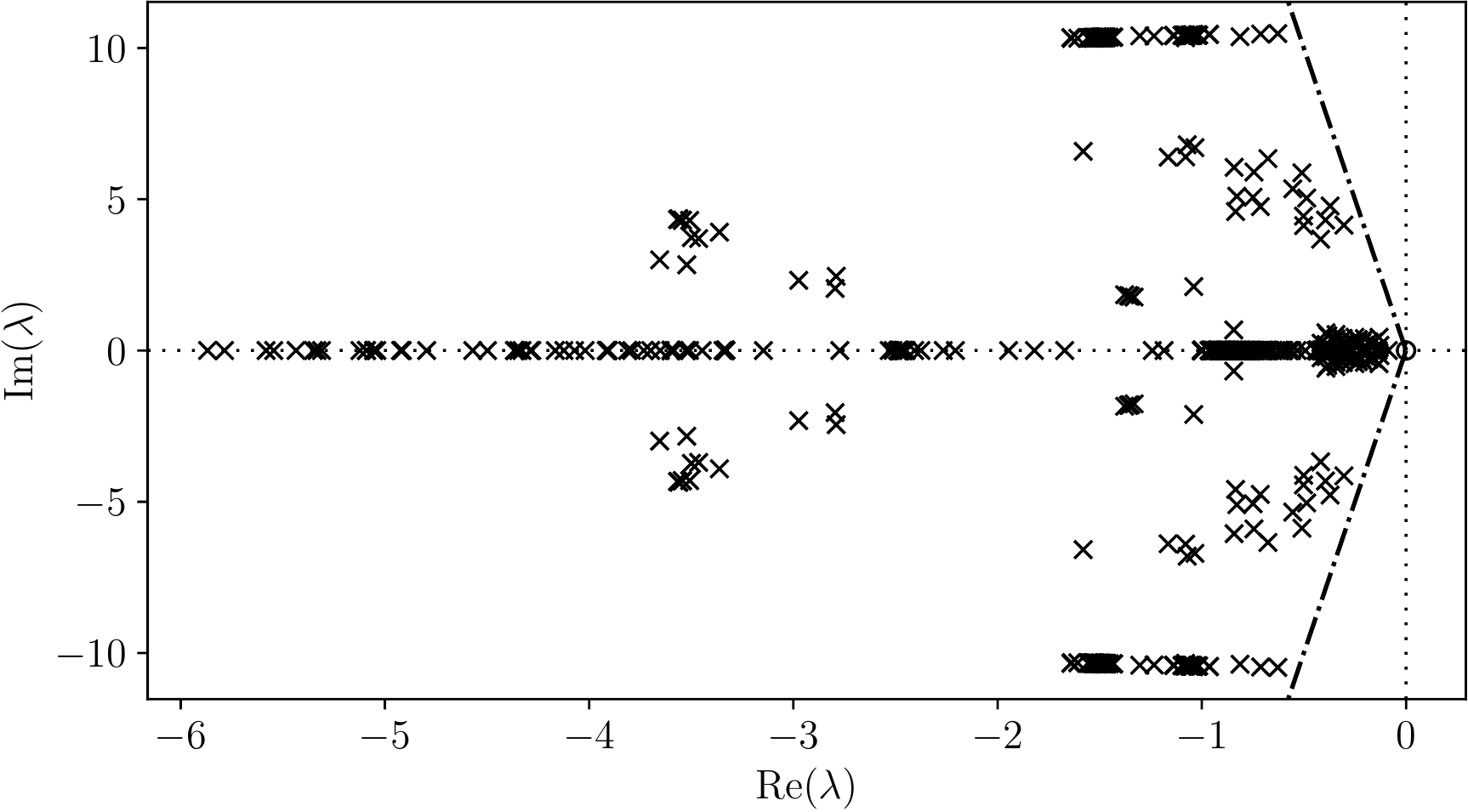}}
      \caption[AIITS: root loci computed with LAPACK]{AIITS: root loci computed with LAPACK.}
      \label{fig:large:irish:lapack}
    \end{center}
  \end{figure}

  The results obtained with Schur
  decomposition methods are presented in Table
  \ref{tab:num_irish_schur}. 
 Both QR and QZ algorithms
 find all $1,443$ finite
  eigenvalues of the system.  For the \ac{gep}, the QZ algorithm also
  finds the additional infinite eigenvalue with its algebraic
  multiplicity. 
  The obtained rightmost eigenvalues are
  the same for both \ac{lep} and \ac{gep}. 
  Since
  LAPACK is the most mature software tool
  among those considered in this section, the accuracy of the
  eigenvalues found with all other libraries is evaluated
  by comparing them with the reference solution computed with LAPACK.
   The system root loci plot
  is shown in Figure~\ref{fig:large:irish:lapack}.  
  Regarding the computational time, it is seen that, for the \ac{lep},
  both LAPACK and the GPU-based MAGMA are very efficient
  at this scale, with MAGMA providing only a marginal speedup. On the other hand, when it comes to solving the \ac{gep} with
  LAPACK's QZ method, scalability becomes a serious issue.
  \begin{table}[ht!]
    \centering
   \renewcommand{\arraystretch}{1.05}
    \caption[AIITS: Schur
      decomposition methods]{AIITS: Schur
      decomposition methods, 
      LEP and GEP.}
    \vspace{1mm}
      \begin{tabular}{l|c|c|c}
        \hline 
        Library & LAPACK & MAGMA & LAPACK\\
        \hline
        Problem & LEP & LEP & GEP \\
        Method & QR\index{QR algorithm} & QR & QZ\index{QZ algorithm} \\ 
        Spectrum & All &All &  All\\
        \hline 
        Time [s] & $3.94$  & $3.54$ & $3,669.77$ \\
        Found & $1,443$ eigs. & $1,443$ eigs. & $8,640$ eigs. \\
        \hline
        LRP eigs. & 
$0.0000$    & $0.0000$ & $0.0000$ \\ 
&$-0.0869$    & $-0.0869$ & $-0.0869$\\ 
&$-0.1276\pm \jj \, 0.1706$    & $-0.1276\pm \jj \, 0.1706$ & $-0.1276\pm \jj \, 0.1706$ \\ 
& $-0.1322\pm \jj \, 0.4353$   & $-0.1322\pm \jj \, 0.4353$ & $-0.1322\pm \jj \, 0.4353$ \\ 
&  $-0.1376 $  & $-0.1376 $ & $-0.1376 $ \\ 
& $-0.1382 $   & $-0.1382 $ & $-0.1382 $ \\ 
& $-0.1386 $   & $-0.1386 $ & $-0.1386 $ \\ 
&  $-0.1390 $  & $-0.1390 $ & $-0.1390 $ \\ 
&  $-0.1391 $  & $-0.1391 $ & $-0.1391 $ \\ 
& $-0.1393 $   & $-0.1393 $ & $-0.1393 $ \\ 
& $-0.1394 $   & $-0.1394 $ & $-0.1394 $ \\ 
\hline
\end{tabular}
\label{tab:num_irish_schur}
\end{table}

\begin{figure}[ht!]
    \begin{center}
      \resizebox{0.85\linewidth}{!}{\includegraphics{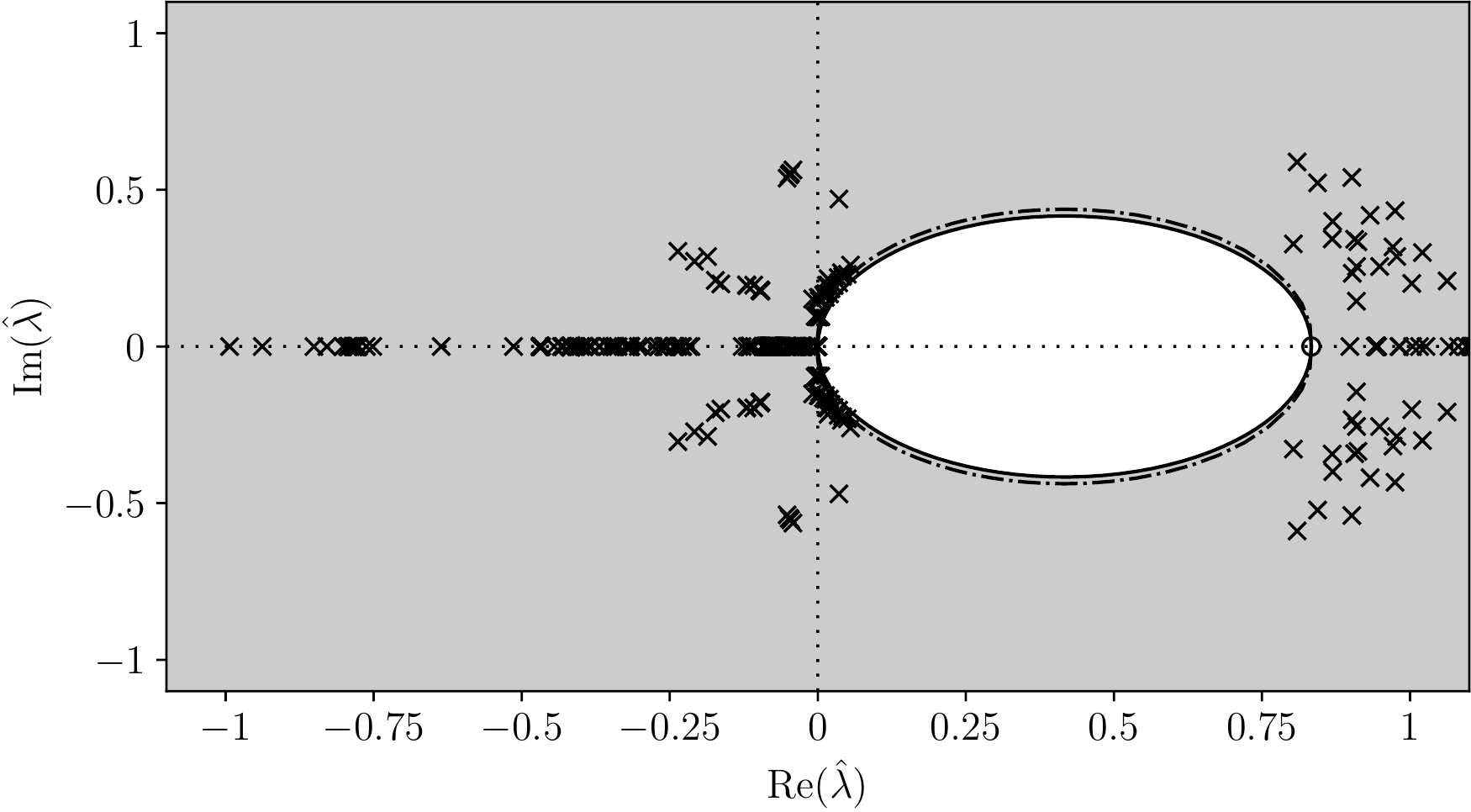}}
      \caption[AIITS: shift \& invert transform image of the spectrum]{AIITS: shift \& invert transform image of the spectrum, $\sigma=-1.2$.}
      \label{fig:moebius:sineg}
    \end{center}
  \end{figure}

\begin{figure}[ht!]
    \begin{center}
      \resizebox{0.85\linewidth}{!}{\includegraphics{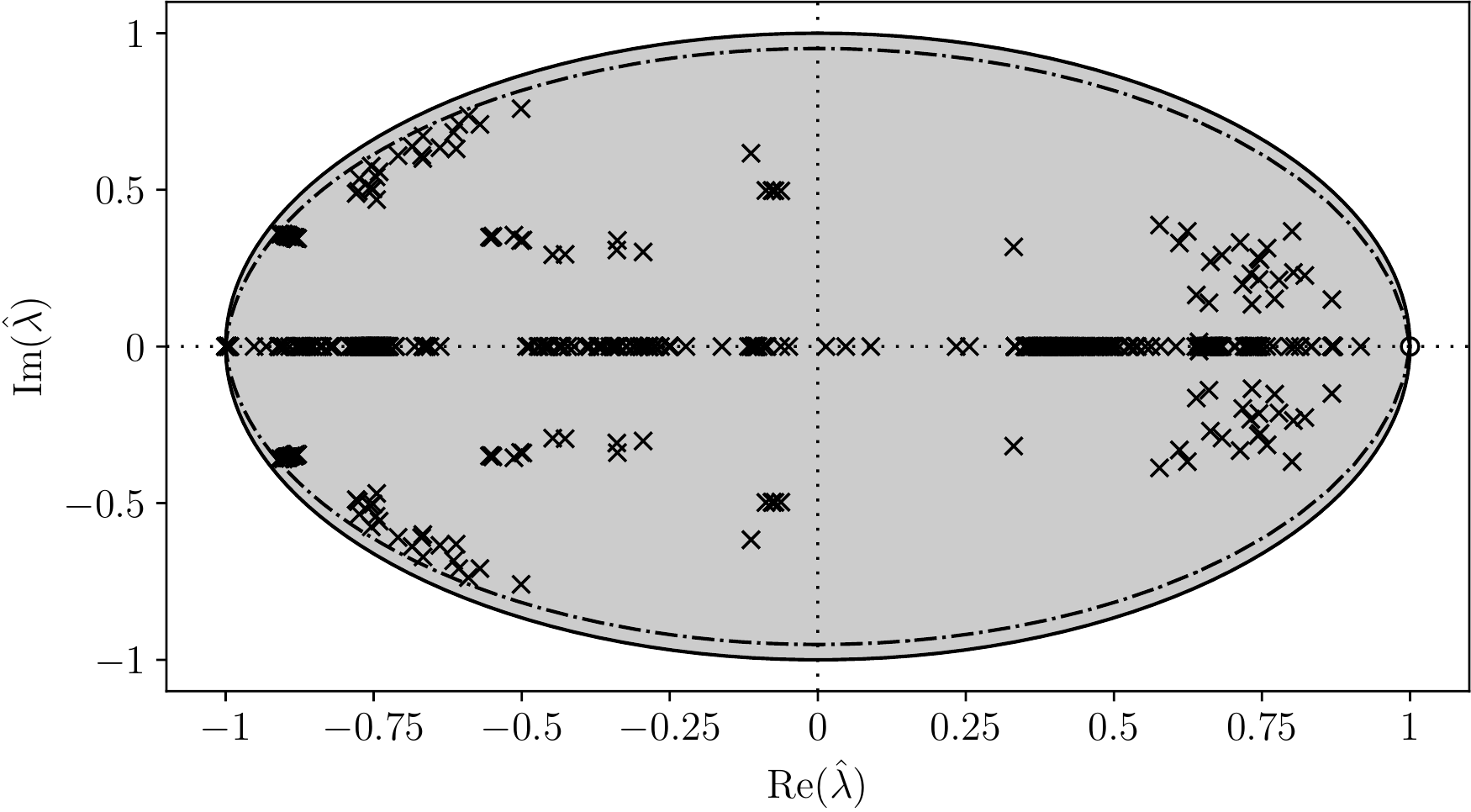}}
      \caption[AIITS: Cayley
        transform image of the spectrum]{AIITS: Cayley
        transform image of the spectrum,
        $\sigma=2$.}
      \label{fig:moebius:cayley}
    \end{center}
  \end{figure}

Figures~\ref{fig:moebius:sineg} and \ref{fig:moebius:cayley} show the spectrum of the \ac{aiits}
  system for a couple of common special M{\"o}bius transforms, in
  particular for the shift \& invert and the
  Cayley transform. 
  In these figures, $\hat \lambda$ denotes an eigenvalue of the transformed pencil. These results
  refer to the
  \ac{lep}, and are obtained using LAPACK.  In each figure, the stable region is shaded, while
  the stability boundary is indicated with a solid line. The
  $5$ \% damping boundary is indicated with a  dash-dotted line. 
  For the shift \& invert transform, the stability boundary is defined
  by the circle with center $c=(1/2\sigma,0)$ and radius
  $\rho=1/2\sigma$. If $\sigma<0$, that is the case of Figure~\ref{fig:moebius:sineg},
  stable eigenvalues are mapped outside the
  circle.  On the other hand, if $\sigma>0$, stable eigenvalues are mapped inside the
  circle.
  If $\sigma=0$, the dual pencil is obtained
  with the corresponding invert transform, and the stable region is
  the full negative right have plane. Finally, Figure~\ref{fig:moebius:cayley} shows the image of the Cayley
  transform \index{Cayley transform} of the \ac{aiits} for
  $\sigma=1.2$.  All stable eigenvalues are located inside the unit
  circle with center the origin.
  
  The implementation of the subspace iteration by
  SLEPc only finds the desired number of LM
  eigenvalues.  However, in the $s$-domain, the relevant eigenvalues
  from the stability point of view
 are not the LM ones, but the ones
  with \ac{LRP} or \ac{SM}.  Especially
  for the \ac{gep}, the \ac{LM} eigenvalue is infinite and, hence, does not
  provide any meaningful information on the system dynamics. 
  For this reason and for the needs of power
system \ac{sssa}, the subspace method and, in general, 
any method that looks for LM eigenvalues, must always be combined with a spectral transform. 
For the needs of this example,
the invert transform is applied and the
pencil of the dual system, i.e.~$z \Asng - \Esng$, is passed to SLEPc.  Then, the method looks for the 50 LM eigenvalues of the dual system, which correspond to the 50 SM eigenvalues of the prime system. With this
setup, the eigenvalues found by the subspace iteration 
for the \ac{gep} are shown in Table~\ref{tab:num_irish_subspace}.  
As it can be seen,
the pair $-0.1322 \pm \jj 0.4353$ is not captured, since its magnitude is larger
than the magnitudes of the 50 SM eigenvalues. To obtain also this pair, one
can customize the spectral transform or simply increase the number of the
eigenvalues to be returned. However, the best setup is not known a priori and
thus, some heuristic parameter tuning is required.  
Finally, the method does not scale well, since solution of the \ac{gep} is completed in $6,807.24$ s.

  \begin{table}[ht!]
    \centering
   \renewcommand{\arraystretch}{1.05}
    \caption[AIITS: subspace
      iteration method]{AIITS: subspace
      iteration method, GEP.}
    \vspace{1mm}
      \begin{tabular}{l|c}
        \hline
        Library & SLEPc  \\ 
        Method &  Subspace \\
        Spectrum & 50 LM  \\
        \hline
        Transform  & Invert \\
        \hline 
        Time [s] & $6,807.24$\\
        Found &  $50$\\
        \hline
        \ac{LRP} eigs. & $-0.0000$  \\
                & $-0.0869$  \\
                & $-0.1276\pm \jj  0.1706$  \\
                & $-0.1376$  \\
                & $-0.1382$  \\
                & $-0.1386$  \\
                & $-0.1390$  \\
                & $-0.1391$  \\
                & $-0.1393$  \\
                & $-0.1394$  \\
                & $-0.1397$  \\
        \hline
      \end{tabular}
    \label{tab:num_irish_subspace}
  \end{table}

  The rightmost eigenvalues found with
  Krylov subspace methods for the \ac{lep} and \ac{gep} are shown in
  Table~\ref{tab:num_irishlep_krylov} and
  Table~\ref{tab:num_irishgep_krylov}, respectively.  For the
  \ac{lep}, ARPACK is set up to find the $50$ \ac{LRP} eigenvalues.
  Although all eigenvalues shown in Table~\ref{tab:num_irishlep_krylov} for ARPACK are actual eigenvalues of the
  system, some of the \ac{LRP} ones are missed. Furthermore, no correct eigenvalues were
  found for the \ac{gep}, since a non-symmetric $\Esng$ is not supported.  In SLEPc methods,
  both for \ac{lep} and \ac{gep} and in
  order to obtain the eigenvalues with good accuracy,
  the option ``Target Real Part'' (TRP) is used, which allows targeting
  eigenvalues with specified real part. In particular, the TRP parameter is set to
  $-0.01$, and a shift \& invert transform with
  $\sigma =-0.01$ is applied.  Both \ac{erd}-Arnoldi and Krylov-Schur methods are able to accurately capture all rightmost
  eigenvalues.  Note that, the eigenvalues
  obtained with SLEPc, when compared to the ones found by
  LAPACK, appeared to be shifted by a constant offset $-\sigma$,
  i.e.~$0.01$ was returned instead of $0$, and so on.  The
  results shown in Table~\ref{tab:num_irishlep_krylov} and Table~\ref{tab:num_irishgep_krylov} take into
  account such a shift by adding $\sigma$ to all output values
  returned by SLEPc.
%
  Finally, the Krylov subspace methods
  by SLEPc appear to be more efficient than ARPACK's
  IR-Arnoldi.  \index{Arnoldi iteration} Compared to Schur
  decomposition methods, \index{Schur decomposition methods} at this
  scale, Krylov methods, although they require some tuning, appear to
  be by far more efficient for the \ac{gep}, but less efficient for
  the \ac{lep}.

\begin{table}[ht!]
  \centering
 \renewcommand{\arraystretch}{1.05}
  \caption[AIITS: Krylov subspace methods, LEP]{AIITS: Krylov subspace methods, LEP.}
  \vspace{1mm}
    \begin{tabular}{l|c|c|c}
      \hline 
      Library & ARPACK & SLEPc & SLEPc \\ 
      Method & IR-Arnoldi & ERD-Arnoldi \index{Arnoldi iteration}                           & Krylov-Schur  \\ 
      Spectrum & 50 \ac{LRP} & $50$ TRP & $50$ TRP \\
      \hline 
      Transform & - & Shift \& invert & Shift \& invert\\
              & &$\sigma=-0.01$ & $\sigma=-0.01$\\
      \hline
      Time [s] & 
      $76.96$ & $17.84$ & $16.58$ \\
      Found  & $26$ eigs. & $54$ eigs. & $55$ eigs. \\
      \hline
      \ac{LRP} eigs.
              & $-0.0000$ &
    {$0.0000$} 
    &
              $0.0000$  
              \\
              & $-0.0869$ & $-0.0869$ & $-0.0869$ \\
              &$-0.1276\pm \jj \, 0.1706$ 
              &
 {$-0.1276\pm \jj \, 0.1706$}
 &
              $-0.1276\pm \jj \, 0.1706$ 
              \\
              &$-0.1322\pm \jj \, 0.4353$ &
{$-0.1322\pm \jj \, 0.4353$}
&
              $-0.1322\pm \jj \, 0.4353$ 
              \\
              &$-0.1615\pm \jj \, 0.2689$ &$-0.1376$ &$-0.1376$ \\
              &$-0.1809\pm \jj \, 0.2859$ &$-0.1382$ &$-0.1382$ \\
              &$-0.2042\pm \jj \, 0.3935$ &$-0.1386$ &$-0.1386$ \\
              &$-0.2172\pm \jj \, 0.2646$ &$-0.1390$ &$-0.1390$ \\
              &$-0.2335\pm \jj \, 0.3546$ &$-0.1391$ &$-0.1391$ \\
              &$-0.2344\pm \jj \, 0.3644$ &$-0.1393$ &$-0.1393$ \\
              &$-0.2503\pm \jj \, 0.4363$
              &
 {$-0.1394$} 
 &
$-0.1394$ %
              \\
      \hline
    \end{tabular}
  \label{tab:num_irishlep_krylov}
\end{table}

\begin{table}[ht!]
\centering
\renewcommand{\arraystretch}{1.05}
  \caption[AIITS: Krylov subspace methods, GEP]{AIITS: Krylov subspace methods, GEP.}
  \vspace{1mm}
  \begin{tabular}{l|c|c}
    \hline
    Library 
    & SLEPc & SLEPc  \\ 
    Method 
    & ERD-Arnoldi \index{Arnoldi iteration} & Krylov-Schur \index{Krylov-Schur method} \\
    Spectrum 
    & $50$ TRP & $50$ TRP \\
    \hline 
    Transform 
    & Shift \& invert& Shift \& invert \\  
    & $\sigma=-0.01$ & $\sigma=-0.01$ \\
    \hline    
    Time [s] 
    & $8.93$ & $7.64$ \\
    Found 
    & $51$ eigs. & $53$ eigs. \\
    \hline
    \ac{LRP} eigs.
    & $0.0000$ &$0.0000$ \\
    &$-0.0869$            &$-0.0869$   \\
    &$-0.1276\pm \jj \, 0.1706$ &$-0.1276\pm \jj \, 0.1706$ \\
    &$-0.1322\pm \jj \, 0.4353$ &$-0.1322\pm \jj \, 0.4353$ \\
    &$-0.1376$    &$-0.1376$ \\
    &$-0.1382$    &$-0.1382$ \\
    &$-0.1386$    &$-0.1386$ \\
    &$-0.1390$    &$-0.1390$ \\
    &$-0.1391$    &$-0.1391$ \\
    &$-0.1393$    &$-0.1393$ \\
    &$-0.1394$    &$-0.1394$ \\
    \hline
  \end{tabular}
\label{tab:num_irishgep_krylov}
\end{table}

\begin{figure}[ht]
  \begin{center}
    \resizebox{0.85\linewidth}{!}{\includegraphics{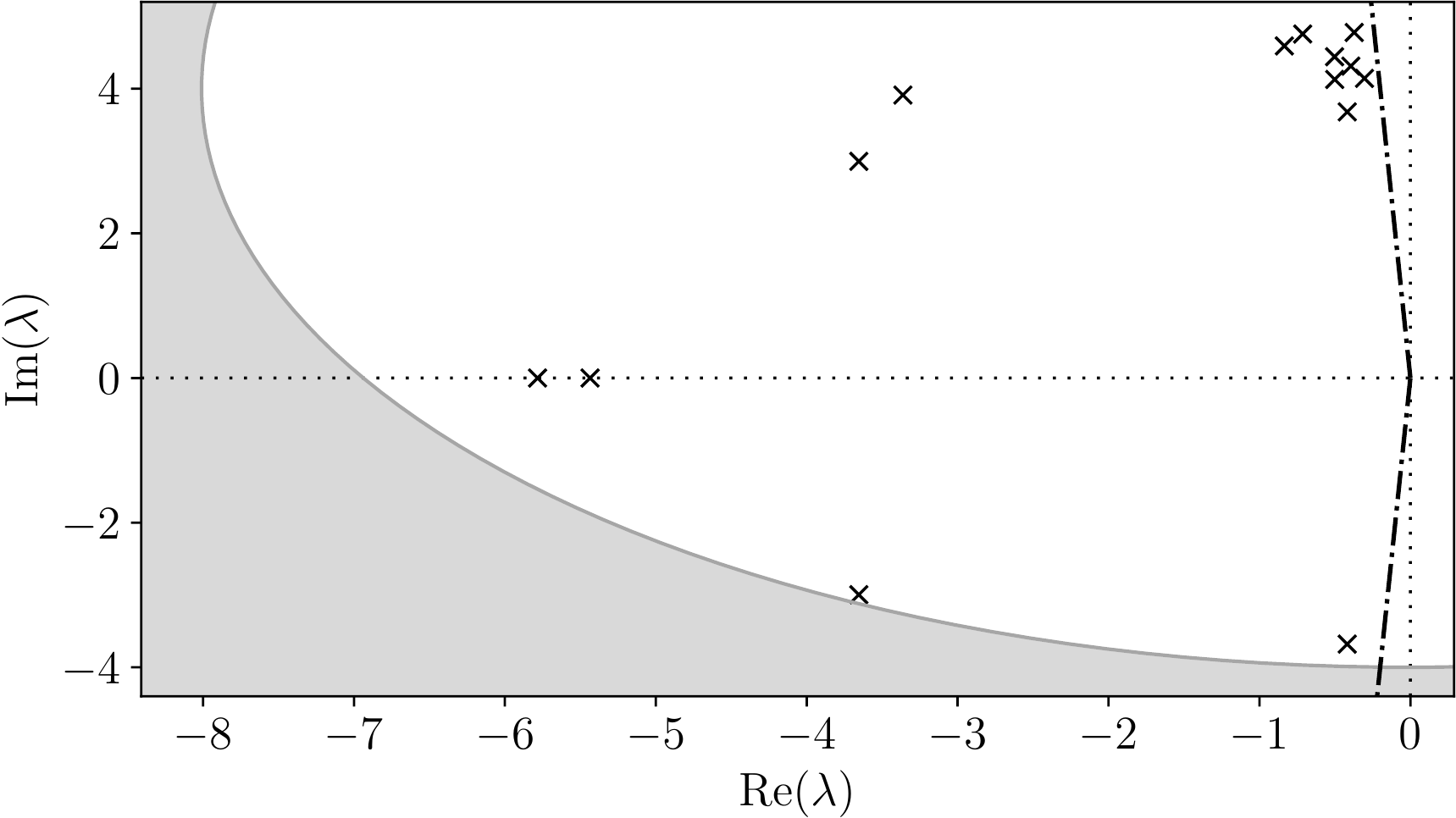}}
    \caption[AIITS: root loci obtained with z-PARES]{AIITS: root loci obtained with z-PARES, LEP.}
    \label{fig:large:irish:contour}
  \end{center}
\end{figure}

The results obtained with
z-PARES' 
\ac{cirr} method are presented in Table~\ref{tab:num_irishlepgep_ci} and Figure~\ref{fig:large:irish:contour}. 
The method is set to look for
solutions in the circle with center the
point $c=(-0.01,4)$ and radius $\rho=8$.
In both cases, the eigenvalues found by z-PARES are actual
eigenvalues of the system, although the eigenvalues found for the
\ac{gep} include noticeable errors, when compared to the results obtained with LAPACK.

\begin{table}[ht!]
 \centering
 \renewcommand{\arraystretch}{1.05}
 \caption[AIITS: contour integration method]{AIITS: contour integration method, LEP and GEP.}
 \begin{tabular}{l|c|c}
\hline
Library 
& \multicolumn{2}{c}{z-PARES} \\ 
Method  
& \multicolumn{2}{c}{CIRR}  \index{Contour integration methods}  \\
Spectrum 
& \multicolumn{2}{c}{$c=(-0.01,4), \rho=8$}  \\
\hline
    Problem & LEP & GEP \\
    \hline
    Time [s] 
    & $10.81$ & $17.10$ \\
    Found 
    & $49$ eigs. & $52$ eigs. \\
    \hline
    \ac{LRP} eigs.
    & $-0.3041 + \jj \, 4.1425$ & $-0.3040 + \jj \, 4.1429$ \\
    & $-0.3720 + \jj \, 4.7773$ & $-0.3715 + \jj \, 4.7774$ \\
    & $-0.3945 + \jj \, 4.3121$ & $-0.3947 + \jj \, 4.3122$ \\
      & $-0.4184\pm \jj \, 3.6794$ 
    & $-0.4187\pm \jj \, 3.6794$ \\
    & $-0.4866 + \jj \, 5.0405$ & $-0.4865 + \jj \, 5.0405$ \\
    & $-0.5011 + \jj \, 4.1276$ & $-0.5007 + \jj \, 4.1274$ \\
    & $-0.5022 + \jj \, 4.4417$ & $-0.5018 + \jj \, 4.4417$ \\
    & $-0.5077 + \jj \, 5.8727$ & $-0.5097 + \jj \, 5.8747$ \\
    & $-0.5555 + \jj \, 5.3444$ & $-0.5542 + \jj \, 5.3436$ \\
    & $-0.6765 + \jj \, 6.3426$ & $-0.6761 + \jj \, 6.3412$ \\
    \hline
  \end{tabular}
  \label{tab:num_irishlepgep_ci}
\end{table}

The most relevant issue is that the eigenvalues obtained
with z-PARES are not the
most important ones for the stability of the system, which means that
critical eigenvalues are missed.  This issue occurs despite the
defined search contour being reasonable.  Of course, there may be some
region for which the critical eigenvalues are captured but, this can
not be known \textit{a priori}.  Regarding the simulation time, the
method for the \ac{aiits} is faster than SLEPc's Krylov subspace
methods for the \ac{lep}, but slower for the \ac{gep}.
The search contour and the location of the characteristic roots found by z-PARES for the \ac{lep} are depicted in 
Figure~\ref{fig:large:irish:contour}.

\subsection{21,177-bus ENTSO-E}
\label{sssa:sec:entsoe}

  This example presents simulation results for a dynamic model of
  the \ac{entsoe}. The system includes
  $21,177$ buses ($1,212 $ off-line); $30,968$ transmission lines and
  transformers ($2,352$ off-line); $1,144$ zero-impedance connections
  ($420$ off-line); $4,828$ power plants represented by $6$-th order
  and $2$-nd order synchronous machine models; and $15,756$ loads
  ($364$ off-line), modeled as constant active and reactive power
  consumption.  Synchronous machines represented by $6$-th order
  models are also equipped with dynamic \ac{avr} and \ac{tg} models.
  The system also includes $364$ \acp{pss}.

  As summarized in Table \ref{tab:num_entsoe_stats}, 
  the system has in
  total $\nx=49,396$ state variables and $\ny=96,770$ algebraic
  variables.  The pencil $s\Esng - \Asng$ has dimensions
  $146,166 \times 146,166$ and the matrix $\Asng$ has $654,950$
  non-zero elements, which represent the $0.003\%$ of the total number
  of elements of the matrix.
\begin{table}[ht!]
\centering
\renewcommand{\arraystretch}{1.05}
  \caption[ENTSO-E: statistics]{ENTSO-E: statistics.}
  \vspace{1mm}
  \begin{tabular}{c|c}
    \hline
    $\nx$ & $49,396$ \\
    $\ny$ & $96,770$ \\
    Dimensions of $\Asng$ & $146,166 \times 146,166$ \\
    Sparsity degree of $\Asng$ [\%] & $99.997$ \\ 
    \hline
  \end{tabular}
\label{tab:num_entsoe_stats}
\end{table}

  Neither the \ac{lep} or \ac{gep} could be solved using Schur
  decomposition methods. \index{Schur decomposition methods} At this
  scale, the dense matrix representation required by LAPACK and
  MAGMA leads to massive memory requirements, and a
  segmentation fault error is returned by the CPU.
  Among the algorithms that support sparse matrices, only
  z-PARES is tested. This, in fact, was the library
  that was able to tackle this large problem on the available hardware.

  The effect of changing the
  search region of z-PARES' \ac{cirr} method on the eigenvalue
  analysis of the \ac{entsoe} is shown in Table
  \ref{tab:num_entsoe_region}.  Interestingly, simulations showed that
  shrinking the defined contour may lead to a marginal increase of the
  computation time. Although not intuitive, this result indicates that
  the mass of the computational burden is mainly determined by the
  large size of the \ac{entsoe}, and that, at this scale, smaller
  subspaces are not necessarily constructed faster by the \ac{cirr}
  algorithm. Regarding the number
  of eigenvalues obtained, using a region that is too small leads, as
  expected, to missing an important number of critical eigenvalues.
  \begin{table}[ht!]
    \centering
  \renewcommand{\arraystretch}{1.05}
      \caption[ENTSO-E: impact of the search region of the CIRR method]{ENTSO-E: impact of the search region of the CIRR method.}
      \vspace{1mm}
      \begin{tabular}{c|c|c|c}
      \hline
        Library & \multicolumn{3}{c}{z-PARES} \\
        Problem & \multicolumn{3}{c}{GEP}     \\
        Method  & \multicolumn{3}{c}{CIRR}        \\
        \hline
        $c$ & $(-0.01,4)$ &  $(-0.01,3)$&  $(-0.01,3)$ \\
        $\rho$ & $8$ & $4$ & $2$ \\
        \hline    
        Time [s] &  $364.85$ & $375.67$ & $378.71$ \\
        Found  & $349$ eigs. & $350$ eigs. & $110$ eigs.\\
      \hline
      \end{tabular}
    \label{tab:num_entsoe_region}
  \end{table}
 
\begin{table}[ht!]
  \centering
 \renewcommand{\arraystretch}{1.05}
    \caption[ENTSO-E: impact of spectral transforms of the CIRR method]{ENTSO-E: impact of spectral transforms of the CIRR method.}
    \vspace{1mm}
    \begin{tabular}{l|c|c|c}
      \hline
      Library & \multicolumn{3}{c}{z-PARES} \\
      Problem & \multicolumn{3}{c}{GEP}     \\
      Method  & \multicolumn{3}{c}{CIRR}        \\
      Spectrum & \multicolumn{3}{c}{$c=(-0.01,4), \rho=8$}  \\
      \hline    
      Transform & - & Invert & Inverted Cayley \index{Inverted Cayley transform} \\
      \hline
      Time [s] &  $364.85$ & $350.82$ & $337.43$ \\
      Found  & $349$ eigs. & $297$ eigs. & $349$ eigs. \\
      \hline
    \end{tabular}
  \label{tab:num_entsoe_st}
\end{table} 
  
  The impact of applying spectral transforms to the matrix pencil $s \Esng - \Asng$ is examined.
  In particular, two transforms are tested. The invert transform, which yields the dual
  pencil $z\Asng - \Esng$; and the inverted Cayley transform,
  i.e.~$s= (z+1)/(\sigma z - \sigma)$, which yields the pencil
  $z(\Esng-\sigma \Asng) - (-\sigma \Asng - \Esng)$.
The results are shown in Table
  \ref{tab:num_entsoe_st}.  Passing the transformed matrices to
  z-PARES provides a marginal speedup to the eigenvalue
  computation. In addition, considering either the prime system or
  the inverted Cayley transform with $\sigma=-1$, results in finding
  the same number of eigenvalues, whereas when the dual system is
  considered a number of eigenvalues is missed.

\section{Concluding Remarks}
\label{sssa:sec:conclusions}

This chapter provides fundamental concepts of power system \ac{sssa} and linear systems of differential equations that are employed throughout the thesis.

The chapter also provides a comprehensive comparison of state-of-art software implementations for the numerical solution of the eigenvalue problems that arise in power systems.  With this regard, the following discussion is relevant.

The main disadvantage of 
dense matrix methods
is that they are computationally
expensive.  In addition, they
generate complete fill-in in general sparse matrices and therefore, cannot be applied to large sparse matrices simply because of massive memory requirements. Even so, LAPACK is the most mature among all computer-based eigensolvers and, as opposed to basically all sparse solvers, requires practically no parameter tuning. For small to medium size problems, the QR algorithm with LAPACK remains the standard and most reliable algorithm for finding
the full spectrum for the conventional \ac{lep}.

%
As for sparse matrix methods,
convergence of vector iteration methods can be very slow, and thus in practice, if
not completely avoided, these algorithms
should be used only for the solution of simple eigenvalue problems. 
With regard to Krylov subspace methods, the main shortcoming of ARPACK's implementation is the lack of support for general, non-symmetric left-hand side coefficient matrices, which is the form that commonly appears when dealing with the \ac{gep} of large power system models. On the other hand, the implementations of
\ac{erd}-Arnoldi and Krylov-Schur by
SLEPc do not have this limitation and 
exploit parallelism while providing good accuracy, although some parameter tuning effort is required.
In addition, for the scale of the \ac{aiits} system and for the \ac{gep}, these methods appear to be by far more efficient than LAPACK.
Moreover, 
the implementation of contour integration by z-PARES is very efficient and can handle systems at the scale of the \ac{entsoe}. The most relevant issue for z-PARES is that, depending on the problem, it may miss some critical eigenvalues, despite the defined search contour being reasonable. Although there may be some parameter settings for 
which this problem does not occur, those cannot be known \textit{a priori}.
\newpage
\chapter{Participation Factors}
\label{ch:pf}

\section{Introduction}
 

Modal participation analysis was first introduced by P{\'e}rez-Arriaga
\textit{et al.} in \cite{arriaga:82_1} and \cite{arriaga:82_2}.  These
studies employed the analytical solution that determines the time
response of a linear time-invariant dynamic system and applied initial
conditions appropriate to define the relative contribution of a system
state in a dynamic mode and \textit{vice versa}.  \acp{pf} were introduced as an
approach to selective modal analysis.    They have
been also utilized in model reduction \cite{book:chow:13}, as well as
in control signal and input placement selection \cite{hsu:87}.
The
properties of \acp{pf} were summarized and extended in
\cite{garofal:O2}.  In \cite{abed:00, abed:09}, the authors
studied the effect of the uncertainty in the initial conditions in the
definition of the \acp{pf}.  
Nowadays, \acp{pf} are 
considered a fundamental tool for power system \ac{sssa}.
Recent efforts have focused on the
modal participation analysis of non-linear systems \cite{tian:18,mili:19}.

Dominant states in lightly damped modes of power systems are typically
the synchronous machine rotor angles and speeds.  The state variables
of poorly tuned controllers, e.g.~the \acp{avr} and \acp{pss}, can
also show high \acp{pf} in critical modes.  Nevertheless, measurement
units installed on the transmission system buses provide information
on the local voltage, frequency and active and reactive power flows,
which in angle and voltage stability studies are modeled as algebraic
variables \cite{milano:10}.  Moreover, these quantities are
typically utilized by \ac{facts} devices as
signals for the implementation of various controllers including
\ac{pod} \cite{abbsvc}.

This chapter provides a tool to study how algebraic variables are coupled
with power system dynamic modes.  
It is precisely recognized that the
algebraic variables of a set of \acp{dae} can
be interpreted as functions of the state variables and, in turn, as \textit{outputs} of the state-space representation
of the power system model.  Until now,
algebraic variables were mostly interpreted either as
\textit{constraints} and thus eliminated when calculating the state
matrix of the system; or as \textit{states with infinitely fast
dynamics} and, as such, their \acp{pf} to system modes were
considered to be null.
The focus is on the \acp{pf} of bus
voltages, frequencies, and power injections;
\ac{rocof} of synchronous machines; \ac{coi} speed of different areas; and any system
parameters.  However, the formulation provided in this chapter is general
and can be extended to any non-linear function of the system states and algebraic variables.


The remainder of the chapter is organized as follows.
Section~\ref{sec:pf} describes the
classical 
modal participation analysis of a power system model.
Section~\ref{sec:pfsing} 
provides 
a new interpretation of the \acp{pf} as eigen-sensitivities -- 
which is derived
from the partial differentiation of the analytical solution of the linearized power system around a valid equilibrium point --
and provides their formulation for a singular dynamical system with eigenvalue multiplicities. 
Based on this interpretation, the proposed approach
to measure the participation of algebraic variables and,
in general, of any function of the variables in power system
modes is presented in Section~\ref{sec:pfalg}.  The case studies are discussed in Section~\ref{pf:tdc:case}.  Finally, conclusions are
drawn in Section~\ref{pf:sec:conclusion}.


\section{Classical Participation Factors}
\label{sec:pf}

\subsection{Definition}

Consider the following system
of \acp{ode}:
\begin{equation}
\label{pf:eq:ode}
 \Dt \xs = \Asng \, \xs \, ,
\end{equation}
where ${\xs} \in \mathbb{R}^{\nd}$ is the vector of state variables and
$\Asng \in \mathbb{R}^{\nd \times \nd}$ is the state matrix. 
System \eqref{pf:eq:ode} can be obtained from \eqref{sssa:eq:sing}
for $\Esng \equiv \bfg I_{\nd}$
and by assuming that no inputs are included. 

Let $\lambda_i$ be an eigenvalue of
$s \bfg I_{\nd} - \Asng$ and all eigenvalues be distinct, i.e.~$\lambda_i \neq \lambda_j$, $i \neq j$, and
  $i,j = 1,2,\ldots,\nd$.
  Let also $\reigv_i$, $\leigv_i$ be the right and left eigenvectors associated to $\lambda_i$, respectively. 
The \ac{pf} 
is defined as the following
dimensionless number:
\begin{equation}
\label{pf:eq:definition}
\PF_{k,i}
= \leigvel_{i,k} \, \reigvel_{k,i} \, ,
\end{equation}
where $\reigvel_{k,i}$ is the $k$-th row element of
$\reigv_i$ and $\leigvel_{i,k}$ is the $k$-th column
element of $\leigv_i$.

The right and
left eigenvectors are usually normalized so that the sum of all \acp{pf} that
correspond to the same eigenvalue equals to $1$ \cite{arriaga:82_1}.  However, this is not always the case \cite{pal:16}. In addition,
the \acp{pf} of a system are typically collected to form the participation matrix
$\PFmat$, which is defined as follows:  
\begin{equation}
\label{pf:eq:PFmat}
\PFmat =  \leigvmat\T \circ \reigvmat \ , 
\end{equation}
where $\circ$ denotes the Hadamard product, i.e.~the element-wise
multiplication; and $\reigvmat$, $\leigvmat$,
are the right and left \textit{modal matrices},
respectively.
That is, the columns of $\reigvmat$ are the right eigenvectors $\reigv_i$ and the rows
of $\leigvmat$ are the left eigenvectors 
$\leigv_i$.

\subsection{Residues}

The \ac{pf} $\PF_{k,i}$ in 
\eqref{pf:eq:definition} basically expresses the relative contribution of $\x_k$ in the structure of the eigenvalue $\lambda_i$, and \textit{vice versa}, but has also 
various other interpretations.  It is also known to represent the sensitivity of an eigenvalue to
variations of an element of the state matrix \cite{arriaga:89} and it has been also viewed as modal energy in the MacFarlane sense
\cite{hamdan:86}.  

In the state space representation,
\acp{pf} can be studied as an important case of residue analysis of the system transfer function
and thus, as joint observabilities/controllabilities of the geometric
approach, which play an important role during the design of control
systems\cite{hamdan:87, garofal:O2}.
Consider the following single-input single-output system:
\begin{equation}
  \label{pf:eq:siso}
  \begin{aligned}
 \Dt \xs &=  \Asng  \, \xs + 
  \bfg b \, \text{u} \, , \\
  {w}& =  \bfg c \,  \xs  \, ,
  \end{aligned}
\end{equation}
where $\bfg b$ is the column vector of the input $\text{u}$; $\bfg c$ is
the row vector of the output $w$.  Then, the residue of the 
transfer function of system
\eqref{pf:eq:siso} associated with the eigenvalue $\lambda_i$
of the pencil 
$s \bfg I_{\nd} - \Asng$ is given by:
\begin{align}
\label{pf:eq:residue}
  \res_i &=  \bfg c \, 
  \reigv_i \, 
  \leigv_i \, \bfg b \, .
\end{align}
The \ac{pf} of the $k$-th state $\x_k$ in $\lambda_i$ can be
viewed as the residue of the transfer function of system \eqref{pf:eq:siso} 
associated with $\lambda_i$, when the input is a perturbation in
the differential equation that defines $\Dt {\x}_k$ and the
output is $\x_k$. Indeed, if
\begin{align}
  \bfg c &=
\begin{bmatrix}
c_{1} & \ldots & c_{k} & \ldots & c_{\nd} \\
\end{bmatrix}
=
\begin{bmatrix}
0 & \ldots & 1 & \ldots & 0 \\
\end{bmatrix}   ,
\nonumber \\
\bfg b\T &=
\begin{bmatrix}
b_{1} & \ldots & b_{k} & \ldots & b_{\nd} \\
\end{bmatrix}\T
=
\begin{bmatrix}
0 & \ldots & 1 & \ldots & 0 \\
\end{bmatrix}\T  ,
\nonumber
\end{align}
equation \eqref{pf:eq:residue} becomes:
\begin{align}
\label{pf:eq:respf}
  \res_i &=  \leigvel_{i,k} \, \reigvel_{k,i} = \PF_{k,i}  \ .
\end{align}
In the case of a multiple-input multiple-output system, the \acp{pf} appear as the diagonal elements of the
emerging residue matrix.  The ability to calculate only a subset of
all residue elements and acquire an approximate but yet accurate
measure of the contribution of system states in system modes (and
\textit{vice versa}), features the physical importance and the
computational efficiency of the \acp{pf}.

\section{Generalized Participation Factors}
\label{sec:pfsing}

\subsection{Formulation}

From the definition of \acp{pf} given in \eqref{pf:eq:definition}, it follows that the main assumptions of classical modal participation analysis are:
\begin{itemize}
    \item All eigenvalues are distinct.
    \item The system is modeled as a set of \acp{ode}, i.e. all eigenvalues are finite.
\end{itemize}

On the other hand, it is common
in the simulation of dynamic models that some eigenvalues are
repeated.  For small size systems, it 
may be possible to avoid multiplicities, e.g. by perturbation of some
parameters.  But this is impractical for
real-world size systems. 
Moreover, for large-scale systems, available algorithms 
are typically able to find a partial solution 
only of the \ac{gep} 
and provided that the matrices are sparse, which implies that 
the system is modeled as a set 
of singular differential equations.  Section~\ref{sssa:sec:numerical} further elaborates on this point.

This section
presents an alternative interpretation of the \acp{pf} as eigen-sensitivities.
In the view of addressing the issues mentioned above,
the focus is on the modal participation analysis of a singular system 
of differential equations with eigenvalue multiplicities.
The proposed approach requires the
solution of the
\ac{gep}, as opposed to the
conventional \ac{lep}, and thus fully exploits the sparsity of Jacobian matrices \cite{dual:17}.  This allows utilizing solvers for eigenvalue
analysis that scale well and are suitable for large real-world
systems. Finally, 
classical \acp{pf} are extracted from the provided formulation as a special case.


Consider system \eqref{sssa:eq:sing} without any inputs, i.e.:
\begin{equation}
\label{pf:eq:sing}
\Esng \, \Dt \xs(t)=\Asng \, \xs(t) \, ,
\end{equation}
where  $\Esng, \Asng \in \mathbb{R}^{\nd \times \nd}$, $\xs(t): \mathbb{R}^+\rightarrow\mathbb{R}^{\nd}$.
%
%
From Theorem \ref{sssa:theorem:sol},
and by applying $\Bsng = \bfg 0_{\nd,\nin}$ in \eqref{sssa:eq:sol}, the
analytical solution of system \eqref{pf:eq:sing} is:
\begin{equation}
\label{pf:eq:sol}
\xs (t)=
\reigvmat_\nf \, e^{\bfg J_\nf t} 
\, \bfb c
\, .
\end{equation}

In order to study the effects of eigenvalue multiplicities in 
\eqref{pf:eq:sing},
\eqref{sssa:eq:sol} has to be rewritten so that
the generalized eigenvectors
appear in the solution of the system explicitly \cite{pfactors2}. 

Firstly, let:
\begin{itemize}
\item $\njb$ be the number of Jordan blocks and $\hat \lambda_i \in \mathbb{C}$, $i=1,2,...,\njb$, be finite eigenvalue that corresponds to the $i$-th Jordan block and $\rjb_i$ be the rank of the block, where $\sum_{i =1}^\njb \rjb_i  = \nf$. 
\item  the infinite eigenvalue have algebraic multiplicity $\ninf$.
\end{itemize}

The following theorem is relevant.

\vspace{2cm}

{\theorem{
Consider system \eqref{pf:eq:sing} with pencil $s\Esng-\Asng$ and
${\rm det}( s \Esng - \Asng)\not\equiv 0$.
.
Let $\hat \lambda_i$, $i=1,2,...,\njb$, be a finite eigenvalue of the pencil, 
where $\njb$ is the number of Jordan blocks.
Let also 
$\rjb_i$ be rank of the corresponding Jordan block, $\sum_{i=1}^\njb \rjb_i=\nf$, $\njb \leq \nf$, and $\reigv_{i}^{[j]}$, $\leigv_{i}^{[j]}$, $j=1,2,...,\rjb_i$, 
denote the $j$-th
right, left, linear independent (generalized) eigenvectors corresponding to the eigenvalue $\hat \lambda_i$, respectively. 
Then \cite{tzou:pfactors}:
\begin{enumerate}[label=(\alph*)]
 \item The solution of \eqref{pf:eq:sing} with initial condition $\xs(0)$ can be written as follows:
\begin{equation}
\xs (t)=\sum_{i=1}^\njb e^{\hat \lambda_i t}\sum_{j=1}^{\rjb_i}\Big(\sum_{k=1}^j t^{k-1}
 \, \leigv_{i}^{[j-k+1]}
 \, \Esng \, \xs (0) \Big) \, \reigv_{i}^{[j]} \, .
\label{sssa:eq:sol3}
\end{equation}
%

\item Let $\x_k(t)$ be the $k$-th element of $\xs(t)$. Then the participation of $\hat \lambda_i$ in $\x_k(t)$, $k=1,2,...,\nd$, is given by:
\begin{equation}
\frac{\partial \x_k(t)}
{\partial e^{\hat \lambda_i t}}
=\sum_{j=1}^{\rjb_{i}}
\Big(
\sum_{\upsigma=1}^j t^{\upsigma-1}
\, 
\leigv_{i}^
{[j-\upsigma+1]} \, \Esng \, 
\xs(0) 
\Big) \, 
\reigvel_{k,i}^{[j]} 
\, ,
\quad
\text{(Participation Factor)}
\label{pf:eq:pfsing}
\end{equation}
where $\reigvel_{k,i}^{[j]}$ is the $k$-th row element of the eigenvector $\reigv_{i}^{[j]}$.
\end{enumerate}
}
\label{pf:theorem:pf}
}

The proof of Theorem~\ref{pf:theorem:pf} is given in Appendix~\ref{app:proofs}.

The following corollary is relevant.

{ \corollary {
Consider system \eqref{pf:eq:sing} with a regular pencil.
Let the finite eigenvalues be either distinct, or with algebraic multiplicity equal to geometric, i.e.~$\rjb_{i}=1$ is the rank of corresponding Jordan block.
Then in Theorem~\ref{pf:theorem:pf}, one has $\njb=\nf$, $\reigv_{i}^{[j]}=\reigv_i$, and:
%

%

\begin{enumerate}[label=(\alph*)]
 \item The solution of \eqref{pf:eq:sing} with initial condition $\xs(0)$ is given by: 
\[
\xs (t)=\sum_{i=1}^\nf \, \leigv_i 
\, \Esng \, \xs(0)\, \reigv_i \, e^{\lambda_it} \, .
\]
\item 
Let $\x_k(t)$ be the $k$-th element of  $\xs (t)$. Then the participation of the $i$-th eigenvalue, $i=1,2,...,\nf$, in $\x_k(t)$, $k=1,2,...,\nd$, is given by:
\begin{equation}
\label{pf:eq:pfs2}
 \frac
{\partial \x_k(t)}
{\partial e^{\lambda_i t}}
=
\bfb w_i \, \Esng \, \xs(0) \, \reigvel_{k,i} \, ,
\quad\text{(Participation Factor)}
\end{equation}
where $\reigvel_{k,i}$ is the $k$-th row element of the eigenvector $\reigv_i$.
\end{enumerate}
}
\label{pf:corollary:singpf}
}

The following remarks are relevant.

{\remark {
Since only the finite eigenvalues appear in \eqref{sssa:eq:sol3}, the
participation matrix of system \eqref{pf:eq:sing} has dimensions
$\nd \times \njb$. Determining the \acp{pf} associated with
the infinite eigenvalue to obtain the full matrix is possible by applying a
spectral transform to system \eqref{pf:eq:sing}.
In particular,
by applying  $z = 1/s$ into \eqref{pf:eq:sing}, one arrives at the dual system 
$\Asng \, \Dt {\hat \xs}= \Esng \, \hat \xs$. 
Let $\x_k(t)$ be the $k$-th element of $\xs (t)$, and $\hat{\x}_k(t)$ be the $k$-th element of  $\hat{\xs}(t)$. Then the participation of the infinite eigenvalue of $s\Esng- \Asng$ in $\x_k(t)$, $k=1,2,...,\nd$, is equal to the participation of the zero eigenvalue of $z\Asng-\Esng$ in $\hat{\x}_k(t)$, $k=1,2,...,\nd$. 
This is a direct result from the duality between \eqref{pf:eq:sing} and its dual system, or, additionally, between their pencils $s\Esng- \Asng$, and $z\Asng-\Esng$ respectively, see \cite{dual:17}.
Note that this discussion, although interesting from a mathematical viewpoint, is not of practical
interest in power system \ac{sssa}, since an
infinite eigenvalue does not represent any particularly meaningful dynamics.
}
\label{pf:remark:inf}
}

{\remark {
Applying appropriate initial conditions in \eqref{pf:eq:pfsing},
i.e.~$\x_k(0)=1$, and $\x_h(0)=0$, $h \neq k$, and imposing
$t \rightarrow 0$, 
allows obtaining the
\acp{pf} in the classical sense \cite{arriaga:82_1}.
Furthermore, 
as already discussed, 
algorithms that are suitable for large sparse matrices allow finding only a partial solution of
the \ac{gep}, typically 
including the most critical
dynamic modes of the system.
This solution
allows determining only the part
of the participation matrix that is associated with the most
critical modes. Therefore, 
by applying the above initial conditions,
the (critical)
participation matrix, i.e.~the part of the participation matrix that
is associated with the most critical eigenvalues of the system, can
be expressed as:
  \begin{equation}
    \label{pf:eq:Pc}
    \PFmatw{\kappa}
    =  \leigvmat_{\kappa}\T \circ 
    (\Esng \, \reigvmat_{\kappa})  \ , 
  \end{equation}
  where 
  $\kappa$, $\kappa \leq \njb$, be the number of the
calculated finite eigenvalues
  and
  $\reigvmat_{\kappa}$, 
  $\leigvmat_{\kappa}$ 
  are the corresponding right and left modal matrices.
}
\label{pf:remark:1}
}

\subsection{Illustrative Example}

As an illustrative example, consider system \eqref{pf:eq:sing} with  
\[ 
\Esng= 
  \begin{bmatrix}
12 & -3&  0&	0  & 0 \\
 4 &  1& -1&	3  & 0 \\
 0 & -4& -5&	1  & 0 \\
 8 &  2& -5&	9  & 0 \\
 0 &  0&  0&	0  & 0 \\
  \end{bmatrix} \, , 
  \, 
\Asng= 
  \begin{bmatrix}
-17 &  8 & -2 &  5 & 3 \\
 -7 & -3 &  3 & -8 & 1 \\
 13 &  9 &  9 &  3 & 1 \\
-12	& -7 & 13 &-22 & 0\\
  1 &  0 &  0 &  0 & 1 \\
  \end{bmatrix}.
\]

The pencil $s\Esng- \Asng$ has $\njb = 2$ finite eigenvalues $\hat\lambda_1=-2$, $\hat\lambda_2=-3$, of algebraic multiplicity $\nfi{1}=2$, $\nfi{2}=1$ and the infinite eigenvalue 
with multiplicity $\ninf=2$. The geometric multiplicity $\gamma_i$ of the finite eigenvalue $\hat \lambda_i$ is found as the dimension of the null space of $\hat \lambda_i\Esng-\Asng$. In this example, $\gamma_1=1$, $\gamma_2=1$.
The right and left eigenvectors of $s\Esng- \Asng$ associated with the finite eigenvalue $\hat\lambda_1 = - 2$ are:

\[
\footnotesize
\reigv_{1}^{[1]} =
\begin{bmatrix}
    0   \\
   -1   \\
   -1   \\
    0   \\
    0   \\
\end{bmatrix},
\
\reigv_{1}^{[2]} =
\begin{bmatrix}
0.0049     \\
-3.282 \cdot 10^7     \\
-3.282 \cdot 10^7     \\
    0   \\
 0.0049    \\
\end{bmatrix},
\
\leigv_{1}^{[1]} =
\begin{bmatrix}
-0.2308 \\
-0.3846 \\
 0.0769\\
 0\\
 1\\
\end{bmatrix}\T ,
\
\leigv_{1}^{[2]} =
\begin{bmatrix}
-0.1426 \\
-0.2376 \\
 0.0475 \\
 0\\
 0.6178\\
\end{bmatrix}\T ,
\]
where $\reigv_{1}^{[2]}$, $\leigv_{1}^{[2]}$ are generalized eigenvectors determined from $(\Asng-\hat\lambda_1 \Esng)\, \reigv_{1}^{[2]} = \Esng \, \reigv_{1}^{[1]}$ and 
$\leigv_{1}^{[2]}\, (\Asng-\hat\lambda_1 \Esng) =  \leigv_{1}^{[1]} \, \Esng$ respectively.
The right and left eigenvectors of $s\Esng- \Asng$ associated with the finite eigenvalue $\hat \lambda_2 = - 3$ are:
\[
\reigv_{2}^{[1]} =
\footnotesize
\begin{bmatrix}
    0   \\
    1   \\
   -0.5 \\
    0   \\
    0   \\
\end{bmatrix},
  \quad
\leigv_{2}^{[1]} =  
  \begin{bmatrix}
    -0.3333  \\
     1       \\
     0.1111  \\
     0       \\
    -0.1111  \\
  \end{bmatrix}\T \, .
\]
The sensitivities 
$\dfrac
{\partial \x_k(t)}
{\partial e^{\hat \lambda_i t}}$ are obtained from Theorem~\ref{pf:theorem:pf} as follows:
\[
 \frac
{\partial \x_k(t)}
{\partial e^{\hat \lambda_i t}}
=\sum_{j=1}^{\rjb_{i}}\Big(\sum_{k=1}^j t^{k-1} \, \leigv_{i}^{[j-(k-1)]} \,
\Esng \, \xs(0) \Big)
\reigvel_{k,i}^{[j]} \, .
\]
For $\hat \lambda_1$ and $\hat \lambda_2$ one has respectively:
\begin{align}
 \frac
{\partial \x_k(t)}
{\partial e^{\hat \lambda_1 t}}
&=\sum_{j=1}^{2}\Big(\sum_{k=1}^j t^{k-1}\leigv_{1}^{[j-(k-1)]} \Esng \xs(0) \Big) \reigvel_{k,1}^{[j]}
 \nonumber \\
&= \leigv_{1}^{[1\, ]}
\,  
\Esng \, \xs(0) \, \reigvel_{k,1}^{[1]}
+ \Big(\sum_{k=1}^2 t^{k-1} \,
\leigv_1^{[2-(k-1)]} \, \Esng \, \xs(0)
\Big) 
\reigvel_{k,1}^{[2]} \nonumber \\
&= \leigv_{1}^{[1]}\,\Esng \, \xs(0) \,\reigvel_{k,1}^{[1]}
+ \leigv_{1}^{[2]} \,\Esng \, \xs(0)\,\reigvel_{k,1}^{[2]} +
t \, \leigv_{1}^{[1]}\,\Esng \, \xs(0)\, \reigvel_{k,1}^{[2]}
\, ,
\nonumber \\
\nonumber \\
 \frac
{\partial \x_k(t)}
{\partial e^{\hat \lambda_2 t}}
&=\leigv_{2}^{[1]} \, \Esng \,  \xs(0) \, \reigvel_{k,2}^{[1]} \, . \nonumber 
\end{align}

Consider $\x_k(0)=1$, and $\x_i(0)=0$, $i\neq k$, which lead to the \acp{pf} related to the system finite modes. One has:

\begin{itemize}
    \item For $\dfrac{\partial \x_1(t)}
{\partial e^{\hat \lambda_i t}}$, one has
$
\xs(0) =
\begin{bmatrix}
    1&0&0&0&0    
\end{bmatrix}\T
$.
Hence,
\begin{align}
\PF_{1,1} &= 0.0130+0.0209t \ , \ 
\PF_{1,2} = 0  \ . \nonumber 
\end{align}

\item For $\dfrac{\partial \x_2(t)}
{\partial e^{\hat \lambda_i t}}$, one has
$
\xs(0) =
\begin{bmatrix}
    0&1&0&0&0    
\end{bmatrix}\T
$.
Hence,
\begin{align}
\PF_{2,1} &= 0.3290+1.0839t \ , \
\PF_{2,2} = 0.6667 \ . \nonumber 
\end{align}

\item For $\dfrac{\partial \x_3(t)}
{\partial e^{\hat\lambda_i t}}$, one has
$
\xs(0) =
\begin{bmatrix}
    0&0&1&0&0   
\end{bmatrix}\T
$.
Hence,
\begin{align}
\PF_{3,1} &= 0.6580+2.1678t \ , \
\PF_{3,2} =  0.3333 \ . 
\nonumber 
\end{align}

\item For $\dfrac{\partial \x_4(t)}
{\partial e^{\hat \lambda_i t}}$, one has
$
\xs(0) =
\begin{bmatrix}
    0&0&0&1&0   
\end{bmatrix}\T
$.
Hence,
\begin{align}
\PF_{4,1} &= 0 \, , \
\PF_{4,2} =  0  \, . \nonumber 
\end{align}

\item For $\dfrac{\partial \x_5(t)}
{\partial e^{\hat \lambda_i t}}$, one has
$
\xs(0) =
\begin{bmatrix}
    0&0&0&0&1   
\end{bmatrix}\T
$.
Hence,
\begin{align}
\PF_{5,1} &= 0 \ , \
\PF_{5,2} =  0 \ . \nonumber
\end{align}

\end{itemize}

Results are summarized in
Table~\ref{tab:pfs}, where $t \rightarrow 0$.
Since $\Esng$, $\Asng$ are $5 \times 5$ matrices and ${\rm rank}(s \Esng - \Asng)=3$, 
there exist $5-3=2$ variables the participation of which to the system finite eigenvalues is zero. These
variables are $\x_4$ and $\x_5$. 
Moreover, Table~\ref{tab:pfs} shows that $\x_3$ is dominant 
in $\hat \lambda_1$, while 
$\x_2$ is dominant in $\hat \lambda_2$.

\begin{table}[!ht]
\centering
\renewcommand{\arraystretch}{1.05}
\caption[Illustrative example: PFs associated to finite modes]{Illustrative example: PFs associated to finite modes.}
\begin{tabular}{c| c| c}
   & $\hat \lambda_1$ & $\hat \lambda_2$ \\
  \hline
  $\x_1$ & 0.0130 & 0 \\
  $\x_2$ & 0.3290 & 0.6667 \\
  $\x_3$ & 0.6580 & 0.3333 \\
  $\x_4$ & 0      & 0   \\
  $\x_5$ & 0      & 0   \\
  \hline
\end{tabular}
\label{tab:pfs}
\end{table} 

\section{Participation Factors of Algebraic Variables}
\label{sec:pfalg}



Consider the explicit \ac{dae} power system model \eqref{sssa:eq:exdaelin} without inputs, i.e.~$\Bdae = \bfg 0_{(\nx+\ny), \nin}$. Let $\xy_k(t)$ be the $k$-th element of $\xys(t)$. Then, the following cases are relevant:

\begin{itemize}
\item $k \leq \nx$, that is $\xy_k$ is a state variable.
Then, substitution of appropriate initial conditions
(see Remark~\ref{pf:remark:1})
gives $\leigv^{[1]}_i \Edae \xys(0)= \leigvel_{i,k}^{[1]}$.
In the special case that the eigenvectors form a complete basis for
the rational vector space of the matrix pencil, which means that all eigenvalues
are either distinct or their algebraic multiplicity is equal with the geometric, one has $\nfi{i}=1$, $\nf=n$, and thus
$\reigvel^{[1]}_{k,i}=\reigvel_{k,i}$, $\leigvel^{[1]}_{i,k}=
\leigvel_{i,k}$ in
\eqref{pf:eq:pfg}.
Substitution in \eqref{pf:eq:pfsing} gives:
\begin{align}
\frac{\partial \xy_k(t)}{\partial e^{\lambda_it}}
=  \leigv^{[1]}_i 
\Edae \, \xys(0)\, \reigvel^{[1]}_{k,i} 
=\leigvel^{[1]}_{i,k} \,
 \reigvel^{[1]}_{k,i} \
=\PF_{k,i} \, .
\label{pf:eq:pfg}
\end{align} 
%
\item 
$\nx < k \leq \nx+\ny$, i.e.~$\xy_k$ is
an algebraic variable.  Then 
  $\leigv_i^{[1]} \, \Edae \, \xys(0)=
  0$. 
  The $\ny$ rightmost columns of $\Edae$ which contain only zero
  elements, impose that the \acp{pf} of the algebraic variables in the
  system finite modes are 
  found to be 
  null. 
  This is a consequence of the fact that
  the coefficients of the first derivatives of the algebraic variables
  are zero, which implies that the algebraic variables introduce only
  infinite eigenvalues to the system.  
  Nevertheless, the algebraic variables
  constrain the system and, in this sense, 
  do participate in the
  system finite modes. 
\end{itemize}

It follows that, for this system, the critical 
participation matrix 
\eqref{pf:eq:Pc} takes 
the following form:
%
\begin{equation}
\label{pf:eq:Pc2}
\PFmatw{\kappa} =  {\leigvmat\T_\kappa} \circ 
(\Esng \, \reigvmat_\kappa) 
= \begin{bmatrix}
  \PFmatw{\bfg x}      \\
  \bfg 0_{\ny,\kappa}\\
\end{bmatrix}  \, , 
\end{equation}
where 
$\PFmatw{\x}  \subset \PFmat$
,
$\PFmatw{\x}  \in \mathbb{C}^{(\nx+\ny) \times \kappa}$. 
The matrix $\PFmatw{\x} $
contains all the information on the dynamics of interest and is the
matrix that is utilized in the remainder of the chapter.  

This section introduces an approach to measure the participation
of algebraic variables in power system dynamic modes, based on the \acp{pf} of the system states \cite{tzou:pfactors}.  These can be algebraic
variables included in the \ac{dae} system model, or, in general, any
algebraic output that is defined as a function of the states and
algebraic variables of the \ac{dae} system.

Let the output vector $\bfg w$, $ \bfg w \in \mathbb{R}^{\nout}$,
be defined as:
\begin{align}
  {\bfg w}&=\bfg h (\bfg x , \bfg y) \, 
  \nonumber,
\end{align}
where
$ \bfg h \, ( \bfg h : \mathbb{R}^{\nx+\ny} \rightarrow \mathbb{R}^\nout)$ is
a non-linear function of $\bfg x$, $\bfg y$.  Then differentiation
around $(\bfg x_o, \bfg y_o)$ yields:
\begin{equation}
  \label{pf:eq:outg}
  {\Delta \bfg w}=
\jacs{h}{x} \, \Delta \bfg x + 
\jacs{h}{y} \, \Delta  \bfg y \, .
\end{equation}
Substitution of \eqref{sssa:eq:algeb} to the last equation gives:
\begin{align}
  \label{pf:eq:out}
  {\Delta \bfg w} &=  \bfg C \, \Delta  \bfg x  \, ,
\end{align}
where $\bfg C = \jacs{h}{x} - \jacs{h}{y} \, \jacsinv{g}{y} \, \jacs{g}{x} $,
$\bfg C \in \mathbb{R}^{\nout \times \nx}$, is the output matrix.

Let ${\Delta w_{\varrho}}$ be the $\varrho$-th system output. Then, the
following expression is a candidate as the \ac{pf} of ${\Delta w_{\varrho}}$ in the mode
$\lambda_i$:
\begin{align}
  \label{pf:eq:pi}
  {\hat \PF_{\varrho, i}}&=\frac{\partial \Delta w_\varrho}{\partial e^{\lambda_i t}} \ .
\end{align}

From the state-space viewpoint, $\hat \PF_{\varrho, i}$ expresses the residue
 (or the joint observability/controllability)  
of the $i$-th mode, when the input is,
exactly as it holds for $\PF_{k, i}$, a perturbation in the differential
equation that defines $\Delta {\Dt x}_k$.  
The output however is
${\Delta w_{\varrho}}$, which can be, in principle, any function of the
system state variables.  The fact that the perturbation that leads from
\eqref{pf:eq:pfsing} and \eqref{pf:eq:pi} to the classical \acp{pf} is the same, is also the reason that
$\hat \PF_{\varrho, i}$ is called \ac{pf}.

{\proposition
{
Let the \ac{pf} ${\hat \PF_{\varrho, i}}$ be the
$\varrho$-th row, $i$-th column element of the participation matrix $\hPFmatw{\hat}{(\bfg  w)}$. Then:
\begin{align}
\label{pf:eq:Pi}
 \hPFmatw{\hat}{(\bfg  w)} &=\bfg C \, \PFmatw{\bfg x}  \, .
\end{align}
}
}

\noindent
\textit{Proof.}  Let $\bfg c_{\varrho}=
\begin{bmatrix}
c_{\varrho, 1} & \ldots &c_{\varrho, \nx} \\
\end{bmatrix}
$
be the $\varrho$-th row of $\bfg C$. 
Then:
\begin{align}
\label{pf:eq:a}
\Delta w_\varrho &= \bfg{c_\varrho}\, \Delta \bfg x = c_{\varrho, 1}\, \Delta x_{1} + c_{\varrho, 2}\, \Delta x_{2} + \cdots + c_{\varrho,\nx} \, \Delta x_{\nx} \ . \nonumber
\end{align}
Partial differentiation over $e^{\lambda_i t}$ leads to:
\begin{align}
   \frac{\partial \Delta w_\varrho}{\partial e^{\lambda_i t}} &= 
   c_{\varrho, 1} \, \frac{\partial \Delta x_1}{\partial e^{\lambda_i t}} + c_{\varrho, 2} \, \frac{\partial \Delta x_2}{\partial e^{\lambda_i t}} + \cdots + c_{\varrho, \nx} \, \frac{\partial \Delta x_\nx}{\partial e^{\lambda_i t}}+ \nonumber \\
   &+\frac{\partial c_{\varrho, 1}}{\partial e^{\lambda_i t}}\,\Delta x_{1} + \frac{\partial c_{\varrho, 2}}{\partial e^{\lambda_i t}}\,\Delta x_{2} + \cdots + \frac{\partial c_{\varrho, \nx}}{\partial e^{\lambda_i t}} \, \Delta x_{\nx} \nonumber \\
  \Rightarrow 
  \hat \PF_{\varrho, i}&= c_{\varrho, 1}\, \PF_{1,i} + c_{\varrho, 2} \, \PF_{2,i} + \cdots + c_{\varrho, \nx}\, \PF_{n,i} \ , \nonumber
\end{align}
where $\dfrac{\partial c_{\varrho, 1}}{\partial e^{\lambda_i t}} = \dfrac{\partial c_{\varrho, 2}}{\partial e^{\lambda_i t}} =\ldots = 
\dfrac{\partial c_{\varrho, n}}{\partial e^{\lambda_i t}} = 0
$, since the elements of $\bfg C$ do not depend on functions of $t$. By applying the same steps for 
all outputs and representing in matrix form, one arrives at \eqref{pf:eq:Pi}.
\hfill
\eop

The main feature of \eqref{pf:eq:Pi} is that it allows defining the
participation matrix not only of the algebraic variables of the
\acp{dae}, but also of any defined output vector that is a function of
the system state and algebraic variables. One has only to specify the
gradients $\jacs{h}{x}$ and $\jacs{h}{y}$ at the operating point, and then
calculate the output matrix $\bfg C$.
The proposed participation matrix $\hPFmatw{\hat}{(\bfg  w)}$ provides meaningful information 
for the system coupling that, to the best of the authors' knowledge, 
has not been exploited in the literature.


{\remark
{
The following special cases for the participation matrix of \eqref{pf:eq:Pi} are relevant:
\begin{enumerate}[label=(\alph*)]

\item \textit{State variables:} If $\bfg w = \bfg x$, the gradients in
  \eqref{pf:eq:outg} become $\jacs{h}{x} = \bfg I_{\nx}$, $\jacs{h}{y} = \bfg 0_{\nout,\ny}$.
  The output matrix is $\bfg C =\bfg I_{\nx}$ and hence the participation
  matrix of the system states is, as expected:
\begin{align}
\hPFmatw{\hat}{(\bfg  x)} &=\PFmatw{\bfg x}  \, .
\end{align}

\item \textit{Algebraic variables:} 
If $\bfg w = \bfg y$, the
gradients in \eqref{pf:eq:outg} become $\jacs{h}{x} = \bfg 0_{\nout,\nx}$,
$\jacs{h}{y} = \bfg I_\ny$.  The output matrix is
$\bfg C =- \jacsinv{g}{y} \, \jacs{g}{x}$.  Thus:
\begin{align}
\label{pf:eq:Piy}
\hPFmatw{\hat}{(\bfg  y)} &=- \jacsinv{g}{y} \, \jacs{g}{x} \, \PFmatw{\bfg x}  \, ,
\end{align}
which is the participation matrix of the algebraic variables in
system modes included in the \ac{dae} model.

\item \textit{Rates of change of state variables:} If the output is defined as $\bfg w = \Dt {\bfg x} = \bfg f(\bfg x, \bfg y)$, the gradients in \eqref{pf:eq:outg} become $\jacs{h}{x} = \jacs{f}{x}$,
  $\jacs{h}{y} = \jacs{f}{y}$. The output matrix is $\bfg C =\bfg A$. Thus:
\begin{align}
\label{pf:eq:Pid}
\hPFmatw{\hat}{(\Dt {\bfg  x})} &= \bfg A \, \PFmatw{\bfg x} \, .
\end{align}

The \ac{rocof} of the
  synchronous machines 
  ($\Dt {\bfg \omega}_{\rm r} $) is a relevant
  case.

\item \textit{Parameters:} Finally, consider the scalar output
  $w = {\eta}$, where ${\eta}$ is a parameter. If ${\eta}$
  appears only in the $j$-th algebraic equation
  $0 = g^j(\bfg x, \bfg y, \eta)$, then  
 the linearization
  of the $j$-th algebraic equation around the operating point yields:
  \begin{align}
    \label{pf:eq:par}
    0 &= \jacs{g^j}{x} \, \Delta {\bfg x}+
        \jacs{g^j}{y}  \, \Delta {\bfg y} +\jac{g^j}{\eta}  \, \Delta {\eta}\ ,
  \end{align}
  where $\jacs{g^j}{x} \in \mathbb{R}^{1 \times \nx}$,
  $\jacs{g^j}{y} \in \mathbb{R}^{1 \times \ny}$ and
  $\jac{g^j}{\eta} \in \mathbb{R}_{\ne 0}$. Solving \eqref{pf:eq:par} for
  $\Delta {\eta}$ and comparing with \eqref{pf:eq:outg}, 
  $\jacs{h}{x} = -\jacs{g^j}{x} \big / \jac{g^j}{\eta}$ and
  $\jacs{h}{y} = -\jacs{g^j}{y} \big / \jac{g^j}{\eta}$.  The participation vector is
  obtained from \eqref{pf:eq:Pi} for
  $\bfg C = (-\jacs{g^j}{x} + \jacs{g^j}{y} \,
  \jacsinv{g}{y} \,
  \jacs{g}{x}) \big / \jac{g^j}{\eta}$.

\end{enumerate}

}
}

Note, finally, that once the eigenvalue analysis is completed and
the modal matrices are known, calculating the proposed participation
matrices involves few matrix multiplications. From the computational
burden viewpoint, the cost of calculating the \acp{pf} is marginal compared
to the eigenvalue analysis.

\section{Case Studies}
\label{pf:tdc:case}

This case study presents two practical applications of the
proposed approach and shows how defining \acp{pf} of algebraic
variables in system modes can help design more effective and robust
controllers.  In particular, Section \ref{pf:sec:kundur} is based on the
well-known two-area system \cite{kundur:94} and shows how the
calculation of \acp{pf} can help select the most effective algebraic
variable to be measured to damp inter-area oscillations.  Section
\ref{pf:sec:irish} utilizes a realistic detailed model of the \ac{aiits} and shows how \acp{pf} can help define the impact
of a given system mode on the network.  This second case study also
serves to discuss the robustness and the scalability of the proposed approach.
Simulations of this section are carried using Dome \cite{vancouver}.


\subsection{Two-Area System}
\label{pf:sec:kundur}

\begin{figure}[ht]
  \begin{center}
    \resizebox{0.9\linewidth}{!}{\includegraphics{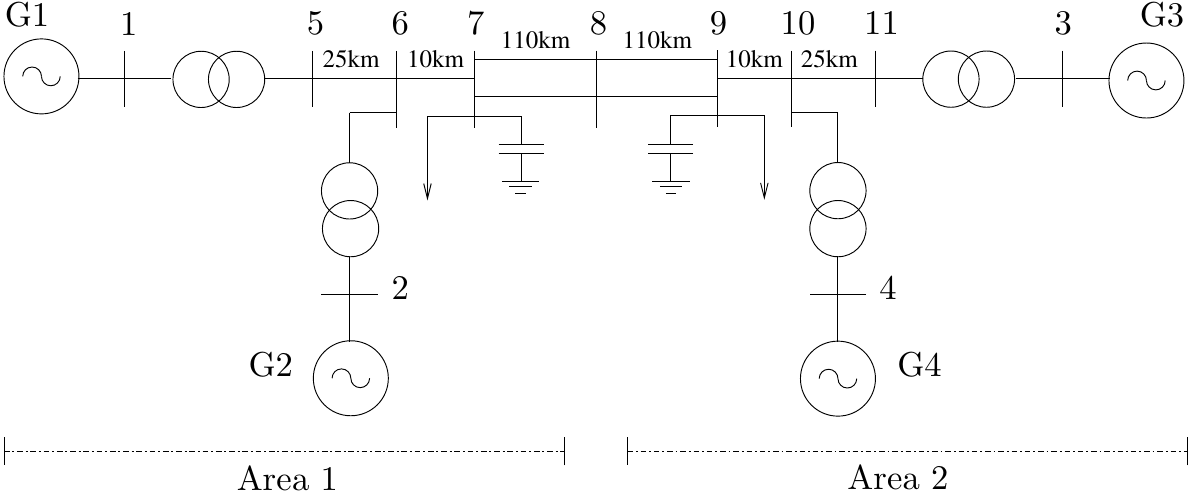}}
    \caption[Two-area four-machine system: single-line diagram]{Two-area four-machine system: single-line diagram.}
    \label{fig:kundur}
  \end{center}
\end{figure}

The two-area system is depicted in Figure~\ref{fig:kundur}.
It comprises two areas connected through a relatively weak
tie; eleven buses and four synchronous machines.
Each generator is equipped with an \ac{avr} of type IEEE DC-1 and a
\ac{tg}. The system feeds two loads connected to buses
7 and 9 and which are modeled as constant active
and reactive power consumption.

The system model has $52$ state variables.  For a system with this
dynamic order, the dense state matrix
$\AS$ can be efficiently calculated and handled.  The most critical modes and the mostly participating
states to these modes are presented in 
Table~\ref{tab:modes}. Area 1
presents a critical local mode $-0.599 \pm \jj 6.604$ with natural
frequency $1.06$ Hz and dominant state the rotor speed $\wGi{2}$.
Area 2 presents a critical local mode as well, which is
$-0.514 \pm \jj6.843$ with natural frequency $1.09$ Hz and dominant
state the rotor speed $\wGi{4}$.  For these modes, the damping
ratio is $> 5\%$.  Finally, the most lightly damped mode is
$-0.096 \pm \jj3.581$, which is an inter-area mode with natural
frequency $0.57$ Hz.  The mostly participating state in the inter-area
mode is the rotor speed $\wGi{3}$.

\begin{table}[ht]
  \centering
  \renewcommand{\arraystretch}{1.05}
  \caption[Two-area system: critical modes]{Two-area system: critical modes.}
  \begin{tabular}{ccccc}
    \hline Mode & ${\rm f}_{\rm n}$ (Hz) & $\zeta$ (\%) & $x$-dom. & $|\PF|_{\rm max}$\\ \hline
    $-0.096 \pm \jj3.581$ & $0.57$ & $2.67$ &$\wGi{3}$ &0.1696\\ 
    $-0.514 \pm \jj6.843$ & $1.09$ & $7.50$ &$\wGi{4}$ &0.2945\\ 
    $-0.599 \pm \jj6.604$ & $1.06$ & $9.04$ &$\wGi{2}$ &0.2530\\ \hline
  \end{tabular}
  \label{tab:modes}
\end{table}

The participation matrix of the algebraic variables for these modes is
calculated from \eqref{pf:eq:Piy}.  
Note that in this section, each
${\hat \PF_{\varrho, i}}$ is divided over the Euclidean norm of the respective
output $\bfg{c}_\varrho$, so that the results are normalized and
comparable according to the geometric approach.  Of course, since the \acp{pf} are a relative measure, one may apply any further normalization, e.g. the maximum or the sum of the values to be equal to 1.  

The following simple test shows how the proposed \acp{pf} of the
algebraic variables are linked to their sensitivities in eigenvalue
changes. Imposing a perturbation in the active power and voltage of
the PV buses 1, 4, changes the most critical mode
$-0.096 \pm \jj3.581$ by $|d \lambda|= 3 \cdot 10^{-5}$. The
calculated eigen-sensitivities ${|d \Delta y_k|} \big / {|d \lambda|}$ are then
compared with the \acp{pf} of the algebraic variables $P_{1}$,
$P_{4}$, $v_{1}$, $v_{4}$, in Table~\ref{tab:sens1_app1}.  As
expected, a highly participating variable in a mode indicates that
this mode is sensitive to small variations of this variable.

\begin{table}[!ht]
\centering
\renewcommand{\arraystretch}{1.05}
\caption[Sensitivity test for the inter-area mode]{Sensitivity test for the inter-area mode, $|d \lambda|= 3 \cdot 10^{-5}$.}
\begin{tabular}{c|c|c|c}
  \hline
$y_k$ (pu) & $|\hat \PF_{k,i}|$  & 
$|d\Delta y_k|$ (pu) & ${|d \Delta y_k|}/{|d \lambda|}$ \\
\hline
  $P_{1}=5.88 $ & 0.3642 & $0.43 \cdot 10^{-3}$ & 14.42 \\
  $P_{4}=7.00$  & 0.9766 & $0.54 \cdot 10^{-3}$ & 18.00 \\
  $v_{1}=1.03$  & 0.0036 & $0.27 \cdot 10^{-4}$ &  0.90 \\
  $v_{4}=1.01$  & 0.0028 & $0.11 \cdot 10^{-4}$ &  0.35 \\
  \hline
\end{tabular}
\label{tab:sens1_app1}
\end{table}


For illustration, consider now the simple example of finding the
participation vector of one system parameter.  Let $P_{7}$ be the
active power consumption of the load connected to bus 7.  Since
$P_{7}$ is also the active power injection at bus 7, 
the following algebraic equation (see Figure~\ref{fig:kundur}) holds:
\begin{align}
  0 &=  v_7v_6(G_{76}\cos(\theta_7-\theta_6)+B_{76}\sin(\theta_7-\theta_6))  \nonumber \\
    &+ v_7 v_8(G_{78}\cos(\theta_7-\theta_8)+B_{78}\sin(\theta_7-\theta_8)) - P_7 \nonumber \\
    &= g(v_{6},v_{7},v_{8},\theta_{6},\theta_{7},\theta_{8}, P_7) 
 \nonumber \ .
\end{align}
Linearization and solving for $\Delta P_7$ yields:
\begin{align}
  \Delta P_7 &= \big(\frac{\partial g}{\partial v_6} 
               \Delta v_6 + \frac{\partial g}{\partial v_7}
               \Delta v_7 + \frac{\partial g}{\partial v_8}  
               \Delta v_8 + \frac{\partial g}{\partial \theta_6} 
               \Delta \theta_6 
             + \frac{\partial g}{\partial \theta_7} 
               \Delta \theta_7 + \frac{\partial g}{\partial \theta_8} 
               \Delta \theta_8 \big) \ \nonumber \, ,
\end{align}
where, the gradients are calculated at
$(v_{6,o},v_{7,o},v_{8,o},\theta_{6,o},\theta_{7,o},\theta_{8,o})$;
and ${\partial g}\big /{\partial P_7}=-1$.  Therefore, $\jacs{h}{x} = \bfg 0_{\nout,\nx}$, and $\jacs{h}{y}$ is the $1 \times \ny$ row vector
which contains the gradients calculated above in the indices of
$v_{6},v_{7},v_{8},\theta_{6},\theta_{7},\theta_{8}$; all other
elements of $\jacs{h}{y}$ are zero.  The output matrix $\bfg C$ is
$\bfg C = - \jacs{h}{y} \, \jacsinv{g}{y} \,
\jacs{g}{x} $,
$\bfg C \in \mathbb{R}^{1 \times \nx}$. The resulting participation
matrix is given by \eqref{pf:eq:Pi}.
 
\begin{table}[ht]
\renewcommand{\arraystretch}{1.05}
\centering
\caption[Two-area system: PFs]{Two-area system: PFs.}
\label{tab:all}
\begin{tabular}{l|c|c|c|c|c|c}
\hline
Mode & \multicolumn{2}{c|}{$-0.096 \pm \jj3.581$} & \multicolumn{2}{c|}{$-0.514 \pm \jj6.843$} & 
\multicolumn{2}{c}{$-0.599 \pm \jj6.604$ }\\
\hline
\hline
Output & Dom. & $|\hat \PF|$ & Dom. & $|\hat \PF|$ & Dom. & $|\hat \PF|$ \\
\hline 
$\bfg v_{\rm B}$    & $v_{11}$     & 0.0192 & $v_{8}$  & 0.0375  & $v_{7}$  & 0.0345 \\
$\bfg \theta_{\rm B}$ &$\theta_{8}$ & 0.1429 & $\theta_{4}$& 0.2385 
& $\theta_{6}$ & 0.2250 \\
$\bfg \omega_{\rm B}$ & $\omega_{8}$ & 0.2065 & $\omega_{10}$ & 0.3247  & $\omega_{6}$ & 0.3113 \\
$\bfg P_{\rm B}$ & $P_{6}$ & 0.1447 & $P_{10}$ & 0.2518 & $P_{6}$ & 0.2719  \\
$\bfg Q_{\rm B}$ & $Q_{11}$ & 0.0258 & $Q_{8}$ & 0.0544 & $Q_{10}$ & 0.0631 \\
$\Dt {\bfg \omega}_{\rm r}$  & 
$\Dt {\omega}_{{\rm r},4}$ & 0.0401 & $\wGi{4}'$ & 0.0917 & $\Dt {\omega}_{{\rm r},2}$ & 0.0539 \\
$\bfg \omega_{\rm CoI}$ & $\omega_{{\rm CoI},2}$ & 0.1700 & $\omega_{{\rm CoI},2}$ & 0.3151 & $\omega_{{\rm CoI},1}$ & 0.3137 \\
\hline            
\end{tabular}
\end{table}

The active ($\bfg P_{\rm B}$) and reactive ($\bfg Q_{\rm B}$) power
injections on all system buses, as well as the \ac{coi} speeds
($\bfg \omega_{\rm CoI}$) of the two areas are defined as outputs and
their \acp{pf} are obtained from \eqref{pf:eq:Pi}.  Correspondingly, the
system bus voltages ($\bfg v_{\rm B}$), angles ($\bfg \theta_{\rm B}$)
and frequencies ($\bfg \omega_{\rm B}$) are included in the algebraic
variables of the \acp{dae}.  Thus, their \acp{pf} are determined from
\eqref{pf:eq:Piy}.  With this aim, ideal frequency estimations of the
system buses are obtained by employing the frequency divider formula, proposed in
\cite{fdf}.  
The formulation of the frequency divider in per units is as follows:
\begin{equation}
  \label{eq:fdf}
  \textbf{B}_{\rm BB} \, \Delta \bfg \omega_{\rm B} = 
  - \textbf{B}_{\rm BG} \,  \Delta \bfg \omega_{\rm r} \, , \nonumber
\end{equation}  
where $\Delta \bfg \omega_{\rm B}$ are the estimated bus frequency
deviations with respect to the reference synchronous speed;
$\Delta \bfg \omega_{\rm r}$ are the synchronous machines rotor speed
deviations; and $\textbf{B}_{\rm BB}$, $\textbf{B}_{\rm BG}$ are system
susceptance matrices that include the internal reactances of the
synchronous machines.  
The accuracy, the numerical
robustness and the computational efficiency of the frequency divider have been discussed in \cite{fdf,pll:18,freqpss:gm}.

The \acp{pf} of the \ac{rocof} of the
synchronous machines ($\Dt{\bfg \omega}_{\rm r}$) are determined
from \eqref{pf:eq:Pid}.
The mostly participating of the above variables in the system critical
modes are summarized in Table~\ref{tab:all}. It is worth observing that the bus
voltages, the reactive power injections and the \ac{rocof} have a low
participation in the system critical modes.  Mostly participating in
the inter-area mode is the bus frequency $\omega_8$.  Similarly, the
bus frequency $\omega_{10}$ is the one mostly participating in the
local mode of Area 2.  Finally, the \ac{coi} speed of Area 1
($\omega_{{\rm CoI},1}$) is the one mostly participating in
$-0.599 \pm \jj6.604$, which is a local mode of this area.

Finally, the calculated \acp{pf} can be utilized to
improve the dynamic behavior of the system.  As already discussed,
the critical mode of the system is the inter-area mode and the mostly
participating variable (Table~\ref{tab:all}) is the bus frequency
$\omega_8$.  A \ac{svc} is installed at bus 8 with a \ac{pod} loop
\cite{abbsvc}.  The \ac{pod} input signal is $\omega_8$.  The \ac{pod}
output is considered as an additional input to the \ac{svc} voltage
reference algebraic equation.  The results are summarized in
Table~\ref{tab:svc}. The eigenvalue analysis shows that, after the
inclusion of the controller, the system is stable and all modes are properly damped.

\begin{table}[ht]
\centering
\renewcommand{\arraystretch}{1.05}
\caption[Impact of SVC-POD 
installation in the critical mode]{Impact of SVC-POD 
installation in the critical mode.}
\begin{tabular}{lcccc}
    \hline  
\ac{svc}-\ac{pod} & Mode & $\zeta$ (\%) \\ \hline
No &$-0.096 \pm \jj3.581$ & $2.67$ \\ 
Yes &$-0.256 \pm \jj3.562$ & $7.16$ \\ 
\hline
  \end{tabular}
  \label{tab:svc}
\end{table}
%

\subsection{All-Island Irish Transmission System}
\label{pf:sec:irish}

This section considers the real-world model of the \ac{aiits} which has been also 
discussed in Section~\ref{sssa:sec:aiits}.
%
%
The dynamic order of the system is 1,480. Eigenvalue analysis
shows that the system is stable when subject to small disturbances.
The system presents both local machine modes and inter-machine modes.
Recall that, a local machine mode refers to a single machine
oscillating against the rest of the system.  On the other hand, an
inter-machine mode refers to a group of machines of the same area
oscillating against each other \cite{kundur:94}.  The remainder of
this section shows two modes with different damping ratios and
natural frequencies.  The examined modes are summarized in
Table~\ref{tab:modesir}.

\begin{table}
 \renewcommand{\arraystretch}{1.05}
 \centering
 \caption[AIITS: examined modes]{AIITS: examined modes.}
 \label{tab:modesir}
  \begin{tabular}{c|c|c|c|c}
 \hline
    Mode Name & \multicolumn{2}{c|}{Mode 1} & \multicolumn{2}{c}{Mode 2} \\
    \hline
    Eigenvalue & \multicolumn{2}{c|}{$-0.586 \pm \jj7.248$} & \multicolumn{2}{c}{$-0.722 \pm \jj4.618$} \\
    \hline
    ${\rm f}_{\rm n}$ (Hz) & \multicolumn{2}{c|}{$1.16$} & \multicolumn{2}{c}{$0.74$} \\
    \hline
    $\zeta$ (\%) & \multicolumn{2}{c|}{$8.06$} & \multicolumn{2}{c}{$15.44$} \\
    \hline 
    Type & \multicolumn{2}{c|}{Local} & \multicolumn{2}{c}{Inter-machine} \\
    \hline
    Dominant States & State & $|\PF|_{\rm max}$ & State & $|\PF|_{\rm max}$ \\
    \hline 
    1-st   & $\dGi{16}$& 0.4456 & $\wGi{2}$& 0.2883   \\
    2-nd & $\wGi{16}$ & 0.4456 & $\dGi{2}$  & 0.2872
    \\
    \hline            
  \end{tabular}
\end{table}

Mode 1 has eigenvalue $-0.586 \pm \jj7.248$, with natural frequency
$1.16$ Hz and damping ratio $8.06\%$.  The dominant states in this
mode are the rotor angle and speed of synchronous generator 16.  The \acp{pf} of these states sum to $0.8912$.  The mode
is local with generator 16 oscillating against the rest of the system.
Mode 2 has eigenvalue $-0.722 \pm \jj4.618$, with frequency $0.74$ Hz
and damping ratio $15.44$~\%.  The mostly participating states are the
rotor speed and angle of  synchronous generator 2.  The
corresponding \acp{pf} sum to $0.5755$.  The natural frequency and the
distribution of the \acp{pf} indicate that this is an inter-machine mode
\cite{kundur:94}.

The Python module \textit{graph-tool} \cite{graphtool:14} is utilized
to generate a graph of the studied network.  The resulting graph has
1,479 vertices, which correspond to the system buses and 1,851 edges,
which correspond to lines and transformers.  Note that the coordinates
of the graph vertices and edges do not represent the actual geography
of the system. The
participation matrices of the bus active power injections are calculated for the examined modes. Then, the
sizes and the colors of the graph vertices are adjusted with respect to the magnitude of the calculated \acp{pf}.

\begin{figure}
    \centering
    \resizebox{0.6\linewidth}{!}{\includegraphics{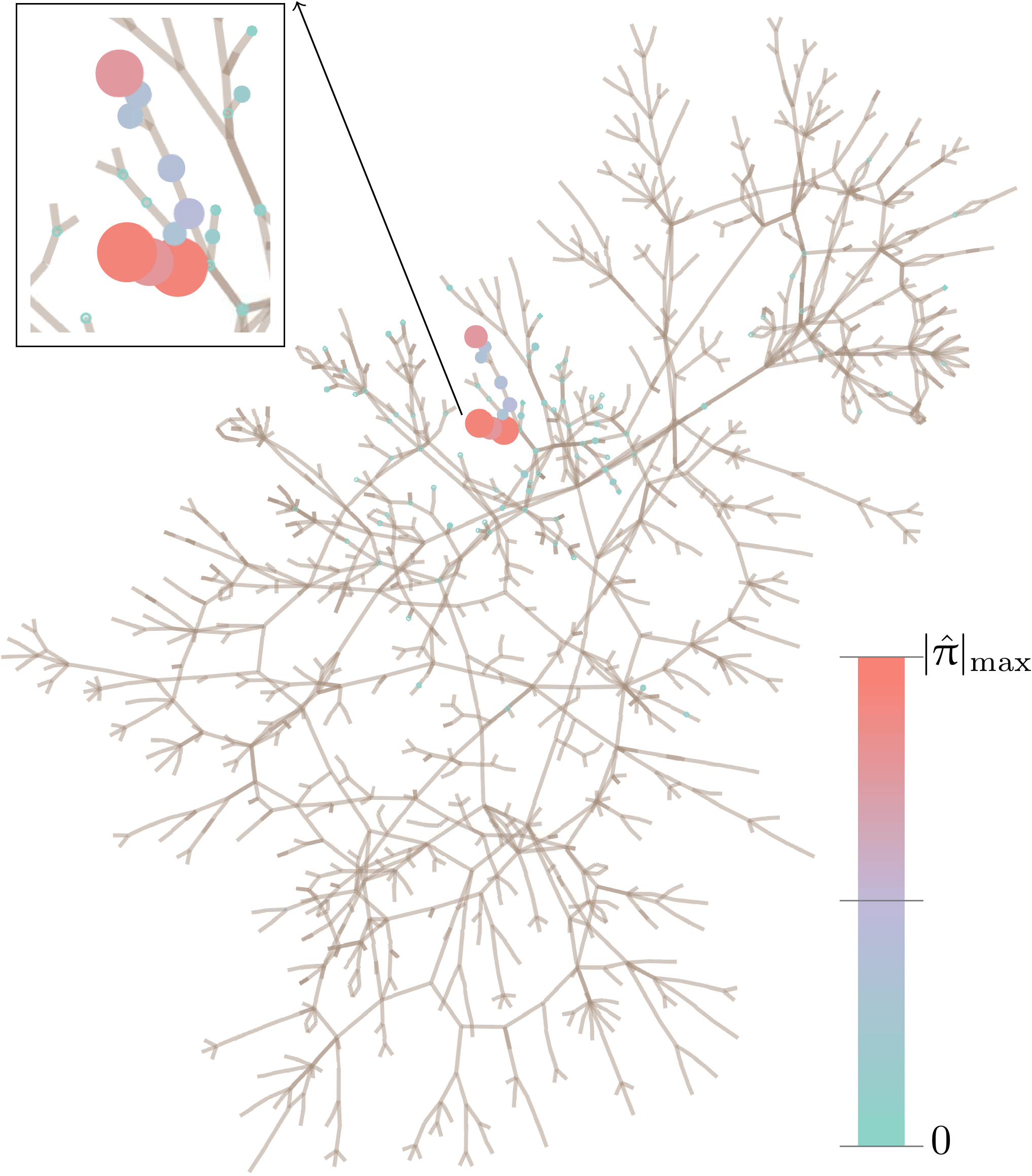}}
    \caption[AIITS: participation of bus active power injections to Mode 1]{AIITS: participation of bus active power injections to Mode 1.}
    \label{fig:graph}
\end{figure}

\begin{figure}
    \centering
    \resizebox{0.6\linewidth}{!}{\includegraphics{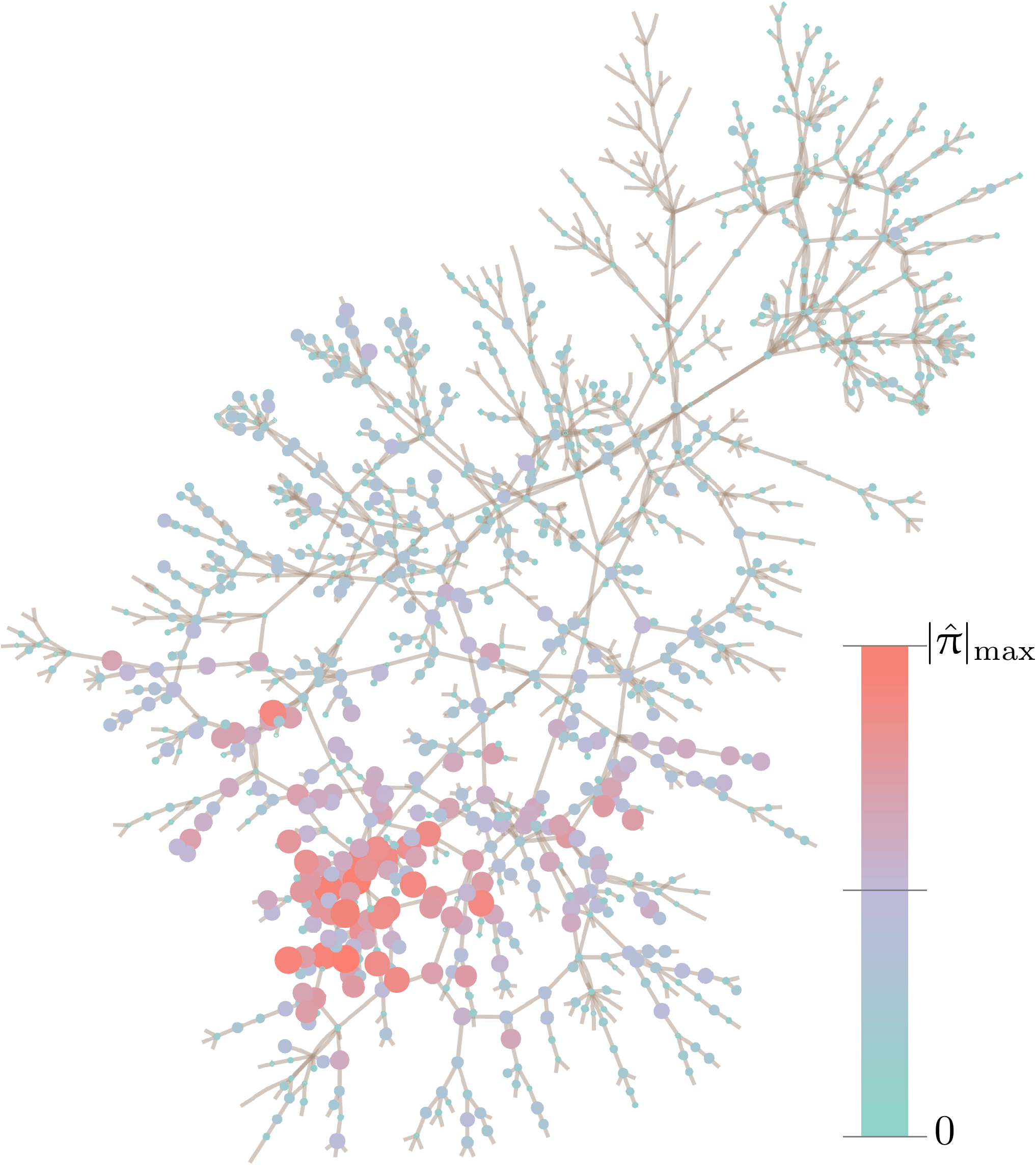}}
    \caption[AIITS: participation of bus active power injections to Mode 2]{AIITS: participation of bus active power injections to Mode 2.}
    \label{fig:graph2}
\end{figure}

The generated graph with the \acp{pf} of all bus active power
injections in the local Mode 1 is illustrated in Figure~\ref{fig:graph}.
The mostly participating active power injection is the one of the bus
552, that is adjacent to generator 16, with
$|\hat \PF|_{\rm max}=0.3218$. The \acp{pf} of all bus active power injections
in the inter-machine Mode 2 is illustrated in Figure~\ref{fig:graph2}.  The mostly participating active power injection is the one of the bus
1,405, that is close to synchronous generator 2, with
$|\hat \PF|_{\rm max}=0.2508$.  Figure \ref{fig:graph2} shows that the lower
frequency oscillations spread over the power system.  In fact, there
are several buses in a large area that have a high participation to
the inter-machine mode.

\section{Conclusions}
\label{pf:sec:conclusion}

The chapter proposes a systematic analytical approach to quantify the
participation of the algebraic variables of a power system model, and
to general of any function of the system variables in the system
modes, through the definition of output vectors of the system's
state-space formulation. The proposed approach, which describes an
alternative interpretation of the \acp{pf} as eigen-sensitivities,
provides a high flexibility, since it allows determining \acp{pf} of
states, algebraic variables, rates of change of system variables, as
well as of system parameters.

Regarding the computational burden of the
participation analysis, once the eigenvalue analysis is completed, the
cost of calculating the proposed \acp{pf} is negligible.
Moreover, 
the proposed approach allows exploiting the sparsity of
the \ac{gep} matrix pencil and can lead to a significant speedup,
provided that a proper eigenvalue solver is employed.

\vspace{-5mm}

\newpage
\chapter{Fractional Order Control}
\label{ch:foc}

\section{Introduction}
\label{foc:sec:intro}

\begin{figure}
\centering
\resizebox{0.35\linewidth}{!}{\includegraphics{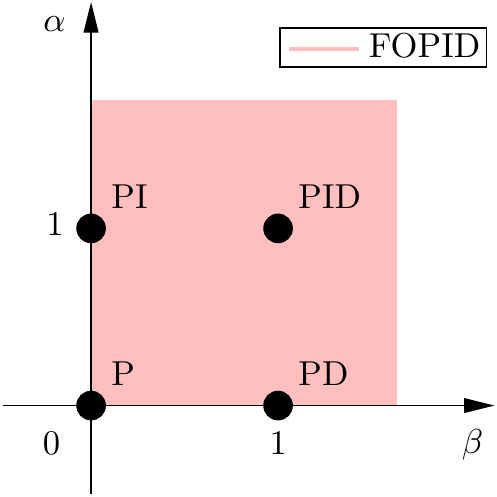}}
\caption[PID vs FOPID: from point to plane]{PID vs FOPID: from point to plane.}
    \label{foc:fig:fopidpl}
\end{figure}

Fractional calculus is the analysis of non-integer order
differentials and integrals.
Although the first discussion on derivatives with
non-integer order dates back to Leibniz \cite{leibniz:1695}, major
studies on fractional calculus started with Liouville \cite{liouville:1832}.  The application of fractional calculus in control was introduced with the definition of the ideal cut-off
characteristic by Bode \cite{bode:1945} and the first systematic study
of the frequency response of \acp{foc} was done by Oustaloup
\cite{oustaloup:1991}.  In \cite{podlubny:book99, podlubny:1999}, Podlubny 
provided a comprehensive analysis of fractional
systems with applications to automatic control, and proposed the use of the \ac{FOPID} controller.
The \ac{FOPID} controller is 
an extension of
the classical \ac{pid}, and is characterized by five parameters:
three gains, namely proportional, integral, and derivative;
and two fractional orders, namely integral ($\alpha$) and derivative ($\beta$).
Employing a \ac{FOPID} extends the four control points of the
\ac{pid} strategy to the plane defined by the fractional orders $\alpha$ and $\beta$ \cite{monje:2010}. 
This is illustrated in Figure~\ref{foc:fig:fopidpl}. 
\acp{foc}
have been applied to various engineering fields, e.g.~heat diffusion
\cite{jesus:2008} and robotic time-delay systems
\cite{lazarevic:2006}.
Recent efforts on \acp{foc} have tackled several issues, such as
modeling and studying the impact of control limiters, see
\cite{PANDEY:2017}, and
variable-order fractional orders, see \cite{Dab}.
Finally, it is relevant to mention here that
fractional calculus
is a promising tool for applications not only in control systems, but in many other applications, for example physics \cite{Hi} and biology \cite{ionescu:17}.


%

%
%

The potential of \acp{foc} for power system applications has not been
discussed until very recently.  Applications include automatic voltage
regulation of synchronous machines
\cite{zamani:2009,tang:2012,pan:2013}; load frequency control
\cite{taher:2014, saxena:2019}; damping control \cite{chaib:2017}; and
voltage control of distributed energy resource systems
\cite{delghavi:2016}.
%
These works mainly focus on the tuning of FOPID controllers through
heuristic algorithms, such as particle swarm \cite{zamani:2009},
chaotic multi-objective \cite{pan:2013}, and imperialist competitive
algorithm \cite{taher:2014}.  Analytical methods employ frequency
response criteria such as the desired gain crossover frequency
\cite{saxena:2019}.
%

From a practical and simulation point of view, fractional dynamics are
typically approximated using appropriate rational order transfer
functions.  Although various techniques have been proposed to define
such transfer functions \cite{vinagre:2000},
the most commonly utilized continuous method is the \ac{ora} \cite{oustaloup:2000}.  
Therefore, the \ac{ora} is the
method considered in this chapter.
The works cited above focus on applications and rely, for
the implementation of \acp{foc}, on proprietary software tools which are
utilized as a {\it black-box}.   This approach is indeed
fostered by the availability of several software tools for the design and simulation of \acp{foc}, see, for example, the Matlab toolboxes
CRONE \cite{cronetool}, Ninteger \cite{ninteger:2004}, and FOMCON
\cite{fomcon:2011}.


The main goal of this chapter is to provide a
systematic study of \acp{foc} for power system applications. 
%
The remainder of the chapter is organized as follows.
Section~\ref{foc:sec:fc} outlines the theory of fractional calculus.
Section~\ref{foc:sec:con} discusses the stability of power systems with
inclusion of \acp{foc}. 
Section~\ref{foc:sec:oust} focuses on the modeling,
computer implementation and tuning of \ac{ora}-based \acp{foc}.   
Two case studies are presented in Sections~\ref{foc:sec:case1} and
\ref{foc:sec:case2}.  Finally, conclusions are drawn in
Section~\ref{foc:sec:conclusion}.

\section{Essentials of Fractional Calculus}
\label{foc:sec:fc}

Fractional calculus is the
analysis of non-integer order differentials and integrals. 
That is, it deals with the problem of extending the
differentiation and integration operators $\frac{d^n}{dt^n}$,
$\int^t_0 d^n(\tau)$, $n \in \mathbb{N}$, for real (or complex) number powers.
There exist several approaches 
that address this problem.
A precise formulation
is given by the \ac{RL} definition.  Consider
a function $\phi : [0, \infty) \rightarrow \mathbb{R}$.  The idea
behind the \ac{RL} definition is to first consider the $n$-fold
integration of $\phi(t)$ and then extend $n \in \mathbb{N}$ to any
$\gamma \in \mathbb{R}^+$.  In its derivative form, the \ac{RL} definition
reads \cite{monje:2010}:

\begin{equation}
  \label{foc:eq:rl}
  \phi^{(\gamma)}(t) = \frac{1}{ \Gamma (\upmu-\gamma) }
  \frac{d^\upmu}{dt^\upmu}
  \Big(  \int^{t}_0 \frac{\phi(\tau)}{(t-\tau)^{\gamma-\upmu+1}} d\tau \Big) \, ,
\end{equation} 
where $\gamma$, $\upmu-1<\gamma<\upmu$, $\upmu \in \mathbb{N}$, is the 
fractional order; and
$\phi^{(\gamma)}(t) = {d^\gamma \phi}/{dt^\gamma}$.
The Laplace transform of \eqref{foc:eq:rl} is:
\begin{equation}
  \label{foc:eq:rll}
  \mathcal{L} \{ \phi^{(\gamma)}(t) \} = 
  s^{\gamma} \Phi(s) - \sum^{\upmu-1}_{j=0} s^j  \phi^{(\gamma-j-1)}(0) \, ,
\end{equation} 
where $s \in \mathbb{C} $ and
$\mathcal{L}\{ {\phi}(t)\}=
\Phi(s)$.
Equation \eqref{foc:eq:rll} requires the knowledge of the fractional order
initial conditions $\phi^{(\gamma-j-1)}(0)$, $j=0,1,\ldots,\upmu-1$.
This raises an issue for engineering systems since, currently, only
integer order initial conditions are well understood and known for
physical variables.  Other properties of the \ac{RL} definition are also
counter-intuitive in the sense of classical differentiation. For
example, the \ac{RL} derivative of a constant function is typically
unbounded at $t = 0$ \cite{podlubny:1999}.

With the aim of meeting the requirements of known physical variables
and systems, \eqref{foc:eq:rl} was revisited by Caputo \cite{caputo:2015}.
Caputo's definition of $\phi^{(\gamma)}(t)$ reads:  
\begin{equation}
  \label{foc:eq:caputo}
  \phi^{(\gamma)}(t) = \frac{1}{ \Gamma (\upmu-\gamma) } \int^{t}_0 \frac{\phi^{(\upmu)}(\tau)}{
    (t-\tau)^{\gamma-\upmu+1} } d\tau \, .
\end{equation} 
The Laplace transform of \eqref{foc:eq:caputo} is:
\begin{equation}
  \label{foc:eq:caputol}
  \mathcal{L} \{\phi^{(\gamma)}(t) \} = 
  s^{\gamma} \Phi(s) - \sum^{\upmu-1}_{j=0} s^{\gamma-j-1} 
  \phi^{(j)}(0)  \, .
\end{equation} 
Equation \eqref{foc:eq:caputol} requires the knowledge of the initial
conditions $\phi^{(j)}(0)$, $j=0,1,\ldots,\upmu-1$, which in this case
are of integer order.  This property is crucial for the solution of
initial value problems.  In fact, for the purpose of
fractional control, that is of concern here, one needs to use a
definition with integer order initial conditions.  
This work utilizes the Caputo definition of fractional
derivative given in \eqref{foc:eq:caputo},
which is more consistent for control
applications and follows the properties of differentiation in the
classical sense.  For example, the Caputo fractional
derivative of a constant function is zero.

In the special case that $0<\gamma<1$,
the Caputo definition \eqref{foc:eq:caputo} reduces to \cite{Bo}: 
\begin{equation}
\label{foc:eq:caputo01}
\phi^{(\gamma)}(t)=\frac{1}{\Gamma (1-\gamma)}\int^t_0\frac{\Dt \phi(\tau)}{(t-\tau)^{\gamma}} d\tau \, .
\end{equation}
%



There are several other definitions of fractional
derivatives/integrals and choosing the appropriate one depends on the
application.  For example, the Gr{\"u}nwald-Letnikov's derivative is
relevant for the numerical solution of fractional differential
equations.  It is important to emphasize that the theory of fractional
calculus applicable to the stability analysis and control of physical
dynamical systems is an active research topic and yet to be fully
understood.  Recent efforts have addressed issues related to Caputo's
formulation, for example its singular kernel for $t=\tau$
\cite{caputo:2015, atangana:2016}.
Finally, note that fractional-order operators are not a straightforward generalization of the classical integer-order operators and, despite the several advances and
interesting recent studies, see e.g.~ \cite{Bo,Das8,Deb,Kac,Liz1,li:2007,Shiri,Su,Wei}, the existing theory of 
fractional differential equations
is far from complete. 

\section{Power System with Fractional Order Control}
\label{foc:sec:con}

This chapter focuses on the modeling and stability of power systems
with inclusion of \acp{foc}.  
Since a general theory of the stability of
non-linear fractional differential equations is not available, we
proceed as follows \cite{foc1}.  This section considers the conditions for the
stability of a linear (or linearized)
singular power system model with inclusion of a \ac{foc}.  These conditions help design the \acp{foc} discussed in Section
\ref{foc:sec:oust}.  However, power system models are non-linear. 
For this
reason, the design of \acp{foc} is checked by solving numerical time domain
simulations of the fully-fledged non-linear model of the system and its
controllers in the case studies presented in 
Section~\ref{foc:sec:casestudies}.  

\subsection{Modeling}

Consider the \ac{dae} linearized power system model \eqref{sssa:eq:singps}, which is repeated here for clarity:
\begin{equation}
\label{foc:eq:singcon}
\Edae \, \Dt{\xys}=\Adae \, \xys+
\Bdae \, 
\Delta 
\bfg u \, ,
\end{equation}
where $\nxy = \nx+\ny$ and $\Edae$, $\Adae \in \mathbb{R}^{\nxy \times \nxy}$,
$\Bdae \in \mathbb{R}^{\nxy \times \nin}$.

Let the vector of the system 
output measurements $\bfg w$, 
$\bfg w \in \mathbb{R}^{\nout}$, be:
\begin{equation}
{\bfg w} = 
\Cdae \, \xys + \Ddae \, \Delta \bfg u \, ,
\label{foc:eq:out}
\end{equation}
%
%
where $\Cdae \in \mathbb{R}^{\nout \times \nxy}$,
$\Ddae \in \mathbb{R}^{\nout \times \nin}$.
%
Then, a multiple-input, multiple-output \ac{foc}
for the system
\eqref{foc:eq:singcon}-\eqref{foc:eq:out},
can be described by a set of fractional 
\acp{dae} as follows:
\begin{equation}
\begin{aligned}
  \bfg E_{c,1} \,  \Dt{\bfg x}_c+ \bfg E_{c,\gamma} \, {\bfg x_c}^{(\gamma)} &= 
  \bfg A_c \, {\bfg x_c}
  + \bfg B_c \, \bfg w \ , \\ 
  \bfg 0_{\nin,1} &= 
  \bfg C_c \, {\bfg x_c}
  + \bfg D_c \, \bfg w - \Delta {\bfg u} \, ,
  \label{foc:eq:focont1} 
 \end{aligned} 
\end{equation}
where $\gamma$ is the controller's fractional order; $\bfg x_c$,
$\bfg x_c \in \mathbb{R}^{\varsigma}$, is the vector of the
controller 
states; $\bfg E_{c,1}, \ \bfg E_{c,\gamma}, \ \bfg A_c \in \mathbb{R}^{\varsigma \times \varsigma}$,
$\bfg B_c \in \mathbb{R}^{\varsigma \times \nout}$,
$\bfg C_c \in \mathbb{R}^{\nin \times \varsigma}$,
$\bfg D_c \in \mathbb{R}^{\nin \times \nout}$.  
It is relevant to mention
that there are \acp{foc} that introduce multiple, distinct fractional
orders.  
Combining \eqref{foc:eq:singcon},
\eqref{foc:eq:out} and \eqref{foc:eq:focont1} yields the closed-loop system
representation.  In matrix form:

\begin{equation}
\begin{aligned}
  \label{foc:eq:clmat}
  \begin{bmatrix}
\bfg E & \bfg 0_{\nxy,\varsigma} & \bfg 0_{\nxy,\nin}\\
\bfg 0_{\varsigma,\nxy} & \bfg E_{c,1} & \bfg 0_{\varsigma,\nin} \\
\bfg 0_{\nin,\nxy} & \bfg 0_{\nin,\varsigma} &  \bfg 0_{\nin,\nin} \\
  \end{bmatrix}
  \hspace{-1.0mm}
  &
  \begin{bmatrix}
 \Dt \xys \\
 \Dt {\bfg x_c} \\
  \Delta \Dt {\bfg u} \\
  \end{bmatrix}
\hspace{-1.5mm} +
\begin{bmatrix}
\bfg 0_{\nxy,\nxy} & \bfg 0_{\nxy,\varsigma} & \bfg 0_{\nxy,\nin}\\
\bfg 0_{\varsigma,\nxy} & \bfg E_{c,\gamma} & \bfg 0_{\varsigma,\nin} \\
\bfg 0_{\nin,\nxy} & \bfg 0_{\nin,\varsigma} &   \bfg 0_{\nin,\nin} \\
  \end{bmatrix}
  \hspace{-1.0mm}
  \begin{bmatrix}
  \xys^{(\gamma)} \\
  \bfg x_c^{(\gamma)} \\
  \Delta \bfg u^{(\gamma)} \\
  \end{bmatrix} \\
  &=
 \begin{bmatrix}
    \bfg A & \bfg 0_{\nxy,\varsigma} & \bfg B\\
   \bfg B_c \bfg C & \bfg A_c & \bfg B_c \bfg D \\
    \bfg D_c \bfg C & \bfg C_c & 
    \bfg D_c \bfg D-\bfg I_{\nin} \\
  \end{bmatrix}
  \begin{bmatrix}
 \xys \\
  {\bfg x_c} \\
  \Delta {\bfg u} \\
  \end{bmatrix}
   \, ,
   \nonumber
 \end{aligned}
\end{equation}
or equivalently,
\begin{equation}
\begin{aligned}
  \bfg M \, \Dt{\bfg \psi}+
  \bfg M_\gamma \, \bfg \psi^{(\gamma)}
&= \bfg A_{\rm cl} \, {\bfg \psi} \, ,
  \label{foc:eq:clsys} 
 \end{aligned} 
\end{equation}
where 
\begin{equation}
 \bfg M = 
 \begin{bmatrix}
\bfg E & \bfg 0_{\nxy,\varsigma} & \bfg 0_{\nxy,\nin}\\
\bfg 0_{\varsigma,\nxy} & \bfg E_{c,1} & \bfg 0_{\varsigma,\nin} \\
\bfg 0_{\nin,\nxy} & \bfg 0_{\nin,\varsigma} &  \bfg 0_{\nin,\nin} \\
  \end{bmatrix}
  \, , \,
\bfg M_\gamma = 
 \begin{bmatrix}
\bfg 0_{\nxy,\nxy} & \bfg 0_{\nxy,\varsigma} & \bfg 0_{\nxy,\nin}\\
\bfg 0_{\varsigma,\nxy} & \bfg E_{c,\gamma} & \bfg 0_{\varsigma,\nin} \\
\bfg 0_{\nin,\nxy} & \bfg 0_{\nin,\varsigma} &   \bfg 0_{\nin,\nin} \\
  \end{bmatrix} \, ,  
  \nonumber
\end{equation}
\begin{equation}
 \bfg A_{\rm cl} = 
 \begin{bmatrix}
    \bfg A & \bfg 0_{\nxy,\varsigma} & \bfg B\\
   \bfg B_c \bfg C & \bfg A_c & \bfg B_c \bfg D \\
   \bfg D_c \bfg C & \bfg C_c & 
\bfg D_c \bfg D-\bfg I_{\nin} \\
\end{bmatrix}
\, , \,
\bfg \psi = 
\begin{bmatrix}
  \xys \\
  {\bfg x_c}  \\
  \Delta {\bfg u} \\
\end{bmatrix} \, ,
\nonumber
\end{equation}
with
$\bfg M, \ \bfg M_\gamma, \ \bfg A_{\rm cl} \in \mathbb{R}^{\rho \times \rho}$, and $\rho=\nxy+\varsigma+\nin$.

\subsection{Stability}

This section studies the stability of the closed-loop system \eqref{foc:eq:clsys}, which is a singular system of differential equations having both first, and fractional order derivatives. With this aim,
the following property of the Caputo fractional derivative is relevant \cite{li:2007}:

{\proposition 
{
Let $\bfg \phi(t)$,
$\bfg \phi(t)\in \mathcal{C}^1[0,T]^{n}$ for some $T>0$, where $\mathcal{C}^1$ denotes the set of continuously differentiable functions.
Then:
\begin{align}
\label{foc:eq:prop1}
  [\bfg \phi^{(a)}(t)]^{(b)}=
[\bfg \phi^{(b)}(t)]^{(a)}=
\bfg \phi^{(a+b)}(t) \ , 
\end{align}
where $a,b\in\mathbb{R^+}$, and $a+b\leq 1$. 
}
}
Note that \eqref{foc:eq:prop1} does not
hold for the \ac{RL} derivative. 

Equation \eqref{foc:eq:clsys} can be rewritten as:
\begin{equation}
\begin{aligned}
  \bfg M \bfg \, \psi^{(\gamma+\beta)}
  +
  \bfg M_\gamma \, \bfg \psi^{(\gamma)}
&= 
  \bfg A_{\rm cl} \, {\bfg \psi}
  \, ,
  \label{foc:eq:clsys2} 
 \end{aligned} 
\end{equation}
where $\gamma+\beta=1$.

Adopting the notation
\begin{align}
\xs_1 &= \bfg \psi \ , \
\xs_2 = \bfg \psi^{(\gamma)} \ ,
\nonumber 
\end{align}
one obtains
$\xs_1^{(\gamma)} = \bfg \psi ^{(\gamma)}
= \xs_2$.
Making use of \eqref{foc:eq:prop1} yields
$\xs_2^{(\beta)}= 
\bfg \psi^{(\gamma+\beta)}$.
Substitution to \eqref{foc:eq:clsys2} gives:
\begin{align}
 \label{foc:eq:clsys3}
\bfg M \, \xs_2^{(\beta)}
+
\bfg M_\gamma \, \xs_2 
&= 
\bfg A_{\rm cl} \, {\xs_1}
\nonumber \ 
\Rightarrow \\
\bfg M \xs_2^{(\beta)}
&=
\bfg A_{\rm cl} {\xs_1} -
\bfg M_\gamma \xs_2 \ .
\end{align} 
Equivalently:
\begin{equation}
\begin{bmatrix}
  \bfg I_{\rho} & \bfg 0_{\rho,\rho}  \\
\bfg 0_{\rho,\rho} & \bfg M \\
\end{bmatrix}
\begin{bmatrix}
\xs_1^{(\gamma)} \\
\xs_2^{(\beta)}  \\
\end{bmatrix}
=
\begin{bmatrix}
\bfg 0_{\rho,\rho} & \bfg I_{\rho}\\
\bfg A_{\rm cl}  & -\bfg M_\gamma \\
\end{bmatrix}
\begin{bmatrix}
  \xs_1\\
  \xs_2\\
\end{bmatrix}  \, .
\label{foc:eq:final}
\end{equation}
%
System \eqref{foc:eq:final} can be rewritten as:
\begin{equation}
\begin{aligned}
  \Efr  \, {\xs}^\Delta
&= 
  \Afr  \, {\xs}
  \, ,
  \label{foc:eq:final2} 
 \end{aligned} 
\end{equation}
or, by recovering the time dependency, as:
\begin{equation}
\label{foc:eq:final2t}
  \Efr \, \xs^\Delta(t)= \Afr \, \xs(t) \, ,
 \end{equation}
where 
\begin{equation}
  \Efr=
\begin{bmatrix}
  \bfg I_{\rho} & \bfg 0_{\rho,\rho}  \\
     \bfg 0_{\rho,\rho} & \bfg M \\
\end{bmatrix}
 , \
\Afr=
\begin{bmatrix}
\bfg 0_{\rho,\rho} & \bfg I_{\rho}\\
\bfg A_{\rm cl}  & 
-\bfg M_\gamma 
\\
\end{bmatrix}  , \
{\xs}=
\begin{bmatrix}
\xs_1\\
\xs_2\\
\end{bmatrix}  
, \ 
{\xs}^\Delta =
\begin{bmatrix}
\xs_1^{(\gamma)} \\
\xs_2^{(\beta)} \\
\end{bmatrix} .
  \nonumber
\end{equation}
We have $\Efr, \Afr \in \mathbb{R}^{r \times r}$,
$\xs:[0,+\infty)\rightarrow\mathbb{R}^{r}$, and
$\beta,\gamma \in (0,1)$, where for simplicity the notation $r = 2\rho$ is used. 

{
\theorem
{
Consider system \eqref{foc:eq:final2t}.  Then its
matrix pencil is given by \cite{foc2}:
\begin{equation}
\begin{aligned}
\label{foc:eq:pencil}
z(s) \,
 \Efr- \Afr \, ,
  \end{aligned}
\end{equation}
where
\begin{equation}
z(s) = \begin{bmatrix}
s^\gamma \bfg I_{\rho} & \bfg 0_{\rho,\rho}\\
\bfg 0_{\rho,\rho} & s^\beta \bfg I_{\rho}\\
\end{bmatrix} \, .
\nonumber
\end{equation}
}
\label{foc:theorem:pencil}
}

The proof of Theorem \ref{foc:theorem:pencil} is given in Appendix \ref{app:proofs}.

Assuming that the pencil \eqref{foc:eq:pencil}
is regular, or equivalently, 
${\rm det}(z \Efr - \Afr) \not\equiv 0$, it can be proven that there always exist solutions of 
\eqref{foc:eq:final2t}. 
Then, similarly to the discussions of Chapter~\ref{ch:sssa}, uniqueness of solutions 
depends on the given initial condition.
The interested reader may find a 
a comprehensive theory for the existence and uniqueness of solutions
of systems in the form of \eqref{foc:eq:final2t} -- with either regular or singular pencil --
in \cite{foc2}.


The eigenvalues of the matrix pencil \eqref{foc:eq:pencil}
provide insight on the stability of system \eqref{foc:eq:final2t}, or
equivalently, of system \eqref{foc:eq:clsys}.
Since the pencil of system 
\eqref{foc:eq:final2t} is
a regular pencil, 
$s \Efr - \Afr$ is also a regular pencil.
Hence and because of the structure of $\Efr$ there exist invariants of
the following type:
\begin{itemize}
\item $\kappa$ finite eigenvalues of algebraic multiplicity $\nfi{i}$,
$i=1,2,...,\kappa$;
\item  an infinite eigenvalue of algebraic multiplicity $\ninf$,
\end{itemize}
where $\sum_{i=1}^\kappa \nfi{i} = \nf$, $\nf+\ninf=\nd$.  


Let 
\[
\leigvmat = \begin{bmatrix}
\leigvmat_{\nf,\gamma}\\
\leigvmat_{p,\beta}\\ 
\leigvmat_{q,\gamma}\\
\leigvmat_{q,\beta}\\ 
\end{bmatrix}
 \, , \ 
 \reigvmat = \begin{bmatrix}
 \reigvmat_{\nf,\gamma}&
 \reigvmat_{\nf,\beta} & 
 \reigvmat_{q,\gamma}&
 \reigvmat_{q,\beta} 
\end{bmatrix} \, ,
\]
where $\leigvmat_{\nf,\gamma} \in \mathbb{C}^{\hat{\nf} \times \nd}$, $\leigvmat_{\nf,\beta} \in \mathbb{C}^{\bar \nf \times \nd}$, $\leigvmat_{\ninf,\gamma} \in \mathbb{C}^{\tilde{\ninf} \times \nd}$, $\leigvmat_{\ninf,\beta} \in \mathbb{C}^{\bar \ninf \times \nd}$, and $\reigvmat_{\nf,\gamma}\in \mathbb{C}^{\nd \times \hat{\nf}}$, $\reigvmat_{\nf,\beta} \in \mathbb{C}^{\nd \times \bar \nf}$, $\reigvmat_{\ninf,\gamma}\in \mathbb{C}^{\nd \times \hat{\ninf}}$, $\reigvmat_{\ninf,\beta} \in \mathbb{C}^{\nd \times \bar \ninf}$. Equivalently, setting: 
\[
\leigvmat_\nf = \begin{bmatrix}
\leigvmat_{\nf,\gamma}\\
\leigvmat_{\nf,\beta} \end{bmatrix} \, , \  \reigvmat_\nf = \begin{bmatrix}
\bfg \reigvmat_{\nf,\gamma}&
 \reigvmat_{\nf,\beta} \end{bmatrix} \, , 
 \]
 \[
\leigvmat_\ninf = \begin{bmatrix}
\leigvmat_{\ninf,\gamma}\\
\leigvmat_{\ninf,\beta} 
\end{bmatrix} \, , \  
\reigvmat_\ninf=
\begin{bmatrix}
\reigvmat_{\ninf,\gamma}&
\reigvmat_{\ninf,\beta} 
\end{bmatrix} \, ,
 \]
one has
\begin{equation}\label{eq8}
\leigvmat = \begin{bmatrix}
\leigvmat_\nf\\ 
\leigvmat_\ninf
\end{bmatrix} \, , \ 
\reigvmat = \begin{bmatrix}
\reigvmat_\nf & \reigvmat_\ninf
\end{bmatrix} \, .
\end{equation}
with $\leigvmat_\nf \in \mathbb{C}^{\nf \times \nd}$, $\leigvmat_\ninf \in \mathbb{C}^{\ninf \times \nd}$, and $\reigvmat_\nf\in \mathbb{C}^{\nd \times \nf}$, $\reigvmat_\ninf \in \mathbb{C}^{\nd \times \ninf}$. 

Employing \eqref{sssa:eq:decomp} for 
$\Afr, \Efr$, and using the notation
$\bfg I_\nf=\bfg I_{\hat{\nf}}  \oplus 
\bfg I_{\bar \nf}$, $\bfg J_p=\bfg J_{\hat{\nf}}  \oplus \bfg J_{\bar \nf}$, one has:
\begin{equation}
\begin{aligned}
 \leigvmat \Efr \reigvmat &=
 \bfg I_{\hat{\nf}}  \oplus \bfg I_{\bar \nf}  \oplus \bfg H_{\hat{\ninf}}  \oplus \bfg H_{\bar \ninf} \, , 
 \\
\leigvmat \Afr \reigvmat &= 
\bfg J_{\hat{\nf}}  \oplus \bfg J_{\bar \nf} \oplus \bfg I_{\hat{\ninf}}  \oplus \bfg I_{\bar \ninf} \, . 
 \end{aligned}
 \nonumber
\end{equation}
%

Finally, the following proposition is relevant:

{\proposition
{ Consider system \eqref{foc:eq:final2t} with a regular pencil. Then:
\begin{enumerate}[label=\alph*)]
\item using the spectrum of the pencil $s \Efr - \Afr$, the general solution is given by:
\begin{equation}
\label{eq23}
\xs(t)=\reigvmat_\nf
\sum_{k=0}^\infty
\begin{bmatrix}
\frac{t^{\gamma k}}{\Gamma(k\gamma+1)}  \bfg I_{\hat{\nf}} &  
\bfg 0_{\hat{\nf}, \hat{\nf}} \\
 \bfg 0_{\bar{\nf},\bar{\nf}} &  \frac{t^{\beta k}}{\Gamma(k\beta+1)}   \bfg I_{\bar{\nf}}
\end{bmatrix}
\bfg J_\nf^k \,  
\textbf{c} \, ,
\end{equation}
where $\bfg J_\nf \in \mathbb{C}^{\nf \times \nf}$ is a Jordan matrix constructed by the finite eigenvalues of the pencil $s \Efr - \Afr$, and their algebraic multiplicity, while $\reigvmat_\nf \in \mathbb{C}^{r \times \nf}$ is a matrix constructed by the linear independent eigenvectors related to the finite eigenvalues of the pencil $s \Efr - \Afr$, and
$\bfb c \in \mathbb{C}^{\nf}$
is a constant vector.
\item the system \eqref{foc:eq:final2t} is asymptotically stable if all eigenvalues $\hat \lambda^*$ of the pencil $s \Efr - \Afr$ satisfy:
\begin{equation}
\label{foc:eq:stability}
|\rm Arg(\hat \lambda^*)| >
\tilde \gamma \, \frac{\pi}{2} \ (rad) 
\, , 
\end{equation}
\end{enumerate}
where $\tilde \gamma=\rm min\left\{\gamma,1-\gamma\right\}$.
}
\label{foc:theorem:sol}
}

The proof of Proposition~\ref{foc:theorem:sol} can be found in \cite{foc2}.  For a more general survey on the stability conditions for systems of fractional differential equations, the interested reader may refer to \cite{petras:2009}.  
%
%
Finally,
for linearized systems, as it is the case of power
systems, the condition \eqref{foc:eq:stability} guarantees stability in a
neighborhood of the operating point utilized to calculate the pencil
\eqref{foc:eq:pencil}.  For this reason, in the case studies
discussed in Sections \ref{foc:sec:case1} and \ref{foc:sec:case2},
numerical integration rather than
\eqref{foc:eq:stability} is utilized to check the stability and the dynamic response
of power systems with inclusion of \acp{foc}.

%

\subsection{Properties of Fractional Order Controllers}

Consider the simple \ac{foc} with transfer function
$\TFUN_{\rm c}(s)=K {s^{\gamma}}$.
Consider the 
frequency response of 
$\TFUN_{\rm c}(s)$, i.e. 
its steady-state response to sinusoidal, periodic input 
signals. In this case, it is $s \in \mathbb{I} $, 
or $s = \jj \omega$. 

{\it Frequency Response:}
The magnitude and phase of $\TFUN_c(s)$ can be written as follows:
\begin{equation}
\begin{aligned}
\label{foc:eq:magph}
{\rm Mag}(\TFUN_{c}(s))\ {\rm (dB)} &= 20 \, {\rm log }|K s^\gamma| 
= 20\, \gamma \, {\rm log }(K \omega) \ , \\
{\rm Arg (\TFUN_{c}(s))} \ {\rm (^\circ)} &= 
{{\rm Arg} (K (\jj \omega)^\gamma)}
= 90 \, \gamma \, .
\end{aligned} 
\end{equation}
Hence, $\TFUN_{ c}(s)$ has a magnitude Bode plot with constant slope of
$20\gamma$~dB/dec, while the phase plot is a horizontal line at
$90\gamma $ degrees.  The \ac{IO} versions of $\TFUN_{ c}(s)$ are obtained for
$\gamma=n$, $n \in \mathbb{Z}$.  Then, from \eqref{foc:eq:magph}, it is
clear that $\TFUN_{ c}(s)$
is an extension of its \ac{IO} versions in frequency domain.  This result
is general, so that all \acp{foc} can be viewed as extensions of the
respective \ac{IO} ones.
 
{\it Robustness:} \acp{foc} have an inherent property of iso-damping, which
implies that the closed-loop system is robust against gain
uncertainties and variations.  Let $\TFUN_o(s)$ be the transfer function of
the open-loop, linearized power system.  Then, the iso-damping
property is defined as:
\begin{equation}
\begin{aligned}
\label{foc:eq:iso}
\Bigg | \frac{d \;{\rm Arg} (\TFUN_{ c}(\jj \omega) \, 
\TFUN_o(\jj \omega))}{d \omega} \Bigg |_{\omega=\omega_{\rm gc}} &=  0 \ ,
\end{aligned} 
\end{equation}
where $\omega_{\rm gc}$ is the system gain crossover frequency.  \eqref{foc:eq:iso} indicates that the system maintains its phase margin
around $\omega_{\rm gc}$.



%

%




\subsection{Examples}

This section provides two illustrative examples
on the \ac{sssa} of power systems with inclusion of \acp{foc}. The first example considers a small linear singular system of differential equations with regular pencil. The second example discusses the damping of the electro-mechanical oscillations of 
the \ac{wscc} 9-bus system
through a \ac{fopss}.

\subsubsection{Illustrative Example}
	
Consider system \eqref{sssa:eq:sing} with:  
\begin{equation}
\begin{aligned}
\Esng  &= 
  \begin{bmatrix*}[r]
4 & 9 & 9 & -2 & 10 & 7 & 3\\ 1 & 5 & 2 & 2 & 3 & 1 & 1\\ 1 & 0 & -2 & -2 & 6 & 4 & 1\\ 5 & -2 & -3 & 18 & 3 & 16 & 2\\ 6 & 8 & 6 & 8 & 6 & 14 & 2\\ 2 & 11 & 3 & 6 & 6 & 2 & 2\\ 4 & 5 & 5 & 6 & 2 & 9 & 1 
  \end{bmatrix*} , \\
\Asng  &= 
  \begin{bmatrix*}[r]
 -15 & -43 & -39 & 4 & -35 & -22 & -5\\ -3 & -19 & -2 & -5 & -12 & -2 & 1\\ -4 & -16 & -30 & 6 & -9 & -1 & -7\\ -25 & -2 & 3 & -72 & -3 & -74 & -4\\ -27 & -32 & -23 & -39 & -24 & -66 & -5\\ -8 & -41 & -15 & -18 & -24 & -8 & -8\\ -18 & -15 & -9 & -29 & -13 & -48 & -1 
  \end{bmatrix*} ,  \nonumber
\ \Bsng = 
  \begin{bmatrix*}[r]
 0 \\
 0 \\ 
 0 \\ 
 0 \\ 
 0 \\ 
 1 \\
 0\\ 
  \end{bmatrix*} .
\end{aligned}  
\end{equation}

The matrix 
pencil $s\Esng-\Asng$
has $\nf=5$ finite, distinct eigenvalues $\lambda_1=-5$,
$\lambda_2=-4$, $\lambda_3=1$,
$\lambda_4=-2$ and $\lambda_5=-3$.
The pencil also has the eigenvalue $\lambda_6 \rightarrow \infty$ with algebraic multiplicity $\ninf=2$. The rightmost eigenvalue of the pencil is $\lambda_3>0$, and thus the system is unstable.

\begin{figure}
    \centering
    \resizebox{0.26\linewidth}{!}{\includegraphics{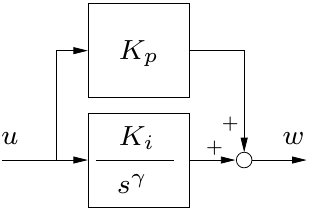}}
    \caption[FOPI block diagram]{FOPI block diagram.}
    \label{fig:fopi}
\end{figure}

Consider that the output of the system is given by \eqref{foc:eq:out},
where:

\begin{equation}
\begin{aligned}
\Csng &= 
  \begin{bmatrix*}[r]
 0 & 0 & 0 & 0 & 0 & 1 & 0\\ 
  \end{bmatrix*} , \ 
\Dsng = 0  \, . \nonumber
\end{aligned}  
\end{equation}
In order to stabilize the system, the following
simple form of
controller \eqref{foc:eq:focont1} is considered:
\begin{equation}
\begin{aligned}
\label{foc:eq:fopim}
   { x_c}^{(\gamma)} &=  K_i \, u \, , \\
   w &= {  x_c} +  K_p  \, u  \, .
\end{aligned}  
\end{equation}
Equation \eqref{foc:eq:fopim} describes a \ac{fopi} controller, where 
$K_p=7$, $K_i=10$, are the proportional and
integral gains, respectively; $\gamma=0.6$
is the controller's fractional order. The block diagram
of the \ac{fopi} is shown in Figure~\ref{fig:fopi}.

Then, the closed-loop system is described by \eqref{foc:eq:clsys}, where
\begin{equation}
\begin{aligned}
 \bfg M &= 
  \setlength{\arraycolsep}{1.5pt}
  \begin{bmatrix*}[r]
4 & 9 & 9 & -2 & 10 & 7 & 3 & 0 & 0\\ 
1 & 5 & 2 & 2 & 3 & 1 & 1 & 0 & 0\\ 
1 & 0 & -2 & -2 & 6 & 4 & 1 & 0 & 0\\ 
5 & -2 & -3 & 18 & 3 & 16 & 2 & 0 & 0\\ 
6 & 8 & 6 & 8 & 6 & 14 & 2 & 0 & 0\\ 
2 & 11 & 3 & 6 & 6 & 2 & 2 & 0 & 0\\ 
4 & 5 & 5 & 6 & 2 & 9 & 1 & 0 & 0\\ 
0 & 0 & 0 & 0 & 0 & 0 & 0 & 0 & 0\\ 
0 & 0 & 0 & 0 & 0 & 0 & 0 & 0 & 0 
  \end{bmatrix*}  , \ 
  \bfg M_\gamma = 
  \setlength{\arraycolsep}{3.1pt}
  \begin{bmatrix*}[r]
0 & 0 & 0 & 0 & 0 & 0 & 0 & 0 & 0 \\
0 & 0 & 0 & 0 & 0 & 0 & 0 & 0 & 0 \\
0 & 0 & 0 & 0 & 0 & 0 & 0 & 0 & 0 \\
0 & 0 & 0 & 0 & 0 & 0 & 0 & 0 & 0 \\
0 & 0 & 0 & 0 & 0 & 0 & 0 & 0 & 0 \\
0 & 0 & 0 & 0 & 0 & 0 & 0 & 0 & 0 \\
0 & 0 & 0 & 0 & 0 & 0 & 0 & 0 & 0 \\
0 & 0 & 0 & 0 & 0 & 0 & 0 & 1 & 0 \\ 
0 & 0 & 0 & 0 & 0 & 0 & 0 & 0 & 0 
  \end{bmatrix*}  , \\ 
\bfg A_{\rm cl} &= 
  \begin{bmatrix*}[r]
-15 & -43 & -39 & 4 & -35 & -22 & -5 & 0 & 0\\ -3 & -19 & -2 & -5 & -12 & -2 & 1 & 0 & 0\\ -4 & -16 & -30 & 6 & -9 & -1 & -7 & 0 & 0\\ -25 & -2 & 3 & -72 & -3 & -74 & -4 & 0 & 0\\ -27 & -32 & -23 & -39 & -24 & -66 & -5 & 0 & 0\\ -8 & -41 & -15 & -18 & -24 & -8 & -8 & 0 & 1.0\\ -18 & -15 & -9 & -29 & -13 & -48 & -1 & 0 & 0\\ 0 & 0 & 0 & 0 & 0 & 10 & 0 & 0 & 0\\ 0 & 0 & 0 & 0 & 0 & 7 & 0 & 1 & -1 
  \end{bmatrix*}  . \nonumber
\end{aligned}  
\end{equation}

The stability of the closed-loop system can be checked
by calculating the eigenvalues of the matrix pencil
$s \Efr - \Afr$, where $\Efr$, $\Afr$, are defined in \eqref{foc:eq:final2t}. 
In this case:

\setcounter{MaxMatrixCols}{65}

\begin{align}
\Efr =
\left[
\fontsize{13}{11}\selectfont
\setlength{\arraycolsep}{3.0pt}
\begin{array}
{ccccccccccccccccccc} 1  & 0 & 0 & 0 & 0 & 0 & 0 & 0 & 0 & 0 & 0 & 0 & 0 & 0 & 0 & 0 & 0 & 0\\ 0 & 1  & 0 & 0 & 0 & 0 & 0 & 0 & 0 & 0 & 0 & 0 & 0 & 0 & 0 & 0 & 0 & 0\\ 0 & 0 & 1  & 0 & 0 & 0 & 0 & 0 & 0 & 0 & 0 & 0 & 0 & 0 & 0 & 0 & 0 & 0\\ 0 & 0 & 0 & 1  & 0 & 0 & 0 & 0 & 0 & 0 & 0 & 0 & 0 & 0 & 0 & 0 & 0 & 0\\ 0 & 0 & 0 & 0 & 1  & 0 & 0 & 0 & 0 & 0 & 0 & 0 & 0 & 0 & 0 & 0 & 0 & 0\\ 0 & 0 & 0 & 0 & 0 & 1  & 0 & 0 & 0 & 0 & 0 & 0 & 0 & 0 & 0 & 0 & 0 & 0\\ 0 & 0 & 0 & 0 & 0 & 0 & 1  & 0 & 0 & 0 & 0 & 0 & 0 & 0 & 0 & 0 & 0 & 0\\ 0 & 0 & 0 & 0 & 0 & 0 & 0 & 1  & 0 & 0 & 0 & 0 & 0 & 0 & 0 & 0 & 0 & 0\\ 0 & 0 & 0 & 0 & 0 & 0 & 0 & 0 & 1  & 0 & 0 & 0 & 0 & 0 & 0 & 0 & 0 & 0\\ 0 & 0 & 0 & 0 & 0 & 0 & 0 & 0 & 0 & 4  & 9  & 9  & -2  & 10  & 7  & 3  & 0 & 0\\ 0 & 0 & 0 & 0 & 0 & 0 & 0 & 0 & 0 & 1  & 5  & 2  & 2  & 3  & 1  & 1  & 0 & 0\\ 0 & 0 & 0 & 0 & 0 & 0 & 0 & 0 & 0 & 1  & 0 & -2  & -2  & 6  & 4  & 1  & 0 & 0\\ 0 & 0 & 0 & 0 & 0 & 0 & 0 & 0 & 0 & 5  & -2  & -3  & 18  & 3  & 16  & 2  & 0 & 0\\ 0 & 0 & 0 & 0 & 0 & 0 & 0 & 0 & 0 & 6  & 8  & 6  & 8  & 6  & 14  & 2  & 0 & 0\\ 0 & 0 & 0 & 0 & 0 & 0 & 0 & 0 & 0 & 2  & 11  & 3  & 6  & 6  & 2  & 2  & 0 & 0\\ 0 & 0 & 0 & 0 & 0 & 0 & 0 & 0 & 0 & 4  & 5  & 5  & 6  & 2  & 9  & 1 & 0 & 0\\ 0 & 0 & 0 & 0 & 0 & 0 & 0 & 0 & 0 & 0 & 0 & 0 & 0 & 0 & 0 & 0 & 0 & 0\\ 0 & 0 & 0 & 0 & 0 & 0 & 0 & 0 & 0 & 0 & 0 & 0 & 0 & 0 & 0 & 0 & 0 & 0 
\end{array}
\right] \nonumber  ,
\end{align}
\begin{align}
\Afr =
\left[
\fontsize{13}{11}\selectfont
\setlength{\arraycolsep}{1.7pt}
\begin{array}
{ccccccccccccccccccc} 0 & 0 & 0 & 0 & 0 & 0 & 0 & 0 & 0 & 1 & 0 & 0 & 0 & 0 & 0 & 0 & 0 & 0\\ 0 & 0 & 0 & 0 & 0 & 0 & 0 & 0 & 0 & 0 & 1 & 0 & 0 & 0 & 0 & 0 & 0 & 0\\ 0 & 0 & 0 & 0 & 0 & 0 & 0 & 0 & 0 & 0 & 0 & 1 & 0 & 0 & 0 & 0 & 0 & 0\\ 0 & 0 & 0 & 0 & 0 & 0 & 0 & 0 & 0 & 0 & 0 & 0 & 1 & 0 & 0 & 0 & 0 & 0\\ 0 & 0 & 0 & 0 & 0 & 0 & 0 & 0 & 0 & 0 & 0 & 0 & 0 & 1 & 0 & 0 & 0 & 0\\ 0 & 0 & 0 & 0 & 0 & 0 & 0 & 0 & 0 & 0 & 0 & 0 & 0 & 0 & 1 & 0 & 0 & 0\\ 0 & 0 & 0 & 0 & 0 & 0 & 0 & 0 & 0 & 0 & 0 & 0 & 0 & 0 & 0 & 1 & 0 & 0\\ 0 & 0 & 0 & 0 & 0 & 0 & 0 & 0 & 0 & 0 & 0 & 0 & 0 & 0 & 0 & 0 & 1 & 0\\ 0 & 0 & 0 & 0 & 0 & 0 & 0 & 0 & 0 & 0 & 0 & 0 & 0 & 0 & 0 & 0 & 0 & 1\\ -15 & -43 & -39 & 4 & -35 & -22 & -5 & 0 & 0 & 0 & 0 & 0 & 0 & 0 & 0 & 0 & 0 & 0\\ -3 & -19 & -2 & -5 & -12 & -2 & 1 & 0 & 0 & 0 & 0 & 0 & 0 & 0 & 0 & 0 & 0 & 0\\ -4 & -16 & -30 & 6 & -9 & -1 & -7 & 0 & 0 & 0 & 0 & 0 & 0 & 0 & 0 & 0 & 0 & 0\\ -25 & -2 & 3 & -72 & -3 & -74 & -4 & 0 & 0 & 0 & 0 & 0 & 0 & 0 & 0 & 0 & 0 & 0\\ -27 & -32 & -23 & -39 & -24 & -66 & -5 & 0 & 0 & 0 & 0 & 0 & 0 & 0 & 0 & 0 & 0 & 0\\ -8 & -41 & -15 & -18 & -24 & -8 & -8 & 0 & 1 & 0 & 0 & 0 & 0 & 0 & 0 & 0 & 0 & 0\\ -18 & -15 & -9 & -29 & -13 & -48 & -1 & 0 & 0 & 0 & 0 & 0 & 0 & 0 & 0 & 0 & 0 & 0\\ 0 & 0 & 0 & 0 & 0 & 10 & 0 & 0 & 0 & 0 & 0 & 0 & 0 & 0 & 0 & 0 & -1 & 0\\ 0 & 0 & 0 & 0 & 0 & 7 & 0 & 1 & -1 & 0 & 0 & 0 & 0 & 0 & 0 & 0 & 0 & 0 
\end{array}
\right] \nonumber  ,
\end{align}

The pencil $s \Efr - \Afr$ has $\hat p=11$ 
distinct finite eigenvalues
$\hat \lambda_{1,2}=1.816 \pm \jj 2.679$,
$\hat \lambda_{3}=-1.130$, 
$\hat \lambda_{4,5}=1.840 \pm \jj 0.996$,
$\hat \lambda_{6,7}=-0.0109 \pm \jj 1.751$,
$\hat \lambda_{8,9}=-0.044 \pm \jj 1.940$,
$\hat \lambda_{10,11}=0.0211 \pm \jj 2.2004$,
and the infinite eigenvalue $\hat   \lambda_{12} \rightarrow \infty$ with algebraic 
multiplicity $\hat q = 7$.
For fractional order $\gamma=0.6$,
$\tilde \gamma=\rm min\left\{0.6, 0.4\right\}=0.4$ in equation \eqref{foc:eq:stability}, and thus,
the closed-loop system is stable
if the arguments 
$\rm Arg(\hat \lambda^*)$
of all finite eigenvalues $\hat \lambda^*$ satisfy:
\begin{equation}
\label{foc:eq:region1}
  |\rm Arg(\hat \lambda^*)| > \frac{\pi}{5}=0.628 \ \rm rad \, .
\nonumber
\end{equation}
\begin{figure}[ht]
    \centering
    \resizebox{0.85\linewidth}{!}{\includegraphics{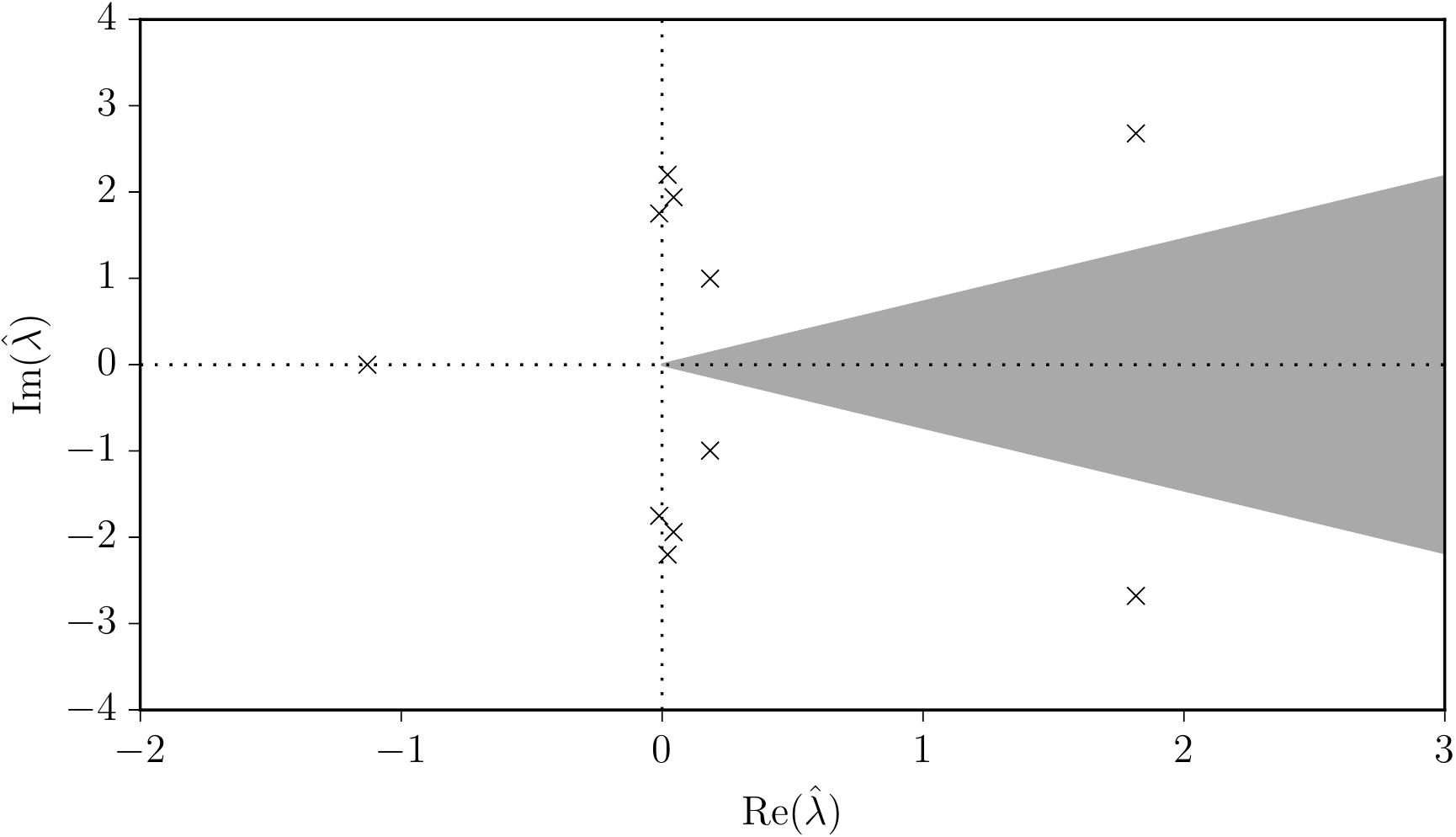}}
    \caption[Numerical example 1: root loci]{Numerical example 1: root loci of $s \Efr - \Afr$. Shaded is the region of instability $|\rm Arg(\hat \lambda^*)| < 0.628$~rad.}
    \label{foc:fig:num_fopss}
\end{figure}

The finite eigenvalues 
of $s \Efr - \Afr$ are illustrated 
in Figure~\ref{foc:fig:num_fopss}. The system is stable, since
all eigenvalues lie in the region given in \eqref{foc:eq:region1}.

\subsubsection{WSCC 9-bus System}

This example is based on the well-known \ac{wscc} 9-bus system, the data of which are provided in \cite{anderson:2003}.  
The system consists of 
3 synchronous machines, 
6 transmission lines, 
3 transformers and 
3 loads, modeled as constant power consumption.  Each machine provides
primary voltage and frequency control through an \ac{avr} and a \ac{tg}, respectively.  The
original system model does not include any fractional dynamics.

Suppose that a \ac{fopss} is installed at the
synchronous machine connected at bus 2. The \ac{fopss} employed has the
following transfer function:
\begin{align}
  \TFUN_{\rm FOPSS} &= K_{\rm w}
\bigg (    \dfrac{T_1 \, s^\gamma +1}{T_2 \,s^\gamma+1} \bigg )^2 \, .
\nonumber 
\end{align}
The controller input is the local rotor speed, while the output is 
an additional input to the algebraic
equation of the local \ac{avr} reference.
The \ac{fopss} can be written in the form of \eqref{foc:eq:focont1}, where:

\begin{equation}
 \bfg E_{c,{\gamma}} = 
 \begin{bmatrix}
    T_2 & 0 & 0 & 0\\
    T_1 & 0 & 0 & 0\\
    0 & 0 & T_2 & 0\\
     0 & 0 & T_1 & 0\\
  \end{bmatrix}
  \nonumber
  \ , \
  \bfg A_{c} = 
 \begin{bmatrix}
    -1 & 0 & 0  & 0\\
    -1 & 1 & 0  & 0\\
     0 & 1 & -1 & 0\\
     0 & 0 & -1 & 1\\
  \end{bmatrix} 
  \ , \
\end{equation}
\begin{equation}
  \bfg B_{c} = 
  \begin{bmatrix}
    K_{\rm w} & 0 & 0 & 0\\
  \end{bmatrix}\T
  \nonumber
   , \,
 \bfg C_{c} = 
 \begin{bmatrix}
    0 & 0 & 0 & 1\\
  \end{bmatrix}
  \nonumber
  , \ \bfg D_c = 0 .
\end{equation}

\begin{figure}[ht]
  \begin{center}
    \resizebox{0.85\linewidth}{!}{\includegraphics{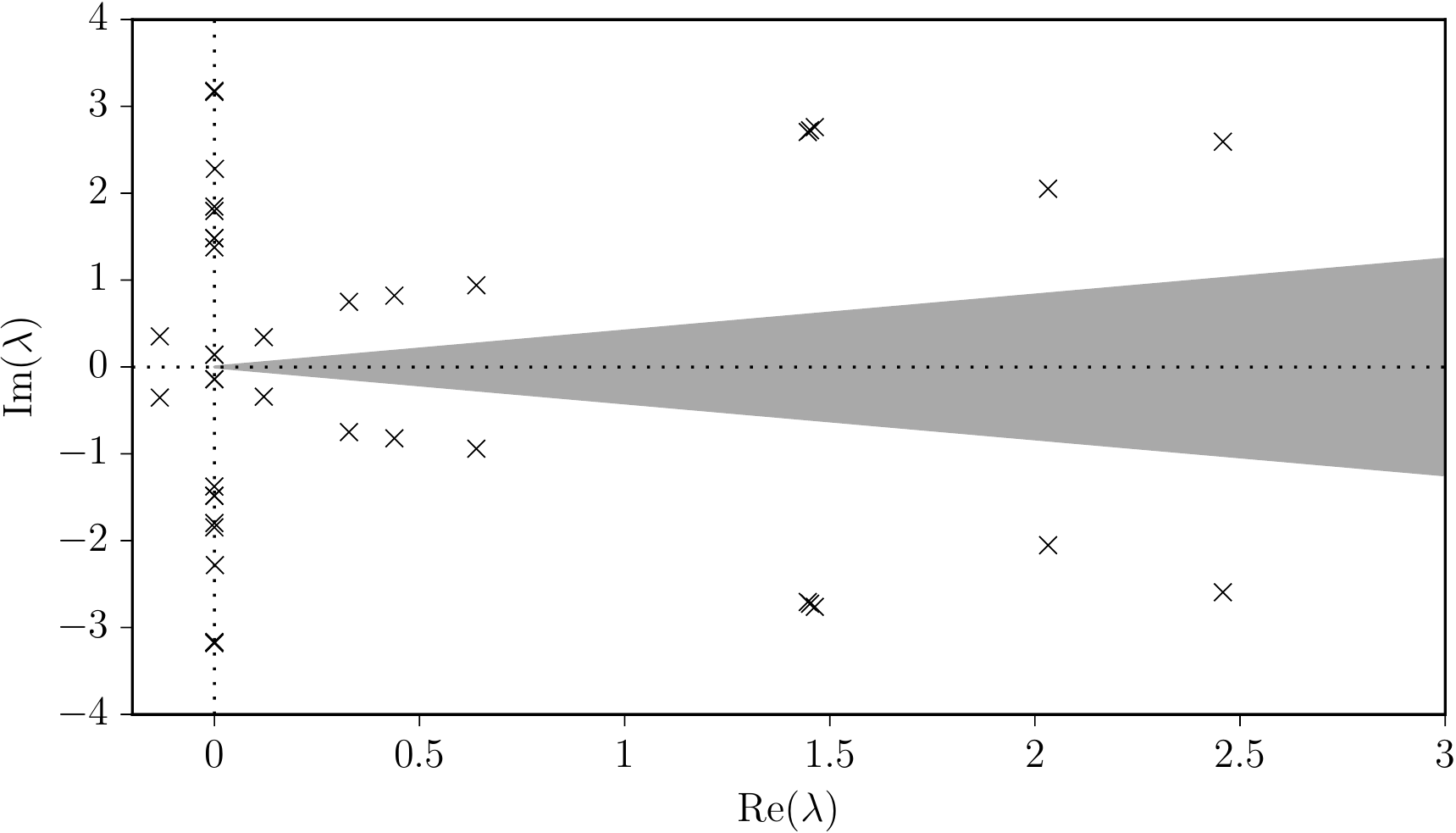}}
    \caption[WSCC system with FOPSS: most critical eigenvalues]{
    WSCC system with FOPSS: most critical eigenvalues. Shaded is the region of instability.}
    \label{fig:locus}
  \end{center}
\end{figure}

Suppose that 
$T_1=0.01$~s, $T_2=0.22$~s,
$\gamma=0.75$. Then, small-signal stability is assessed by calculating
the eigenvalues of \eqref{foc:eq:pencil}. From
\eqref{foc:eq:stability}, the system is stable if all
finite eigenvalues $\lambda^*$ satisfy:
\begin{equation}
\label{foc:eq:region2}
  |{\rm Arg}(\lambda^*)| > \tilde \gamma \, \frac{\pi}{2}
  = 0.393 \ \rm
  rad \, ,
  \nonumber
\end{equation}
where $\tilde \gamma = {\rm min}\left\{0.75, 0.25\right\}=0.25$.
The most critical eigenvalues of the closed-loop system are shown in
Figure~\ref{fig:locus}, where the shaded region is unstable. As it can
be seen, the system with the \ac{fopss} is in this case stable.

%


\section{Oustaloup's Recursive Approximation}
\label{foc:sec:oust}

\subsection{Formulation}

%
The theoretical analysis based on fractional calculus is essential for
a better understanding of ``ideal'' \acp{foc} and hence, for a robust \ac{foc}
design.  In practice, however, the implementation of \acp{foc} is typically
done by approximating the fractional derivatives and integrals with
rational transfer functions.  Although this is an
important aspect of \acp{foc} implementation, some studies omit mentioning
what approximation technique and/or parameters they use, effectively
forcing the adoption of a black-box approach.   In the following the \ac{ora} method, which is arguably the most common
continuous approximation technique, is employed.  The generalized \ac{ora} of a
fractional derivative of order $\gamma$ is defined as
\cite{monje:2010}:
\begin{align}
\label{foc:eq:oust}
s^\gamma \approx \omega_h^{\gamma} \,
\prod_{k=1}^{N}
\frac{s+\omega_k'}{s+\omega_k}
\ , 
\end{align} 
where 
\begin{align}
\omega_k' &= \omega_b
\, \omega_v^{(2k-1-\gamma)/N} \, ,
\nonumber 
\\
\omega_k &= \omega_b \, \omega_v^{(2k-1+\gamma)/N} \, ,
\nonumber 
\\
\omega_v &= \sqrt{\frac{\omega_h}{\omega_b}} \, .
\nonumber 
\end{align}

\begin{figure}[ht!]
  \begin{center}
  \resizebox{0.85\linewidth}{!}{\includegraphics{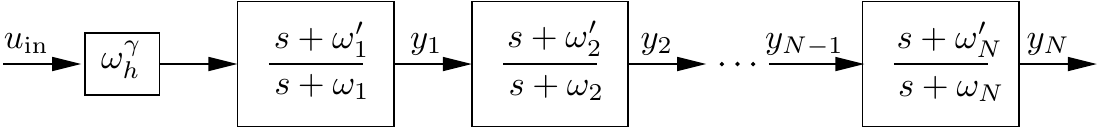}}
    \caption[Oustaloup's recursive approximation block diagram]{Oustaloup's recursive approximation block diagram.}
    \label{fig:oustaloup}
  \end{center}
\end{figure}

In the above expressions,
$[\omega_b,\omega_h]$ is the frequency range for which the
approximation is designed to be valid; $N$ is the order of the
polynomial approximation; The term ``generalized" implies that, in
\eqref{foc:eq:oust}, $N$ can be either even or odd \cite{monje:2010},
while the term ``recursive" implies that the values of $\omega_k'$,
$\omega_k$ result from a set of recursive equations
\cite{oustaloup:2000}.  The block diagram of \ac{ora} is shown in
Figure~\ref{fig:oustaloup}.

\begin{figure}[ht!]
\begin{center}
    \resizebox{0.85\linewidth}{!}{\includegraphics{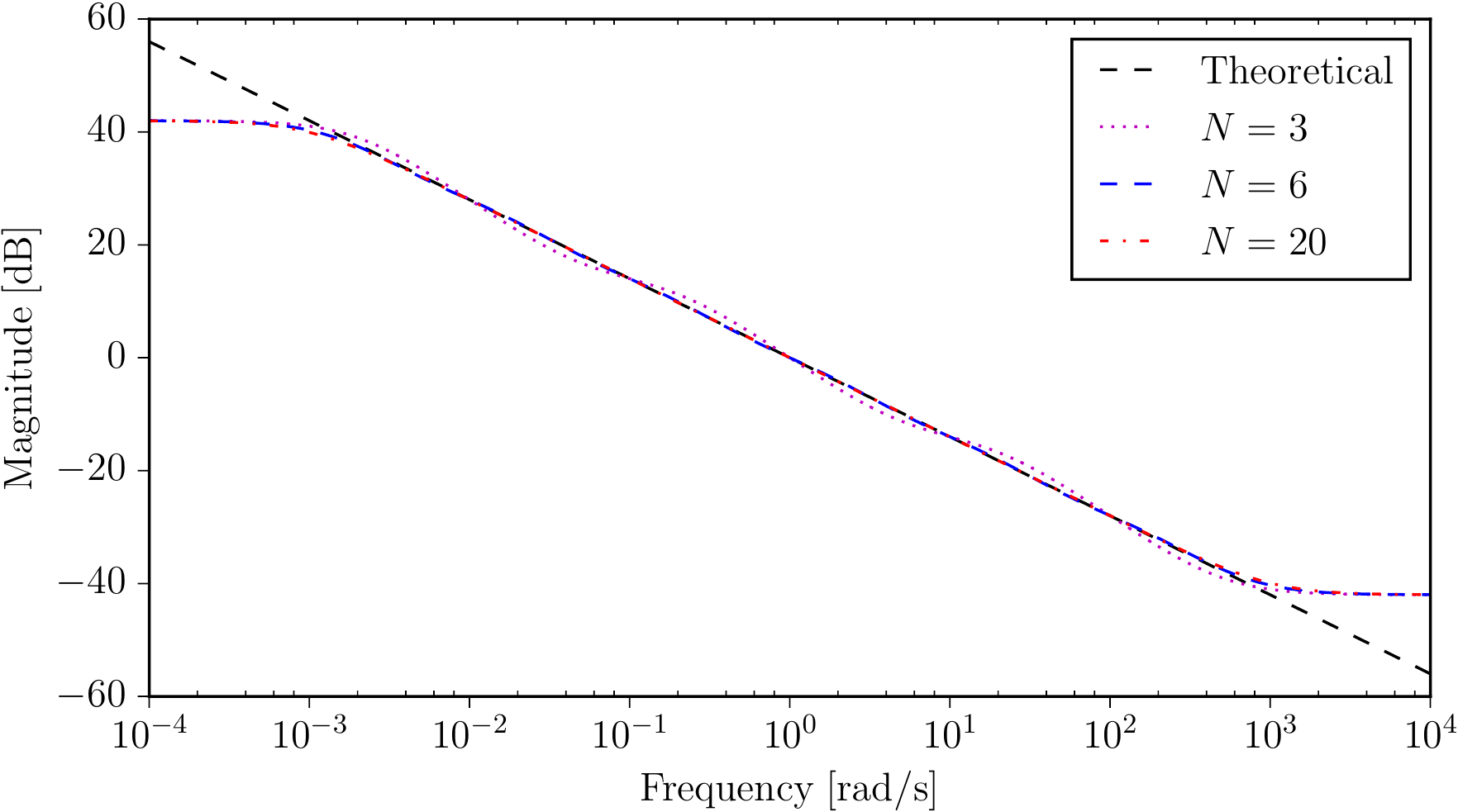}}
\end{center}
\begin{center}
\vspace{0.2cm}
\resizebox{0.85\linewidth}{!}{\includegraphics{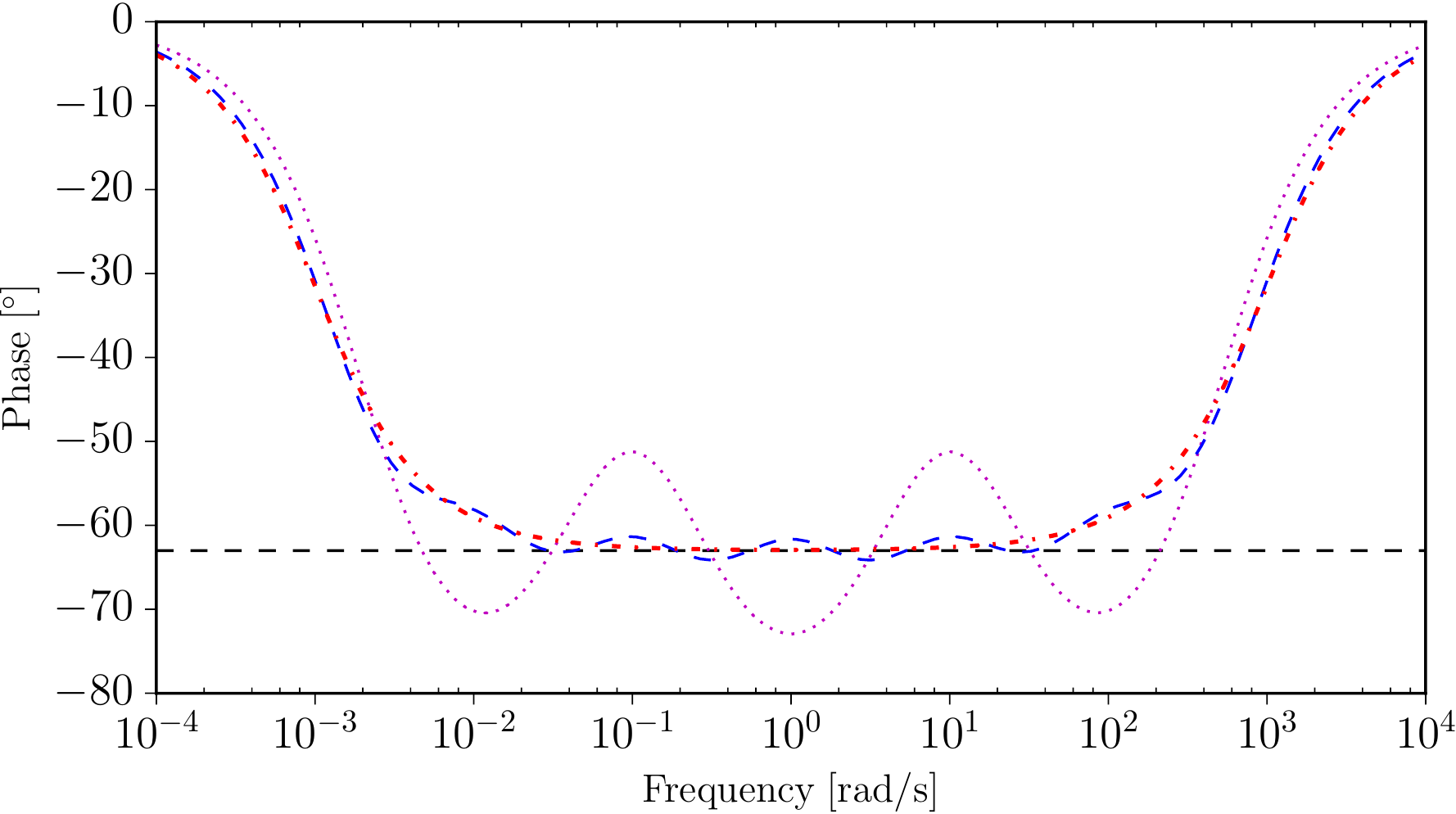}}
\end{center} 
\caption[Bode plot of the ORA]{Bode plot of the ORA for different approximation orders $N$
($\gamma = -0.7$, $[\omega_b,\omega_h]=[10^{-3},10^{3}]$~rad/s).}
\label{fig:bode1}
\end{figure}

Figure~\ref{fig:bode1} compares the theoretical frequency response of
$s^{-0.7}$, which is given by \eqref{foc:eq:magph} for $K=1$, with the
respective plots provided by \acp{ora} of different dynamic orders.  This
simple example shows the typical behavior of the \ac{ora}: the approximation
is more accurate for higher dynamic orders and for frequencies closer
to the middle of the interval $[\omega_b,\omega_h]$.

A final remark on equation \eqref{foc:eq:oust} is that the \ac{ora} of $s^\gamma$ is
typically accurate enough for fractional orders that satisfy
$0\leq |\gamma| \leq 1$.  For higher fractional orders, accuracy can
be maintained by implementing $s^\gamma$ as a multiplication of a
suitable integer order block and a fractional order block, as follows:
\begin{align}
  \label{foc:eq:separ}
  s^\gamma = s^n \, s^{\gamma-n}, \ n \in \mathbb{Z}, 
  \ (\gamma-n) \in [0,1] \ .
\end{align} 

\subsection{DAE Model}

In time-domain, the \ac{ora} can be described by the following set of
explicit \acp{dae}:
%
\begin{equation}
\begin{aligned}
\label{foc:eq:oustfin}
\Dt {\bfg \chi}_{{\scriptscriptstyle \rm E}} &= \bfg A_{{\scriptscriptstyle \rm E,ORA}} \, \bfg \chi_{{\scriptscriptstyle \rm E}}+ \bfg B_{{\scriptscriptstyle \rm E,ORA}} \, u_{\rm in} 
\,
\\
0 &= y_N - \bfg C_{{\scriptscriptstyle \rm E,ORA}} \, \bfg \chi_{{\scriptscriptstyle \rm E}} + \omega_h^{\gamma} \, u_{\rm in}  \, , 
\end{aligned} 
\end{equation}
where $\bfg \chi_{{\scriptscriptstyle \rm E}} = [ \chi_{{{\scriptscriptstyle \rm E}},1} \ \ \chi_{{{\scriptscriptstyle \rm E}},2} \cdots \ \chi_{{{\scriptscriptstyle \rm E}},N} ]\T $ is the \ac{ora} state vector; and
\begin{equation}
\begin{aligned}
\bfg A_{{\scriptscriptstyle \rm E,ORA}} &= 
  \footnotesize
  \begin{bmatrix}
-\omega_1 &    &   & &   \\
\omega'_2-\omega_2 & -\omega_2 &  &   &  \\
\omega'_3-\omega_3 & \omega'_3-\omega_3 & -\omega_3 &  &  \\
\vdots &  \vdots &  \vdots &  \ddots &  \\ 
    \omega'_N-\omega_N & \omega'_N-\omega_N & \cdots  & \omega'_N-\omega_N &  -\omega_N \\
  \end{bmatrix}
  \, , \nonumber \\
  \bfg B_{{\scriptscriptstyle \rm E,ORA}} &= 
  \footnotesize
  \begin{bmatrix}
    \omega_h^{\gamma} \, (\omega'_1-\omega_1) & \omega_h^{\gamma} \, (\omega'_2-\omega_2) & 
    \cdots & 
    \omega_h^{\gamma} \, (\omega'_N-\omega_N) \\
  \end{bmatrix}\T \, ,  \nonumber \\
  \bfg C_{{\scriptscriptstyle \rm E,ORA}} &= 
  \footnotesize
  \begin{bmatrix}
    1 & 1 & \cdots & 1 \\
  \end{bmatrix}\T \, . 
  \nonumber
  \end{aligned}
\end{equation}
The dimensions of $\bfg A_{{\scriptscriptstyle \rm E,ORA}}$, $\bfg B_{{\scriptscriptstyle \rm E,ORA}}$, $\bfg C_{{\scriptscriptstyle \rm E,ORA}}$,
are ${N \times N}$, ${N \times 1}$ and ${1 \times N}$,
respectively. 
%
As discussed in Chapter~\ref{ch:sssa},
an alternative way to describe a dynamic model is by using
a semi-implicit 
\ac{dae} formulation.
With this formulation, the \ac{ora} can be written as:
\begin{align}
\Dt \chi_{{\scriptscriptstyle \rm I},1} &= -\omega_1 \,\chi_{{\scriptscriptstyle \rm I},1} + 
\omega_h^{\gamma} \, u_{\rm in}    \nonumber \\
\Dt \chi_{{\scriptscriptstyle \rm I},2} - \Dt \chi_{{\scriptscriptstyle \rm I},1} &= \omega'_1 \, \chi_{{\scriptscriptstyle \rm I},1} -\omega_2 \, \chi_{{\scriptscriptstyle \rm I},2}    \nonumber \\
 & \vdots  \nonumber \\
\Dt \chi_{{\scriptscriptstyle \rm I},N}- \Dt \chi_{{\scriptscriptstyle \rm I},N-1} 
 &= \omega'_{N-1} \, \chi_{{\scriptscriptstyle \rm I},N-1} -
 \omega_{N} \, \chi_{{\scriptscriptstyle \rm I},N} \nonumber \\
-\Dt \chi_{{\scriptscriptstyle \rm I},N} &= \omega'_{N} \, \chi_{{\scriptscriptstyle \rm I},N} 
- y_N \nonumber \, ,
\end{align} 
where $\bfg \chi_{\scriptscriptstyle \rm I} = [ \chi_{{\scriptscriptstyle \rm I},1} \ \ \chi_{{\scriptscriptstyle \rm I},2} \cdots \ \chi_{{\scriptscriptstyle \rm I},N} ]\T $ is the \ac{ora} state vector; and, in matrix notation,
\begin{equation}
\begin{aligned}
\label{foc:eq:oustfinsemi}
\bfg E_{{\scriptscriptstyle \rm I,ORA}} \Dt \xys_{\scriptscriptstyle \rm I} &= \bfg A_{{\scriptscriptstyle \rm I,ORA}} \, \xys_{\scriptscriptstyle \rm I} + \bfg B_{{\scriptscriptstyle \rm I,ORA}} \, u_{\rm in} \, , 
\end{aligned} 
\end{equation}
where $\xys_{\scriptscriptstyle \rm I} = [ \bfg \chi_{\scriptscriptstyle \rm I} \ \ y_N]\T $ and
\begin{equation}
  \bfg E_{{\scriptscriptstyle \rm I,ORA}} = 
  \footnotesize
  \begin{bmatrix}
 1 &   & &       & & \\
-1 &  1   &  &       &  & \\
 & \ddots &  \ddots &  & & \\ 
 	 &  & -1 & 1  & & \\    
 &   &  & -1 & 1 &     \\
  &   &  &  & -1 & 0    \\
  \end{bmatrix} 
  \ , \ 
  \normalsize
  \bfg B_{{\scriptscriptstyle \rm I,ORA}} = 
  \footnotesize
  \begin{bmatrix}
    \omega_h^{\gamma}  \\
    0  \\
    \vdots \\
    0  \\  
   0  \\   
   0  \\
  \end{bmatrix} \ , \nonumber 
\end{equation}
\begin{equation}
  \bfg A_{{\scriptscriptstyle \rm I,ORA}} = 
  \footnotesize
  \begin{bmatrix}
-\omega_1 &   & &       &  &\\
\omega'_1 & -\omega_2 &  &     &   &\\
 & \ddots &  \ddots &  &  \\ 
 	 &  & \omega'_{N-2} & -\omega_{N-1}  &  &\\    
 &   &  & \omega'_{N-1}  & -\omega_N &\\
  &   &  &  & \omega'_N    & -1   \\
  \end{bmatrix}
  \ , \nonumber 
\end{equation}
where the dimensions of $\bfg E_{{\scriptscriptstyle \rm I,ORA}}$, $\bfg A_{{\scriptscriptstyle \rm I,ORA}}$, $\bfg B_{{\scriptscriptstyle \rm I,ORA}}$,
are ${(N+1) \times (N+1)}$, 
${(N+1) \times (N+1)}$ 
and ${(N+1) \times 1}$,
respectively. 

In \eqref{foc:eq:oustfin}, the total number of non-zero elements of the
coefficient matrices is $\Theta(N^2)$, whereas in
\eqref{foc:eq:oustfinsemi} it is $\Theta(N)$.  As a result, the
proposed semi-implicit model is sparser 
than the explicit one.  Moreover,
\eqref{foc:eq:oustfinsemi} prevents the input $u_{\rm in}$ to propagate
through the equations to the output.


\subsection{Steady State Error}
\label{foc:sec:sse}

Consider the simple \ac{foc}
with transfer function 
$\TFUN_c(s) = K {s^{\gamma}}$.
By approximating $s^{\gamma}$ from
\eqref{foc:eq:oust}, $\TFUN_c(s)$ can be written
as:
\begin{align}
\label{foc:eq:expan}
\TFUN_c(s) & \approx K \, \omega_h^{\gamma}
\,
\prod_{k=1}^{N}
\frac{s+\omega_k'}{s+\omega_k} 
= \frac{c_1 \, s^N + c_2 \, s^{N-1} + \cdots + c_N }
{d_1\,s^N + d_2 \, s^{N-1} + \cdots + d_N} 
\, ,
\end{align} 
with
\begin{equation}
\begin{aligned}
\label{foc:eq:cNdN}
c_N &=K \, \omega_h^{\gamma} \,  \prod_{k=1}^{N} \omega'_k 
=
K \, \omega^\gamma_h \, \omega^N_b
\, \omega_v^{( N-\gamma) }
 \, , \\
d_N &= \prod_{k=1}^{N} \omega_k 
=\omega^N_b \, \omega_v^{(N+\gamma ) }
 \, ,
\end{aligned} 
\end{equation}
where the expressions for $\omega_k$ and $ \omega'_k $ have been substituted 
as in \eqref{foc:eq:oust} and $\sum^N_{k=1} = \frac{N(N+1)}{2}$.
From \eqref{foc:eq:expan}, one can deduce that the 
controller's unity feedback closed-loop steady state error 
$e(t \rightarrow \infty)$ for
an arbitrary input $U(s)$ is:
\begin{align}
\label{foc:eq:sserror}
e(t \rightarrow \infty) 
&=  \lim_{s \rightarrow 0} 
\frac{s \, U(s)}{1+\TFUN_c(s)} \nonumber \\ &= 
\frac{d_N }{c_N + d_N} \lim_{s \rightarrow 0} {s \, U(s)}
\nonumber  \\
&= \frac{1}{K+\omega^\gamma_b} \lim_{s \rightarrow 0} {s \, U(s)} \, ,
\end{align} 
where $\omega_v = \sqrt{\omega_h/\omega_b}$.  The
steady-state error in \eqref{foc:eq:sserror} depends on $K$,
$\omega^\gamma_b$, and the applied input $U(s)$.  Evaluating the
controller's unit step input response yields $U(s)=1/s$ and
$e_{\rm step}(t \rightarrow \infty) = {1}/{(K+\omega^\gamma_b)}$. Considering
$\gamma<0$ yields that an \ac{ora}-based \ac{FO} integral controller is not
perfect tracking.  This result is 
not consistent with the theoretical
behavior of $K s^\gamma$, $\gamma<0$, which has
$e_{\rm step}(t \rightarrow \infty) = 0$.  However, the design of an
almost perfect-tracking \ac{foc} is possible with appropriate selection
of \ac{ora} parameters.

\subsection{Parameters Selection}
\label{foc:sec:param}

While the value of $N$ is 
usually constrained due to
computational concerns, most studies that consider \ac{ora}-based \acp{foc} in
power systems provide a rather arbitrary selection of the range of
frequencies $[\omega_b,\omega_h]$.  This section discusses the tuning
of \ac{ora} parameters and provides an empirical rule that simplifies the
design of \acp{foc}.

\begin{figure}[ht!]
  \begin{center}
    \resizebox{0.85\linewidth}{!}{\includegraphics{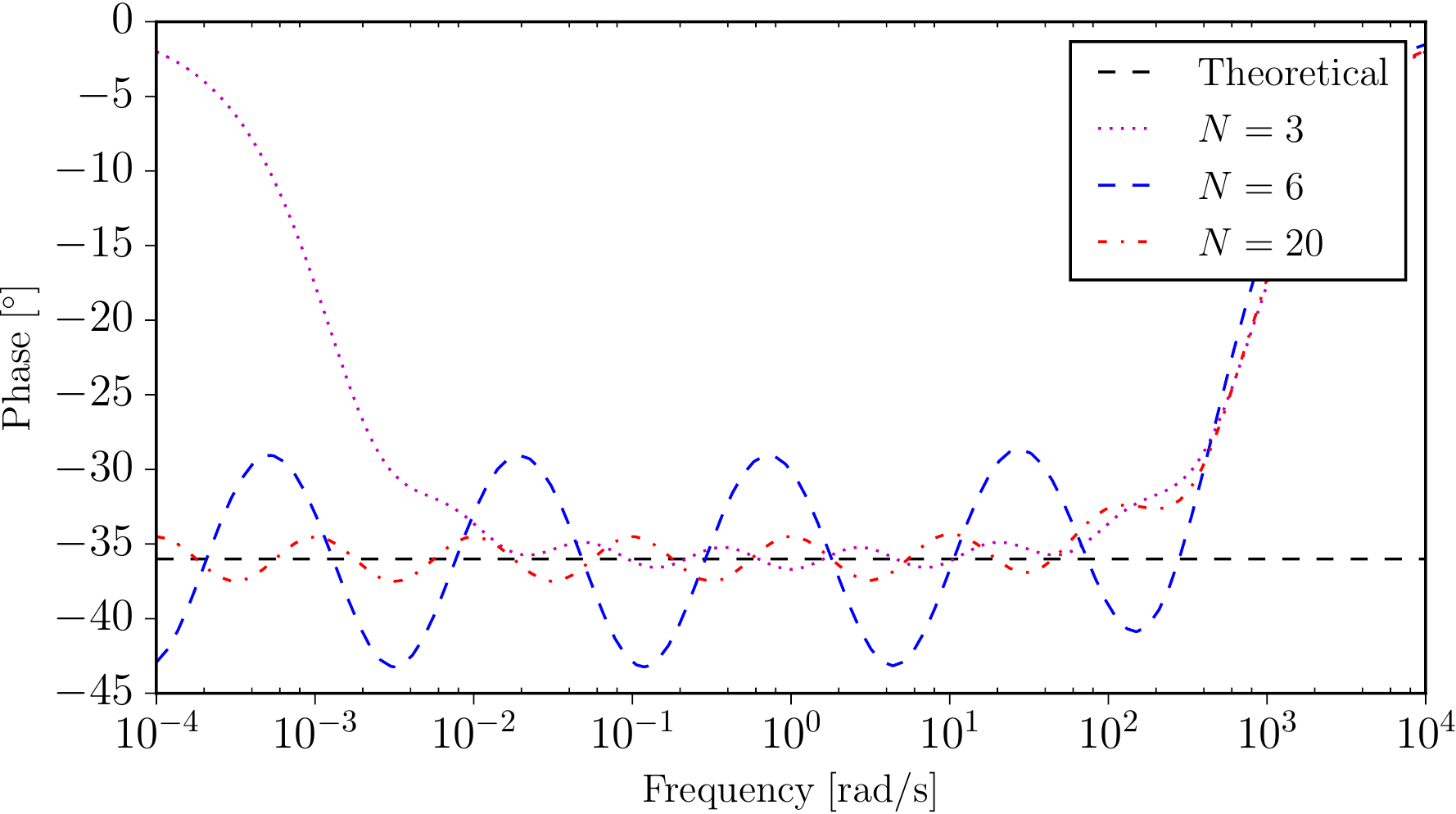}}
  \end{center} 
  \caption[ORA frequency response]{Effect of $\omega_b$ on ORA frequency response
($\gamma = -0.4$, $\omega_h=10^{3}$~rad/s).}
\label{fig:bode2}
\end{figure}

\begin{itemize}
\item \textit{Low frequency $\omega_b$:} A very small $\omega_b$ reduces the steady state error in
\eqref{foc:eq:sserror}.  However, a poor choice can significantly degrade
the phase fitting of \ac{ora}.  An example is shown in
Figure~\ref{fig:bode2}, where $\omega_b$ is varied from $10^{-3}$
to $10^{-8}$~rad/s.
\item \textit{High frequency $\omega_h$:} A very high $\omega_h$ may increase the system gain margin.  Large
gains lead to fast response and stability enhancement, as well as to 
elimination of steady state errors.  However, increasing excessively
the speed of the system response may trigger closed-loop resonant
points.  Note that such resonant points can remain undetected if they
stem from unmodeled high frequency dynamics.
\item \textit{Approximation order $N$:}
The phase fitting degradation caused by the decrease of $\omega_b$ can
be compensated by increasing the dynamic order $N$, e.g.~from $7$ to
$11$ (see Figure~\ref{fig:bode2}).  Increasing $N$ has an impact on the computational
complexity, which can be a serious constraint, especially if multiple
filters are required and if a large system (like real-world
power systems) is studied.  Another possible problem of a very high
$N$ is that multiple poles are placed very close to each other and
close to the imaginary axis.  For digital filters, such a pole-placement
may affect the discretization process, with multiple poles being
mapped on the unity circle, due e.g.~to rounding errors.
\end{itemize}

Control parameters need to provide an adequate compromise among
accuracy, computational burden and performance.  A good practice is to
limit the range of $[\omega_b,\omega_h]$ to the frequencies of the
dynamics of interest.  This also avoids unexpected resonances, as
discussed above.  Then, given a range let's say $\omega_b = 10^{-\upnu_b}$ and
$\omega_h = 10^{\upnu_h}$, $\upnu_b, \upnu_h \in \mathbb{N}$, a choice that provides
a very good compromise is $N = \upnu_b + \upnu_h$, with $N \geq 4$.

\section{Case Studies}
\label{foc:sec:casestudies}

\subsection{WSCC 9-bus System}
\label{foc:sec:case1}

This section presents three 
power system applications of \acp{foc}.  
(i) a \ac{FO} integral controller for secondary
frequency regulation; (ii) a \ac{FO} lead-lag controller for
primary frequency regulation of an \ac{ess}; 
and (iii) the
voltage regulation provided by a \ac{statcom} with inclusion of multiple \ac{fopi} controllers. 
In these three examples, the pre-disturbance equilibrium of the fractional \ac{dae} model is stable, i.e.~condition
\eqref{foc:eq:stability} holds. The focus is on
time-domain simulations carried to 
discuss the dynamic performance of \ac{ora}-based
\acp{foc} and check the system stability under
large disturbances. In all cases, the system is numerically integrated using the implicit trapezoidal method. A brief description of implicit integration of power systems is provided in Section \ref{osda:sec:tdi}.
Examples of this section are based on the \ac{wscc} 9-bus system. 
All simulation results are obtained with Dome.



\subsubsection*{Automatic Generation Control}
\label{foc:sec:agc}

In this example, an \ac{agc}, that coordinates the three generators and
provides secondary frequency regulation, is included in the \ac{wscc} system.
The \ac{agc} measures the \ac{coi} frequency
($\omega_{\rm CoI}$) and produces a dynamic active power
signal ($P_{\rm s}$), which is sent to the synchronous
generator turbine governors, and is proportional to their droops.  The
power order ($P_{{\rm ord},{i}}$) received by the $i$-th governor is:
\begin{align}
  \label{foc:eq:agc}
  P_{{\rm ord},{i}} &= \frac{\mathcal{R}_i}{\mathcal{R}_T} \, P_{\rm s} \, ,
\quad \quad 
i = \{1, 2, 3\} \, , 
\end{align}
where $\mathcal{R}_i$ is the $i$-th \ac{tg} droop constant; and
$\mathcal{R}_T = \mathcal{R}_1 + \mathcal{R}_2 + \mathcal{R}_3$.
The simplest model of an \ac{agc} assumes an integral controller. The
differential equation that describes the dynamic behavior of a \ac{foi} \ac{agc} is:
\begin{align}
  \label{foc:eq:foagc}
P_{\rm s}^{(\gamma)}&= K_{i} \,
 (\omega^{\rm ref} - \omega_{\rm CoI})\ , 
\end{align}
where $K_{i}$ is the \ac{foi}-\ac{agc} gain; $\omega^{\rm ref}$ is the reference
angular frequency; and $\gamma$ is the order of integration.  The \ac{IO}
version of this controller (I-\ac{agc}) is obtained for $\gamma=1$.

It is of interest to compare the performance of the I-AGC and the \ac{ora}-based \ac{foi}-\ac{agc}.
With this aim, a three-phase fault is considered at bus 4 occurring at
$t=3$~s.  After $80$~ms, the line that connects buses 4 and 5 trips
and the fault is cleared.

The parameters of both controllers are tuned by optimizing the \ac{coi} frequency profile through
trial-and-error. The I-\ac{agc}
gain is $K_i = 15$, while the parameters of the \ac{foi}-\ac{agc} are $K_i = 50$ and $\gamma = 0.7$.

\begin{figure}[ht]
  \begin{center}
    \resizebox{0.85\linewidth}{!}{\includegraphics{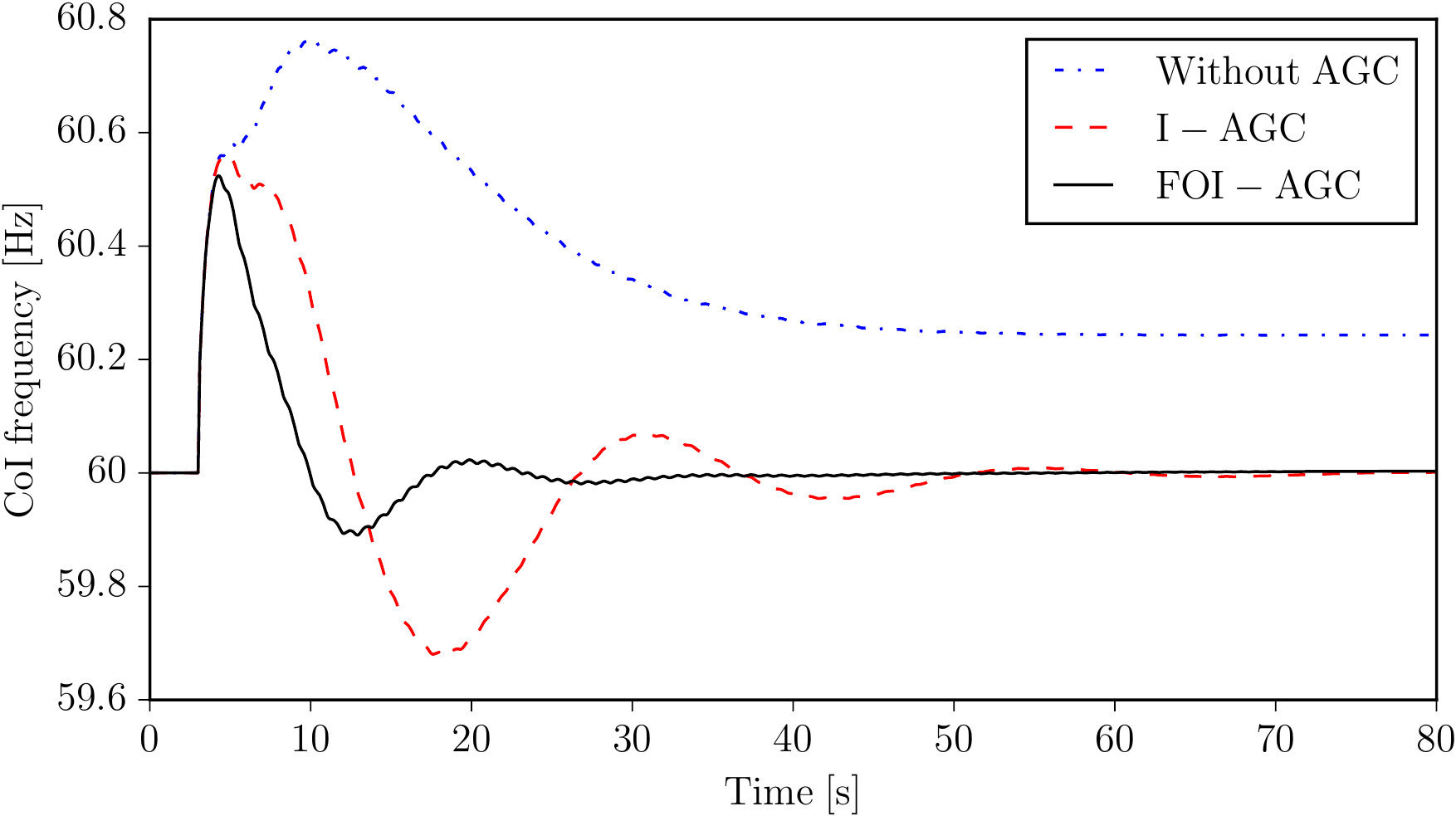}}
    \caption[WSCC system with AGC: CoI frequency]{WSCC system with AGC: CoI frequency.}
    \label{fig:agc1}
  \end{center}
\end{figure}

Taking into account the discussion in
Section~\ref{foc:sec:param}, the \ac{ora} parameters are set to
$[\omega_b, \omega_h] = [10^{-3}, 10^1]$~rad/s, $N=4$.
Figure~\ref{fig:agc1} shows the \ac{coi} frequency response of the
system without \ac{agc}; with I-\ac{agc}; with \ac{foi}-\ac{agc}.  The \ac{foi}-\ac{agc} improves
significantly the dynamic response of the frequency of the system.
Note that, with the selected parameters, the \ac{foi}-\ac{agc} achieves
practically a perfect-tracking behavior.

\subsubsection*{Energy Storage System}
\label{foc:sec:ess}

\begin{figure}[ht]
  \begin{center}
    \resizebox{0.7\linewidth}{!}{\includegraphics{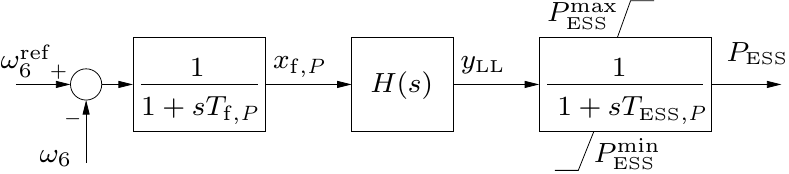}}
    \caption[Active power flow of simplified ESS model]{Active power flow of simplified ESS model.}
    \label{fig:ess}
  \end{center}
\end{figure}

In this example, a converter-interfaced \ac{ess} is
installed at bus 6 of the 9-bus system. A simplified model is employed
to describe the \ac{ess} dynamics.  Figure~\ref{fig:ess} shows the block
diagram of the \ac{ess} active power control.  The \ac{ess} measures the
local frequency at bus 6 and regulates its active power $P_{\sss {\rm ESS}}$
to provide frequency support.  The frequency error
$\omega^{\rm ref}_6 - \omega_6 $ is filtered.
$T_{{\rm f},\it{P}}$ is the time constant of the applied filter and
$x_{\rm f,\it{p}}$ is the filtered signal 
as well as the input of the
frequency control transfer function $\TFUN(s)$.  Finally, $T_{{\sss {\rm ESS}}, \it{P}}$ is the time
constant of the \ac{ess} active power dynamics.  The interested reader can
find more details on the employed \ac{ess} model in
\cite{milano_ortega:2019}.

In this example, $\TFUN(s)$ is assumed to be a \ac{FO} lead-lag controller defined as:
\begin{equation}
  \label{foc:eq:follgs}
  \begin{aligned}
  \TFUN(s) = K \, \frac{T_{\rm 1} \, s^\gamma +1}{T_{\rm 2} \, s^\gamma +1 } \, .
  \end{aligned} 
\end{equation}
The
equations that describe the \ac{FO} lead-lag are:
\begin{equation}
  \label{foc:eq:foll}
  \begin{aligned}
    T_{2} \, x^{(\gamma)}_{\sss \rm LL} &=K \, x_{{\rm f},\it{P}}-x_{\sss \rm LL} \, ,  \\
    y_{\sss \rm LL} &= x_{\sss \rm LL} + T_1 \, x^{(\gamma)}_{\sss \rm LL} \, ,  
  \end{aligned} 
\end{equation}
where $x_{\sss \rm LL} $ is the controller's state.  The \ac{IO} version of this
controller (\ac{IO} lead-lag) is obtained for $\gamma=1$.

\begin{figure}[ht]
  \begin{center}
    \resizebox{0.85\linewidth}{!}{\includegraphics{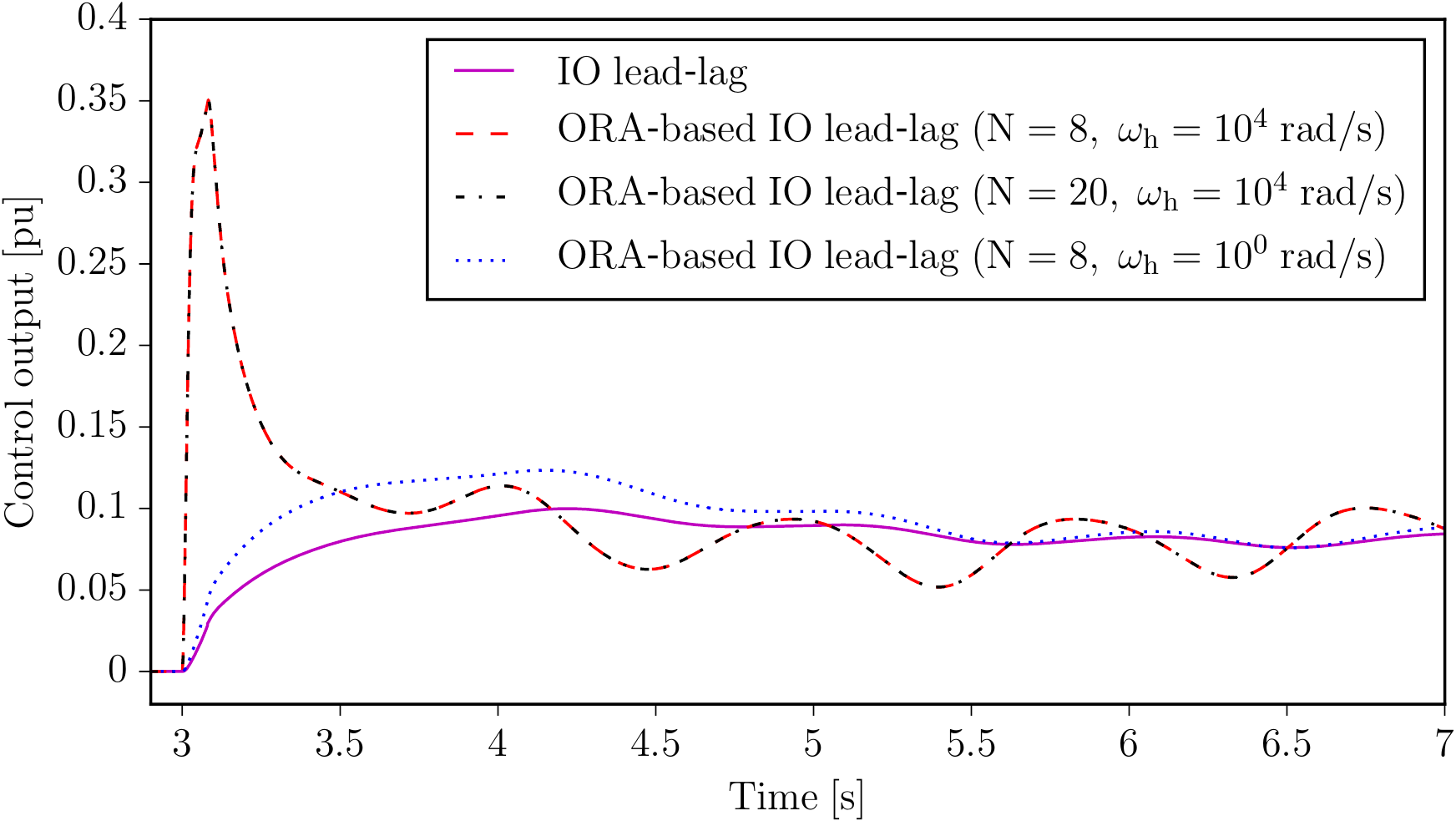}}
    \caption[WSCC system with ESS frequency control]{WSCC system with ESS frequency control: lead-lag
      parameters: $T_{1}=2$~s, $T_{2}=0.01$~s, $K=10$. ORA parameters:
      $\omega_b=10^{-4}$~rad/s, $\gamma=1$.}
    \label{fig:ess1}
  \end{center}
\end{figure}

We consider the same disturbance examined at the previous example
(fault at bus 4 cleared after $80$~ms).  Two implementations of the IO
lead-lag are compared, namely, the \ac{IO} lead-lag and the \ac{ora}-based IO
lead-lag controller for $\gamma=1$,
$[\omega_b, \omega_h]=[10^{-4},10^4]$~rad/s.  The results are shown in
Figure~\ref{fig:ess1}.  The value $\omega_{h}=10^4$~rad/s is high enough
to trigger a closed-loop high frequency resonant point, which
significantly impacts the control output.  Figure~\ref{fig:ess1} also shows that,
while it is independent from the approximation order, the overshoot
can be avoided by properly reducing the value of $\omega_h$.

Next, the dynamic performance of the \ac{IO} lead-lag is compared with two
\ac{ora}-based \ac{FO} lead-lags, namely ${\rm FOLL}1$ and ${\rm FOLL}2$, which
have different tuning.  The parameters of the three controllers are
shown in Table \ref{tab:esspars}.  For comparison, the gain and time
constants of ${\rm FOLL}1$ have been set equal to the ones of the \ac{IO}
lead-lag.  In this case, only the order $\gamma$ needs to be tuned.
In general, however, the control parameters of a \ac{foc} are not directly
mapped onto those of its \ac{IO} version and should be retuned.
${\rm FOLL}2$ represents the retuned controller. 
 
To tune ${\rm FOLL}2$, 
$T_1=2$~s is fixed and
the rest of the parameters are selected by optimizing the local bus frequency profile through trial-and-error. 

\begin{figure}[ht]
  \begin{center}
    \resizebox{0.85\linewidth}{!}{\includegraphics{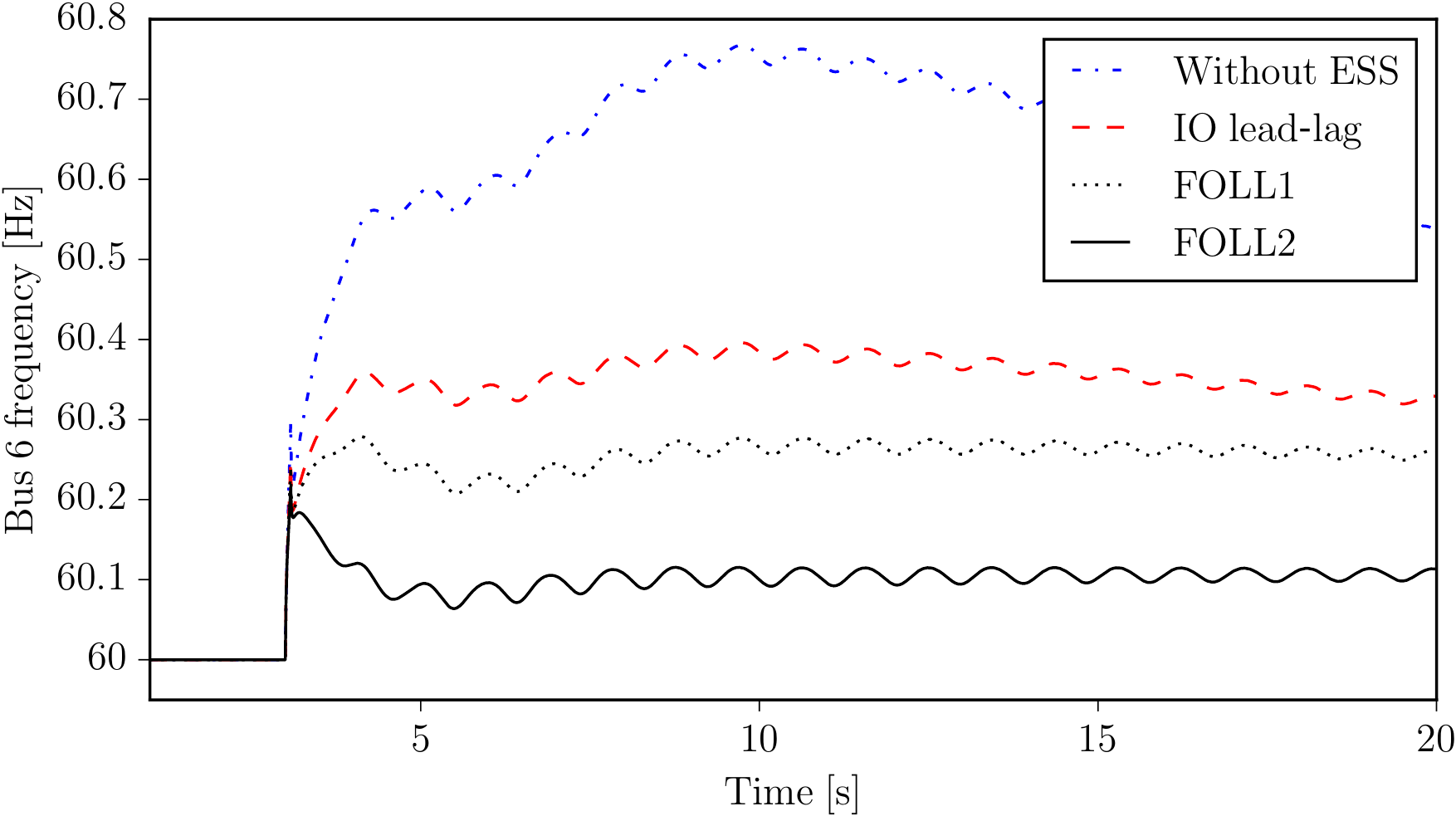}}
    \caption[WSCC system with ESS frequency control]{WSCC system with ESS frequency control: ORA parameters:
      $[\omega_b, \omega_h] = [10^{-3}, 10^1]$~rad/s, $N=4$.}
    \label{fig:ess2}
  \end{center}
\end{figure}

The response of the frequency at bus 6 is shown in
Figure~\ref{fig:ess2}.  Shifting the fractional order $\gamma$
(${\rm FOLL}1$) allows reducing both the frequency overshoot and the
steady state error of the local bus frequency.  Retuning all control
parameters leads to a further performance improvement (${\rm FOLL}2$).


\begin{table}[ht]
  \centering
  \renewcommand{\arraystretch}{1.3}
  \caption[Parameters of the ESS lead-lag frequency controllers]{Parameters of the ESS lead-lag frequency controllers.}
  \begin{tabular}{l|l}
    \hline 
    \ac{IO} lead-lag &  $T_{1}=2$~s, $T_{2}=0.01$~s, $K=20$,  \\
    ${\rm FOLL}1$ &  $T_{1}=2$~s, $T_{2}=0.01$~s, $K=20$, $\gamma=0.3$ \\
    ${\rm FOLL}2$ &  $T_{1}=2$~s, $T_{2}=0.005$~s, $K=60$, $\gamma=0.2$ \\
    \hline 
  \end{tabular}
  \label{tab:esspars}
\end{table}

\subsubsection*{STATCOM}
\label{foc:sec:statcom}

In this example, a \ac{statcom} connected to bus 8 provides
reactive power support.  The reactive power
variations provided by the \ac{statcom} rely on the control of a \ac{vsc}.  The \ac{vsc} is represented by an average value
model.  It consists of an AC/DC converter, an AC-side high
voltage/medium voltage transformer, and a DC-side condenser.  The \ac{vsc}
parameters are given in \cite{ahsan:2018}.

\begin{figure}[!ht]
  \begin{center}
    \resizebox{1\linewidth}{!}{\includegraphics{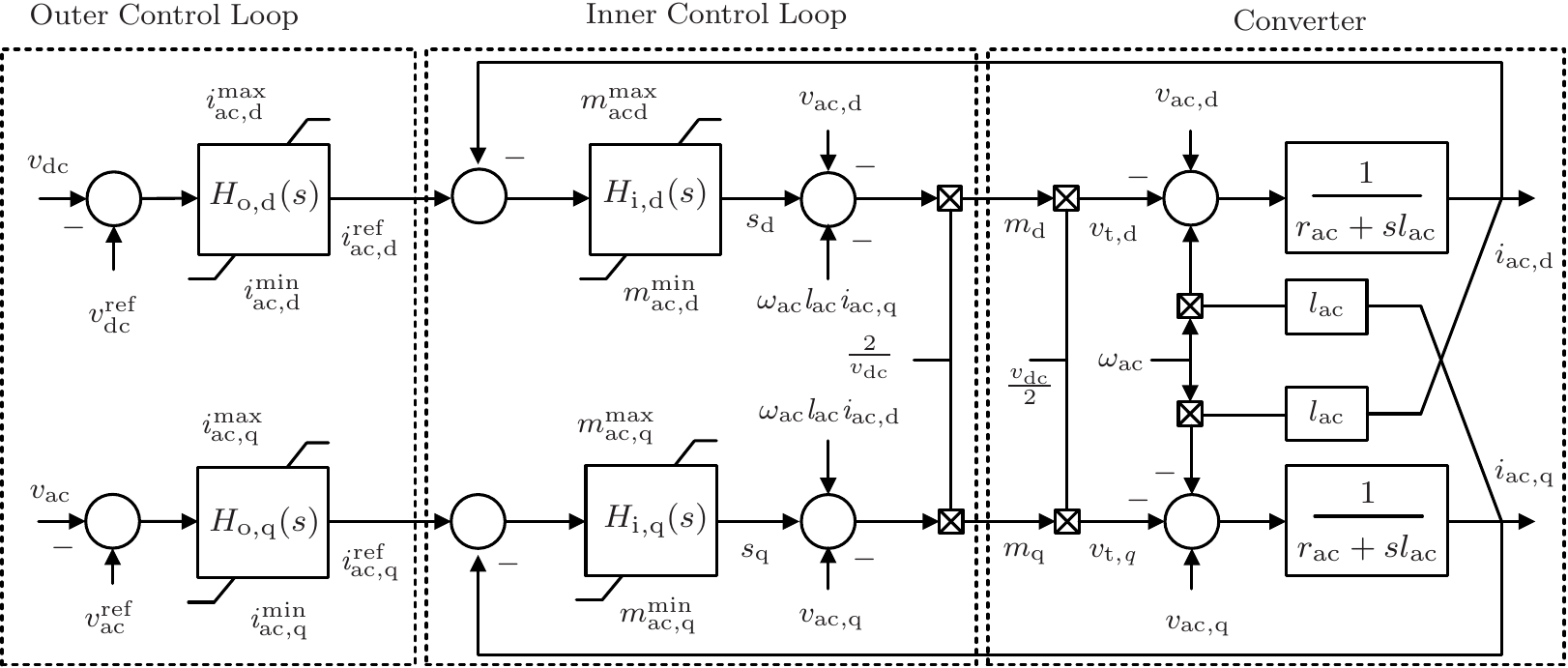}}
    \caption[VSC outer and inner control in \dqax-frame]{VSC outer and inner control in \dqax-frame.}
    \label{fig:vsccon}
  \end{center}
\end{figure}

The \ac{vsc} is controlled by employing a vector-current control strategy.
The control is based on a \dqax-axis 
reference frame and a phase-locked loop refers all phases to
the AC side voltage phasor angle \cite{amirnaser}.  The block diagram
of the considered vector-current control is depicted in
Figure~\ref{fig:vsccon}.  The \dax- and \qax- axis current components are
decoupled by the inner control loop, through the controllers
$\TFUN_{{\rm i},\dax}(s)$ and $\TFUN_{{\rm i},\qax}(s)$, respectively.  In the
\ac{statcom} configuration, the outer control loop utilizes the \dax-axis and \qax-axis
current components to provide regulation of the DC and AC voltages,
through the controllers $\TFUN_{{\rm o},\dax}(s)$ and $\TFUN_{{\rm o},\qax}(s)$,
respectively.

$\TFUN_{{\rm i},\dax}(s)$, $\TFUN_{{\rm i},\qax}(s)$,
$\TFUN_{{\rm o},\dax}(s)$ and $\TFUN_{{\rm o},\qax}(s)$, are assumed to be \ac{fopi} controllers.  The equations that describe
the behavior of the \ac{fopi} are:
\begin{equation}
  \begin{aligned}
    \label{foc:eq:fopi}
    x^{(\gamma)}_G &= K_{i} \, u_G
    \ , \\
    y_G &= K_{p} \, u_G + x_G
    \ , 
  \end{aligned} 
\end{equation}
where $K_p$ and $K_i$, are the proportional and integral gains,
respectively; $x_G$, $y_G$, are the state and output variable of the
controller, respectively; and $u_G$ is the controller input.  The \ac{IO} version of this controller, i.e. the classical \ac{PI} controller, is
obtained for $\gamma=1$.

To study the impact of the \ac{statcom} voltage regulation, a
stressed operating condition of the \ac{wscc} system is considered.  With this aim, the consumed power is increased by $60\%$
compared to the base case.  Then, for the purpose of transient
analysis, an additional $15\%$ consumption increase of the load
connected at bus 8 is considered, occurring at $t=3$~s.  The system
response is compared for the three following scenarios: without the
\ac{statcom}; with the \ac{statcom} connected and all four controllers modeled
as classical \acp{PI}; with the \ac{statcom} connected and the four controllers modeled as
\ac{ora}-based \acp{fopi}.

The values of the \ac{statcom} control parameters are shown in
Table~\ref{tab:statcompars}. 
The inner control loop parameters are tuned based on the pole cancellation technique as in
\cite{ahsan:2019}, while the outer
control loop parameters are tuned 
by optimizing the local bus voltage
profile though trial-and-error. 

Regarding the \ac{ora} parameters of the \acp{fopi}, we
have set the frequency range at $[10^{-3}, 10^{2}]$ rad/s for the
inner control loop; at $[10^{-4}, 10^{1}]$~rad/s for the outer control
loop.  The dynamic order is $N=5$ for all \ac{fopi} controllers.

\begin{table}[ht]
  \centering
  \renewcommand{\arraystretch}{1.2}
  \caption[Parameters of the STATCOM controllers]{Parameters of the STATCOM controllers.}
  \begin{tabular}{l|rr|rrr}
    \hline
    & \multicolumn{2}{c|}{\ac{PI}} & \multicolumn{3}{c}{\ac{fopi}} \\
    &$K_p$ & $K_i$  & $K_p$ & $K_i$ & $\gamma$ \\ 
    \hline
    $\TFUN_{{\rm i},\dax}(s)$ &$0.2$ & $20$ &$0.2$ & $20$&  $0.20$ \\
    $\TFUN_{{\rm i},\qax}(s)$& $0.2$& $20$&$0.2$ & $20$& $0.25$ \\
    $\TFUN_{{\rm o},\dax}(s)$& $50$& $25$&$50$& $25$& $0.40$ \\
    $\TFUN_{{\rm o},\qax}(s)$& $2.3$& $6$&$2.3$& $80$& $0.50$ \\
    \hline 
  \end{tabular}
  \label{tab:statcompars}
\end{table}

\begin{figure}[ht]
  \begin{center}
    \resizebox{0.85\linewidth}{!}{\includegraphics{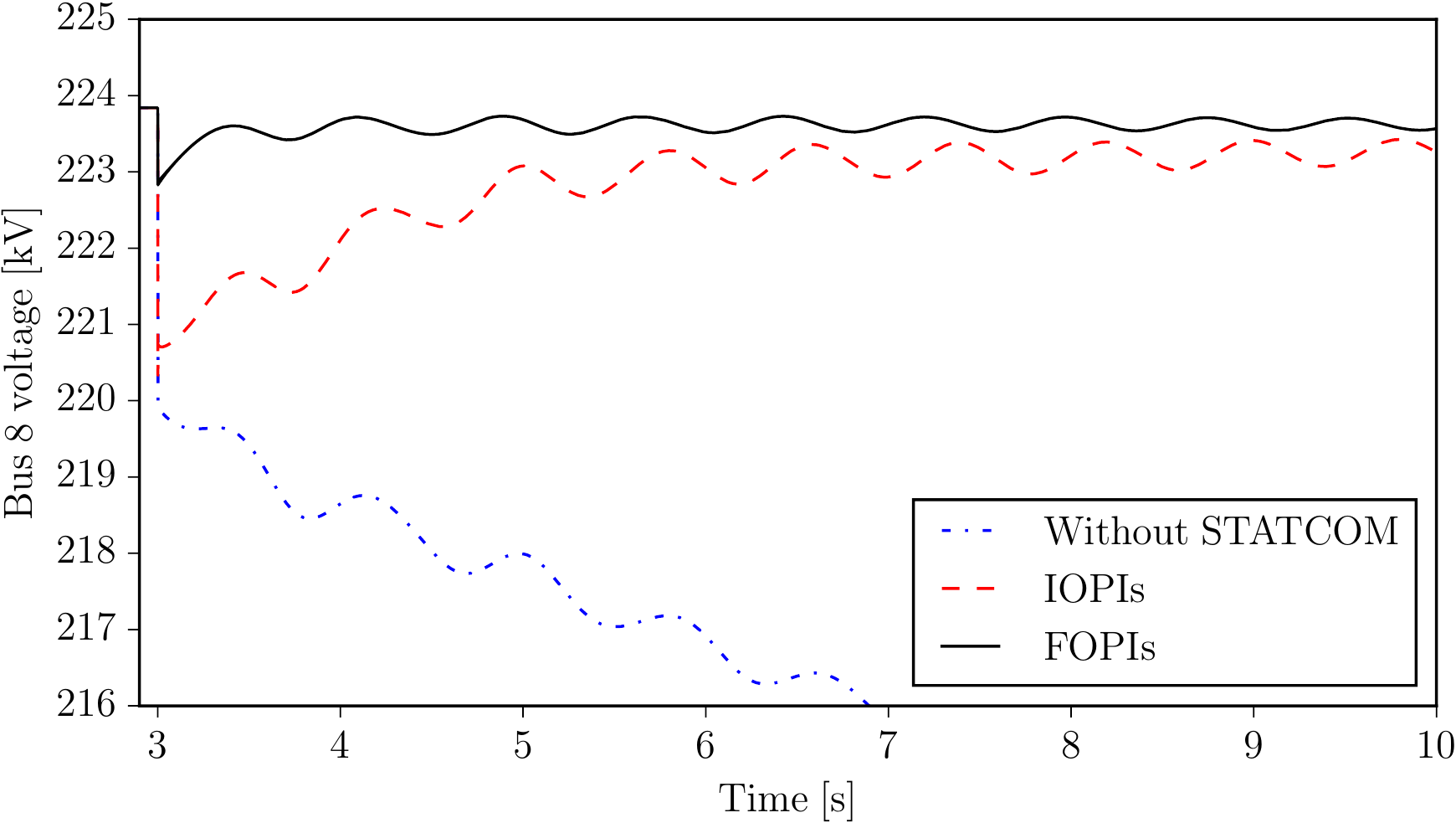}}
    \caption[WSCC system with STATCOM: voltage at bus 8]{WSCC system with STATCOM: voltage at bus 8.}
    \label{fig:statcom}
  \end{center}
\end{figure}

Simulation results are presented in Figure~\ref{fig:statcom}. The use of multiple \acp{fopi} for \ac{statcom} voltage regulation
is able to provide a significant improvement to the local voltage response.

\vspace{-3mm}

\subsection{All-Island Irish
Transmission System}
\label{foc:sec:case2}

This section presents simulation results based on a detailed model of the
\ac{aiits}, which has been described in the case study of Section~\ref{sssa:sec:aiits}.
In addition to the components described in 
Section~\ref{sssa:sec:aiits}, in the system
is connected also an I-\ac{agc}.

For the needs of this case study, 
the \ac{aiits} model has been validated by utilizing the frequency
data from a severe event that occurred in the real system \cite{ahsan:gm2019}.  
The examined event refers to the tripping -- on
the 28-th of February 2018 -- of the \ac{vsc}-based \ac{hvdc} link 
\ac{ewic}
that connects the \ac{aiits} with the Great Britain
transmission system.  At that moment, Ireland was exporting 470 MW to
Great Britain. Following the loss of the \ac{ewic}, the frequency in the
Irish grid showed a peak of 50.42~Hz, which led to the triggering of
over-frequency protections and wind farm active power generation
curtailment.

\begin{figure}[ht]
  \begin{center}
    \resizebox{0.85\linewidth}{!}{\includegraphics{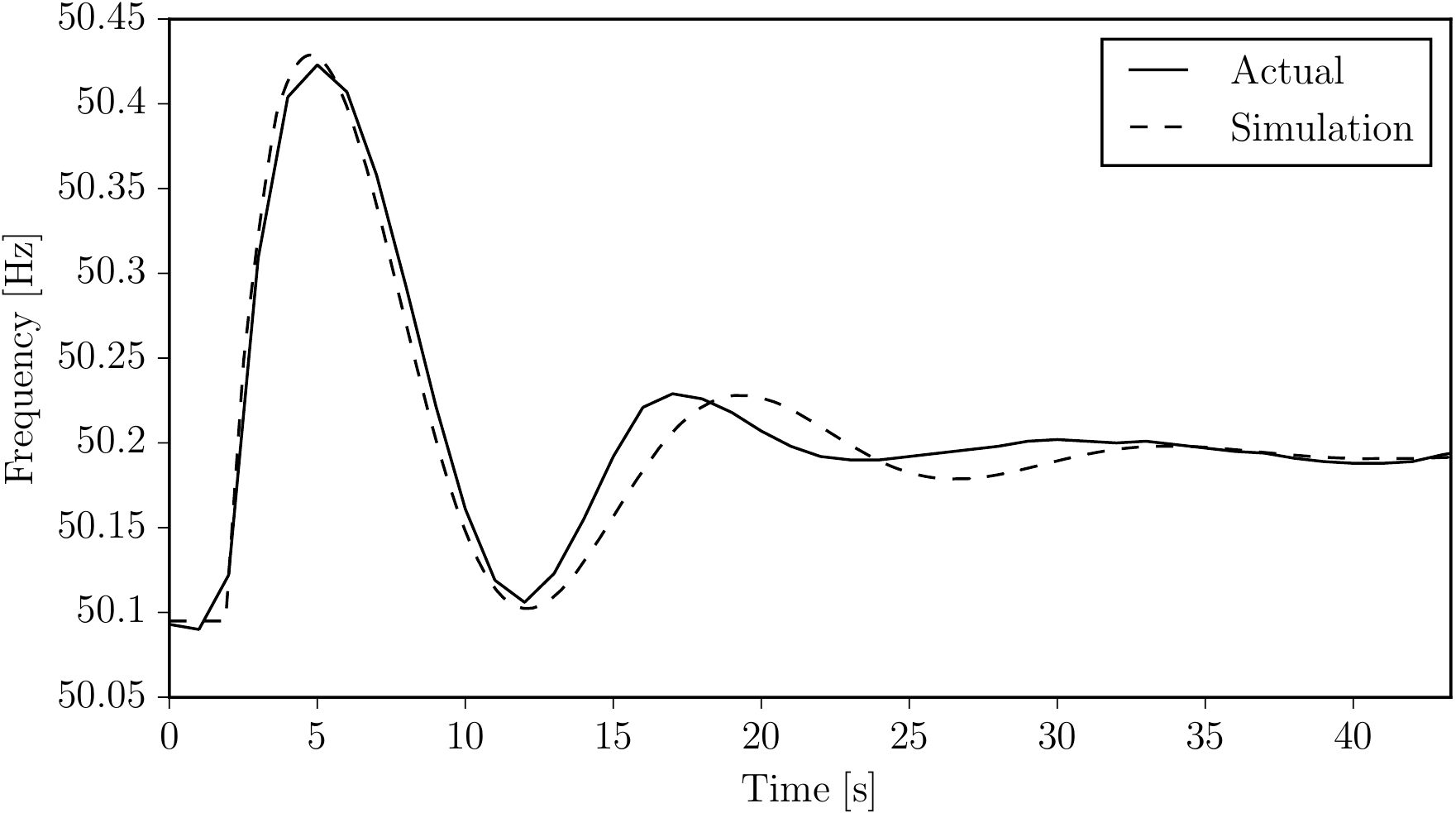}}
    \caption[AIITS: frequency response following the loss of EWIC]{
    AIITS: frequency response following the loss of EWIC.}
    \label{fig:val_irish}
  \end{center}
\end{figure}

A comparison of the actual system response and the one simulated with
\dome is shown in Figure~\ref{fig:val_irish}. As it can be seen, the
simulated transient closely follows the real system behavior.

%

\begin{figure}[ht]
  \begin{center}
    \resizebox{0.85\linewidth}{!}{\includegraphics{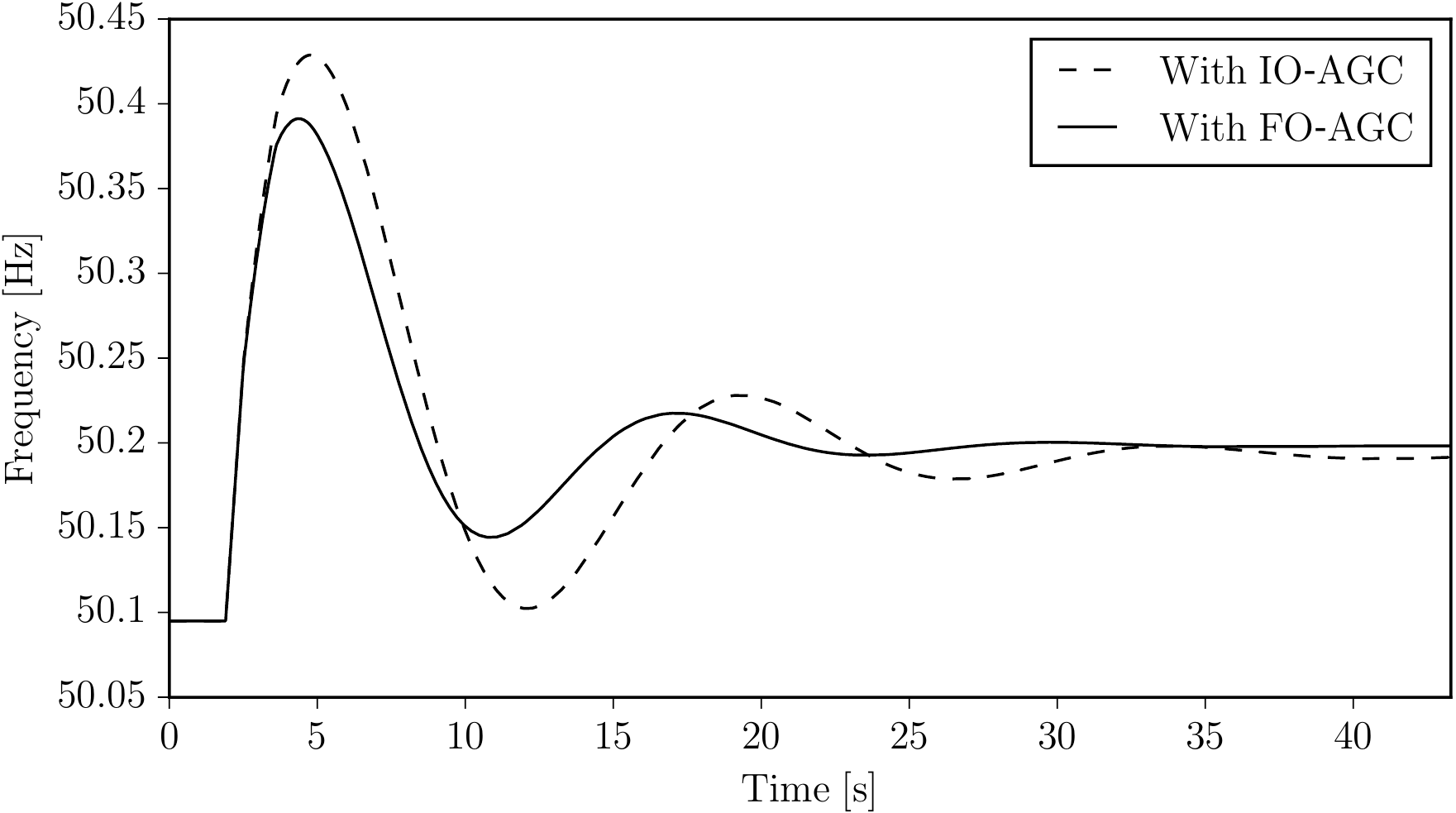}}
    \caption[AIITS: impact of 
    FO-AGC on frequency response]{
    AIITS: impact of 
    FO-AGC on frequency response.
    }
    \label{fig:foagc_irish}
  \end{center}
\end{figure}

We examine the impact of \ac{foc} on the 
secondary frequency regulation of the system. To this aim, the I-\ac{agc} is substituted with the 
\ac{foi}-\ac{agc} model described by \eqref{foc:eq:foagc}. The parameters of the \ac{foi}-\ac{agc} are tuned to $K_i = 500$, $\gamma = 0.15$.
The \ac{ora} parameters are 
$[\omega_b, \omega_h] = [10^{-3}, 10^1]$~rad/s, $N=4$.
Figure~\ref{fig:foagc_irish} shows the frequency response of the system with I-\ac{agc} and \ac{foi}-\ac{agc}. The \ac{foi}-\ac{agc} is able to improve the frequency regulation of the \ac{aiits}.

\section{Conclusions}
\label{foc:sec:conclusion}

The chapter studies the theory, stability analysis, computer
implementation and practical design aspects of \acp{foc} for power system
applications.  It provides a comprehensive theory on fractional
calculus for control, as well as a detailed description of \ac{ora}-based
\acp{foc}.  
In all considered examples, the proposed \acp{foc} are shown to
perform better than the conventional \ac{IO} versions while requiring only a little
additional tuning effort.  This is a general result that shows the
potential of \acp{foc} for power system
applications.


\newpage
\chapter{Time-Delay-based Control}
\label{ch:tdc}

\section{Introduction}
\label{tdc:sec:intro}



Time delays appear in many control systems mainly because it takes
time to measure/acquire information, formulate a decision based on
this information, and implement the decision to achieve a particular
control mission.  Delays arise in many applications, such as in
network control systems when sending/receiving information between
physical locations \cite{Sipahi2011_CSM,Abidi:2016-IJRNC}; in
connected vehicle models due to delays in communication/sensing lines
and human reaction times \cite{Helbing:2004-PRE,Orosz:2010-PTRSL}; and in the dynamics of multi-agent systems
\cite{Olfati:2005-ACC,qiao2016consensus}.

Since delays are in general a source of poor performance and
instability, many studies have focused on the fundamentals of
explaining these characteristics within a control theoretic approach
\cite{Gopalsamy:1992-BOOK, Bellman:1963-BOOK, Hale:1993-BOOK}.  Along
these lines, stability theory has been developed to address the
peculiarities of systems with delays and these results were more
recently combined with powerful convex optimization tools to study the
stability of and design controllers for time-delay systems, see,
e.g.~\cite{Fridman2016}.

While most results in the literature treat delays as undesirable,
there is also a large amount of work that has focused on the
advantages of having delays in a closed-loop setting. In these
studies, the goal is to incorporate delays intentionally into the
closed-loop and systematically analyze the dynamics to show that for
certain delays and controllers, the closed-loop dynamics can behave
more desirably based on certain metrics, such as response time
\cite{Ramirez:2015-ISA, Kokame:2001-TAC,Ulsoy:2015-JDSMC}.  A simple
``delay-based'' controller is the one in which a derivative of a
signal $\dot{x}(t)$ is approximated using a first order Euler's
approximation: 
\begin{equation}
\dot{x}(t) \approx \frac{x(t)-x(t-\tau)}{
\tau} \, ,
\label{tdc:eq:euler}
\end{equation}
where $\tau>0$ is the delay \cite{Kokame:2001-TAC}.

Delay-based controllers have a rich history with many promising
directions \cite{Smith:1959,Pyragas:1992-PLA,Kallmann:1940}. Recent
studies have focused on analytical tractability. This is a challenging
effort since delays cause infinite dimensional system dynamics, study
of which cannot be performed using standard tools available for
finite-dimensional systems. A remedy to this was proposed by utilizing some salient
features of algebraic geometry on a class of delay systems, and 
deriving analytical formulae that prescribe how to tune the delays
and control gains to achieve a desired performance from these systems
\cite{Ramirez:2015-ISA,Ramirez:2017-SICON,Ramirez:2016-TAC}. These
results have been recently extended to distributed control of multi-agent
systems with the goal to achieve fast consensus of agents
\cite{Ramirez:2018-CYBERNETICS}.

Despite the aforementioned advances, benefits of utilizing time delays
as part of controllers are yet to be fully explored in engineering
applications.  In electric power engineering, the vast majority of
studies have emphasized only the destabilizing effects of time delays, e.g. 
in \acp{wadc}, see, 
\cite{heydt:2002,  heydt_stochastic:08, delay1}.   Some studies have focused
on modeling of delays that arise in a \ac{wams} \cite{wang:2012,muyang:wams}, while others have explored
numerical methods for the stability analysis of power systems with
inclusion of delays \cite{multiple:16,li:2019}.  Only very
recently were delays in power systems viewed as tunable control
design parameters \cite{asghari:2018,roy:2019}.

In light of the above discussion, there exists an
opportunity to connect the recent results in time-delay systems
literature toward improving the stability of power systems \cite{delaydamp}.  The main
goal of this chapter is to systematically assess the impact of the
structure and control parameter settings of delay-based \acp{pss} on the
small-signal stability and in particular on the damping
characteristics of power system electromechanical oscillations.

The remainder of the chapter is organized as follows.  Section
\ref{sec:tds} describes a comprehensive treatment for the stability
analysis of small and large scale time-delay systems.  Section
\ref{sec:omib} provides analytical results on the \ac{omib} power system.
Section \ref{tdc:sec:case} discusses a case study based on the IEEE
standard 14-bus system model.  Finally, conclusions are drawn in
Section \ref{tdc:sec:conclusion}.

\section{Spectral Analysis of Time-Delay Systems}
\label{sec:tds}

This section provides first some preliminaries on the spectral properties of
\ac{lti} systems with time delay.  This is followed
by further discussions on a benchmark second-order time-delay system.
Then, for this system, the conditions that have to be
satisfied to guarantee stability independently from the magnitude of
the delay are rigorously deduced. Finally, this section shows how delay-independent stability enables
``connected'' stability regions.

\subsection{Preliminaries}
\label{subsec:pre}

Since the study is concerned with the dynamics following small perturbations, it is relevant to provide here a concise discussion on the stability
properties of linear systems affected by time delays.  Given that the
focus is on time-invariant systems, consider the following \ac{lti} system:
\begin{equation}
  \label{eq:lti}
  \dot{\bfg x}(t) = \bfg A_0 \, \bfg x(t) + \bfg A_1 \, \bfg x(t-\tau) \, ,
\end{equation}
where $\bfg A_0$ and $\bfg A_1$ are matrices with constant entries,
delay is denoted by $\tau\ge0$, and $\bfg x$,
$\bfg x \in \mathbb{R}^{\nx}$, is the state vector.  System
\eqref{eq:lti} is a set of linear functional differential equations of retarded type, also known as linear
\acp{dde}.  Moreover, this system is of
retarded type, i.e.~the highest derivative of the state is not
influenced by the delay term.

To assess exponential stability of system \eqref{eq:lti}, one must
study its characteristic roots, which are the zeros of the system
characteristic function given by:
\begin{equation}\label{eq:CE}
\poly(s,\tau) = \det(s \bfg I_n-\bfg A_0-\bfg A_1 e^{-\tau s}) \, ,
\end{equation}
where the delay appears in
the exponents as per Laplace transform. Due to the presence of the
exponential function, this equation is not in polynomial form in $s$, and is often called
a \textit{quasi-polynomial}
\cite{Niculescu:2001}.

For a given delay $\tau$, system \eqref{eq:lti} is
exponentially stable if and only if all its characteristic roots
have negative real parts. That is, for all $\lambda$
satisfying $\poly(\lambda,\tau) = 0$,
${\rm Re}(\lambda^\ast)<0$
holds \cite{Stepan:1989-BOOK}. While in principle stability definition is not different from
that for \acp{ode}\footnote{This is mainly
  because the spectrum of `retarded' type \ac{lti} systems exhibit similar
  characteristics as those of ordinary differential equations
  \cite{Stepan:1989-BOOK}. }, computing $\lambda$ to assess stability
is challenging due to the transcendental exponential terms in
$\poly(s,\tau)$ that arise due to the delay $\tau$. This is because these terms
bring about infinitely many characteristic roots, computation of which
is prohibitive \cite{Sipahi2011_CSM}.

A remedy to the above issue is to recognize that the characteristic
roots of the system vary on the complex plane in a continuum as the
delay parameter changes in a continuum \cite{Datko:1978-QAM}.  Hence,
the only way the system may become unstable is that a characteristic
root (or a pair of roots) touches the imaginary axis of the complex
plane at $s=\jj \omega$, $\omega\in\mathbb{R}^+$.  That is, whenever $\poly(\jj \omega,\tau)=0$
for some $\omega\ge0$ and $\tau$, the system ``may be'' in transition from
stability to instability, or vice versa\footnote{Note that it is
  necessary, but not sufficient, that the system has at least one root
  on the imaginary axis for its transition from stable to unstable
  behavior.  For sufficiency, the system must be stable for
  $\tau-|\varepsilon|$, $|\varepsilon|\ll 1$.}.

{\definition
{
Consider that $\lambda_{i,i+1} = \alpha_i \pm \jj \beta_i$ define a pair of 
roots of \eqref{eq:CE}. Then, the system is called:
\begin{itemize}
\item $\sigma$-stable, if $\forall \, i \in \mathbb{N^*}$,
  $\alpha_i < -\sigma$, where $\sigma>0$ is a prescribed exponential
  decay rate \cite{Ramirez:2015-ISA}.
\item $\zeta$-stable, if $\forall \, i \in \mathbb{N^*}$,
  $\dfrac{-\alpha}{\sqrt{\alpha^2+\beta^2}}<\zeta$, where $\zeta$ is a
  prescribed dominant oscillation damping ratio.
\end{itemize}
%
}
}

In contrast to the concept of $\sigma$-stability, which has been employed in several studies,  
the term $\zeta$-stability is, to the best of the author's knowledge, first introduced for the needs of this thesis and, in particular, to the aim of studying the structure of the delay-gain parameter space 
and the damping characteristics of the system discussed in the case study of Section~\ref{tdc:sec:case}.

\subsection{Analytical Study of Second-Order LTI Systems}
\label{subsec:lti}

This section presents some salient stability characteristics of second-order
\ac{lti} systems, namely, a subset of the systems described by equation
\eqref{eq:lti}.  These results are critical to establish the stability
features of the \ac{omib} power system.

\subsubsection{System Description}

Consider the \ac{lti} system:
\begin{align}
  \label{eq:lti1}
  {\ddot x(t)} + c_1 \, {\dot x(t)} + c_2 \, x(t)  &= - u(t) \, ,
\end{align}
where $c_1,c_2,  \in \mathbb{R}$ and $u(t)$ is a scalar input.
Next, let $u(t)$ be defined as a delay-based controller.  Specifically,
$u$ is designed as \ac{pr} controller:
\begin{align}
  \label{eq:pr1}
  u(t) = \Gain_p \,
  {\dot x(t)} - \Gain_r \, {\dot x}(t- \tau_r)
  \, ,
\end{align}
where $\Gain_p$, $\Gain_r$, are the proportional, retarded gains,
respectively; and $\tau_r \geq 0$ is a constant delay.  Combining
\eqref{eq:lti1} and \eqref{eq:pr1}, and taking the Laplace transform
of the arising dynamics leads to the closed-loop system characteristic
equation $q(s,\tau_r,\Gain_r) = 0$, where
\begin{align}
  \label{eq:charquasi}
  q(s,\tau_r,\Gain_r) &= s^2 + (c_1+ {  \Gain_p}) \, s + c_2- { \Gain_r}\, s \, e^{-s \tau_r} \, ,
\end{align}
is the system characteristic equation.

\subsubsection{Stability Analysis}

In order to study the $\sigma$-stability of system \eqref{eq:lti1}, the change of variable
$s \rightarrow (s-\sigma)$ is applied to \eqref{eq:charquasi}. This yields the following
quasi-polynomial:
\begin{align}
  \label{eq:quasi2}
  \tilde q(\sigma, s,\tau_r,\Gain_r) = \tilde q_0(\sigma,s)
  +\tilde q_1(\sigma,s) \, \Gain_r \,  e^{\sigma \tau_r} \, e^{-s \tau_r} \, ,
\end{align}
where 
\begin{align}
  \tilde q_0(\sigma,s) &= (s-\sigma)^2 + (c_1 + {  \Gain_p})(s-\sigma) + c_2 \, , \nonumber \\
  \tilde q_1(\sigma,s) &= - (s-\sigma) \, . \nonumber
\end{align}
Recall that the roots of the characteristic equation change
continuously with respect to variations of system parameters and time
delays. The system can thus change from stable to unstable, and vice versa, only if a root (or a pair of roots) crosses the imaginary axis of the complex
plane. Hence, the $\sigma$-stability of \eqref{eq:lti1} can be
assessed by finding the set of crossing points
$(\tau_r^{\rm cr}, \Gain_r^{\rm cr})$, that satisfy:
\begin{align}
  \label{eq:crcond1lti}
  \tilde q(\sigma, \jj , \tau_r^{\rm cr},\Gain_r^{\rm cr}) = 0 \, ,
\end{align}
where $s=\jj \omega$.  The set
$(\tau_r^{\rm cr}, \Gain_r^{\rm cr})$ can be determined by considering the
magnitude and the argument of \eqref{eq:crcond1lti}, as follows
\cite{Ramirez:2016-TAC}:
%
\begin{align}
  \label{eq:arg1lti}
  \tau_r^{\rm cr} &= \frac{1}{\jj \omega} \Big ( {\rm Arg}(\tilde q_1(\sigma,\jj \omega)) -
{\rm Arg}(\tilde q_0(\sigma,\jj \omega)) 
+ \frac{\pi}{2}(4\upmu+\upnu+1) \Big ) \, ,   \\
  \Gain_r^{\rm cr} &= \upnu \, e^{-\sigma \tau_r^{\rm cr}} \Bigg|
     \frac{\tilde q_0(\sigma, \jj \omega)}{\tilde q_1(\sigma,\jj \omega)} \Bigg | \, ,
 \label{eq:mag1lti}
 \end{align}
%
where $\upnu = \pm 1$, $\upmu = 0,\pm 1,\pm 2,\ldots$. Equations
\eqref{eq:arg1lti} and \eqref{eq:mag1lti}, allow tracing the domains of
stability that correspond to specified exponential decay rates,
i.e.~the \textit{$\sigma$-stability map} in the $(\tau_r, \Gain_r)$ space.

Finally, note that if the time-delayed state in \eqref{eq:lti1} is not
utilized, i.e.~$\Gain_r=0$, then the closed-loop system behavior is
determined by the polynomial $\tilde q_0(\sigma, s)$.  In this case,
dissipative terms included in the system are defined by the
coefficient of $s$ corresponding to the first derivative of the state:
\begin{align}
  \label{eq:frictionlti}
  c = c_1 +  {\Gain_p} \, .
\end{align}
Here, the coefficient $c_1$ defines the damping of the open-loop
system oscillatory mode, while $\Gain_p$ defines the amount of non-delayed
artificial damping introduced by the \ac{pr} controller.

\subsubsection{Delay-Independent Stability}

Under certain conditions, \ac{lti} systems with delays can remain stable no matter how large/small the delays are. This phenomenon, known as delay-independent stability, offers the advantage of rendering the dynamics robustly stable against the delays. 

The system \eqref{eq:lti1} is stable regardless the magnitude of the time delay $\tau_r$ provided that certain conditions on the gain $\Gain_r\in\mathcal{K}$,
$\mathcal{K} \subset \mathbb{R}$, 
are satisfied.
For a given set $\mathcal{K}$, a necessary condition for delay
independent stability is that the roots of the system characteristic
equation never cross the imaginary axis, or equivalently:
\begin{align}
  \label{eq:dis}
  q(\jj \omega,\tau_r,\Gain_r) &\neq 0 \ , \ \forall \tau_r \geq 0, 
                    \ \forall \Gain_r \in \mathcal{K} \, .
\end{align}
Using \eqref{eq:frictionlti} in \eqref{eq:dis} yields:
\begin{align}
  \label{eq:dis1lti}
  & -\omega^2 + c \, \jj \omega + c_2 - {  \Gain_r } \, \jj \omega \, e^{-\jj \omega \tau_r}
    \neq  0 
    \nonumber  \\ 
  \Rightarrow \quad &
    \frac{- \omega^2 + c \, \jj \omega + c_2}{{  \Gain_r } \, \jj \omega}  \neq 
    e^{-\jj \omega \tau_r}
    \nonumber  \\ 
  \Rightarrow \quad &
    \dfrac{c}{  \Gain_r }
    + \jj \dfrac{ 1 }{  \Gain_r }( \omega-\frac{c_2}{ \omega}) \neq 
    e^{-\jj \omega \tau_r}
    \, .
\end{align}
Note that the real part of \eqref{eq:dis1lti} does not depend on
$w$, and thus, in the complex plane, the left hand side defines the
vertical line with abscissa $c/ \Gain_r$. In addition,
$e^{-\jj \omega \tau_r}$ defines in the complex plane a unit circle 
centred at $(0,0)$, regardless of the value of the delay $\tau_r$.
Then, the critical condition for delay independent stability is
that the line $c/\Gain_r$ is tangent to the unit circle.
Equivalently:
\begin{align}
  \label{eq:dis2}
  \dfrac{c}{  \Gain_r}
  &= \pm 1
    \quad \Rightarrow \quad 
    c = \pm {   \Gain_r }
    \, .
\end{align}
From equation \eqref{eq:dis2}, the following cases are of interest \cite{delaydamp}:
\begin{itemize}
\item If $c=-  \Gain_r^0<0$, $\Gain_r =\Gain_r^0 > 0$, the system is delay
  independent unstable in $\mathcal{K}=(-\Gain_r^0,\Gain_r^0)$.  Moreover, since $c<0$, the system is unstable around the origin of the $\tau_r$-$\Gain_r$ plane. Hence, even if stable regions exist, these regions are guaranteed to be disconnected.
\item If $c=0$, there are no delay independent stable or unstable
  regions.
\item If $c= \Gain_r^0>0$, $\Gain_r = \Gain_r^0 > 0$, the system is delay independent
  stable in $\mathcal{K}=(-\Gain_r^0,\Gain_r^0)$.  The existence of a delay
  independent stable region around the zero gain guarantees that
  there is a large connected stable domain in the $\tau_r$-$\Gain_r$
  plane. This feature is very important for two reasons: {(i)} there is the possibility that the dynamics can be characterized by high exponential decay rates for large delay values, {(ii)} the presence of a delay-independent stable region indicates that there exists at least one large, ``connected'' stable region from zero to infinite delay.
\end{itemize}

Notice that delay independent stable/unstable region is symmetric
with respect to the gain $\Gain_r$.

\subsection{Linear Large-Scale Time-Delay Systems}
\label{subsec:numlarge}

For a second-order \ac{lti} system with \ac{pr} control, such as the one
discussed above, one can analytically identify the parameter regions  with
specified exponential decay rates, as well as the conditions for delay
independent stability.  However, real-world dynamical systems are
larger in size and much more complex. Capturing the impact of delays on the
behavior of large system models can be achieved only by carrying out a
numerical analysis. Nevertheless, such studies must be carefully guided by the analytical understanding of small scale dynamical systems. This is the approach utilized below.

This section describes next how to assess the stability of large scale linear time-delay
systems.  To this aim, system \eqref{eq:lti} is extended to include
multiple delays $\tau_i$. The resulting \ac{lti} dynamical system is described
through the following set of \acp{dde}:
\begin{equation}
  \label{eq:ltimul}
  \Dt{\bfg x}(t) = \bfg A_0 \, \bfg x(t) + \sum^{\rho}_{i=1} \bfg A_i \, \bfg x(t-\tau_i) \, ,
\end{equation}
where $\tau_i \geq 0$, $i=1,2,\ldots,\rho$, The characteristic
matrix of \eqref{eq:ltimul} has the following form \cite{multiple:16}:
\begin{align}
  \label{eq:charmatmul}
  s  \bfg{I}_{\nx} - \bfg{A}_0
  -\sum^{\rho}_{i=1} \bfg A_i \, e^{-{s \tau_i}}  \, .
\end{align}
Since \eqref{eq:charmatmul} is transcendental, it has infinitely many
eigenvalues, and only an approximation of the solution is possible.
Different approaches have been proposed to overcome this problem
\cite{multiple:16}.  In this chapter, the \ac{dde} system \eqref{eq:ltimul}
is transformed to a formally equivalent set of \acp{pde}, which has infinite dimensions.  The \ac{pde} system is
then reduced to a finite dimensional problem through Chebyshev discretization \cite{bellen:2000,Breda:2015-BOOK}.  If $N_C$ is the number of points
of the Chebyshev differentiation matrix \cite{multiple:16}, then
discretization leads to an approximate linear matrix pencil in the form:
\begin{align}
  \label{eq:approxpencil}
  s  \bfg{I}_{nN_C} - \bfg{M} \, ,
\end{align}
where the matrix $\bfg M$ has dimensions $nN_C \times nN_C$.
The spectrum of \eqref{eq:approxpencil} -- which can be found using
any common numerical method, e.g.~the QR algorithm \cite{Francis} -- represents an
approximate spectrum of \eqref{eq:charmatmul}.
The Chebyshev discretization technique has been successfully applied
to single and multiple time-delay systems, e.g.~to power systems with
constant and stochastic delays affecting damping controllers
\cite{muyang:wams,tzounas2018}.

After the above analysis is complete, one can reveal the most critical
eigenvalue(s), by comparing the damping ratios $\zeta_i$ of all
computed eigenvalues. This work is concerned with the
parametric analysis in a delay versus control gain space. The above
analysis therefore allows building a map of specified dominant
oscillation damping ratio $\zeta$.  In the remainder of the chapter, this
map is referred to as the \textit{$\zeta$-stability map}.

\subsection{Non-Linear Large-Scale Time-Delay Systems}
\label{sec:pssssa}

Consider the non-linear \ac{dae}
power system model \eqref{sssa:eq:sidae}, where for simplicity, but without loss of generality, no inputs are included and
$\bfg T = \bfg I_\nx$,
$\bfg R = \bfg 0_{\ny,\nx}$,
see \eqref{sssa:eq:explcon}. This system can be written as:
\begin{equation}
  \label{tdc:eq:dae}
  \begin{aligned}
    \dot{\bfg x} &= \bfg f( \bfg x, \bfg y) \, , \\
    \bfg 0_{\ny,1} &= \bfg g(\bfg x, \bfg y) \, .
  \end{aligned}
\end{equation}
%

The presence of time delays, for example, in control loops, changes
the set of \acp{dae} \eqref{tdc:eq:dae} into a set of \acp{ddae}.  
Inclusion of time delays 
in \eqref{tdc:eq:dae} yields the following system:
\begin{equation}
  \label{eq:ddae1}
  \begin{aligned}
    \dot{\bfg x} &= \bfg f( \bfg x, \bfg y, \bfg x_d, \bfg y_d) \, , \\
    \bfg 0_{\ny,1} &= \bfg g(\bfg x, \bfg y, \bfg x_d , \bfg y_d) \, ,
  \end{aligned}
\end{equation}
where $ \bfg x_d$, $ \bfg x_d \in \mathbb{R}^{{\nx}_d}$, and $ \bfg y_d$,
$\bfg y_d \in \mathbb{R}^{{\ny}_d}$ are the delayed state and algebraic
variables, respectively.  Suppose that the system includes a single
constant delay $\tau$.  Then, one has:
\begin{equation}
  \label{eq:xdyd}
  \begin{aligned}
   \bfg  x_{d} &= { \bfg x}(t-\tau) \, , \\ 
  \bfg  y_{d} &= { \bfg y}(t-\tau) \, ,
  \end{aligned}
\end{equation}
where $t$ is the current time. 

The equilibrium of \eqref{eq:xdyd} is defined in the time interval $[-\tau,0]$. This implies the assumption that a time equal to $\tau$ has to elapse before a valid equilibrium of the system is reached. 
Although there is no theoretical upper bound to $\tau$,
the aforementioned assumption may render the 
consideration of a very large value impractical. 
A limit case example is a \ac{pss} whose control signal is affected by an infinite delay. In principle, infinite time has to pass before the equilibrium of a system with inclusion of such \ac{pss} is obtained. The issue is resolved by modeling only delays that lie in the time-scale of the dynamics of interest. Variables affected by delays that are much larger than the time-constants of the system dynamics are, in fact, irrelevant to the model under study which, in turn, can be conveniently modified to disregard such variables. In the above limit case example, the delayed \ac{pss} does not have any effect on the system and thus, it can be simply disregarded from the system model. In this thesis, only delays that lie in the same time scale with the rest of the system dynamics 
are considered.

When $\bfg y_d$
does not appear in the algebraic equations of \eqref{eq:ddae1}, this leads
to the index-1 Hessenberg form of \acp{ddae}:
\begin{equation}
  \label{eq:ddae}
  \begin{aligned}
    \dot{\bfg x} &= \bfg f( \bfg x, \bfg y, \bfg x_d, \bfg y_d) \, , \\
    \bfg 0_{\ny,1} &= \bfg g(\bfg x, \bfg y, \bfg x_d) \, .
  \end{aligned}
\end{equation}
Model \eqref{eq:ddae} is adopted instead of \eqref{eq:ddae1}, since it
allows simplifying the form of the characteristic equation of the
corresponding linearized system, while being adequate for the
applications considered in this chapter. The interested reader can find
a detailed study on the \ac{sssa} for
non-index 1 Hessenberg form systems of \acp{ddae} in \cite{hessenberg:16}.

For sufficiently small disturbances, and for the purpose of \ac{sssa}, see
Chapter~\ref{ch:sssa},
\eqref{eq:ddae} can be linearized around a valid stationary point, as
follows:
\begin{align}
  \label{eq:ddaelinf}
  \Delta\dot{\bfg x} &=  \jacs{f}{x}\Delta \bfg x + \jacs{f}{y} \Delta \bfg y +
\jacs{f}{x_d}\Delta \bfg x_d + \jacs{f}{y_d} \Delta \bfg y_d \, , \\
  \label{eq:ddaeling}
  \bfg  0_{m,1} &= \jacs{g}{x} \Delta \bfg x +\jacs{g}{y} \Delta \bfg y +
    \jacs{g}{x_d} \Delta \bfg x_d \, ,
\end{align}
where $\jacs{f}{x}$, $\jacs{f}{y}$,
$\jacs{g}{x}$, $\jacs{g}{y}$, are the Jacobian
matrices of the delay-free variables; and $\jacs{f}{x_d}$,
$\jacs{f}{y_d}$, $\jacs{g}{x_d}$, are the Jacobian
matrices of the delayed variables of \eqref{eq:ddaelinf} and \eqref{eq:ddaeling}.

In the linearized system \eqref{eq:ddaelinf}, \eqref{eq:ddaeling}, the
algebraic variables $\Delta \bfg y$, $\Delta \bfg y_d$ can be
eliminated, under the assumption that $\jacs{g}{y}$ is not
singular. Substitution of \eqref{eq:ddaeling} into \eqref{eq:ddaelinf}
yields:
\begin{equation}
  \begin{aligned}
    \label{eq:ddaelin}
    {\Delta \dot {\bfg x}}(t) = \bfg A_0 \, \Delta \bfg x(t) 
    +
    \bfg A_1 \, \Delta \bfg x(t-\tau) 
    + \bfg A_2 \, \Delta \bfg x(t-2\tau) \, ,
  \end{aligned}
\end{equation}
where 
\begin{equation}
  \begin{aligned}
    \bfg A_0 &= \jacs{f}{x}
    -\jacs{f}{y}\,  \jacsinv{g}{y} \, 
    \jacs{g}{x} \, ,
    \\
\bfg A_1 &= \jacs{f}{x_d}
-\jacs{f}{y} \, \jacsinv{g}{y} \, \jacs{g}{x_d}-\jacs{f}{y_d} \, \jacsinv{g}{y} \, \jacs{g}{x} \, , \\
    \bfg A_2 &= -\jacs{f}{y_d} \, \jacsinv{g}{y} \,  \jacs{g}{x_d} \, .
    \nonumber 
  \end{aligned}
\end{equation}
Applying the Laplace transform in \eqref{eq:ddaelin} yields the
following, quasi-polynomial characteristic matrix:
\begin{align}
  \label{eq:charmat}
  s \bfg I_{\nx}
  - \bfg A_0 
  - \bfg A_1 
  e^{-s \tau}
  - \bfg A_2
  e^{-2 s \tau} \, .
\end{align}
Note that the form of the characteristic matrix \eqref{eq:charmat}
can be retrieved from \eqref{eq:charmatmul} for $\rho=2$ and $\tau_2=2\tau_1$.

\section{One-Machine Infinite-Bus System}
\label{sec:omib}

\begin{figure}[ht!]
  \centering
  \resizebox{0.5\linewidth}{!}{\includegraphics{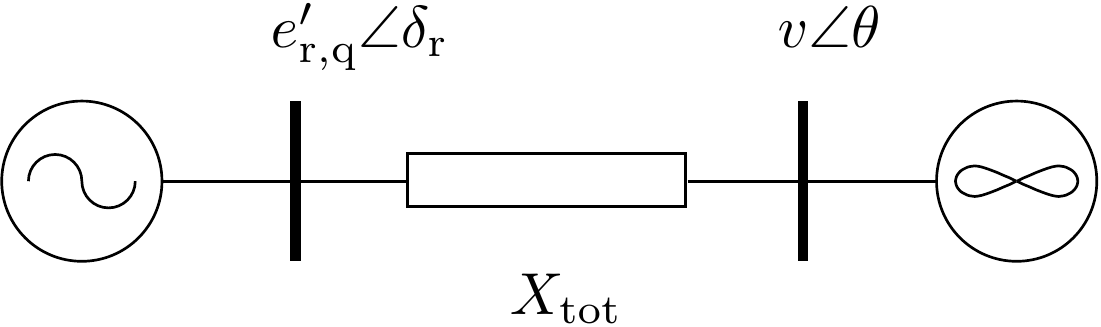}}
  \caption[OMIB system: single-line diagram]{OMIB system: single-line diagram.}
  \label{fig:omib}
\end{figure}

Consider the simple example of the \ac{omib} system shown in
Figure~\ref{fig:omib}.  This section first describes the classical
machine model and then includes in such a model a simplified \ac{pss} with a
\ac{pr} control, i.e.~with two input signals, one
instantaneous and one delayed.

\subsection{Classical Model} 

The classical per-unit model of this system is as follows \cite{kundur:94}:
\begin{equation}
  \begin{aligned}
    \dot \dG &={\Omega_b}(\wG-1)  \, ,  \\
    M \dot \wG &=
    P_{\rm m} -  P_{\rm e} (\delta)
    - D(\wG - 1) \, , 
    \label{eq:omibeqs}
  \end{aligned}
\end{equation}
where $\dG$, $\wG$, are the rotor angle and the rotor speed of
the synchronous machine, respectively; $P_{\rm m}$ and $P_{\rm e}$ are the
mechanical, electrical power output of the machine, respectively.  In
addition, $M$ is the machine mechanical starting time; $D$ is the machine
rotor damping coefficient; 
and $\Omega_b$ is the nominal synchronous angular frequency in
rad/s.  

The electrical power $P_{\rm e}$ is described by the following
non-linear expression:
\begin{align}
  \label{eq:pe}
  P_{\rm e} (\dG) = \frac{e'_{{\rm r},\qax} v }{X_{\rm tot}}
  {\rm sin}(\dG -\theta) \, ,
\end{align}
where $v$, $\theta$, are the (constant) voltage magnitude and
angle at the infinite bus; $e'_{{\rm r},\qax}$ is the internal electromotive
force of the synchronous machine, which is taken as constant, by
assuming an integral \ac{avr}.  $X_{\rm tot}$
is the total reactance, comprising the machine transient reactance
($X'_{\dax}$) and the line reactance ($X$), where the latter is referred
to the machine power base.

Defining the system state vector as $ \left[ \dG \ \ \wG \right]\T $,
making use of \eqref{eq:pe} and linearizing \eqref{eq:omibeqs} around
a valid equilibrium $\left[ \dGi{o} \ \ \wGi{o} \right]\T$ yield:
\begin{align}
  \label{eq:delta}
  \Delta {\dot \dG } &= \Omega_b \Delta \wG \, ,\\
  \Inert \Delta {\dot \wG } &= -\frac{e'_{{\rm r},\qax} v {\rm cos}(\dGi{o} - \theta) }{ X_{\rm tot}} 
  \Delta \dG - D \Delta \wG \label{eq:omega} \, ,
\end{align}
where $\Delta \dG = \dG - \dGi{o}$ and
$\Delta \wG = \wG - \wGi{o}$. Equations
\eqref{eq:delta} and \eqref{eq:omega} can be rewritten as a second-order
\ac{lti} system:
\begin{align}
  \label{eq:olomib}
  \Delta {\ddot \dG} 
  +
  d \Delta {\dot \dG } +
  b \Delta \dG
  &=  0 \, ,
\end{align}
where $\Delta \dG \equiv x$ and
\begin{equation}
  b = \dfrac{\Omega_b e'_{{\rm r},\qax} v {\rm cos}(\dGi{o}) }{\Inert X_{\rm tot}},
  \quad d = \dfrac{D}{\Inert} \, .
\end{equation}

\subsection{Power System Stabilizer with PR Control}

\begin{figure}[ht!]
  \centering
  \resizebox{0.35\linewidth}{!}{\includegraphics{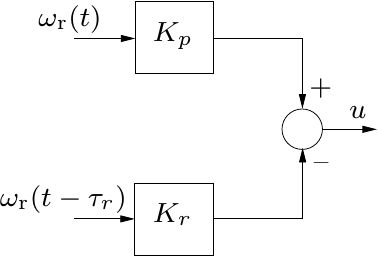}}
  \caption[PR control based PSS diagram]{PR control based PSS diagram.}
  \label{fig:conscheme1}
\end{figure}

In its simplest form, the \ac{pss} measures the machine rotor speed,
i.e.~$\Dt \dG = \wG$, and introduces a fictitious damping into
the swing equation \eqref{eq:delta}. The linearized closed-loop
system can therefore be written as:
\begin{align}
  \Delta {\ddot \dG} 
  + d \, \Delta {\dot \dG } +
  b \, \Delta \, \dG
  &=  - u(\Delta \dot \dG)
  \label{eq:omib2} \, .
\end{align}
The damping controller is modeled here as a proportional \ac{pss} with two
control channels, one with and one without delay.  The \ac{pss} diagram is
shown in Figure~\ref{fig:conscheme1}.

Dual-channel \acp{pss} have been employed in the past, e.g.~as
decentralized-hierarchical schemes for wide-area stabilizing control
\cite{kamwa:2001}.  The dual-channel \ac{pss} output is described as:
\begin{equation}
  \label{eq:prpss}
  u = \Gain_p \, \Delta {\dot \dG} - \Gain_r \, \Delta {\dot \dG}(t- \tau_r) \, .
\end{equation}
Merging \eqref{eq:delta}, \eqref{eq:omib2} and \eqref{eq:prpss} leads
to the following closed-loop system representation:
\begin{align}
  \Delta {\ddot \dG} +
  (d + \frac{\Gain_p}{\Omega_b}) \, 
  \Delta {\dot \dG } +
  b \, \Delta \dG - \frac{\Gain_r}{\Omega_b} \,  \Delta {\dot \dG}(t- \tau_r)
  &= 0
\label{eq:clsys}
\, ,
\end{align}
which is exactly in the form of \eqref{eq:lti1}-\eqref{eq:pr1}. Applying the Laplace transform and substituting the initial conditions
$\Delta {\dG}(0)=\Delta {\dot \dG}(0)=0$, yields the following
characteristic quasi-polynomial:
\begin{align}
  \label{eq:quasiomib}
  q(s,\tau_r,\Gain_r) &= s^2 + (d + \frac{\Gain_p}{\Omega_b}) \, s +
b - \frac{\Gain_r}{\Omega_b} \, s \, e^{-s \tau_r} \, .
\end{align}
Comparing the quasi-polynomial \eqref{eq:quasiomib} with the one in \eqref{eq:charquasi}, one has $c_1=d$, $c_2=b$, $ =\Omega_b^{-1}$.
Therefore, the analysis of $\sigma$-stability and the conditions for
delay independent stability can be studied through the derivations of
Section~\ref{subsec:lti}.  The amount of friction included in the
delay-free \ac{omib} system is according to \eqref{eq:frictionlti}:
\begin{align}
  \label{eq:frictionomib}
  c = d + \frac{\Gain_p}{\Omega_b} \, .
\end{align} 
The critical condition for which the \ac{omib} system is delay independent
stable is that $\Omega_b c/\Gain_r$ is tangent to the unit circle.
Equivalently, one has:
\begin{align}
  \label{eq:dis2omib}
  \dfrac{\Omega_b c}{\Gain_r}
  &= \pm 1
    \quad \Rightarrow \quad
    c = \pm 
    \dfrac{\Gain_r }{\Omega_b}
    \, .
\end{align}

\subsection{Illustrative Example}
\label{subsec:example}

We provide a numerical example on the closed-loop \ac{omib} system. Let
$e'_{{\rm r}, \qax}=1.22$~pu, $v=1$~pu, $\theta=0$~rad, $P_{\rm m}=1$~pu,
$X_{\rm tot} = 0.7$~pu. Then, the initial value $\dGi{o}$ of the
rotor angle is given by:
\begin{align}
  \label{eq:delta0}
  \dGi{o}= {\rm arcsin} 
  \left (\dfrac{P_{\rm m}  X_{\rm tot}}{v e'_{{\rm r},\qax} } \right ) \, .
\end{align}
The examined equilibrium is hence $[ 0.61 \, , \, 1 ]\T$.  Let also
$\Inert=5$~MW~s/MVA, and $\Omega_b=100\pi$~rad/s ($50$~Hz system).  Then,
$b = 89.756$~pu in \eqref{eq:olomib}.  The following sections discuss the
$\sigma$-stability map of the system for the three cases of negative,
zero and positive values of $c$.

\subsubsection*{Case 1} 

\begin{figure}[ht!]
\centering
\resizebox{0.85\linewidth}{!}{\includegraphics{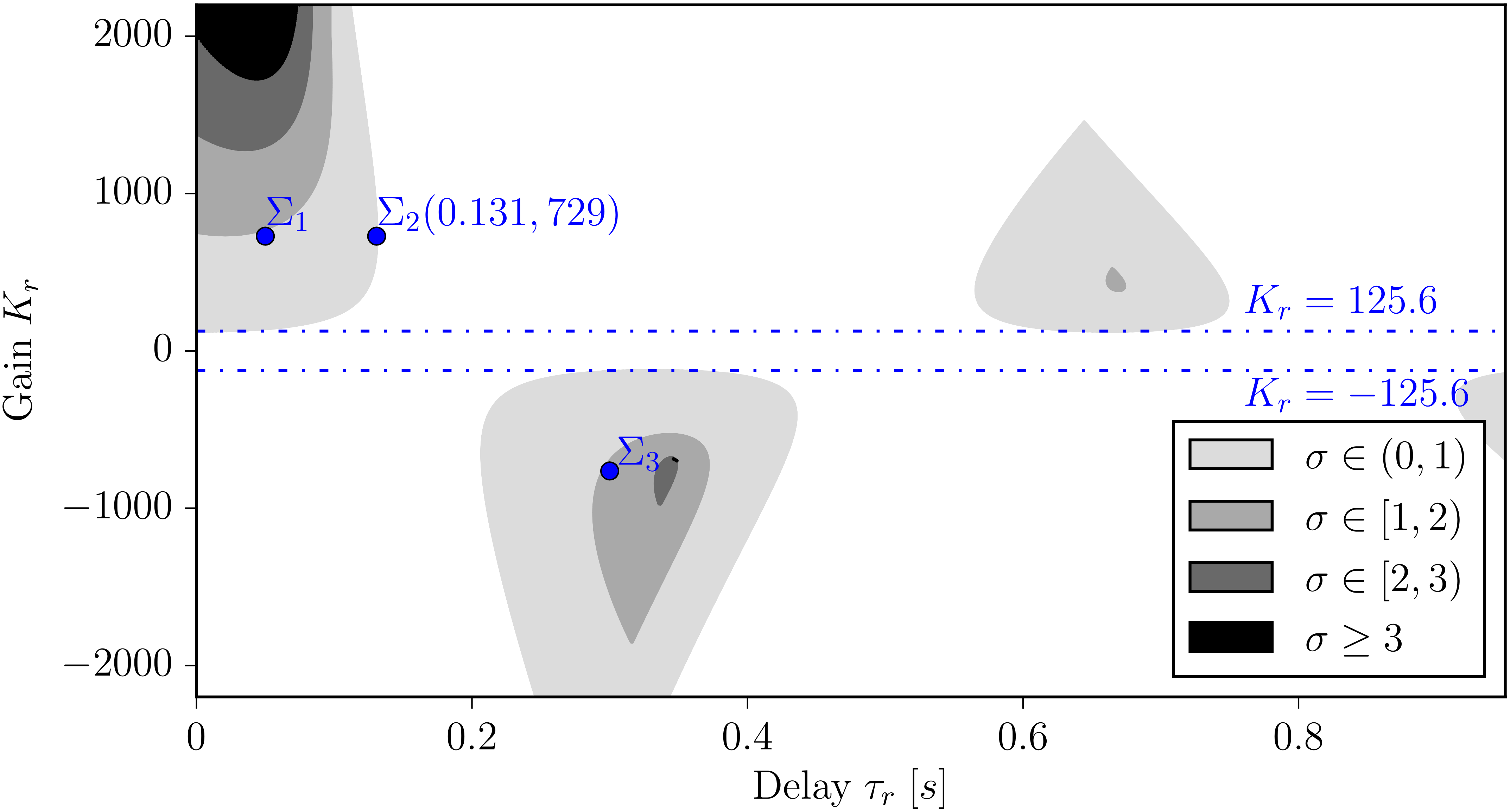}}
\caption[Closed-loop linearized OMIB system: stability map]{
Closed-loop linearized OMIB system: $\sigma$-stability map in the $\tau_r$-$\Gain_r$ plane, $c = -0.4$.}
\label{fig:map1}
\end{figure}

For $c=-0.4<0$, the stability map is shown in Figure~\ref{fig:map1}.
The map has a symmetric delay independent unstable region obtained for
$\Gain_r \in (-125.6, 125.6)$. In addition, \ac{pr} control can stabilize the
system, provided that the delay is $\tau_r<0.131$~s and a proper
$\Gain_r>0$ is selected (see e.g.~point $\point_1(0.05,729)$).

There also exist stable regions of the map in Figure~\ref{fig:map1} for delays
higher than $0.131$~s.  For example, the system is stable around the
point $\point_3(0.30, -763.4)$.  Note, however, that obtaining the
equilibrium of a delayed system implies that a time equal to the
maximum delay included in the system has elapsed but, meanwhile, the
system may have been already rendered unstable. Indeed,
Figure~\ref{fig:map1} indicates that there is no path to $\point_3$ without
crossing the system stability boundary, which implies that the system
necessarily becomes unstable before actually reaching $\point_3$.

The effect of crossing the stability boundary of the
closed-loop \ac{omib} system is illustrated with a time domain simulation.
Suppose that the non-linear system \eqref{eq:omibeqs} with the inclusion
of the \ac{pr} controller \eqref{eq:prpss}, operates around the stable
equilibrium defined by the point $\point_1$ of Figure~\ref{fig:map1}.

The system is numerically integrated considering a small noise on the
measurement of the \ac{omib} rotor speed.  The noise is a normal process
with zero mean and standard deviation of $0.0002$.  The noise amplitude
is set to a small value with the purpose of showing the dynamics of
the system in a neighbourhood of the equilibrium point.  At $t=2$~s,
the gain and delay are switched to $\Gain_r=-765$ and $\tau_r=0.3$,
respectively, so that the system is set at the new equilibrium point $\point_3$.

\begin{figure}[ht!]
  \begin{center}
  \resizebox{0.85\linewidth}{!}{\includegraphics{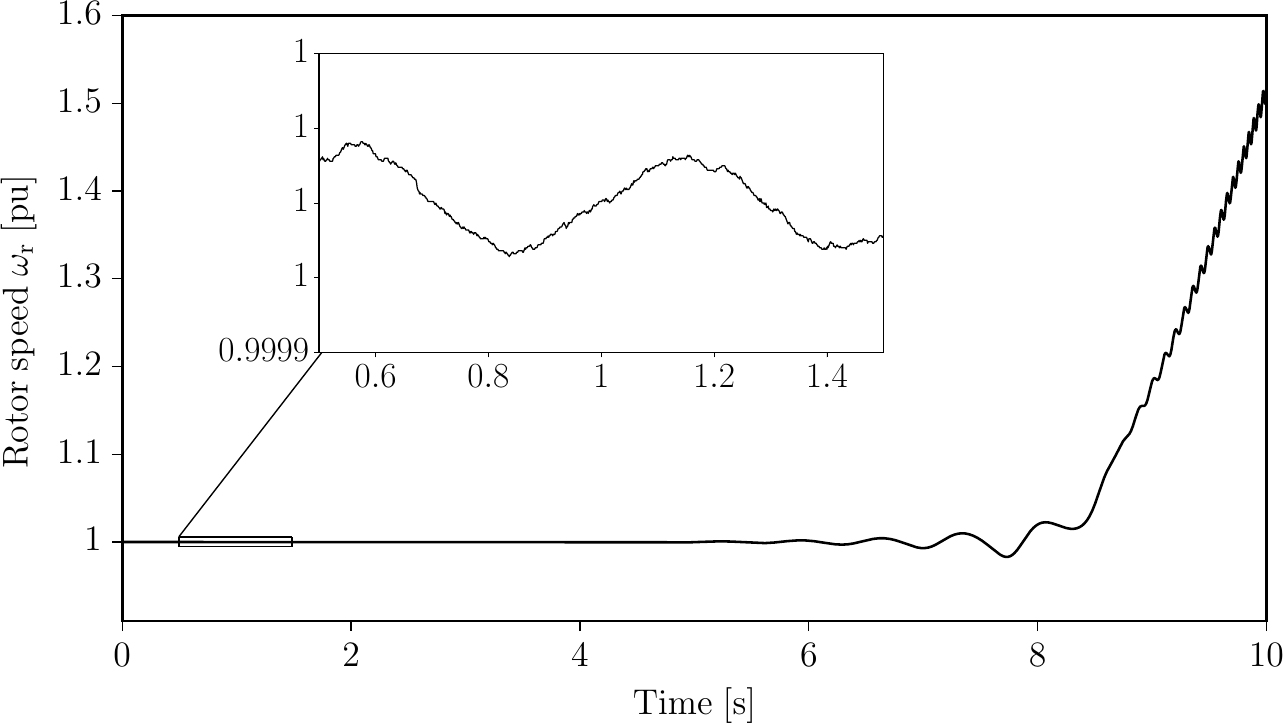}}
  \end{center} 
  \caption[Closed-loop non-linear OMIB system with noisy rotor speed
measurement]{Closed-loop non-linear OMIB system \eqref{eq:omibeqs} with noisy rotor speed
    measurement: the equilibrium is switched from $\point_1$ to $\point_3$ at
    $t=2$~s.}
  \label{tdc:fig:trans}
\end{figure}

Figure~\ref{tdc:fig:trans} shows the simulation result, and indicates that,
as expected, attempting to jump to a different, not connected stable
region by crossing the stability boundary during a transient, renders
the system unstable. Thus, $\point_3$ is an example of infeasible
stationary point, and thus, the delay margin of the system is $0.131$~s.

\subsubsection*{Case 2}

\begin{figure}[ht!]
  \centering
  \resizebox{0.85\linewidth}{!}{\includegraphics{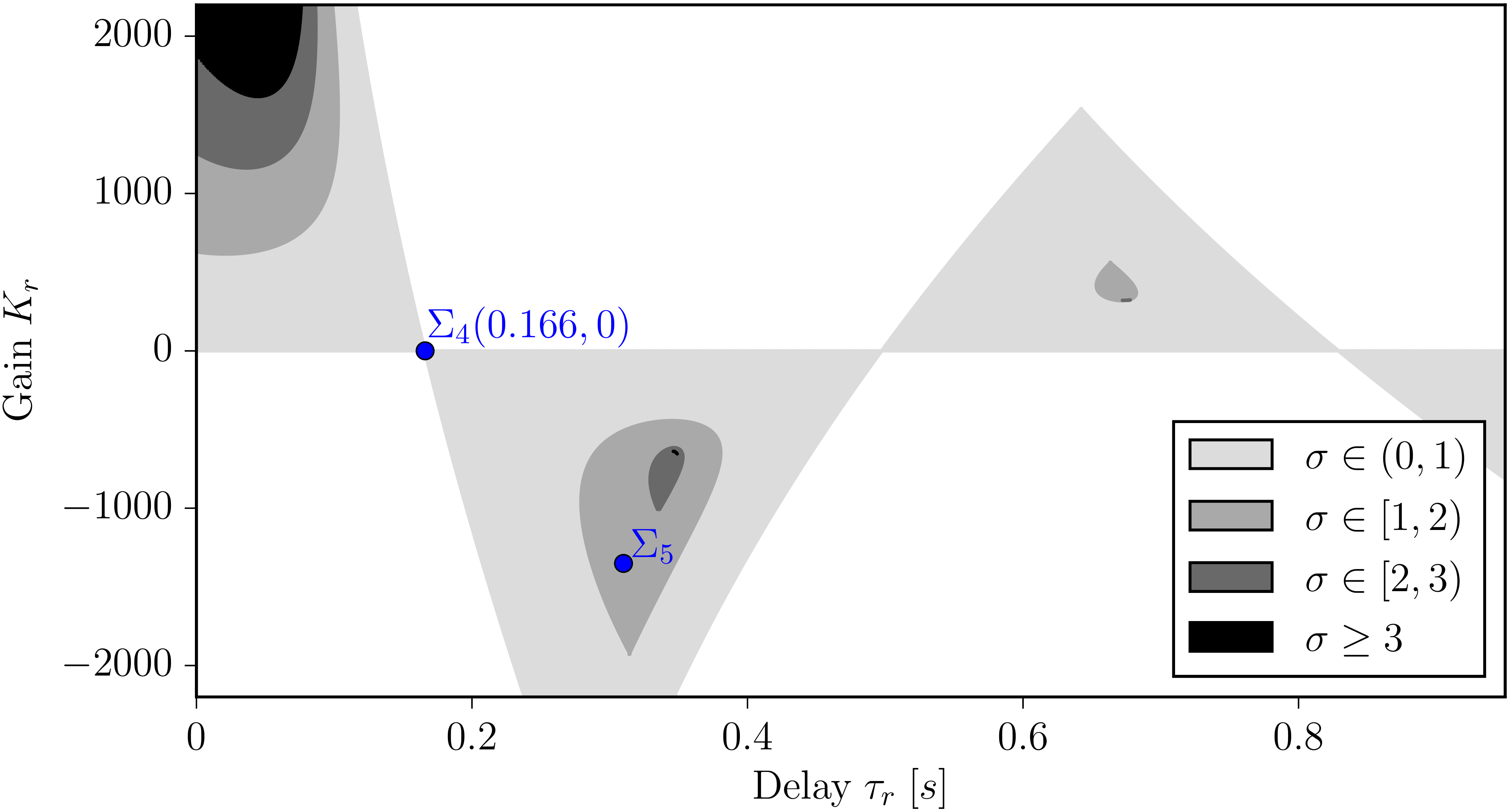}}
  \caption[Closed-loop linearized OMIB system: stability map]{Closed-loop linearized OMIB system: $\sigma$-stability map in the
    $\tau_r$-$\Gain_r$ plane, $c = 0$.}
  \label{fig:map2}
\end{figure}

The $\sigma$-stability map for $c=0$ is presented in
Figure~\ref{fig:map2}.  In this case, the stability of the system
depends on the magnitude of the delay, regardless of the value of the
gain $\Gain_r$.  In fact, the horizontal line $\Gain_r=0$ comprises
bifurcation points.  The delay-free closed-loop system is stable for
$\Gain_r>0$ and unstable for $\Gain_r<0$.  Provided that a proper positive
$\Gain_r$ value is selected and that $\tau_r<0.166$~s (see point $\point_4$),
the delayed system is stable.

There also exist stable regions for $\tau_r>0.166$~s. For example, the
system is small-signal stable around $\point_5$.  However, similarly to the
discussion of Case~1, the system will likely lose stability
before actually reaching e.g.~$\point_5$.  An exception occurs if the
system crosses $\point_4$, which is a bifurcation point that connects two
stable regions.  In this scenario, the first order information
provided by the linearized system in Figure~\ref{fig:map2} is
inconclusive on the feasibility of operating at $\point_5$.

\subsubsection*{Case 3}

\begin{figure}[ht!]
  \centering
  \resizebox{0.85\linewidth}{!}{\includegraphics{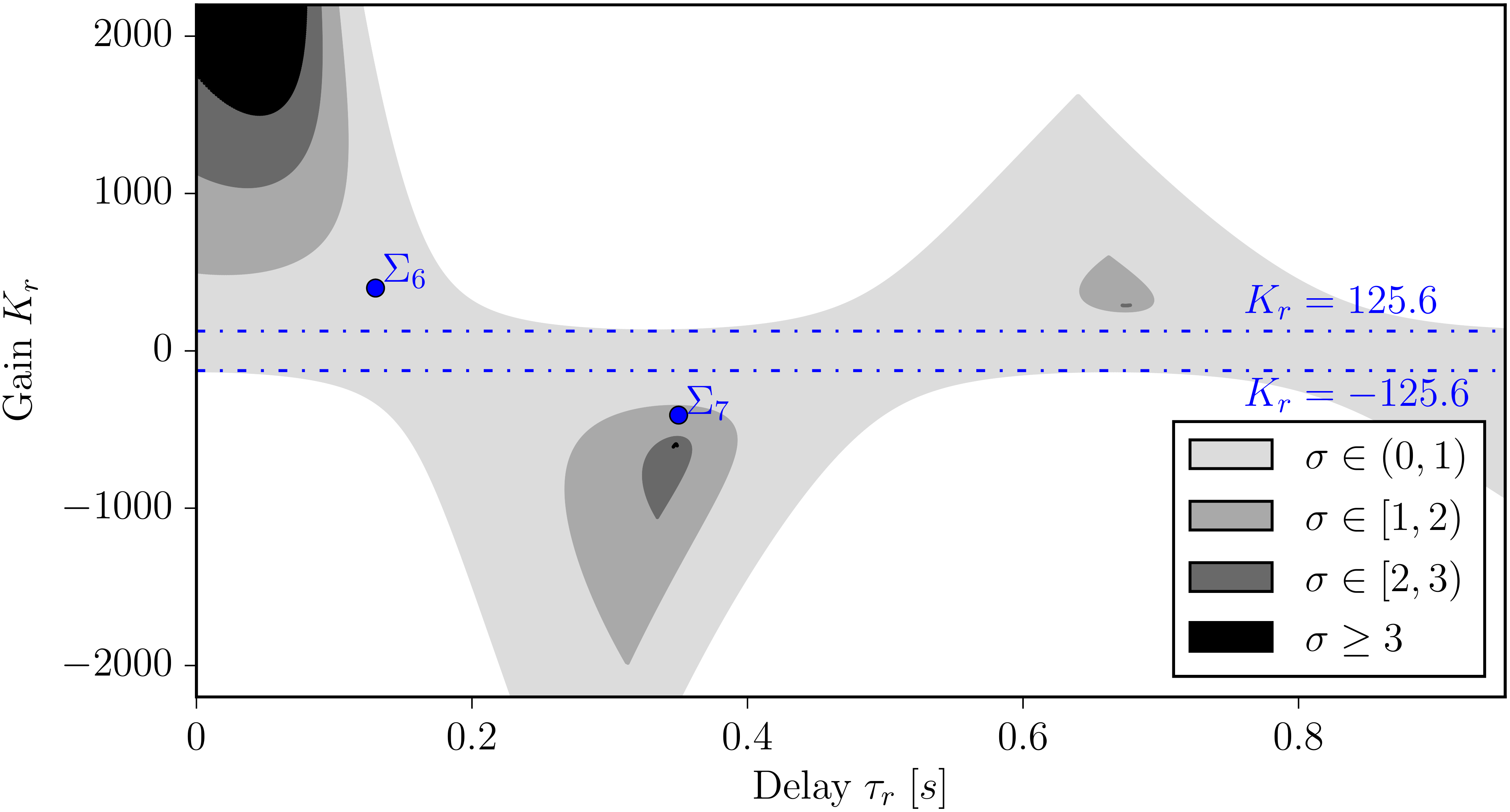}}
  \caption[Closed-loop linearized OMIB system: stability map]{Closed-loop linearized OMIB system: $\sigma$-stability map in the
    $\tau_r$-$\Gain_r$ plane, $c = 0.4$.}
  \label{fig:map3}
\end{figure}

The stability map for $c=0.4>0$ is shown in Figure~\ref{fig:map3}. In
this case, the stable region is compact. For $\Gain_r \in (-125.6, 125.6)$
the system is stable regardless of the magnitude of the delay $\tau_r$.
Moreover, all points of Figure~\ref{fig:map3} with $\sigma>0$ represent
stable and feasible stationary points of the linearized \ac{omib} system. For example,
such points are $\point_6(0.13,400)$ and $\point_7(0.35,-410)$.

\begin{figure}[ht!]
  \centering
  \resizebox{0.85\linewidth}{!}{\includegraphics{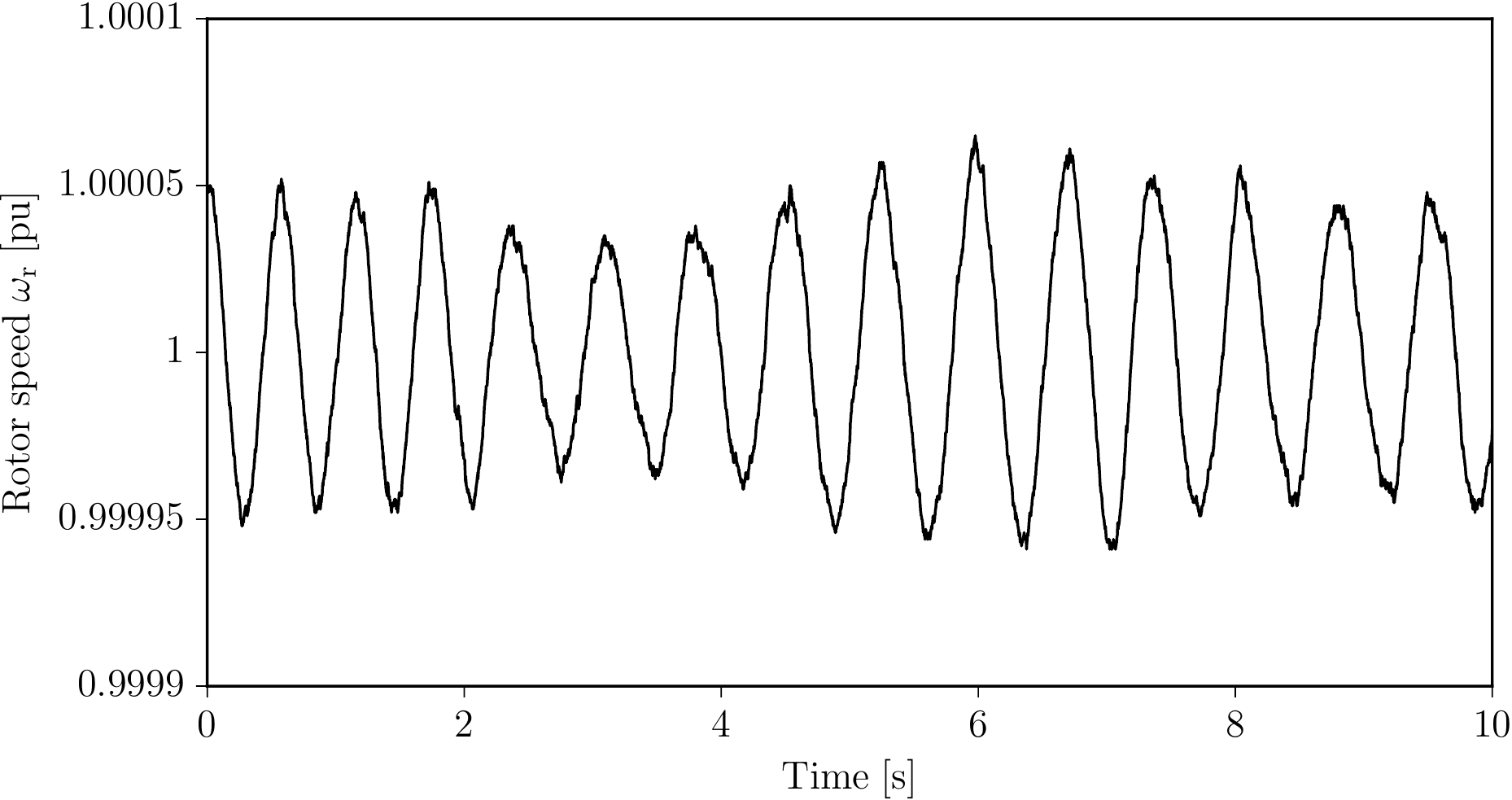}}
  \caption[Closed-loop non-linear OMIB system with noisy rotor speed
    measurement]{Closed-loop non-linear OMIB system with noisy rotor speed
    measurement: the equilibrium is switched from $\point_6$ to
    $\point_7$ at $t=2$~s.}
  \label{tdc:fig:trans2}
\end{figure}

The results of a time domain simulation, including the same noise model on the
rotor speed measurement as in Case~1, are shown in Figure~\ref{tdc:fig:trans2}. At $t=2$~s, the system
equilibrium is switched from $\point_6$ to $\point_7$. The trajectory shows that the machine maintains
synchronism.

Overall, proper design of the \ac{pss} given by the \ac{pr} law \eqref{eq:prpss}
allows unifying the $\sigma$-stable regions, and thus allows one to
operate the \ac{omib} system under the presence of large delays.  In
particular, this is achieved by properly adjusting the control
parameter $\Gain_p$ which introduces delay-free artificial damping to the
system.

Finally, the delay $\tau_r$ in this example is assumed to be a fully
controlled parameter.  However, the above discussion is relevant also
for systems with inherent delays.  For the sake of example, consider
again point $\point_6$ of Figure~\ref{fig:map3}. Suppose that the
corresponding delay, i.e.~$0.13$~s, represents an uncontrolled
physical phenomenon, e.g.~the latency of a measurement transmitted
through a communication system.  In power systems, this situation
describes, for example, the behavior of a wide area measurement system
\cite{muyang:wams}.  In such a scenario, the parameter $\tau_r$ can be
adaptively adjusted to add an artificial delay, which ensures that the
system under the total delay $0.13+\tau_r$ always operates at a region
of high exponential decay rate.  Along these lines, see, for example,
the idea of delay scheduling in \cite{Olgac:2005}.

\section{Case Study: IEEE 14-bus System}
\label{tdc:sec:case}

\begin{figure}[ht!]
  \centering
  \resizebox{1\linewidth}{!}{\includegraphics{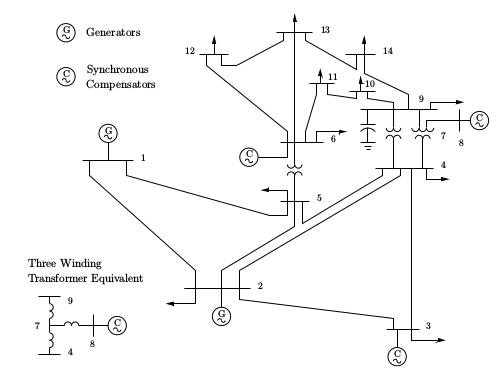}}
  \caption[IEEE 14-bus system: single-line diagram]{IEEE 14-bus system: single-line diagram.}
  \label{fig:ieee14}
\end{figure}

We next study the stability characteristics of the IEEE 14-bus system. The single-line diagram of this system is 
depicted in Figure~\ref{fig:ieee14}.  The system consists of fourteen
buses, five synchronous machines, twelve loads, twelve transmission
lines and four transformers.  All machines are equipped with \acp{avr}.  The static and dynamic data of the system
can be found in \cite{milano:10}.  Simulations in this section are
carried out using Dome.

\begin{figure}[ht!]
  \centering
  \resizebox{0.9\linewidth}{!}{\includegraphics{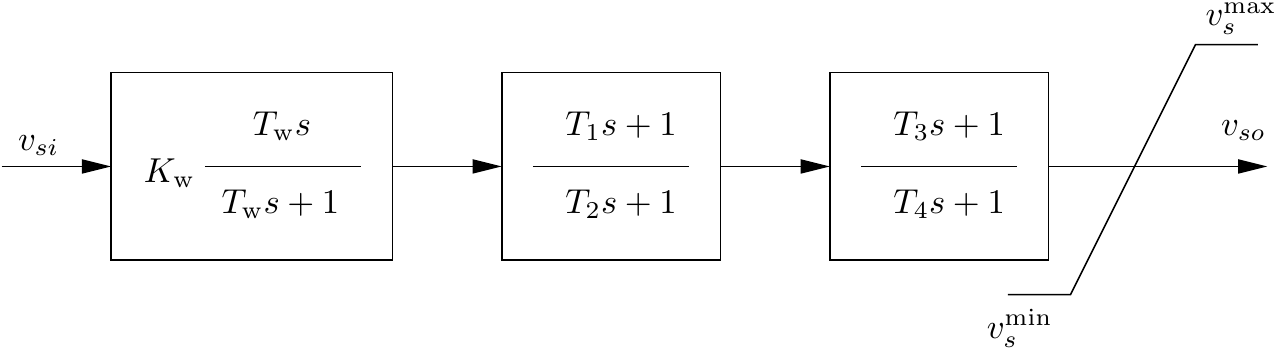}}
  \caption[Power system stabilizer block diagram]{Power system stabilizer block diagram.}
  \label{fig:pss2}
\end{figure}

Without any \ac{pss} installed to the system, \ac{sssa} shows that the rightmost
pair of eigenvalues is $0.3522 \pm \jj 9.12$, and thus, the system is
unstable around the examined equilibrium. A \ac{pss} is utilized to stabilize the system.  The \ac{pss} model
employed in this section is described by the following \acp{dae}:
\begin{equation}
  \begin{aligned}
    T_{\rm w} \dot v_{1} &= -\Gain_{\rm w} \, v_{si}-v_1  \, , \nonumber \\
    T_2 \dot v_{2} &= \Big (1-\frac{T_1}{T_2} \Big )(\Gain_{\rm w} \, v_{si}+v_1  )-v_2 \, ,
    \\
    T_4 \dot v_{3} &= \Big  (1-\frac{T_3}{T_4} \Big) \Big (v_2+\frac{T_1}{T_2}(\Gain_{\rm w} \, v_{si}+v_1) \Big )-v_3    \, , \\
    0 &= v_3+\frac{T_3}{T_4} \Big (v_2+\frac{T_1}{T_2}(\Gain_{\rm w} \, v_{si}+v_1) \Big )-v_{so} \, ,\\
  \end{aligned}
\end{equation}
where $v_1$, $v_2$, $v_3$ are the \ac{pss} state variables; $T_{\rm w}$, $T_1$,
$T_2$, $T_3$, $T_4$ are time constants; $\Gain_{\rm w}$ is the \ac{pss} gain.  In
addition, the input $v_{si}$ is the local rotor speed, which,
depending on the examined scenario, may be delayed or not. Finally,
the output signal $v_{so}$ is an additional input to the local \ac{avr}
reference, so that the \ac{pss} provides damping of electromechanical
oscillations through excitation control. The \ac{pss} block diagram is
depicted in Figure~\ref{fig:pss2}.

To study the effect of time-delayed damping control on the small
signal stability of the IEEE 14-bus system,
two damping control configurations are compared, namely, a conventional \ac{pss} with
delayed input signal; and a \ac{pss} that consists of two channels, one
delayed and one non-delayed.  In both cases, the damping controller is
installed at the \ac{avr} of the synchronous machine connected at bus 1.

The impact of time delay in each case is evaluated by means of
constructing the $\zeta$-stability map in the delay-control gain
space. For each point of the plane, an eigenvalue analysis is carried
out by applying the Chebyshev discretization technique (see
Section~\ref{subsec:numlarge}).  The spectrum of the approximate
matrix pencil is calculated using the QR algorithm with LAPACK
\cite{lapack}.  Then, comparison among the eigenvalues allows
obtaining the most poorly damped one determining the $\zeta$-stability.

\subsection*{Standard PSS with Delayed Input Signal}
\label{sec:standard}

The employed \ac{pss} model is as shown in Figure~\ref{fig:pss2}. The control
input signal is considered to be the delayed local rotor speed
measurement:
\begin{align}
  \label{eq:pssin}
  v_{si} &= \wGi{1}(t - \tau) \, ,
\end{align}
where $\tau$ is an intentional constant delay.  The \ac{pss} time constant values are
summarized in Table~{\ref{tab:psspars}}.

\begin{table}[ht!]
  \centering
  \renewcommand{\arraystretch}{1.3}
  \caption[IEEE 14-bus system: PSS parameters]{IEEE 14-bus system: PSS parameters.}
  \begin{tabular}{l}
    \hline 
    $T_{1}=T_3=0.28$~s, $T_{2}=T_{4}=0.02$~s, $T_{\rm w}=10$~s  \\
    \hline 
  \end{tabular}
  \label{tab:psspars}
\end{table}

\begin{figure}[ht!]
  \centering
  \resizebox{0.85\linewidth}{!}{\includegraphics{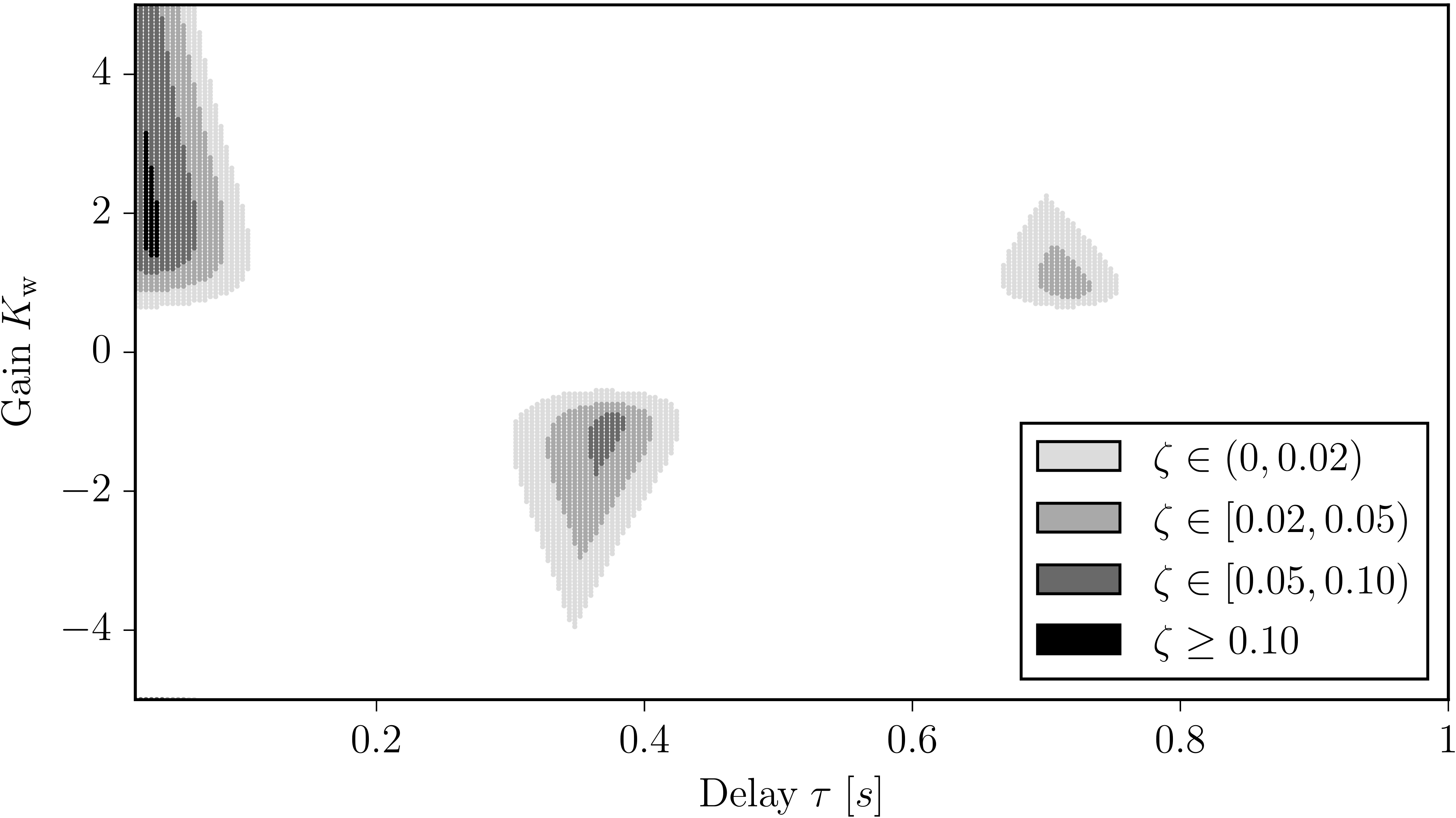}}
  \caption[IEEE 14-bus system: stability map]{IEEE 14-bus system: $\zeta$-stability map in the
$\tau$-$\Gain_{\rm w}$ plane.}
  \label{fig:ieee14damp1}
\end{figure}

The dynamic order of the system is $54$.  Setting the number of points
of the Chebyshev differentiation matrix to $N_C=10$, $540$
eigenvalues are found in total.  The system $\zeta$-stability map in the
$\tau$-$\Gain_{\rm w}$ plane is shown in Figure~\ref{fig:ieee14damp1}.  The map consists of distinct and not compact stable regions,
which stems from the fact that, without the \ac{pss}, the system is
unstable.  For $\Gain_{\rm w} \in (-0.55,0.65)$, the system is unstable
regardless of the magnitude of the delay.  The delay margin of the
system is $0.104$~s and is obtained for $\Gain_{\rm w} = 1.5$. Thus, operation
under the presence of a large delay, e.g.~$0.35$~s, is infeasible.

\subsection*{Dual-channel PSS}
\label{sec:dual}

In the \ac{omib} system example of Section~\ref{sec:omib}, a compact stable
region in the delay-control gain plane can be achieved by employing a
\ac{pr}-based \ac{pss} scheme, tuned to operate the system at a point with good
damping characteristics.

\begin{figure}[ht!]
  \centering
  \resizebox{0.65\linewidth}{!}{\includegraphics{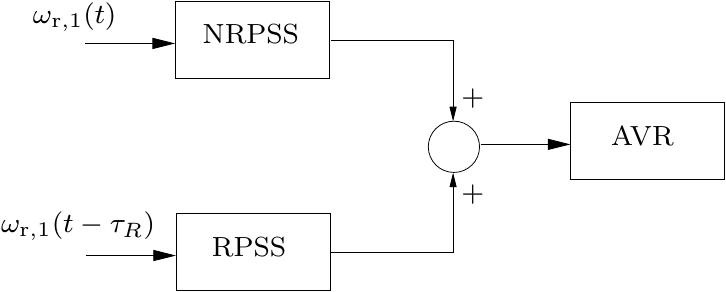}}
  \caption[Dual-channel PSS configuration]{Dual-channel PSS configuration.}
  \label{fig:conscheme}
\end{figure}

We apply the same principle in the IEEE 14-bus system. To this aim, we
test a \ac{pss} with two control channels: first channel is not delayed;
second channel is delayed.  The examined dual-channel \ac{pss}
configuration is shown in Figure~\ref{fig:conscheme}.

The first channel, namely Not Retarded \ac{pss} (NRPSS), is tuned to render
the non-delayed system small-signal stable.  The control input of
NRPSS is the local rotor speed $\wGi{1}(t)$.  The second channel,
namely Retarded \ac{pss} (RPSS), tunes the delay dynamics so that the
system operates at a point with good damping characteristics.  The
input signal of the RPSS is the delayed rotor speed
$\wGi{1}(t-\tau_{R})$, where $\tau_{R} \geq 0$ is the magnitude of
the delay.  The time constants of both NRPSS and RPSS are as
summarized in Table~\ref{tab:psspars}.  In addition, $\Gain_{{\rm w},P}$ and
$\Gain_{{\rm w},R}$ denote the gains of NRPSS and RPSS, respectively.  An
analogy between the dual-channel \ac{pss} configuration and the PR-based
\ac{pss} of the \ac{omib} system example of Section~\ref{sec:omib} is given in
Table~\ref{tab:analogy}.

\begin{table}[ht!]
  \centering
  \renewcommand{\arraystretch}{1.3}
  \caption[Analogy between the dual-channel PSS
    and the PR controller]{Analogy between the examined dual-channel PSS configuration
    and the PR controller of Section~\ref{sec:omib}.}
  \begin{tabular}{l|ll}
    \hline 
    System    & OMIB & IEEE 14-bus  \\
    \hline
    Non-retarded control & Proportional $\Gain_p$ & NRPSS \\
    Retarded control & $\Gain_r$, $\tau_r$ & RPSS \\
    \hline 
  \end{tabular}
  \label{tab:analogy}
\end{table}

The NRPSS gain is tuned so that the system without delayed control is
small-signal stable.  For $\Gain_{{\rm w},P}=5$, $\Gain_{{\rm w},R}=0$, \ac{sssa} shows that
the rightmost pair of eigenvalues is $-0.1376 \pm \jj 0.0203$.  The most
poorly damped pair is $-0.5171 \pm \jj 7.2516$, which yields a damping
ratio $0.071$.

\begin{figure}[ht!]
  \centering
\resizebox{0.85\linewidth}{!}{\includegraphics{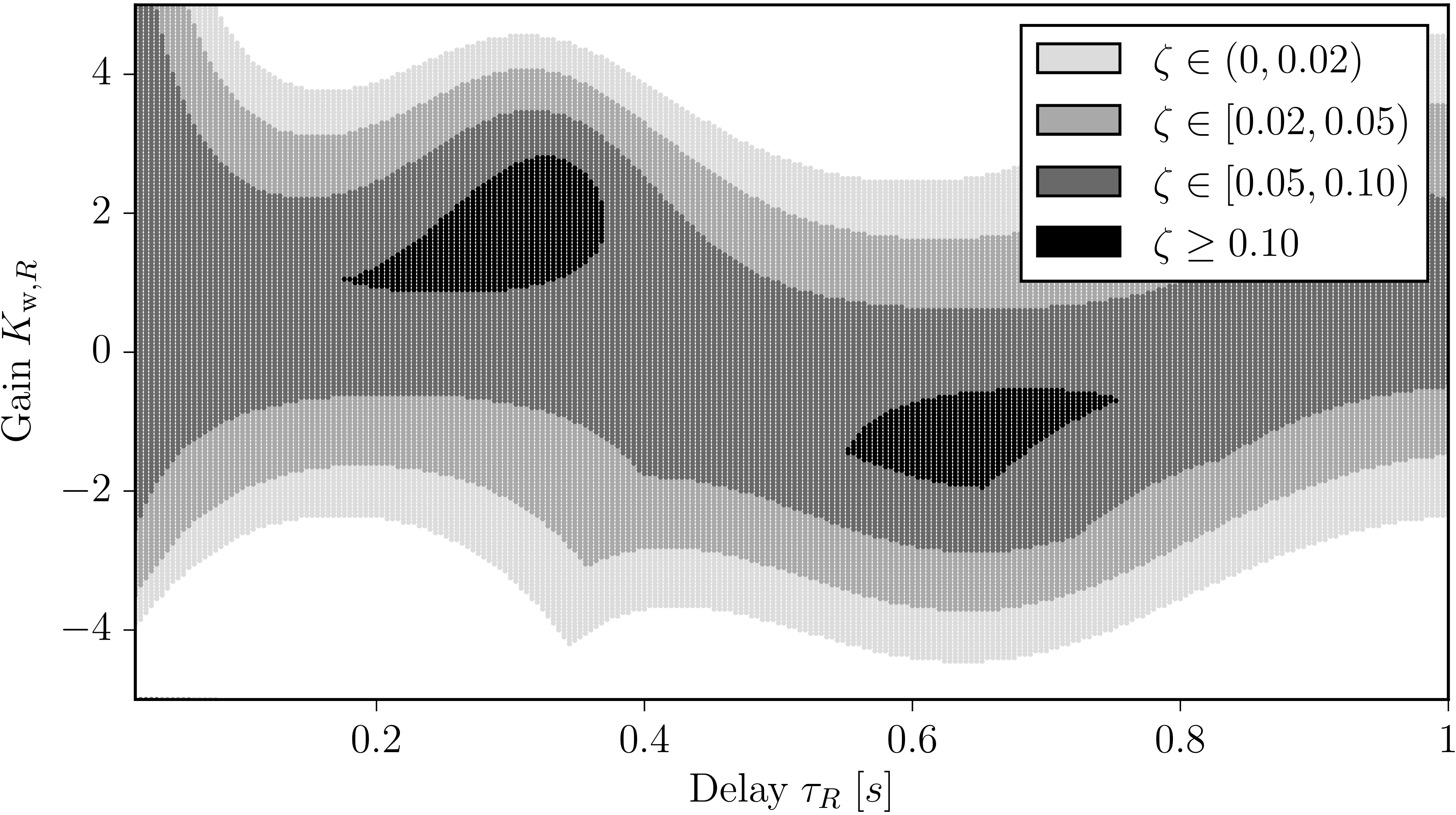}}
\caption[IEEE 14-bus system: stability map]{IEEE 14-bus system: $\zeta$-stability map in the
$\tau_R$-$\Gain_{{\rm w},R}$ plane.}
\label{fig:ieee14damp2}
\end{figure}

Considering $\Gain_{{\rm w},P}=5$, the $\zeta$-stability map of the
system is constructed in the $\Gain_{{\rm w},R}-\tau_{R}$ plane.  In this case, the dynamic
order of the system is $57$ and, using $N_C=10$, $570$ eigenvalues are in total calculated to obtain each point of the map.
The resulting map, presented in Figure~\ref{fig:ieee14damp2}, shows that
the stable region is compact, while the area with
$\Gain_{{\rm w},R} \in (-2.4,2.5)$ is delay independent stable.  In
Figure~\ref{fig:ieee14damp2}, maximum damping is $0.178$ and is achieved
for $\tau_R=0.34$~s, 
i.e.~a relatively
large delay value.

\section{Remarks}
\label{discussiom}

Delays arising in power system damping controllers are typically not
tunable but inherent, i.e.~they represent measurement and/or
communication latencies.  Although studying the impact of inherent
delays is not the main focus of this chapter, this section 
briefly discusses their relevance to the results presented above.

Regarding the standard \ac{pss} with delayed input signal of
Section~\ref{sec:standard}, assuming that $\tau$ is inherent does not
change the structure of the stability map in
Figure~\ref{fig:ieee14damp1} but only changes the interpretation of the role of delay.  Since in this case $\tau$ is not tunable,
the small delay margin of $0.104$~s may be a severe stability issue.
For inherent delays that have a small magnitude, a delay-dependent
design of a standard \ac{pss} allows
increasing the delay margin and avoid instability.

In fact, in Figure~\ref{fig:ieee14damp1}, the region of the highest damping
$\zeta \geq 0.10$ is obtained for a non-zero delay value.  The
closed-loop loci related to the critical system mode have an angle of
departure closer to $180^\circ$ when $\tau=0.03$~s.  In other words,
the phase shift introduced by the \ac{pss} is optimal when a small delay is present.  

In the case of the dual-channel \ac{pss} of Section~\ref{sec:dual}, if the
delay is inherent, the structure of Figure~\ref{fig:ieee14damp2} implies
that if one introduces a proper artificial delay on top of the
inherent delay, the system can be led to a region of better damping
characteristics.  This extra delay can be introduced, for example, by
a properly designed controller that adjusts both the delay and gain
values following a stable path, through consecutive quasi-steady state
shifts of the system equilibrium.

\section{Conclusions}
\label{tdc:sec:conclusion}

This chapter presents new results on time-delayed damping control of
power system synchronous machine electromechanical oscillations.  The
chapter focuses on the delay-control gain space, and studies the
stability boundaries, as well as the relationship between the
existence of delay-independent stability and connected stability domains.
Connected stable regions are obtained by employing a \ac{pss} with two
control channels and indicate that best damping characteristics may be
achieved for large delay values. 

The \ac{omib} system with inclusion of a \ac{pss} is a relevant example of power system
model that allows an analytical assessment of its stability when
delays are considered. This chapter shows the conditions for
which the stability of the linearized \ac{omib} equations is guaranteed
independently from the magnitude of the delay, and present how system response time 
as measured by the concept of $\sigma$-stability can be understood in view of recent results \cite{Ramirez:2015-ISA}.

On the plane of controller gain vs. intentional delay, the linearized equations of the \ac{omib} system typically exhibit stable
  regions that are separated by unstable regions.  This however does not
  allow tuning the ``non-linear'' dynamics to operate in separate
  stable regions as this would require the non-linear dynamics to first cross through an unstable region. This practically-relevant aspect of delayed dynamical
  systems is addressed by presenting the conditions under which the stable region
  of the \ac{omib} system can be all ``connected" so that the non-linear
  dynamics can be tuned for any settings inside this region.
  Finally, the analytical results based on the benchmark \ac{omib} system are extended through numerical methods and the concept of
  $\zeta$-stability.
 Specifically, the IEEE 14-bus system model serves to illustrate how to achieve improved damping characteristics for a set
  of controller gains and intentional delays, and how to achieve a fully
  connected stability region to be able to fully explore the parameter space, without introducing instability.
\newpage
\chapter{One-Step Delay Approximation}
\label{ch:osda}

\section{Introduction}


As discussed in Chapter~\ref{ch:tdc}, introducing time delays in a set of \acp{dae} turns it into a set of
functional \acp{dae} of retarded type,
also known as
\acp{ddae}. \acp{ddae} are typically
employed to model physical time delays. 
In power systems, apart from
their application to automatic control, mostly due to the inevitable latencies that are present in measurement 
and communication systems, see Chapter~\ref{ch:tdc},
delays have been also considered to study the effect of long transmission lines \cite{venk:94}. 
In addition, time delays 
are inherent to phenomena occuring in many other engineering applications, such as circuit and microwave theory \cite{bellen:99, minasian:06}. 

\begin{figure}[ht]
  \begin{center}
    \resizebox{0.7\linewidth}{!}{\includegraphics{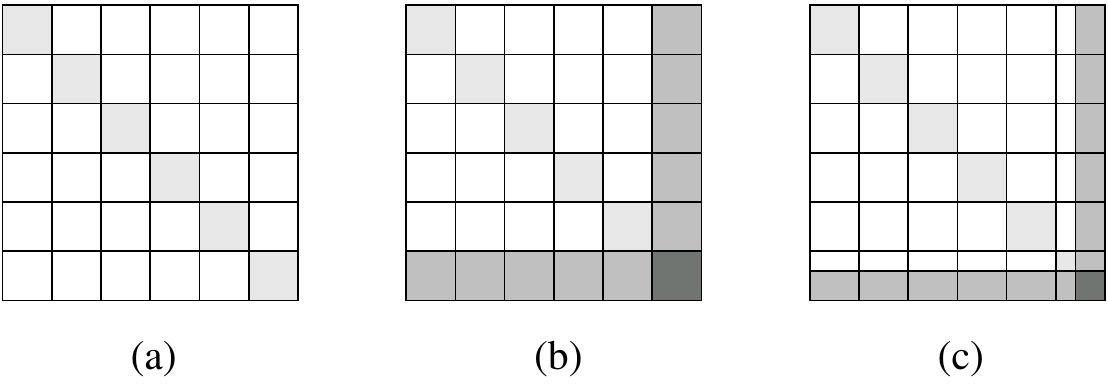}}
    \caption[Types of Jacobian matrices]{Types of Jacobian matrices: (a) block-diagonal matrix of
    a fully decoupled system; (b) coupled system ordered to exploit
    the BBD structure; and (c) coupled system with ordered BBD
    structure and extra decoupling obtained through fictitious
    delays.}
    \label{fig:bbd}
    \vspace{-2mm}
  \end{center}
\end{figure}

A property of constant delays is that the Jacobian 
elements with respect to retarded variables are null.  This feature
has been utilized for the simulation of \acp{emt}
that include long transmission lines \cite{WatsonEMTP} or control
systems \cite{317667, 1583706}.  The sending- and receiving-end variables of
long overhead lines, in fact, are decoupled by the transmission delay
and, hence, sections of circuits connected through long lines are
naturally decoupled.  In \cite{1583706}, on the other hand, the
control system is solved at the previous step of the \ac{emt} circuit
equations which, \textit{de facto}, introduces a delay in the control
equations.  This allows ordering the Jacobian matrix of the \acp{dae} with
a block diagonal structure (see Figure~\ref{fig:bbd}.a).  Each block can
be handled separately at each time step -- which is of the same order
of the delay, i.e.~$\mu$s -- and allows exploiting parallelization
techniques.


The effort of developing parallel algorithms in \acp{emt} simulations stems
from the fact that simulations of this type are slow for large systems
\cite{uriante:13}.  For systems with the same number of buses,
simulations based on quasi-steady state phasors and electromechanical
models are much faster.  However, the \ac{tdi} of
large power systems requires iteratively solving stiff non-linear
hybrid \acp{dae}, which is still a time-consuming task to complete.  The
time required to complete a $N-1$ contingency analysis, in fact, can
be a critical constraint, e.g.~for on-line dynamic security assessment
(see, for example, Chapter 15 of \cite{grigsby:2007}).

The \acp{dae} for transient stability analysis are naturally coupled
through the admittance matrix of the grid, which is modeled with a
set of algebraic equations, as well as by secondary frequency
controllers, and generally do not include delays.  This leads to a
\ac{bbd} structure of the Jacobian matrix (see
Figure~\ref{fig:bbd}.b) \cite{Shahidehpour:2003}.
The \ac{bbd} structure can be enforced in any set of \acp{dae} through
\textit{diakoptics}, i.e.~by introducing additional algebraic
equations \cite{kron:1963, URIARTE2012146}.  These equations increase
the order of the system but tend to increase sparsity and, in some
cases, may also speed up the factorization of the Jacobian matrix of
the system.  A technique conceptually similar to diakoptics, called
MANA, namely Modified Augmented Nodal Analysis, has been utilized in
unbalanced power flow analysis \cite{6624122} and \acp{emt} \cite{8274156}
but has no clear application for single-phase equivalent phasor-based
transient stability models.

The main idea of this chapter is that, if one includes fictitious delays
in the power system transient stability model, the Jacobian matrix can
be further decoupled (see Figure~\ref{fig:bbd}.c) without increasing the
system's order, thus increasing sparsity and reducing the
computational burden of numerical methods.  In this vein, in
\cite{VCutsemCOI}, the authors proposed the use of the \ac{coi} at the previous time step to decouple the equations of
the rotor angles of the synchronous machines.  In \cite{VCutsemCOI},
the ``slightly'' delayed \ac{coi} was tested on a 4-bus system using the
\ac{itm} with time step $0.01$~s and showed
not to affect the system transient response.

This chapter proposes a systematic way to implement and evaluate the idea of including \emph{one-step-delays}
to a \ac{dae} model for transient stability analysis. 
In order to ensure that the inclusion of fictitious one-step delays in the non-linear DAE power system model does not have a noticeable impact on the system trajectories,  the following crucial aspects of the technique are addressed: 
%
(i)  only variables that do not contribute to critical dynamic modes of the system are delayed.  Such variables are systematically identified based on the values of their residues at the frequency range of the dynamics of interest; and
(ii)  given a set of selected variables, a proper upper bound of the integration time-step is
established. 
 This is done by recognizing that the one-step-delay approximation in a coupled system can be formally studied as a set of \acp{ddae} \cite{osda}.

The remainder of the chapter is organized as follows.  Section
\ref{osda:sec:tdi} recalls a conventional implicit \ac{tdi} scheme for power
systems.  Section \ref{osda:sec:del} discusses the proposed approach to
one-step-delay approximation.  Section \ref{osda:sec:select} discusses how
to select the variables of a \ac{dae} model to be delayed.  Section
\ref{osda:sec:maxdel} provides a method to calculate the maximum admissible
delay for a given \ac{ddae} model.  The case study is discussed in Section~\ref{osda:sec:case}.
Conclusions are drawn in Section
\ref{osda:sec:conclusion}.
    
\section{Implicit Integration of Power Systems}
\label{osda:sec:tdi}

Consider the non-linear \ac{dae}
power system model \eqref{sssa:eq:sidae}, where for simplicity, but without loss of generality, no inputs are included and 
$\bfg T = \bfg I_\nx$,
$\bfg R = \bfg 0_{\ny,\nx}$,
see \eqref{sssa:eq:explcon}. This system can be written as:
\begin{equation}
  \label{osda:eqdae}
  \begin{aligned}
    \dot{\bfg x} &= \bfg f( \bfg x, \bfg y) \, , \\
    \bfg 0_{\ny,1} &= \bfg g(\bfg x, \bfg y) 
    \, .
  \end{aligned}
\end{equation}
Equations of system
\eqref{osda:eqdae} are \textit{stiff}, i.e. some numerical methods, when employed for the solution of \eqref{osda:eqdae}, are unstable. This happens for two reasons: 

\begin{enumerate}[label=(\alph*)]
    \item the time constants of
the differential equations typically span multiple time scales;
\item and the algebraic equations can be viewed as \textit{infinitely fast}
differential equations associated with zero time constants.
\end{enumerate}

There exist both explicit and implicit methods for the numerical integration of \eqref{osda:eqdae}.
Explicit numerical methods are known to 
be impractical for the integration of this system, since for stiff problems they require very small time steps which in turn leads to poor performance.
Thus, a more common approach to numerically integrate
\eqref{osda:eqdae} is to use an implicit method with a direct solver.
Employing an implicit method allows a simultaneous solution of both
state and algebraic variables \cite{stott:1979}, and requires the
solution of the following set of non-linear equations:
\begin{equation}
  \label{osda:eqpqeqs}
  \begin{aligned}
    \bfg 0_{\nx,1} &= \bfg \phi(\bfg x,\bfg y, h) \, , \\
    \bfg 0_{\ny,1} &= \bfg \psi(\bfg x,\bfg y, h) \, ,
  \end{aligned}
\end{equation}
where $\bfg \phi$, ($\bfg \phi$ :
$\mathbb{R}^{(\nx+\ny)} \rightarrow \mathbb{R}^{\nx}$) and $\bfg \psi$,
($\bfg \psi$ : $\mathbb{R}^{(\nx+\ny) } \rightarrow \mathbb{R}^{\ny}$) are
non-linear functions that depend on the differential and algebraic
equations, respectively, as well as on the applied implicit method.
The update of the state and algebraic variables at each time step can
be expressed as follows:
\begin{equation}
    \begin{bmatrix}
      \bfg x^{[i+1]}(t+h) \\
      \bfg y^{[i+1]}(t+h)
    \end{bmatrix} =
    \begin{bmatrix}
      \bfg x^{[i]}(t+h) \\
      \bfg y^{[i]}(t+h)
    \end{bmatrix}+ 
    \begin{bmatrix}
      \Delta \bfg x^{[i]} \\
      \Delta \bfg y^{[i]}
    \end{bmatrix} \, ,
  \end{equation} 
  where $h$ is the time step length; $\bfg x^{[i]}(t+h)$ denotes the
  vector $ \bfg x$ at the $i$-th iteration of time $t+h$.  The
  increments $\Delta \bfg x^{[i]}$, $\Delta \bfg y^{[i]}$ are obtained
  by employing the Newton method, as follows:
\begin{equation}
    \begin{bmatrix}
      \Delta \bfg x^{[i]} \\
      \Delta \bfg y^{[i]}
    \end{bmatrix} = -
    \begin{bmatrix}
      \bfg A^{[i]}
    \end{bmatrix}^{-1}
    \begin{bmatrix}
      \bfg \phi^{[i]} \\
      \bfg \psi^{[i]}
    \end{bmatrix} \, ,
\end{equation}
where $\bfg A^{[i]}$, $\bfg A^{[i]} \in \mathbb{R}^{(\nx+\ny) \times
  (\nx+\ny)}$, is defined as:
\begin{equation}
  \label{osda:eqnumjac1}
  \bfg A^{[i]} = 
  \begin{bmatrix}
    \bfg \phi_{\bfg x}^{[i]}  & \bfg \phi_{\bfg y}^{[i]} \\
    \bfg \psi_{\bfg x}^{[i]} & \bfg \psi_{\bfg y}^{[i]}
  \end{bmatrix} \, ,
\end{equation}
where $\bfg \phi_{\bfg x}^{[i]}$, $\bfg \phi_{\bfg y}^{[i]}$, $ \bfg \psi_{\bfg
  x}^{[i]}$, $ \bfg \psi_{\bfg y}^{[i]}$ are the Jacobian matrices of
$\bfg \phi$ and $\bfg \psi$ at the $i$-th iteration of time $t$.

Among the various implicit numerical methods utilized by power system
software tools to define equations \eqref{osda:eqpqeqs}, for simplicity
but without lack of generality, this chapter considers only one, namely the \ac{itm}, which is a well-known and widely utilized $A$-stable integration
scheme particularly adequate to handle \ac{dae} stiffness. 
%
The \ac{itm} leads to the following form of \eqref{osda:eqpqeqs}:
\begin{equation}
  \label{osda:eqitm}
  \begin{aligned}
    \bfg 0_{\nx,1} &= \bfg \phi^{[i]}= 
    \bfg x^{[i]} -
    \bfg x({t-h})-0.5 \, h\,  \big (\bfg f^{[i]} +
    \bfg f({t-h}) \big ) \, , \\
    \bfg 0_{\ny,1} &= \bfg \psi^{[i]} = \bfg g^{[i]} \, ,
  \end{aligned}
\end{equation}
where $h$ is the time step length; 
$\bfg f^{[i]}=\bfg f(\bfg x^{[i]},
\bfg y^{[i]})$, $\bfg g^{[i]}=\bfg g(\bfg x^{[i]}, \bfg y^{[i]})$ and
$\bfg f({t-h})=\bfg f(\bfg x({t-h}), \bfg y({t-h}))$.  The
Jacobian matrix \eqref{osda:eqnumjac1} at the $i$-th iteration of time $t$ is defined as:
\begin{equation}
  \label{osda:eqitmjac}
  \begin{aligned}
    {\bfg \phi_{\bfg x}^{[i]}} &= \bfg I_{\nx} - 0.5 \, h \,  \bfg f_{\bfg x}^{[i]} \, , \\ 
    {\bfg \phi_{\bfg y}^{[i]}} &=-0.5\, h\, 
    \bfg f_{\bfg y}^{[i]} \, , \\
    {\bfg \psi_{\bfg x}^{[i]}} &= \bfg g_{\bfg x}^{[i]}  \, , \\
    {\bfg \psi_{\bfg y}^{[i]}} &= \bfg g_{\bfg y}^{[i]} \, , 
  \end{aligned}
\end{equation}
where $\bfg f_{\bfg x}^{[i]}$, $\bfg f_{\bfg y}^{[i]}$, $\bfg g_{\bfg
  x}^{[i]}$, $\bfg g_{\bfg y}^{[i]}$ are the Jacobian matrices of the \acp{dae}.    
    
\section{One-Step-Delay Approximation}
\label{osda:sec:del}


Assume that some variables -- how to
select such variables is discussed later in this chapter -- of the \ac{dae} system \eqref{osda:eqdae} are
substituted with their values at the previous time step.  This
system can be formally studied as a system of non-linear \acp{ddae} with
a constant delay, as follows:
\begin{equation}
 \label{osda:eqddae}
  \begin{aligned}
    \dot{\bfg x}  &= \bfg {\tilde f}( \bfg x, \bfg y, \bfg x_d, \bfg y_d)
    \, , \\
    \bfg 0_{\ny,1} &= \bfg {\tilde g}(\bfg x, \bfg y, \bfg x_d, \bfg y_d) \, ,
  \end{aligned}
\end{equation}
where $\bfg x_d$, $ \bfg x_d \in \mathbb{R}^{\nx_{d} }$, $\bfg y_d$, $
\bfg y_d \in \mathbb{R}^{\ny_{d} }$, are the delayed state and algebraic
variables, respectively, as follows:
\begin{equation}
  \label{osda:eqab}
  \begin{aligned}
    \bfg x_{d} &= {\bfg x}(t-h) \, , \\ 
    \bfg y_{d} &= {\bfg y}(t-h) \, .
  \end{aligned}
\end{equation}
Note that \eqref{osda:eqddae} is an approximation of \eqref{osda:eqdae}.  The
delay $h$, in fact, is fictitious as it does not model any physical
phenomenon.

The numerical integration of \eqref{osda:eqddae} requires the solution of
the following set of non-linear equations \cite{delay1}:
\begin{equation}
  \label{osda:eqpqeqsddae}
  \begin{aligned}
    \bfg 0_{\nx,1} &= \bfg {\tilde \phi}(\bfg x, \bfg y, \bfg x_{d}, \bfg y_{d}, h) \, , \\
    \bfg 0_{\ny,1} &= \bfg {\tilde \psi}(\bfg x, \bfg y, \bfg x_{d}, \bfg y_{d}, h) \, .  
  \end{aligned}
\end{equation}

If at the $i$-th iteration of time $t$, $\bfg {\tilde f}^{[i]}_{{\bfg
    x}}$, $\bfg {\tilde f}^{[i]}_{{\bfg y}}$, $\bfg {\tilde
  g}^{[i]}_{{\bfg x}}$, $\bfg {\tilde g}^{[i]}_{{\bfg y}}$, are the
delay-free and $\bfg {\tilde f}^{[i]}_{{\bfg x}_d}$, $\bfg {\tilde
  f}^{[i]}_{{\bfg y}_d}$, $\bfg {\tilde g}^{[i]}_{{\bfg x}_d}$, $\bfg
{\tilde g}^{[i]}_{{\bfg y}_d}$, are the Jacobian matrices of the
delayed variables of system \eqref{osda:eqddae}, then the following
identities apply:
\begin{equation}
  \label{osda:eqcompjacs}
  \begin{aligned}
    \bfg {f}^{[i]}_{\bfg x} &= \bfg{\tilde f}^{[i]}_{\bfg x}+\bfg{\tilde f}^{[i]}_{\bfg x_d} \, , \\
    \bfg {f}^{[i]}_{\bfg y} &= \bfg{\tilde f}^{[i]}_{\bfg y}+\bfg{\tilde f}^{[i]}_{\bfg y_d} \, , \\
    \bfg {g}^{[i]}_{\bfg x} &= \bfg{\tilde g}^{[i]}_{\bfg x}+\bfg{\tilde g}^{[i]}_{\bfg x_d} \, , \\
    \bfg {g}^{[i]}_{\bfg y} &= \bfg{\tilde g}^{[i]}_{\bfg y}+\bfg{\tilde g}^{[i]}_{\bfg y_d} \, . \\
  \end{aligned}
\end{equation}
The main difference between \eqref{osda:eqpqeqsddae} and \eqref{osda:eqpqeqs}
is that the Jacobian matrices of \eqref{osda:eqpqeqsddae} do not include
the terms that depend on $\bfg x_{d}$ and $\bfg y_{d}$, as these
variables are ``constants'' at time $t$.  One has:
\begin{equation}
  \label{osda:eqnewton1}
  \begin{bmatrix}
    \Delta \bfg x^{[i]} \\
    \Delta \bfg y^{[i]} \\
  \end{bmatrix} = -
  \begin{bmatrix}
    \bfg {\tilde A}^{[i]}
  \end{bmatrix}^{-1}
  \begin{bmatrix}
    \bfg {\tilde \phi}^{[i]} \\
    \bfg {\tilde \psi}^{[i]} \\
  \end{bmatrix} \, ,
\end{equation}
where
\begin{equation}
  \label{osda:eqnumjac1delf}
  \bfg {\tilde A}^{[i]} = 
  \begin{bmatrix}
    \bfg {\tilde \phi}_{\bfg x}^{[i]} & \bfg {\tilde \phi}_{\bfg y}^{[i]}  \\
    \bfg {\tilde \psi}_{\bfg x}^{[i]} & \bfg {\tilde \psi}_{\bfg y}^{[i]}
  \end{bmatrix} \, .
\end{equation}
The terms $\bfg{\tilde \phi}_{\bfg x}^{[i]}$, $\bfg{\tilde \phi}_{\bfg
  y}^{[i]}$, $\bfg{\tilde \psi}_{\bfg x}^{[i]}$, $\bfg{\tilde \psi}_{\bfg
  y}^{[i]}$ are the delay-free Jacobian matrices of $\bfg {\tilde
  \phi}^{[i]}$ and $\bfg {\tilde \psi}^{[i]}$.  For a detailed description
on the modifications required by the \ac{itm} in order to
integrate a set of \acp{ddae} with inclusion of more general (time-varying
and state-dependent) delays, the interested reader may refer to
\cite{delay1}.
Since matrix $\bfg {\tilde A}$ is composed only of the delay-free
Jacobian matrix elements of $\bfg {\tilde \phi}$ and $\bfg {\tilde \psi}$,
$\bfg {\tilde A}$ is sparser than $\bfg {A}$.
The scope of this chapter is to 
take advantage of the fact that approximating \eqref{osda:eqdae} with
\eqref{osda:eqddae} leads to a sparser Jacobian matrix.

\section{Selection of Variables to be Delayed}
\label{osda:sec:select}



Inclusion of fictitious time delays in a set of \acp{dae} introduces an
inevitable approximation in its transient response.  It is thus
crucial to identify the variables and the equations that, if subject
to a small variation, do not lead to a significant change in the
system trajectories.  With this regard, delaying variables that are
slower than the dynamics of main interest, causes a smaller variation
in the system trajectories.
%


Another aspect is the position of the selected elements in the
Jacobian matrix.  Removing elements that introduce dense rows/columns
in the Jacobian leads not only to a sparsity increase, but also to
decoupling of the system equations, which in turn
allows exploiting state-of-the-art algorithms that parallelize the
factorization.  Such algorithms usually exploit the specific
formulation of current-injection power system models and the
admittance matrix to take advantage of the \ac{bbd} structure of the
Jacobian matrix \cite{Fong1978,FabozziLNewton}.  Exploiting
parallelization, 
however, is out of the scope of this chapter.  Thus the
general \ac{dae} model  \eqref{osda:eqdae} is considered.


\subsection{Systematic Selection of Variables}
\label{osda:subsec:var}


This section provides a systematic small-signal based method to select the delayed variables $\bfg x_d$ and $\bfg y_d$ of a model,
based on the geometric approach \cite{hamdan:87,freqpss:powertech}.
The geometric approach has been widely employed in control design
to provide a measure for {(i)}
the observability of a dynamic mode from
a signal; {(ii)} the controllability of a mode from a control input placement.
The product of these two measures provides the joint observability/controllability index. 
The smaller this index is, 
the less the examined mode is affected by the specific  signal-control input set. 
In the following, the geometric approach is utilized to determine the sensitivity of
system modes to variations of all non-zero elements of the \ac{dae}
Jacobian matrices.

%
%
%
%
Differentiating \eqref{osda:eqdae} around an equilibrium point yields:
\begin{equation}
  \label{osda:eq:lin}
  \begin{aligned}
    \Delta \dot{\bfg x} &= \jacs{f}{x} \, \Delta \bfg x + \jacs{f}{y} \, \Delta \bfg y \, , \\
    \bfg 0_{\ny,1} &= \jacs{g}{x} \, 
      \Delta \bfg x + \jacs{g}{y} \, \Delta \bfg y \, .
  \end{aligned}
\end{equation}
Elimination of $\Delta \bfg y$ leads to $\Delta \dot {\bfg{x}} = {\AS \Delta \bfg x}$, where the matrix
$\AS=\jacs{f}{x} - \jacs{f}{y} 
\, 
\jacsinv{g}{y} \, \jacs{g}{x}$
%
has $\nx$ finite eigenvalues 
$\lambda_1, \lambda_2, \ldots, \lambda_\nx$. 
For rotor angle stability studies, eigenvalues of interest are those that define oscillatory modes with natural frequency ${\rm f}_{{\rm n}, i} \in [0.1, 2]$~Hz \cite{arriaga:82_2}.
In the remainder of the chapter, these modes (and the
respective eigenvalues) are referred to as \textit{relevant}.

A perturbation is introduced into \eqref{osda:eq:lin} as follows:
\begin{equation}
  \label{osda:eqperturb}
  \begin{aligned}
    \Delta \dot {\bfg x} &= { \jacs{f}{x} \, \Delta \bfg x} + {\jacs{f}{y} \, \Delta \bfg y} + {\bfg {B}_{f} \, \Delta \bfg u_{f}}
    \, , \\ 
    \bfg 0_{\ny,1} &= {\jacs{g}{x} \, \Delta \bfg x} + \, \jacs{g}{y}
      \Delta \, \bfg y + {\bfg {B}_g \Delta \, \bfg u_g} \, ,
  \end{aligned}
\end{equation}
where $\Delta \bfg u_{f} \in \mathbb{R}^{\nx}$,
$\Delta \bfg u_g \in \mathbb{R}^{\ny}$ are the perturbation vectors of
the differential, algebraic equations, respectively; and
$\bfg {B}_{f}$, $\bfg {B}_g$ are the perturbation matrices associated
with $\Delta \bfg u_{f}$ and $\Delta \bfg u_{g}$, respectively.
Eliminating $\Delta \bfg y$ from \eqref{osda:eqperturb} yields:
\begin{equation}
  \label{osda:eqss}
  \begin{aligned}
    \Delta \Dt{\bfg x} &= {\AS \, \Delta \bfg x} + {\bfg B_{f} \,
      \Delta \bfg u_{f}} - \jacs{f}{y} \, \jacsinv{g}{y}
{\bfg {B}_g \, \Delta \bfg u_g} \, .
  \end{aligned}
\end{equation}
Considering zero perturbation matrices in \eqref{osda:eqss}, as discussed
in \cite{hamdan:87}, the output matrices of the
state and algebraic variable variations can be defined as $\bfg C_x = \bfg I_{\nx}$,
$\bfg C_y = - \jacsinv{g}{y} \jacs{g}{x}$, respectively.

Let $\jacs{f}{x}(\mu_f , \nu_x)$, be the $\mu_f$-th row, $\nu_x$-th column element of $\jacs{f}{x}$;
$\jacs{f}{y}(\mu_{f} , \nu_y)$
the 
element of $\jacs{f}{y}$;
$\jacs{g}{x}(\mu_g , \nu_x)$ the
$\mu_g$-th row, $\nu_x$-th column element of $\jacs{g}{x}$; and
$\jacs{g}{y}(\mu_g , \nu_y)$ the $\mu_g$-th row, $\nu_y$-th
column element of $\jacs{g}{y}$.
Then, the 
geometric controllability/observability
(${\rm gco}$)
measures of 
$\lambda_i$
from the Jacobian matrix elements of the system are determined as follows:

\begin{equation}
  \label{osda:eqgco:fx}
  {\rm gco}({\bfg f_{\bfg x}}(\mu_{f} , \nu_x)) =
  \frac{|\bfg c_{x,\mu_{f}} \, \reigv_i \, \leigv_i \, \bfg b_{f,\nu_x}|}
       {||\reigv_i || \, ||\bfg c_{x,\mu_{f}} || \,
         ||\leigv_i || \, ||\bfg b_{f,\nu_x} ||} \, , 
\end{equation}
 where $\bfg c_{x,\mu_{f}}$ is the $\mu_{f}$-th row of $\bfg C_x$; $\bfg b_{f,\nu_x}$ is the $\nu_x$-th column of $\bfg B_{f}$; $|\cdot|$ and $||\cdot||$
 denote the modulus and Euclidean norm, respectively;
\begin{equation}
  \label{osda:eqgco:fy}
  {\rm gco}({\bfg f_{\bfg y}}(\mu_{f} , \nu_y)) =
  \frac{|\bfg c_{y,\mu_{f}} \, \reigv_i \, \leigv_i \, \bfg b_{f,\nu_y}|}
       {||\reigv_i || \, ||\bfg c_{y,\mu_{f}} || \,
         ||\leigv_i || \, ||\bfg b_{f,\nu_y} ||} 
  \, , 
\end{equation}
 where $\bfg c_{y,\mu_{f}}$ is the $\mu_{f}$-th row of $\bfg C_y$; $\bfg b_{f,\nu_y}$ is the $\nu_y$-th column of $\bfg B_{f}$;
\begin{equation}
  \label{osda:eqgco:gx}
  {\rm gco}({\bfg g_{\bfg x}}(\mu_g , \nu_x)) =
  \frac{|\bfg c_{x,\mu_g} \, \reigv_i \, \leigv_i \bfg b_{g,\nu_x}|}
{||\reigv_i || \, ||\bfg c_{x,\mu_g} || \,
||\leigv_i || \, ||\bfg b_{g,\nu_x} ||}
  \, , 
\end{equation}
 where $\bfg c_{x,\mu_g}$ is the $\mu_g$-th row of
 $\bfg C_x$; $\bfg b_{g,\nu_x}$ is the $\nu_x$-th column of $\bfg B_g$; and
\begin{equation}
  \label{osda:eqgco:gy}
  {\rm gco}({\bfg g_{\bfg y}}(\mu_g , \nu_y)) =
  \frac{|\bfg c_{y,\mu_g} \, \reigv_i \, \leigv_i \, \bfg b_{g,\nu_y}|}
       {||\reigv_i || \, ||\bfg c_{y,\mu_g} || \,
         ||\leigv_i || \, ||\bfg b_{g,\nu_y} ||} 
  \, , 
\end{equation}
where $\bfg c_{y,\mu_g}$ is the $\mu_g$-th row of the output matrix
$\bfg C_y$; $\bfg b_{g,\nu_y}$ is the $\nu_y$-th column of the
perturbation matrix $\bfg B_g$.


%

Expressions \eqref{osda:eqgco:fx}-\eqref{osda:eqgco:gy}
allow selecting the elements of the Jacobians of \eqref{osda:eq:lin} that
can be delayed and thus can be eliminated from the matrices
$\jacs{f}{x}$, $\jacs{f}{y}$, $\jacs{g}{x}$ and
$\jacs{g}{y}$.  
Specifically, elements of such matrices that have low $\rm gco$ values for all relevant modes of the system, do not noticeably impact
the dynamic behavior of the system. Therefore, a candidate to be delayed is any element whose
$\rm gco$ value is below a given threshold $\rm gco_{\rm max}$.
Note that $\jacs{f}{x}$, $\jacs{f}{y}$, $\jacs{g}{x}$ and
$\jacs{g}{y}$ are stored as sparse matrices, hence only non-zero
elements are considered for the analysis above, which leads to an
efficient implementation.

\subsection{Illustrative Examples}  
\label{osda:subsec:examples}
 
The criteria described above are further discussed through some
illustrative examples, which are based on well-known devices and
models utilized in transient stability analysis.  In particular, this chapter
considers devices and controllers that are slow and/or couple several
variables of the system.



\subsubsection{Center of Inertia}



The algebraic variable of the \ac{coi} speed ($\omega_{\rm CoI}$) is
defined by the following algebraic equation:
\begin{align}
  \label{osda:eqcoi2}
  g_{(\omega_{\rm CoI})} :=0 &= \omega_{\rm CoI} -
  \sum_{i = 1}^{\upkappa}\frac{\Inert_i}{\Inert_T} \wGi{i} \ ,
\end{align}
where $\wGi{i}$, $i = 1,2,\ldots,\upkappa$, is the state of the speed
of the $i$-{th} machine; $\Inert_i$ is the mechanical starting time of the
$i$-{th} machine; and $\Inert_T = \Inert_1 + \Inert_2 + \ldots + \Inert_{\upkappa}$.
The \ac{coi} speed is used as a reference in the differential equations of
the generator rotor angles:
\begin{align}
  \label{osda:eqangle}
 f_{(  \Dt \delta_{{\rm r},i} )} := \Dt \delta_{{\rm r},i} =
 \Omega_b \big(
 \wGi{i}-\omega_{\rm CoI} \big) \, ,
\end{align}
where $\Omega_b$ is the angular frequency base.  The \ac{coi} provides the
``average" frequency trend of the system and thus represents a
relatively slow dynamic.
Delaying $\wGi{i}$ and $\omega_{\rm CoI}$ in \eqref{osda:eqcoi2},
\eqref{osda:eqangle}, respectively, allows removing the elements
$\partial g_{(\omega_{\rm CoI})} \big / \partial \wGi{i}$ and
$\partial f_{(\Dt \delta_{{\rm r},i})} \big / \partial \omega_{\rm CoI}$, 
which
constitute a dense row in $\jacs{g}{x}$ and a dense column in
$\jacs{f}{y}$, respectively.

\subsubsection{Turbine Governor}
The action of some \acp{tg} can be significantly slow,
as compared to primary damping and voltage controllers and hence,
adding one-step delays in some \ac{tg} \ac{dae} models, e.g. the ones described
in \cite{milano:10}, leads to increased sparsity without jeopardizing
the \ac{tdi} accuracy. On the other hand, since \ac{tg} variables typically do
not constitute dense segments in the Jacobian matrix, the increased
sparsity does not come with significant decoupling.

%
%

\subsubsection{Automatic Generation Control}

The \ac{agc} is used to provide secondary
frequency regulation to the power system. Consider a simplified
continuous \ac{agc} model that measures the \ac{coi} frequency and produces a
dynamic active power command ($P_{\rm s}$) which is distributed to the
machine \acp{tg} proportionally to their droops \cite{1601717}.  The
algebraic variable of the power order received by the $i$-th \ac{tg} is
defined by the following algebraic equation:
\begin{align}
  \label{osda:eqagc}
  g_{ (P_{{\rm ord},{i}} )} :=
  0 &= P_{{\rm ord},{i}}- \frac{\mathcal{R}_i}{\mathcal{R}_T} P_{\rm s} \, ,
\end{align}
where $P_{{\rm ord},{i}}$ is the \ac{tg} power order; $\mathcal{R}_i$ is the droop
constant; and $\mathcal{R}_T = \mathcal{R}_1 + \mathcal{R}_2 + \ldots + \mathcal{R}_{\upkappa}$.  Delaying
$P_{{\rm ord},{i}}$ in \eqref{osda:eqagc} removes
$\partial g_{(P_{{\rm ord},{i}})} / \partial P_{\rm s}$, which forms a
dense column in $\jacs{g}{x}$, while accuracy is not impacted,
because of the \ac{agc} slow action.

\subsubsection{Secondary Voltage Regulation}

The \ac{svr} model employed in this chapter is
based on the scheme proposed in the grid code of the Italian system.
For a detailed description of this scheme, the interested reader may
refer to \cite{sart:11}.  The \ac{svr} mainly consists of two control
levels.  The external loop receives the voltage measurement of a
selected pilot bus and computes the vector $\bfg Q_{\rm ref}$ that
represents reactive power limits for the participating to the \ac{svr}
generators.  $\bfg Q_{\rm ref}$ is compared with the actual reactive
power generation vector $\bfg Q$ and the error
$\bfg Q_{\rm err} = \bfg Q_{\rm ref} - \bfg Q$ is further processed by a
dynamic decoupling matrix $\bfg {D}$.  The produced vector is finally
sent to the \acp{grpr}.  Each \ac{grpr} is
basically a \ac{PI} control, the output of which is considered as input to
the voltage reference of the generator's \ac{avr}.
The dynamic behavior of the the $i$-th \ac{grpr} state variable $x_{r,i}$
is given by the PI differential equation:
\begin{align}
  \label{osda:eqxri}
  f_{ (x_{r,i})} := \Dt x_{r,i} = K_{I} \, {\bfg {D}_i} \, \bfg Q_{\rm err} \, ,
\end{align}
where $K_I$ is the integral gain of the \ac{grpr}; $\bfg {D}_i$ is
the $i$-th row of $\bfg {D}$.  
Delaying $\bfg Q_{\rm err}$ in \eqref{osda:eqxri} allows eliminating
$\partial f_{(x_{r,i})} / \partial \bfg Q_{\rm err}$, which constitutes
a dense block of columns and rows in $\jacs{f}{y}$. The accuracy
of the integration is maintained, due to the relatively slow time
scale of the \ac{svr} action.





\section{Maximum Delay / Time Step}
\label{osda:sec:maxdel}

This section presents a technique 
based on \ac{sssa}, which
for a selected
set of $\bfg x_d$ and $\bfg y_d$, estimates the maximum admissible delay $h_{\rm max}$
that allows keeping the errors between the
original \acp{dae} and the modified \acp{ddae} below a threshold.
%
%
To this aim, first one has to solve the eigenvalue problem of the linearized
delayed system.  Linearizing \eqref{osda:eqddae} around a valid operating
point yields:
\begin{equation}
  \begin{aligned}
    \label{osda:eq:linddae}
    \Delta \dot{\bfg x} &=
    \tilde {\bfg f}_{\bfg x} \, \Delta \bfg x +
    \tilde {\bfg f}_{\bfg y} \, \Delta \bfg y +
    \tilde {\bfg f}_{\bfg x_d} \, \Delta \bfg x_d +
    \tilde {\bfg f}_{\bfg y_d} \, \Delta \bfg y_d \, ,
    \\
    \bfg 0_{\ny,1} &=
    \tilde {\bfg g}_{\bfg x} \, \Delta \bfg x +
    \tilde {\bfg g}_{\bfg y} \, \Delta \bfg y +
    \tilde {\bfg g}_{\bfg x_d} \, \Delta \bfg x_d +
    \tilde {\bfg g}_{\bfg y_d} \, \Delta \bfg y_d \, .
  \end{aligned}
\end{equation}
Eliminating the algebraic variables from \eqref{osda:eq:linddae} is
possible under the assumption that $\bfg {\tilde g}_y$ is not
singular, as follows:
\begin{align}
  \Delta  \dot{\bfg x} = \bfg {A}_0 \, \Delta \bfg x + 
  \bfg A_1 \, \Delta \bfg x_d + 
  \sum^{\infty}_{k=2} \big (\bfg A_k \, \bfg x(t- k \,h)
  \big ) \, ,
 \label{osda:eq:01k}
 \end{align}
 where $\bfg {A}_0$ is the delay-free system matrix; 
 $\bfg A_k$, $k \geq 2$, are the delayed system matrices.
%
%
%
%
%
A rigorous proof of \eqref{osda:eq:01k}, as well as the condition under
which the series in \eqref{osda:eq:01k} converges, are provided in
\cite{hessenberg:16}.  The series typically converges rapidly as $k$
increases and, thus, it is acceptable to assume a finite maximum value
for $k$, say 
$\rho$,
in the summation of \eqref{osda:eq:01k}, and hence,
the characteristic matrix of \eqref{osda:eq:01k} can be approximated with
the following pencil:
\begin{align}
  \label{osda:eqchareq}
  s \, \bfg{I}_{\nx} - \bfg{A}_0
  -\sum^{\rho}_{k=1} \big (e^{-s k h} \bfg A_k \big) \, .
\end{align}

Based on the above, the following proposition on the
continuity of the eigenvalues of \eqref{osda:eqchareq} is
relevant.

{
\proposition
{
Let $\hat \lambda$ 
be an eigenvalue of \eqref{osda:eqchareq} with
multiplicity 
$\alpha$.  
There exists a constant $\hat{\epsilon}$ such
that for all $\epsilon>0$ satisfying $\epsilon<\hat{\epsilon}$, there
is a number $\xi>0$ such that the pencil:
\begin{align}
  \label{osda:eqcontin}
 s \, \bfg{I}_{\nf} - (1+\xi) \Big ( \bfg{A}_0
    -\sum^{\rho}_{k=1} e^{-s k (1 + \xi) h} \bfg A_k
    \Big ) \, ,
\end{align}
where 
\begin{align*}
  & \xi {h} \in \mathbb{R},  \ \ \ \ \ 
  \ ||\xi{h}||<\xi,  \ {h}+\xi {h} \geq 0 \, , \nonumber \\
  & \xi \bfg A_k \in \mathbb{R}^{\nx \times \nx},  \ 
  ||\xi \bfg A_k||_2<\xi \, , \, k=0, 1, \dots, \rho \, , \nonumber
\end{align*}
has exactly $\alpha$ 
eigenvalues in the disk:
$\{ s \in
\mathbb{C}:|s-\hat \lambda|<\epsilon \}$.  
The notation $||
\cdot ||_2$ implies the induced matrix 2-norm.
}
\label{osda:proposition:roots}
}

The proposition above states that the characteristic roots of a delayed
system behave
continuously with respect to variations of system matrices and delays \cite{niculescu}.

In the proposed scheme, the modes of the time-delay system are viewed
as approximations of the modes of the delay-free system.  Let
$\lambda_i$ 
and $\hat{\lambda}_i$ 
be the $i$-th rightmost, 
non-null
eigenvalues of the delay-free and the delayed system, respectively.
The associated relative error is:
\begin{align}
  \label{osda:eqrelerr}
  \eta_{i} = \frac{|\hat \lambda_i - \lambda_i|}{|\lambda_i|} \ .
\end{align}
The limit case $h=0$ leads to 
$\eta_{i} = 0 $, 
$\ 0 \leq i \leq n$; and, for $h>0$, $\eta_{i} \ge 0$.  
Assigning a maximum admissible error, say $\eta_{\max}$, allows
finding the delay upper bound $h_{\max}$, as follows:
\begin{align}
  \label{osda:eqrelerrmax}
  \eta_{\max} \ge \frac{|\hat \lambda_i(h_{\max}) - \lambda_i|}{|\lambda_i|} \, ,
  \quad \forall i=1,2, \dots, \nu, \ \nu \leq n \, .
\end{align}

The calculation of $h_{\rm max}$ requires to find the eigenvalues of
\eqref{osda:eqchareq}, which implies solving a non-linear, transcendental
characteristic equation.  Transforming \eqref{osda:eqchareq} into a linear
pencil is possible by using a partial differential equations representation of the system, which however, has infinite dimensions.
A reduced set of eigenvalues can be found
by employing Chebyshev discretization, a brief description of which was given in Section
\ref{subsec:numlarge}.
 
The following remarks are relevant.

{
\remark
{ \textit{(Delay vs time step):}
In general, the time step of the numerical integration is determined based on the fastest dynamics of the system, whereas the variables that are delayed in this work are typically associated with slow dynamics.  This means that the time step is always smaller than the time scale of delayed variables.
Regarding the magnitude of the time delay, if the delay is greater than the integration time step, an extra, undesirable approximation is introduced into the system,
as Proposition \ref{osda:proposition:roots} indicates that the difference between
the \acp{dae} and \acp{ddae} are the smaller, the smaller is the delay.  On the
other hand, for delays smaller than the time step, the numerical integration has to
interpolate the delayed values, which introduces an additional source
of error in the trajectories of the \acp{ddae}.  In general, handling
delays smaller than the step size is an open research topic, as it
creates difficulties even for special integration methods for stiff
\acp{dde} \cite{Guglielmi2001}.  For the reasons above, in the proposed
formulation, the delay is always equal to the time step.
}
}

{
\remark
{
\textit{(Stiffness)}:
Apart from the approximation introduced with the delay, the maximum
step $h_{\max}$ is also constrained by the stiffness of the \acp{ddae} and
the numerical integration method.
%
In the following, the system is integrated using the
\ac{itm}.


}
}

{
\remark
{
\textit{(Computational burden)}:
The approach presented in 
Section~\ref{osda:subsec:var} and 
Section~\ref{osda:sec:maxdel} is based
on \ac{sssa}, which is valid around an equilibrium. \ac{sssa} based techniques in this chapter are used to capture a feature of the power system model that is ``robust'', i.e. does not substantially change by varying the operating point. Hence, the analysis can be carried out only once per network. 
Some references have addressed a similar problem.  For example, see the discussion on the participation matrix and identification of relevant state variables in \cite{arriaga:82_2}; and the use of \ac{sssa} techniques for non-linear dynamic model reduction in \cite{book:chow:13}. 
}
}
\section{Case Studies}
\label{osda:sec:case}
 
Two power system models are considered in this section.
In particular, Section~\ref{osda:subsec:39bus}
is based on the \newengland
system and employs the discussions of Sections \ref{osda:sec:select} and 
\ref{osda:sec:maxdel}
for selection of variables and
estimation of the maximum admissible time step.
Then, Section~\ref{osda:subsec:entsoe}
considers a 
21,177-bus model of the \ac{entsoe}.
This system is large enough to allow
properly discussing the impact of the proposed approach on
the convergence and the 
computational 
burden of the \ac{tdi}.
 
 
\subsection{\newengland System} 
\label{osda:subsec:39bus}
 
This section presents simulation results based on the IEEE 39-bus system, also known as the New England 10-machine system. 
Detailed static and dynamic data of the \newengland system can be found in \cite{milano_ortega:2019}.
It consists of $10$ synchronous generators, all represented by $4$-th
order (two-axis) models  \cite{milano:10}; $34$ transmission lines; $12$ transformers; and $19$ loads. 
Each generator is equipped with \ac{avr}, \ac{tg} and \ac{pss},
and thus provides primary voltage, primary frequency and damping control,
respectively, to the system. 
In this chapter, all generators are assumed to participate to secondary frequency and voltage control
through \ac{agc} and \ac{svr} schemes, respectively.  Note that the \ac{coi} speed is
used as angular frequency reference of the generators. In total,
the system has $141$ state variables 
and $253$ algebraic variables. 
 
The state matrix $\bfg A$ has $141$
finite eigenvalues, $48$ of which have natural frequencies that fall
in the range $[0.1, 2]$~Hz and are thus considered \textit{relevant}
eigenvalues for the analysis carried out below.
Table \ref{tab:jacnnz} shows the \ac{nnz} elements of
the Jacobian matrix of the original system.  The full $394 \times 394$
Jacobian matrix $\bfg A$ has $1,704$ non-zero elements, which
corresponds to density $1.098 \%$.
\begin{table}[ht!]
  \renewcommand{\arraystretch}{1.15}
  \centering
  \caption[\newengland system: NNZ Jacobian elements of the original DAE system]{\newengland system: NNZ Jacobian elements of the original DAE system.}
  \label{tab:jacnnz}
  \begin{tabular}{cccccc}
    \hline            
    $\jacs{f}{x}$  & $\jacs{f}{y}$ 
 &  $\jacs{g}{x}$& $\jacs{g}{y}$ & Total & Density (\%)\\
    \hline
    $281$&$271$ &$140$ &$1,012$ &$1,704$& $1.098$\\
    \hline  
  \end{tabular}
\end{table}

\begin{figure}[ht!]
  \begin{center}
  \resizebox{0.85\linewidth}{!}{\includegraphics{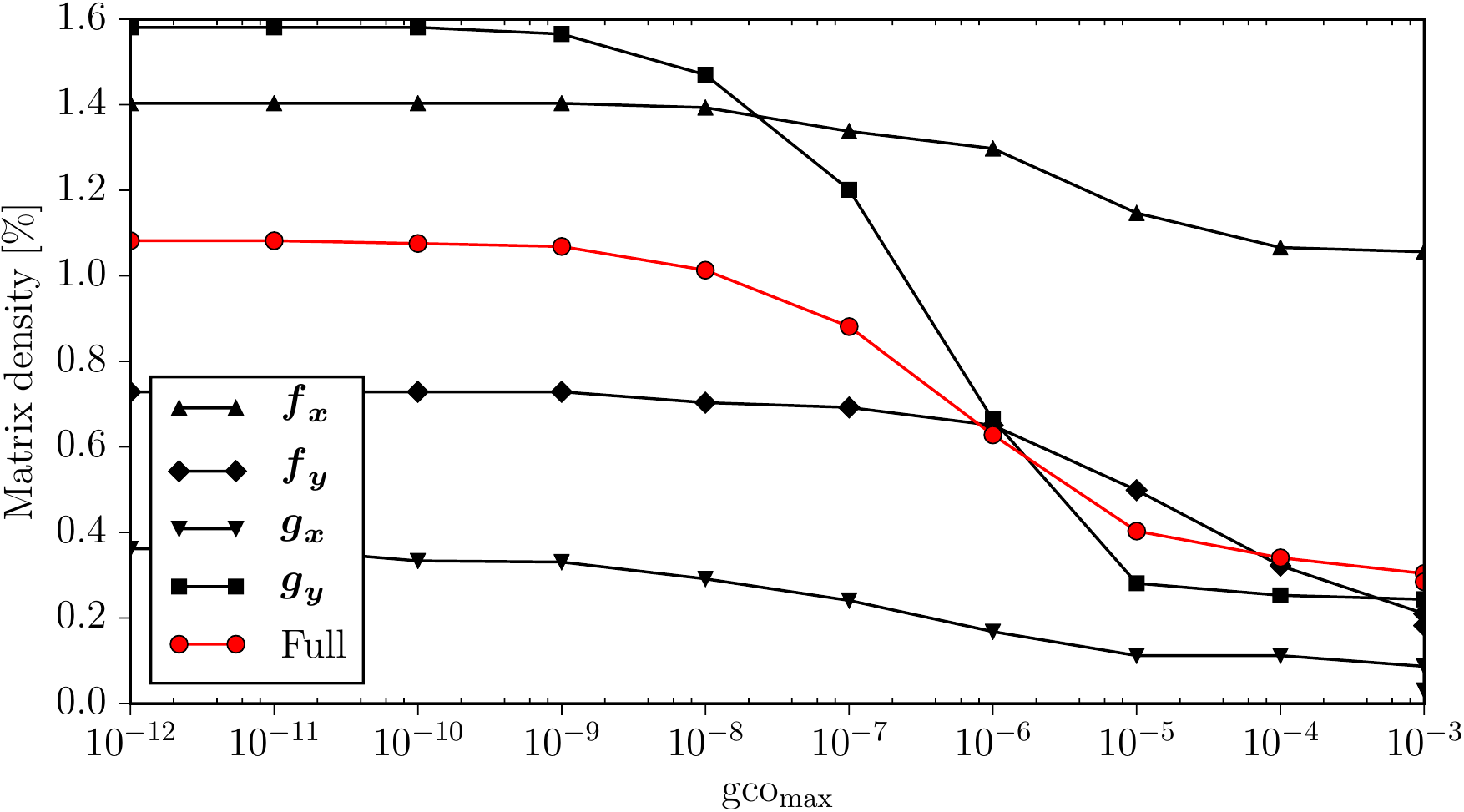}}
    \caption[\newengland system: density of Jacobians]{
    \newengland system: density of Jacobians as $\rm gco_{\max}$ varies.}
    \label{fig:dens}
  \end{center}
\end{figure}
The method discussed in
Section~\ref{osda:subsec:var} is applied to the \newengland system. 
The effect of the selected
threshold $\rm gco_{\max}$ on the density of the system Jacobian
matrices is shown in Figure~\ref{fig:dens}.  As expected, the higher
$\rm gco_{\max}$, the more elements are selected and the sparser
the delayed Jacobian matrices become.
 
\begin{table*}[ht!]
  \renewcommand{\arraystretch}{1.05}
  \centering
  \caption[\newengland system without delays: rightmost eigenvalues]{\newengland system without delays: rightmost eigenvalues.}
  \label{osda:tab:eigs}
  \begin{tabular}{c}
  \hline
    $-0.00782$  
    \\
    $-0.01400\pm \jj0.03721$
    \\
    $-0.02000$  
    \\
    $-0.02890$  
    \\
    $-0.02998$ 
    \\
    $-0.04009$
    \\
    $-0.04368$
    \\
    $-0.05554$
    \\
    $-0.05776$     
    \\
    $-0.06160$      
    \\
    $-0.06179$   
    \\
    $-0.06312$    
    \\
    $-0.08362\pm \jj0.02745$
    \\
    $-0.10001$      
    \\
    $-0.10002$      
    \\
    \hline   
  \end{tabular} 
\end{table*} 

\begin{table*}[ht!]
  \renewcommand{\arraystretch}{1.05}
  \centering
  \caption[\newengland system with one-step delay: rightmost eigenvalues]{\newengland system with one-step delay: rightmost eigenvalues, $\rm gco_{\max} = 10^{-10}$.}
  \label{osda:tab:eigdel}
  \begin{tabular}{c|c|c}
    \hline
     \multicolumn{1}{c|}{$h=0.01$~s}
    &\multicolumn{1}{c|}{$h=0.2$~s}  
    &\multicolumn{1}{c}{$h=0.24$~s}\\
    \hline
    \hline
    $\hat \lambda_i$ 
    & $\hat \lambda_i$ 
    & $\hat \lambda_i$ 
    \\
    \hline 
    $-0.00782$        
    & $-0.00783$            
    & $-0.00784$            
    \\
    $-0.01400\pm \jj0.03720$
    & $-0.01399\pm \jj0.03719$
    & $-0.01397\pm \jj0.03717$
    \\
    $-0.02000$            
    & $-0.02000$            
    & $-0.02000$            
    \\
    $-0.02890$ 		      
    & $-0.02890$ 		      
    & $-0.02890$ 		      
    \\
    $-0.02998$ 		      
    & $-0.02998$ 		      
    & $-0.02998$ 		      
    \\
    $-0.04009$            
    & $-0.04009$            
    & $-0.04009$            
    \\
    $-0.04368$            
    & $-0.04368$            
    & $-0.04368$            
    \\
    $-0.05554$            
    & $-0.05553$            
    & $-0.05553$            
    \\
    $-0.05776$            
    & $-0.05776$            
    & $-0.05776$            
    \\
    $-0.06160$            
    & $-0.06160$            
    & $-0.06160$            
    \\
    $-0.06179$            
    & $-0.06179$            
    & $-0.06179$           
    \\
    $-0.06312$            
    & $-0.06312$            
    & $-0.06312$            
    \\
    $-0.08364\pm \jj0.02748$
    & $-0.08388\pm \jj0.02770$
    & $-0.08426\pm \jj0.02804$
    \\
    $-0.10001$            
    & $-0.10001$            
    & $-0.10001$            
    \\
    $-0.10002$            
    & $-0.10002$            
    & $-0.10002$            
    \\
    \hline   
  \end{tabular} 
\end{table*}
 
\begin{table*}[ht!]
  \renewcommand{\arraystretch}{1.05}
  \centering
  \caption[\newengland system with one-step-delay: relative errors of rightmost eigenvalues]{\newengland system with one-step-delay: relative errors of rightmost eigenvalues, $\rm gco_{\max} = 10^{-10}$.}
  \label{osda:tab:err2}
  \begin{tabular}{c|c|c}
    \hline
     \multicolumn{1}{c|}{$h=0.01$~s}
    &\multicolumn{1}{c|}{$h=0.2$~s}  
    &\multicolumn{1}{c}{$h=0.24$~s}\\
    \hline
    \hline
    $\eta_i$\,(\%) 
    & $\eta_i$\,(\%) 
    & $\eta_i$\,(\%) 
    \\
    \hline 
    $0.00$
    & $0.13$ 
    & $0.26$ 
    \\
    $0.03$
    & $0.06$ 
    & $0.13$ 
    \\
    $0.00$
    & $0.00$ 
    & $0.00$ 
    \\
    $0.00$
    & $0.00$ 
    & $0.00$
    \\
    $0.00$
    & $0.00$ 
    & $0.00$
    \\
    $0.00$
    & $0.00$ 
    & $0.00$
    \\
    $0.00$
    & $0.00$ 
    & $0.00$
    \\
    $0.00$
    & $0.02$ 
    & $0.02$
    \\
    $0.00$
    & $0.00$ 
    & $0.00$
    \\
    $0.00$
    & $0.00$ 
    & $0.00$
    \\
    $0.00$
    & $0.00$ 
    & $0.00$
    \\
    $0.00$
    & $0.00$ 
    & $0.00$
    \\
    $0.04$
    & $0.41$ 
    & $0.99$
    \\
    $0.00$
    & $0.00$
    & $0.00$ 
    \\
    $0.00$
    & $0.00$ 
    & $0.00$
    \\
    \hline   
  \end{tabular} 
\end{table*}

For example, consider $\rm gco_{\rm max} = 10^{-10}$.  
In this case, selected
variables include
variables of \acp{tg}; rotor speeds that appear in the equation of the \ac{coi}; variables of the \ac{svr}. 
The method suggests first variables of slow acting devices, which is also consistent
to the discussion of
Section~\ref{osda:sec:select}.
The $35$ rightmost
eigenvalues of the system without and with inclusion of different delays are shown in Tables~\ref{osda:tab:eigs}
and \ref{osda:tab:eigdel}.
The relative errors of the system for these
eigenvalues are calculated according to \eqref{osda:eqrelerr} 
and 
results are presented in 
Table~\ref{osda:tab:err2}.  If
$h=0.01$~s, all relative eigenvalue errors are below $0.05 \%$. The
relative eigenvalue errors increase for larger delays. According to the discussion of Section~\ref{osda:sec:maxdel}, if the maximum
relative error is $\eta_{\max}=1$~\%, then $h_{\max}=0.24$~s.
Finally, as illustrated in Figure~\ref{fig:dens}, constantly increasing $\rm gco_{\max}$ leads to more and more variables being selected, which gradually limits
the value of $h_{\max}$.  
However, following from 
\eqref{osda:eqgco:fx}-\eqref{osda:eqgco:gy},
variables that inherently define relevant modes are consistently not selected.

The geometric approach can provide an insight of the system structure in a systematic and
model-agnostic way,  
unlike for example, the methods
proposed in \cite{Fong1978} and \cite{FabozziLNewton}.
This feature is particularly important for modern power systems where converter-interfaced devices can change, in a future not too far away, the overall dynamic response of the system. 
Still, it is common that variables of
a 
conventional power system \ac{dae} model are well-known. Then, $\bfg x_d$, $\bfg y_d$
can be selected based on the user's experience, and thus
without applying a systematic method.
The variables that if delayed, do not change or change in a negligible way the overall dynamic behavior of the system, are typically the ones
with significantly slower dynamic response as compared to the critical modes of the system. Selected variables are thus naturally decoupled by the critical dynamics of the system due to their different time scale.
With this regard, two comments are relevant. 
First, for any set of selected elements,
$h_{\rm max}$ is not
known a priori, so it can be still estimated according to
the method described in Section~\ref{osda:sec:maxdel}.
Second, while selecting $\bfg x_d$, $\bfg y_d$, the user should take into account that, how slow a variable
actually is depends on the state matrix 
$\bfg A$ and, in turn, on the parameters of the examined system.  For example,
consider again the example of the \ac{coi}.
Differentiation
of \eqref{osda:eqcoi2} yields:
\begin{align}
  \label{osda:eqrococoi}
  \dot \omega_{\rm CoI}
  &= \sum_{i = 1}^{\upkappa}\frac{\Inert_i}{\Inert_T}  \,
  \dot {\omega}_{{\rm r},i} \, ,
\end{align}
where $ \dot {\omega}_{{\rm r},i}$ is given by the well-known swing equation:
\begin{align}
  \label{osda:eqswing}
\dot {\omega}_{{\rm r},i} &= \frac{1}{\Inert_i} ( \torque_{{\rm m},i}-\torque_{{\rm e},i}
  - D_i \, ( \wGi{i} - \omega_{\rm CoI}) ) \, ,
\end{align}
where $\torque_{{\rm m},i}$, $\torque_{{\rm e},i}$ are the mechanical and
electrical torque, respectively; $D_i$ is the damping coefficient of the $i$-{th} machine.
Substitution of \eqref{osda:eqswing} to \eqref{osda:eqrococoi} gives:
\begin{align}
  \label{osda:eqrococoi2}
  \dot \omega_{\rm CoI} 
  &=   
  \frac{1}{\Inert_T} \big( \torque_{{\rm m},T}-\torque_{{\rm e},T} -
  \sum_{i = 1}^{\upkappa} 
  D_i \, ( \omega_i - \omega_{\rm CoI})    \big)
  \ ,
\end{align}
where $\torque_{{\rm m},T}=\torque_{{\rm m},1}+\torque_{{\rm m},2} +
\ldots+\torque_{{\rm m},\upkappa}$; and $\torque_{{\rm e},T}=\torque_{{\rm
    e},1}+\torque_{{\rm e},2} + \ldots+\torque_{{\rm e},\upkappa}$.
A characteristic of the 39-bus system is that $\Inert_1 \gg \Inert_i$, $i \neq
1$ ($\Inert_1 = 1000$~MWs/MVA, while the second larger mechanical starting
time is $\Inert_{10} = 84$~MWs/MVA).  In this case, the rate of change of
$\omega_{\rm CoI}$ is still slow
(see Section~\ref{osda:subsec:examples}), but, as seen from  \eqref{osda:eqrococoi2}, its rate of change is comparable
with that of $\wGi{1}$. 
%

Delayed variables are associated with secondary controllers or ``slow'' variables such as the center of inertia.  The dynamic response of these variables cannot change even for relatively big changes of the operating point and topology of the system.  As a matter of fact, one could select \textit{a priori} most of these variables.  
However, the eigenvalue analysis provides a systematic approach that can cope with any system setup and any device and controller.  This feature is particularly important for modern power systems where converter-interfaced devices can change, in a future not too far away, the overall dynamic response of the system.


%
\begin{figure}[ht!]
\centering
  \begin{center}
    \resizebox{0.85\linewidth}{!}{\includegraphics{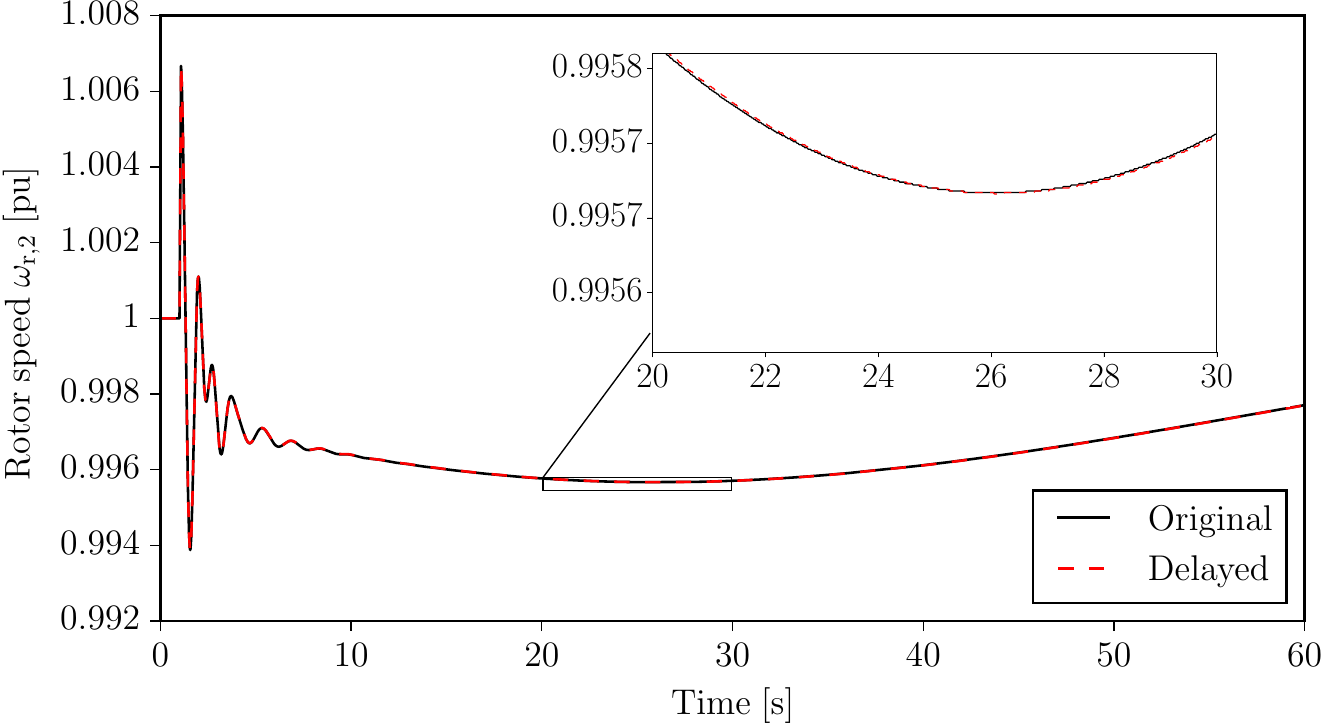}}
    \subcaption{$h=0.02$~s.}
    \vspace{0.5cm}
     \resizebox{0.85\linewidth}{!}{\includegraphics{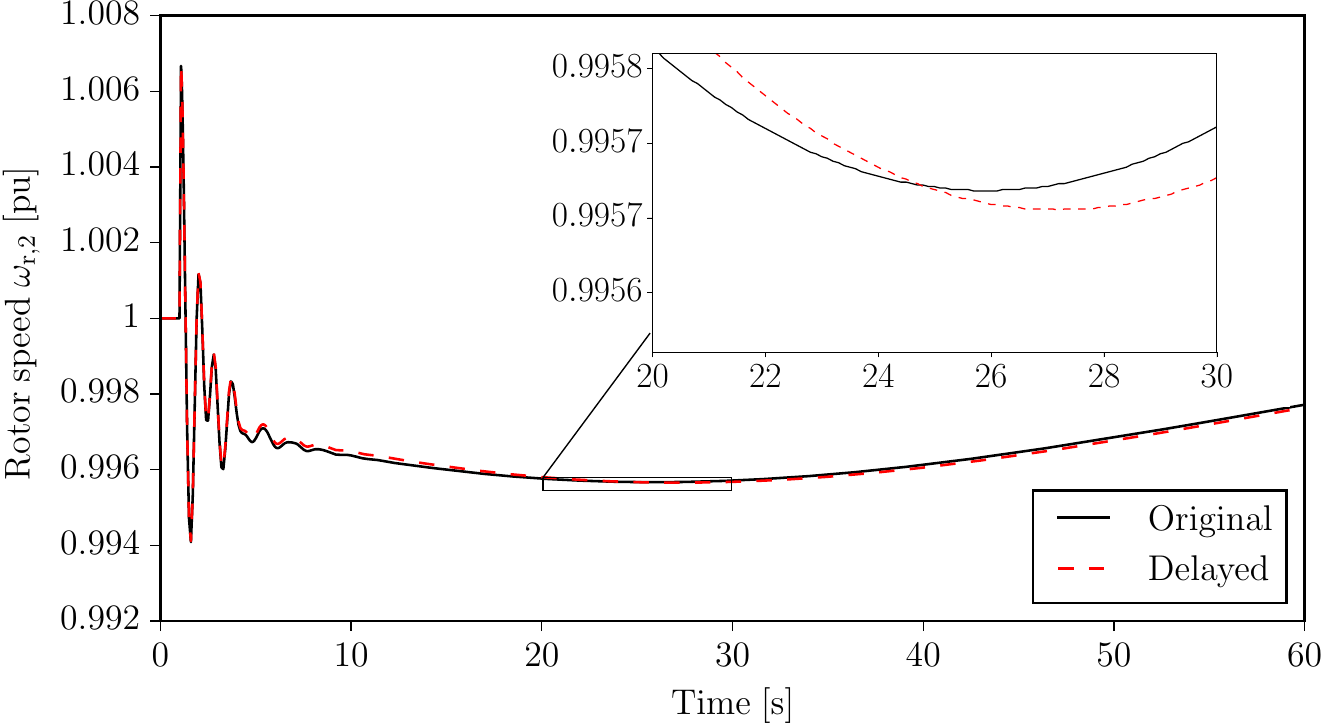}}
    \subcaption{$h=0.10$~s.}
    %
  \end{center} 
  \caption[\newengland system: transient following a three-phase fault]{\newengland system: transient following a three-phase fault.}
  \label{fig:trans}
\end{figure}

The next example discusses the effect of
the one-step delay approximation in the transient response of the
39-bus system by carrying out non-linear time domain simulations.  
With this aim and according to the discussion
of Section~\ref{osda:subsec:examples},
we eliminate the dense segments 
$\partial g_{(\omega_{\rm CoI})} / \partial
\wGi{i}$, $\partial f_{(\dot {\dGi{i}})} / \partial \omega_{\rm
  CoI}$, $\partial g_{(P_{{\rm ord},{i}})} / \partial P_{\rm s}$,
$\partial f_{(x_{r,i})} / \partial \bfg Q_{\rm r}$, that arise from
\eqref{osda:eqcoi2}-\eqref{osda:eqxri}.
We simulate the transient following a three-phase fault applied at bus 6 at $t=1$~s. The fault is cleared after $80$~ms by tripping the
transmission line that connects buses $5$ and $6$.  The system is
numerically integrated using the \ac{itm}.  Figure~\ref{fig:trans} shows
the transient behavior of the rotor speed of generator 2 for
integration step sizes $h=0.02$~s and
$h=0.1$~s.  The larger $h$ is, the larger is the mismatch between the two trajectories. 
In both plots though, 
the trajectory of the \ac{ddae} system closely follows the original trajectory, as expected. 


It is relevant to check the accuracy of the proposed one-step delay technique under different operating conditions and contingencies. In addition to the operating condition considered above
(from here and on referred as the base case), two other operating conditions are considered, namely, $10$\% and $20$\% increase in the total power consumption of the system.  For each operating point, the transient response of the system is examined. Two different disturbances are considered: first, the three phase fault applied at bus 6 described above; and second, the loss of the load connected to bus 39 at $t=1$~s, which leads to a $1.109$~GW decrease in the power consumption of the system.  In all scenarios, the delayed variables do not change and are the ones used to plot the base case in Figure~\ref{fig:trans}.

The response of the rotor speeds of the \ac{dae} system are compared with the respective speed trajectories obtained by integrating the \ac{ddae} system.  Each system is simulated for $100$~s and for two time step sizes, $h=0.02$~s and $h=0.10$~s.  The maximum absolute rotor speed trajectory errors are summarized in Table~\ref{tab:scen39bus}.  As expected, the proposed technique shows high accuracy for all considered operating conditions and disturbances.

\begin{table}[ht!]
  \renewcommand{\arraystretch}{1.3}
  \centering
  \caption[\newengland system: Maximum absolute rotor speed trajectory mismatches induced by the proposed method]{\newengland system: Maximum absolute rotor speed trajectory mismatches induced by the proposed method.}
  \label{tab:scen39bus}
  \begin{tabular}{cl|cl}
    \hline        
Operating & Applied & $h=0.02$~s & $h=0.10$~s\\
condition & disturbance & -- max. error & -- max. error \\
\hline
\hline
Base case & Fault at bus 6  & $6.0 \cdot 10^{-6}$ & $8.3 \cdot 10^{-5}$ \\
& Bus 39 load trip & $9.0 \cdot 10^{-6}$ & $4.2 \cdot 10^{-5}$ \\
\hline
$+10$\% load & Fault at bus 6  &  $6.0 \cdot 10^{-6}$ & $4.8 \cdot 10^{-4}$  \\
& Bus 39 load trip & $1.2 \cdot 10^{-5}$ & $5.7 \cdot 10^{-5}$ \\
\hline
$+20$\% load  & Fault at bus 6 &  $7.0 \cdot 10^{-6}$ & $5.8 \cdot 10^{-4}$ \\
& Bus 39 load trip & $1.7 \cdot 10^{-5}$ & $8.4 \cdot 10^{-5}$ \\
\hline
  \end{tabular}
\end{table}

\subsection{21,177-bus ENTSO-E} 
\label{osda:subsec:entsoe}

This subsection presents simulation results on a dynamic model of the
\ac{entsoe} transmission system, which has been also discussed in Section~\ref{sssa:sec:entsoe}.
In addition to the model components considered in Section~\ref{sssa:sec:entsoe}, the system examined 
in this section also includes \ac{agc} and \ac{svr} mechanisms,
which provide secondary frequency and voltage control, respectively,
to different areas of the system.
%
In total, the system has 
$49,930$ state variables and $97,304$ algebraic variables.
The full Jacobian matrix has dimensions
$147,234 \times 147,234$ and $1,226,492$ non-zero elements, which
yields a density degree of $0.0057$~\%.

In order to show the impact of the one-step delay approximation on the
accuracy, number of factorizations and computational burden of the \ac{tdi}, the dense segments that arise from
\eqref{osda:eqcoi2}-\eqref{osda:eqxri} are eliminated, leading to a sparser
and less coupled model.  The \ac{nnz} Jacobian elements of the original and
delayed system are summarized in Table~\ref{tab:jacentsoe}.
\begin{table}[ht!]
  \renewcommand{\arraystretch}{1.15}
  \centering
  \caption[ENTSO-E system: NNZ Jacobian elements]{ENTSO-E system: NNZ Jacobian elements.}
  \label{tab:jacentsoe}
  \begin{tabular}{lrrr}
    \hline        
    System & \ac{nnz} Jacobian elements 
    & Density (\%) & Relative difference \\
    \hline
    Original & $1,226,492$ & $0.0057$ \\
    Delayed & $936,871$ & $0.0043$ & $-23.61$~\%\\
    \hline
  \end{tabular}
\end{table}

\begin{figure}[ht!]
  \begin{center}
  \resizebox{0.85\linewidth}{!}{\includegraphics{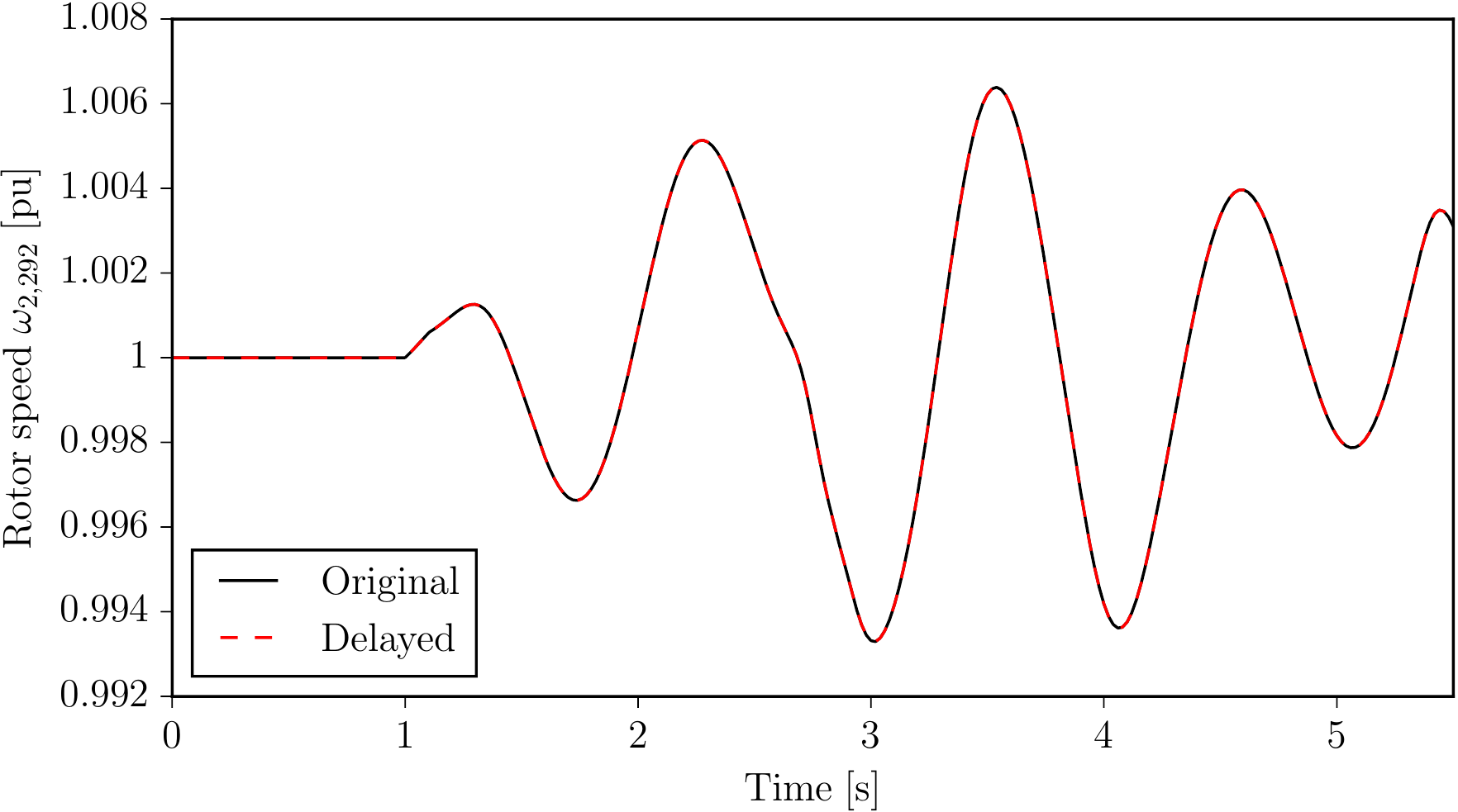}}
  \subcaption{$h=0.02$~s.} 
  \vspace{0.5cm}
  \resizebox{0.85\linewidth}{!}{\includegraphics{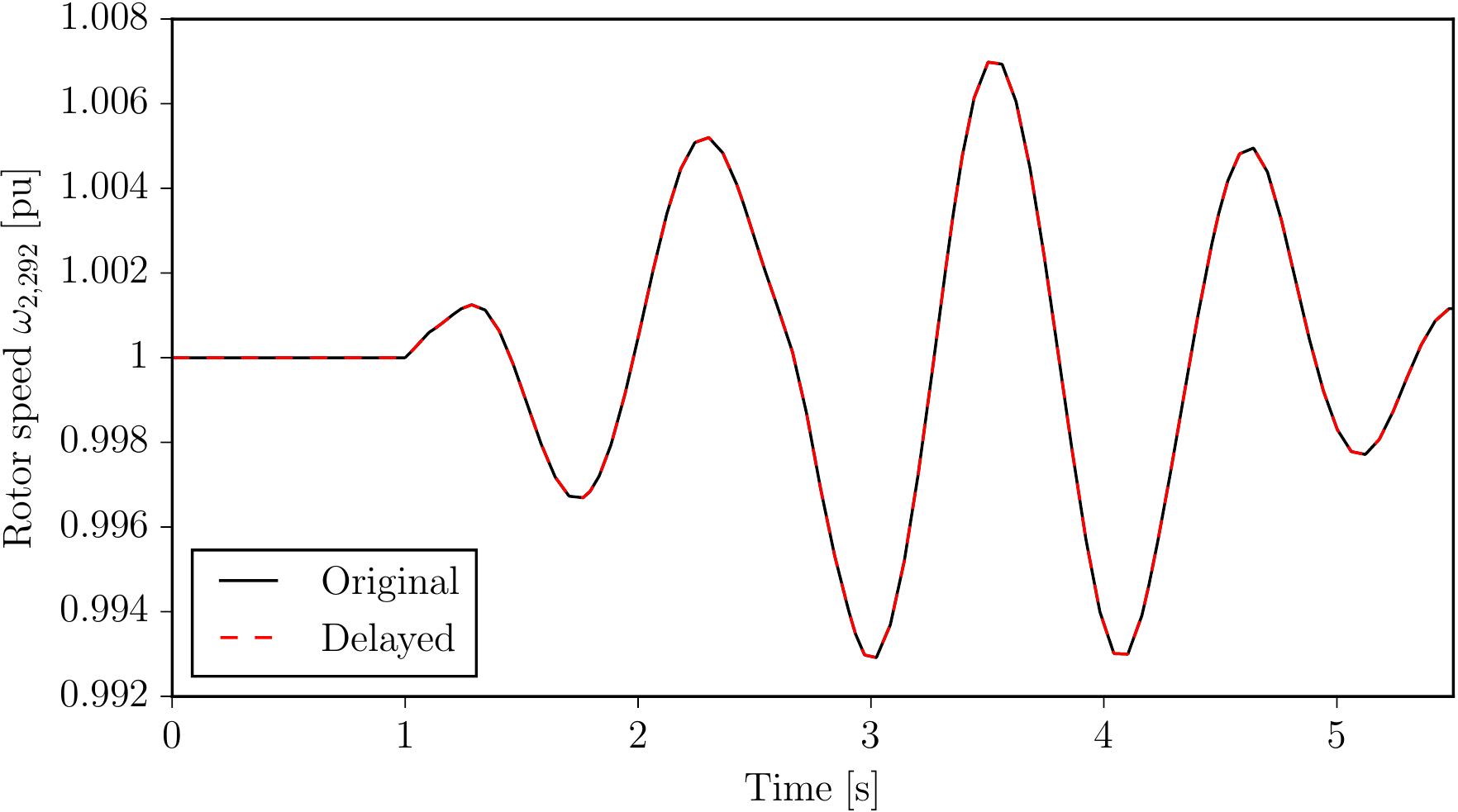}}
  \subcaption{$h=0.06$~s.}
  \end{center} 
  \caption[ENTSO-E system: transient following a three-phase fault]{ENTSO-E system: transient following a three-phase fault.}
  \label{fig:entsoe1}
\end{figure}

We consider a 
three-phase fault at bus
$12,921$, occuring at $t=1$~s.
The fault is
cleared after $100$~ms.
The response of the rotor speed of the
synchronous generator 
connected at bus $2,292$
during the first seconds following the fault, is shown in Figure~\ref{fig:entsoe1}
for two different time step sizes. 
The difference between the two trajectories is very small,
which indicates that accuracy is maintained. In
particular, the maximum absolute mismatch between the two trajectories
for the cases shown in Figure~\ref{fig:entsoe1} are: (a) $1.0 \cdot 10^{-6}$, 
(b) 
$7.0 \cdot 10^{-6}$. 

The impact of the one-step delay approximation on the number of factorizations of the \ac{tdi} is examined next. 
Following a disturbance, 
the system shows a transient
and, provided that the trajectory is stable, finally reaches a stationary point. 
While in steady state, the \ac{itm} requires exactly one factorization for each time step, both for the original and the delayed system. 
Hence, any noticeable differences in the number of factorizations required 
by the original and the delayed system occur during the first seconds 
following the disturbance.

\begin{figure}[ht!]
  \begin{center}
  \resizebox{0.85\linewidth}{!}{\includegraphics{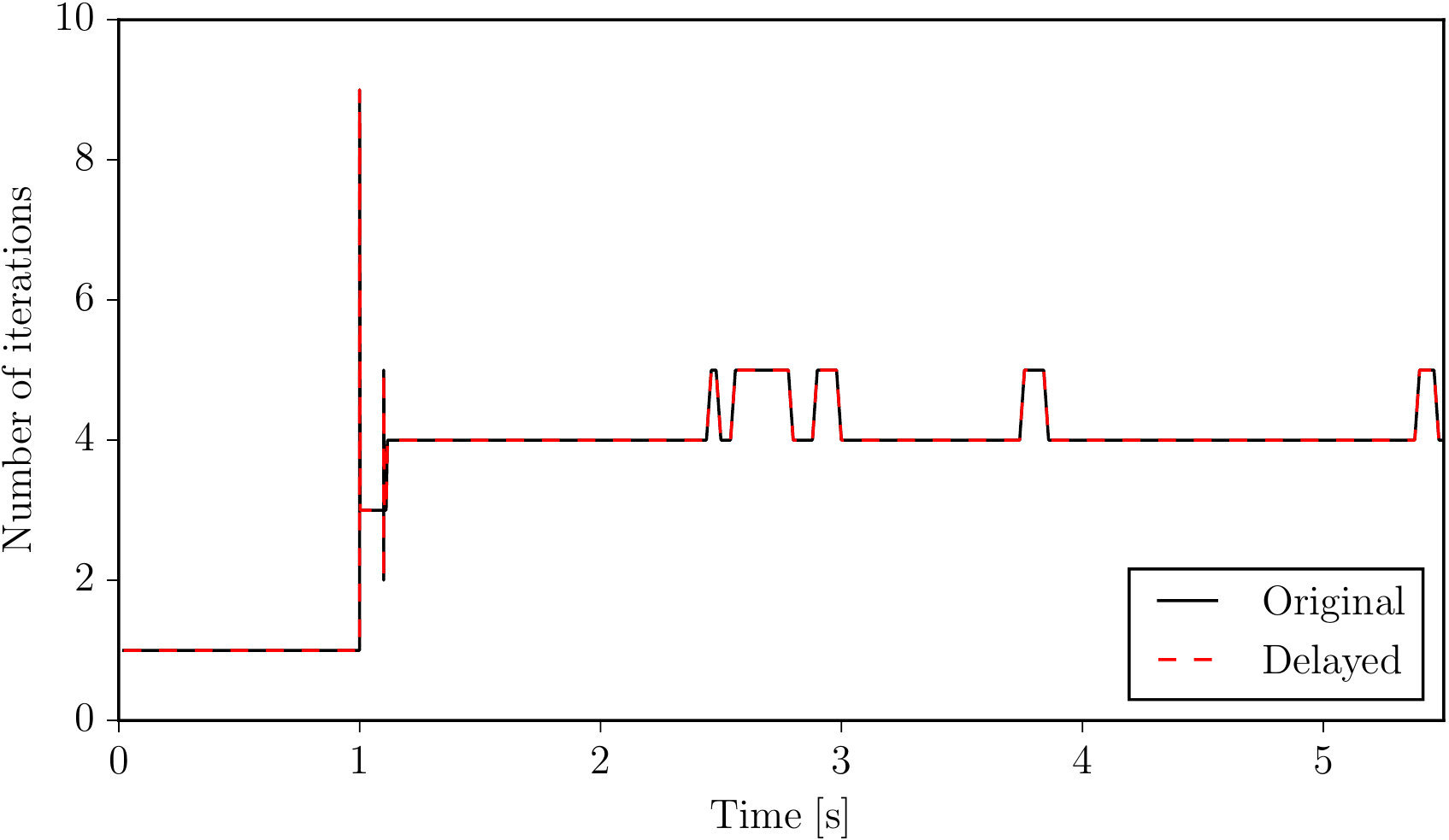}}
  \subcaption{$h=0.02$~s.} 
  \vspace{0.5cm}
  \resizebox{0.85\linewidth}{!}{\includegraphics{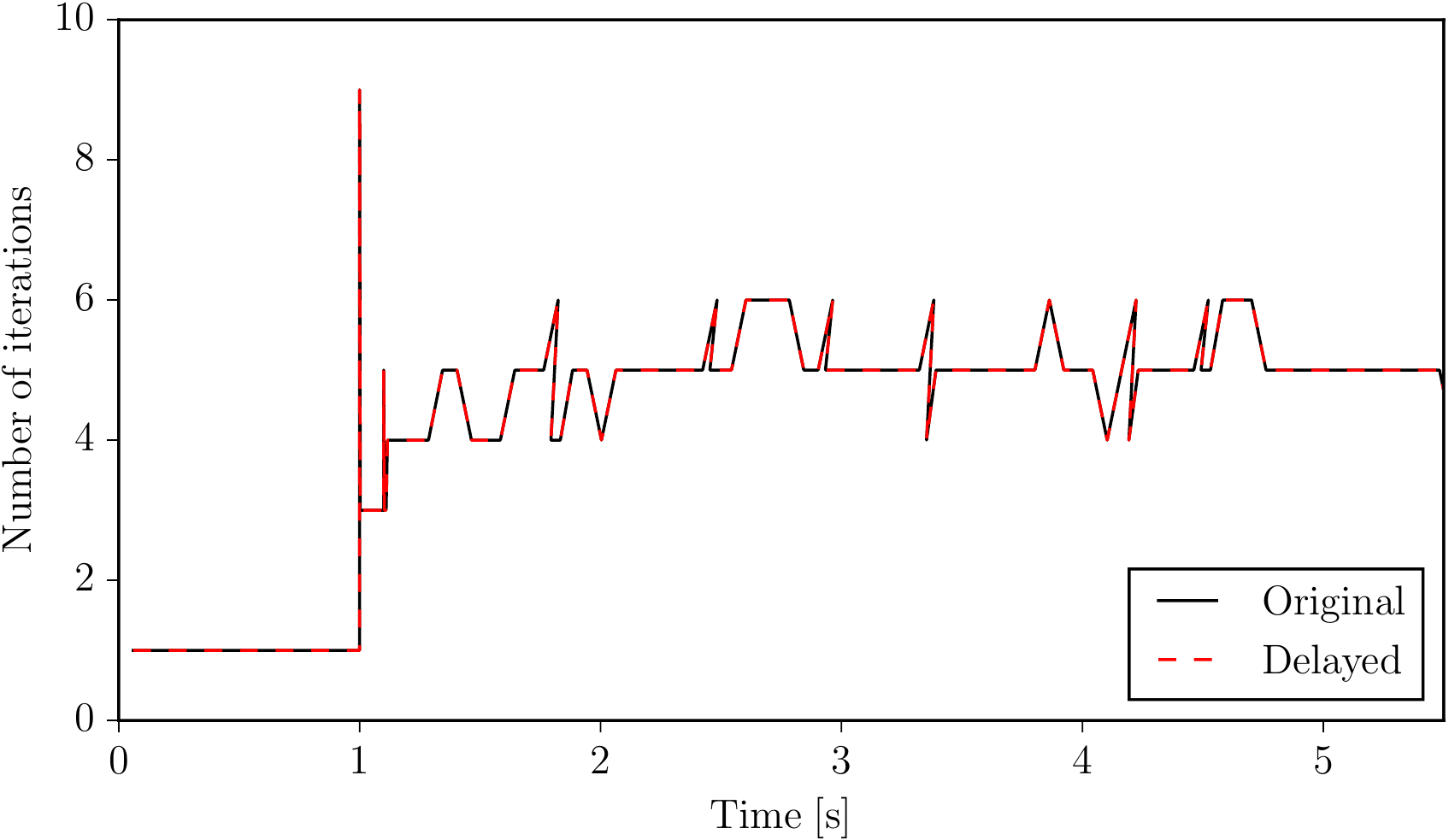}}
  \subcaption{$h=0.06$~s.}
  \end{center} 
  \caption[ENTSO-E system: number of Newton iterations]{ENTSO-E system: number of Newton iterations.}
  \label{fig:entsoe2}
\end{figure}

The number of factorizations
required by the original and the delayed system during the first seconds following the three-phase fault, are shown in
Figure~\ref{fig:entsoe2}.
Since
the increments of the variables at each time step are updated according to the standard Newton method (see Section~\ref{osda:sec:tdi}), 
the number of factorizations at each time step is equal to the number of Newton iterations. The original 
and the delayed system require 
in this case
the same number of factorizations at each time step to converge. This indicates that the approximation does not jeopardize the convergence. 

Finally, it is relevant to evaluate the effect of the one-step delay approximation on
the computational burden of the \ac{tdi}. The method reduces the coupling
of the \ac{entsoe} system and facilitates the potential application of
techniques that factorize decoupled blocks of the Jacobian matrix in
parallel. In turn, enabling parallelization leads to a significant
speedup of the simulation.  However, as already stated, the goal of
this chapter is to provide a technique for decoupling and sparsity
increase rather than applying parallel techniques. Hence, the original and delayed \ac{entsoe} systems are compared in terms of
computational effort required for a non-parallel numerical
integration.

The full Jacobian matrix without
introducing delays requires $0.245$~s per each factorization, in
average, on a $8 \times$ $3.5$~GHz Intel Xeon CPU desktop computer,
while the Jacobian matrix of the delayed system requires $0.223$~s,
which corresponds to a speedup of $9.04$~\%.


Apart from the factorization speed-up, one has also to evaluate
whether the delayed system requires more or less iterations than the
original system to solve the Newton method for each point of the time
domain integration. 
With this regard, 
Figure~\ref{fig:entsoe2} has already shown an example where
the two systems require at each point the same number of iterations. 
In addition, several cases have been carried out considering a variety
of contingencies and time steps and it has been found out that the proposed
technique is able to reduce the simulation time in range from 
$5$ to $20 \%$. 

For the sake of example, consider the three-phase fault at bus 2,292
discussed above.  The system is integrated for $7$~s.  With a time
step $h=0.02$~s, the original system completes the numerical
factorization in $298.63$~s, while the delayed system in $262.14$~s,
which corresponds to a speedup of $12.22$~\%.


    
    The proposed one-step-delay technique is agnostic with respect to the integration scheme utilized for the \ac{tdi}.  For this reason, the proposed approach can be coupled with any other numerical technique to speed up time domain simulation software. Hence, even if the speed-up provided by the proposed formulation \textit{per se} is not huge, it can be combined with other techniques.
   %
   Moreover, reducing the computational burden is not the only benefit of the proposed one-step delay technique.  A relevant feature is that it increases the decoupling of system variables.  This leads to a sparser and more decoupled system Jacobian matrix.  The latter is a feature that is expected to be beneficial to further speed up the time domain analysis if combined with parallelization techniques.



\section{Conclusions}
\label{osda:sec:conclusion}

The chapter presents a systematic approach to exploit delays to reduce
the coupling of the equations of conventional \ac{dae} models of power
systems for transient stability analysis.  With this aim, the chapter
discusses how to select the variables of a power system \ac{dae} model that
can be delayed and provides an estimation of the maximum admissible
time delay so that simulation accuracy is maintained. 
This analysis has to be carried out only once per network.
Numerical
simulations support the theoretical appraisal
of the proposed
approach and show its accuracy, convergence and computational burden.

\newpage
\chapter{Conclusions and Future Work}
\label{ch:conclusions}

This thesis proposes novel
\ac{sssa}-based techniques with application to modal analysis, robust control, and numerical integration of power systems.
The objective of this chapter is to summarize the main conclusions of the thesis and support directions for future work.

\begin{itemize}
    \item 
\textit{Modal analysis}: The study on modal analysis shows that the classical assumptions made when computing \acp{pf}, namely that the system is modeled as a set of \acp{ode} and that all eigenvalues are distinct, are not binding.
In fact, considering a singular system of differential equations with eigenvalue multiplicities, allows extracting a generalized expression of \acp{pf} in the form of eigen-sensitivities, from which
\acp{pf} in the classical sense arise as a special case. 
Moreover, prior to this work, algebraic variables of a power system model were either eliminated or treated as states with infinitely fast dynamics and, as such, their \acp{pf} to finite modes were
considered to be null.
The proposed formulation shows that it is possible to quantify the \acp{pf} of algebraic variables of a power system, and
in general of any function of the system variables, through the definition of appropriate input/output vectors of the system's state-space model. 

Future work will focus on studying the 
effect of network constraints on the design of control schemes based on the  
proposed modal analysis.
A control signal/actuator selection
that is based on the \acp{pf} of line power flows and takes into account information on the 
current/power capacity limits of transmission lines is a relevant example. Such information is readily available to system operators and thus it can be effectively included to the analysis.

Moreover, the proposed formulation of \acp{pf}
can be extended to include the analysis 
of systems of 
fractional differential equations. Such analysis
will allow measuring the coupling between the variables of a power system models and its fractional order dynamics, such as the ones introduced by \acp{foc} discussed in Chapter~\ref{ch:foc}. 
Moreover, extension of the proposed modal participation analysis is also relevant for systems of \acp{ddae}. 
This will allow an efficient assessment of control signals impacted by time delays, including wide-area controllers and controllers with intentional delays, such as the ones described in 
Chapter~\ref{ch:tdc}. In this case, the
calculation of \acp{pf} will be a challenging task, since the existence of an infinite dimensional spectrum makes it difficult to determine the coupling between eigenvectors and system variables in an accurate and efficient way.


\item
\textit{Fractional Control}: The contributions of the thesis are a systematic study of controllers based on fractional calculus and a technique to carry out eigenvalue analysis 
and assess the small-signal stability of power systems with inclusion of exact fractional dynamics of Caputo type. 
Furthermore, the properties of \ac{ora} are investigated and, through time domain simulations, it is shown that \acp{foc}
perform better than their conventional \ac{IO} versions for synchronous machine \ac{agc}, 
\ac{ess} frequency control, and 
\ac{statcom} voltage control,
while they require only a small additional tuning effort. 

A relevant extension of the work on \acp{foc} is the study of practical
aspects, such as potential modeling and stability issues introduced with the inclusion of control saturation limits. 
With this regard, a preliminary study on \ac{foc} control limits for power system applications can be found in \cite{foc:ifac}.
Furthermore, time domain simulations in Chapter~\ref{ch:foc} are based on the widely employed \ac{ora}, but there exist also other methods that approximate fractional order dynamics, see e.g. \cite{chen:06}. To the best of our knowledge, a systematic study that provides an eigenvalue-based comparison of approximation methods for fractional order dynamics with application to power systems is still missing and it is worth consideration.


\item
\textit{Delay-based Control}: The study on time-delayed control
focuses on the stability boundary of power systems with
delay-based \acp{pss}, as well as on the
relationship between the
existence of delay-independent stability and connected stability domains in the delay-control gain space.
Analytical results based on the \ac{omib} system, as well as a numerical analysis 
of the IEEE 14-bus system,
illustrate that in contrast to their bad reputation, large delays
may achieve best damping characteristics, provided that the \ac{pss} is properly configured.
It is shown that one such possible configuration is 
to employ a two-channel \ac{pss}, which permits a fully connected stability region in the delay-gain space. Then, non-linear dynamics can be tuned 
to achieve high damping, 
by fully exploring the parameter space and without introducing instabilities.

A possible future work direction is the design of an adaptive control scheme so that the delay and gain are automatically tuned,
following a stable path and through consecutive quasi-steady state
shifts of the system equilibrium. This is particularly interesting in case that part of the delay is inherent, i.e.~represents measurement and/or
communication latency. In this case, high damping can be achieved by
adding an artificial controlled delay on top of the inherent delay. 
This is a novel approach, whose effectiveness can be assessed in comparison with other techniques used for mitigating the
destabilizing effect of communication delays, such as delay compensation methods \cite{muyang:compensation:mdpi,muyang:powertechmilan}.

\item
\textit{Numerical Integration}:
Finally, this work proposes an ``one-step delay" approximation technique for the numerical integration of the \acp{dae} utilized to study power system transient stability, and provides a
first evaluation of its accuracy, convergence and computational burden.
Chapter~\ref{ch:osda} shows that the proposed approach 
allows exploiting delays equal to the time step of the numerical integration, in order to reduce
the coupling of the equations of conventional power system \ac{dae} models for transient stability analysis. \ac{sssa}-based techniques are discussed for the
selection of the variables of a power system \ac{dae} model that
can be delayed, as well as for the estimation of the maximum admissible time delay, so that the accuracy of trajectories is not compromised. 

The next step is to embed the proposed one-step delay approximation technique in algorithms that apply state-of-the art parallelization techniques.
This will typically require to exploit a current injection-based power system model formulation, in order
to take advantage of the \ac{bbd} structure of 
the corresponding Jacobian matrix, see 
e.g.~\cite{Fong1978,FabozziLNewton}.

\end{itemize}

\begin{appendix}
\renewcommand{\chaptername}{\Alph{chapter}}
\newpage

\cleardoublepage
\phantomsection
\addcontentsline{toc}{chapter}{Appendices}
\thispagestyle{empty}
\vspace*{\stretch{1}}
\begin{center}
\Huge \textbf{Appendices}
\vfill
\end{center}

\normalsize
\pagebreak
\chapter{Proofs}
\label{app:proofs}

\section{Proof of Theorem~\ref{sssa:theorem:sol}}

By substituting the transformation
\begin{equation}
\xs(t)=\reigvmat \, \bfg \xi(t)
    \label{pf:eq:transf}
\end{equation}
into \eqref{sssa:eq:sing}, and by multiplying by $\leigvmat$, one obtains:
\begin{equation}
\leigvmat 
\Esng \, \reigvmat \, \Dt {\bfg \xi}(t)=
\leigvmat \, \Asng \, \reigvmat \, \bfg \xi(t) +
\leigvmat \, \Bsng \, \textbf{u}(t)
\, .
\label{pf:eq:ind1}
\end{equation}

Let $\reigvmat_{\nf}$, $\reigvmat_\ninf$ be the matrices that contain
all right eigenvectors of the finite, and infinite eigenvalues
respectively. Then by setting $ \bfg \xi = [\bfg \xi_\nf \ \ \bfg \xi_\ninf]\T $,
$\reigvmat= [\reigvmat_\nf \ \  \reigvmat_\ninf]
$, 
with $\bfg \xi_\nf \in \mathbb{C}^{\nf}$,
$\bfg \xi_\ninf \in \mathbb{C}^{\ninf}$, and using \eqref{sssa:eq:decomp}, we
arrive at two subsystems of \eqref{pf:eq:ind1}:
\begin{align}
\Dt {\bfg \xi}_\nf(t) &= \bfg J_\nf \, \bfg \xi_\nf(t)
+ \leigvmat_{\nf} \, 
\Bsng \, \textbf{u}(t)
\, , \nonumber \\
\bfg H_\ninf \,
\Dt {\bfg \xi}_\ninf(t) &= \bfg \xi_\ninf(t) 
+ \leigvmat_{\ninf} \,
\Bsng \, \textbf{u}(t) \, . 
\nonumber
\end{align}
The first subsystem has solution:
\begin{align}
\label{sssa:eq:solfinite}
    \bfg \xi_\nf(t)&= e^{\bfg J_\nf t} \, 
    \bfb c
    + \int_0^\infty
  e^{\bfg J_\nf(t-\kappa)}
 \, \leigvmat_{\nf} 
 \, \Bsng 
  \, \textbf{u}(\kappa) 
  \, d\kappa
    \, ,
\end{align}
where $\bfb c = \bfg \xi_\nf(0)$ is a constant vector.
For the second subsystem, let $\ninf_*$ be the index of the nilpotent
matrix $\bfg H_\ninf$, i.e.~$\bfg H_\ninf^{\ninf_*}=\bfg 0_{\ninf,\ninf}$.  Then following matrix equations are obtained: 
\begin{align}
\bfg H_\ninf \, \Dt {\bfg \xi}_\ninf(t) &= 
\bfg \xi_\ninf(t) 
+ \leigvmat_{\ninf} 
\, \Bsng \, \textbf{u}(t)
\nonumber \\
\bfg H_\ninf^2 \, \ddot {\bfg \xi}_\ninf(t) &= \bfg H_\ninf \, \Dt {\bfg \xi}_\ninf(t)
+ \bfg H_\ninf \,  \leigvmat_{\ninf} 
\, \Bsng \,
\Dt {\textbf{u}}(t)
\nonumber \\
& \ \vdots \nonumber \\
\bfg H_\ninf^{\ninf_*-1}\, \bfg \xi^{(\ninf_*-1)}_\ninf(t) &=
\bfg H_\ninf^{\ninf_*-2}\, \bfg \xi_\ninf^{(\ninf_*-2)}(t)
+ \bfg H_\ninf^{\ninf_*-2} \, \leigvmat_{\ninf} \,
\Bsng \, \textbf{u}^{(\ninf_*-2)}(t)
\nonumber \\ 
\bfg H_\ninf^{\ninf_*}\bfg \xi^{(\ninf_*)}_\ninf(t) &=
\bfg H_\ninf^{\ninf_*-1}\bfg \xi_\ninf^{(\ninf_*-1)}(t) 
+ \bfg H_\ninf^{\ninf_*-1} \leigvmat_{\ninf} \,
\Bsng \, \textbf{u}^{(\ninf_*-1)}(t)
\nonumber \, .
\end{align}
By taking the sum of the above equations, the
solution for the second subsystem is:
\begin{equation}
\label{pf:eq:sub2}
   \bfg \xi_\ninf(t)=
   - \sum^{\ninf_*-1}_{i=0}
   \bfg H_\ninf^i \leigvmat_{\ninf} 
\, \Bsng \, 
{\textbf{u}}^{(i)}(t) \, .
\end{equation}
Using the solutions \eqref{sssa:eq:solfinite} and \eqref{pf:eq:sub2} in \eqref{pf:eq:transf}, ones gets:
\begin{equation}
\label{sssa:eq:transf2}
\xs(t) =  
\begin{bmatrix}
\reigvmat_\nf & \reigvmat_\ninf
\end{bmatrix}
\begin{bmatrix}
\bfg \xi_\nf(t)\\
\bfg \xi_\ninf(t)
\end{bmatrix}
\, ,
\end{equation}
or equivalently,
\begin{equation}
\xs(t) =  
 \reigvmat_\nf \, e^{\bfg J_\nf t} 
 \, \bfb c
 +
\reigvmat_\nf
 \int_0^\infty
  e^{\bfg J_\nf(t-\kappa)}
 \, \leigvmat_{\nf} 
  \, \Bsng \, 
  \, \textbf{u}(\kappa) 
  \, d \kappa
 -  \reigvmat_\ninf
 \sum^{\ninf_*-1}_{i=0}
   \bfg H_\ninf^i \leigvmat_{\ninf} 
    \, \Bsng \, 
{\textbf{u}}^{(i)}(t)
 \, , \nonumber
\end{equation}
%
which is the general solution 
\eqref{sssa:eq:sol}.
\hfill
\eop


\section{Proof of Theorem~\ref{pf:theorem:pf} }

\begin{enumerate}[label=(\alph*), leftmargin=*]

\item
By using the transformation $\xs (t)=
\reigvmat \, \bfg \xi (t)$, from the proof of Theorem~\ref{sssa:theorem:sol},
and in particular from 
\eqref{sssa:eq:transf2}, one has for $\Bsng = \bfg 0_{\nd,\nin}$:
\begin{equation}
\label{pf:eq:transf3}
\xs (t)=\reigvmat_\nf \, \bfg \xi_\nf(t) \, .
\end{equation}

Let $\leigvmat_\nf$, $\leigvmat_\ninf$ be the matrices that contain
all left eigenvectors of the finite, and infinite eigenvalues of the pencil 
$s \Esng-\Asng$, respectively. 
Then by using the notation
$\leigvmat= [\leigvmat_{\nf} \ \  \leigvmat_\ninf]\T$, 
and making use of \eqref{sssa:eq:decomp}, there exist $\leigvmat_\nf$, 
$\reigvmat_\nf$, such that
$\leigvmat_\nf \,\Esng \,
\reigvmat_\nf=\bfg I_\nf$. 
Multiplying \eqref{pf:eq:transf3} by $\leigvmat_\nf \, \Esng$ yields:
\[
\leigvmat_\nf \, \Esng \, \xs(t) =
\leigvmat_\nf \, \Esng \, \reigvmat_\nf \, 
\bfg \xi_\nf(t) \, ,
\]
or, equivalently,
\[
\bfg \xi_\nf(t)=\leigvmat_\nf \, \Esng \, \xs(t) \, .
\]
Hence:
\begin{equation}
\label{pf:eq:solxi_0}
\bfg \xi_\nf(0)= \leigvmat_\nf \, \Esng \, \xs(0) \, .
\end{equation}
Substitution of \eqref{pf:eq:solxi_0} into the general solution
\eqref{pf:eq:sol} gives:
\begin{equation}
\label{pf:eq:solxi2}
\xs(t) = \reigvmat_\nf \, e^{\bfg J_\nf t} \, \leigvmat_\nf  \,\Esng \, \xs(0) \, .
\end{equation}
The matrices $\reigvmat_\nf$, $\leigvmat_\nf$
can be written as:
\begin{align}
\label{pf:eq:vp}
\reigvmat_\nf &=
\big [
\reigv_{1}^{[\rjb_1]} \ldots \
\reigv_{1}^{[2]} \ 
\reigv_{1}^{[1]} \  \ldots \ \reigv_{\njb}^{[\rjb_\njb]} 
\ldots \ 
\reigv_{\njb}^{[2]} \ 
\reigv_{\njb}^{[1]} 
\big ] \, , 
\\
\leigvmat_\nf &=
\big [
\leigv_{1}^{[\rjb_1]} \ldots \
\leigv_{1}^{[2]} \
\leigv_{1}^{[1]} \  \ldots \ \leigv_{\njb}^{[\rjb_\njb]} 
\ldots \ 
\leigv_{\njb}^{[2]} \ 
\leigv_{\njb}^{[1]}
\big ]\T \, , 
\label{pf:eq:wp}
\end{align} 
where $\reigv_{i}^{[j]}$,
$\leigv_{i}^{[j]}$,
$j=1,2,...,\rjb_i$, 
linear (generalized) independent right, left eigenvectors of 
$\hat \lambda_i$, 
$i=1,2,...,\njb$, respectively.

The Jordan matrix $\bfg J_{\nf}$ has the following form:

\[
\bfg J_{\nf}  :=\bfg J_{\rjb_1} ({\hat \lambda_1 }) \oplus  \dots  \oplus \bfg J_{\rjb_{\njb}}(\hat  \lambda_{\njb } ) \, ,
\]
where
   \[
  \bfg J_{\rjb_{i}} ({\hat  \lambda_i })=
  \left[\begin{array}{ccccc}
  \hat \lambda_i  & 1 & \dots&0  & 0  \\
   0 & \hat \lambda_i  &   \dots&0  & 0  \\
    \vdots  &  \vdots  &  \ddots  &  \vdots  &  \vdots   \\
   0 & 0 &  \ldots& \hat \lambda_i & 1\\
   0 & 0 & \ldots& 0& \hat \lambda_i 
   \end{array}\right] \in {\mathbb{C}}^{\rjb_{i}\times \rjb_{i} } \, ,\quad i=1,2,...,\njb \, , 
\]
is the Jordan block that corresponds to the
eigenvalue $\hat \lambda_i$.
The matrix exponential of $\bfg J_\nf t$, denoted as
$e^{\bfg J_\nf t}$, is defined as:

\begin{equation}
e^{\bfg J_\nf t}  :=e^{\bfg J_{\rjb_1} ({\hat \lambda_1 })t} \oplus  \dots  \oplus e^{\bfg J_{\rjb_{\njb}}({\hat \lambda_\njb})t} \, ,
\label{app:eq:jordanexp}
\end{equation}
where
   \[
   e^{\bfg J_{\rjb_i} ({\hat \lambda_i }) t}=
  \left[\begin{array}{ccccc}
   e^{\hat \lambda_it}  &  e^{\hat \lambda_it} t & \dots & e^{\hat \lambda_it}\frac{t^{{\rjb_i}-1}}{({\rjb_i}-1)!}  & e^{\hat \lambda_it}\frac{t^{{\rjb_i}}}{{\rjb_i}!}  \\
   0 & e^{\hat \lambda_i t} &   \dots&
e^{\hat \lambda_it}\frac{t^{{\rjb_i}-2}}{({\rjb_i}-2)!} & e^{\hat \lambda_it} \frac{t^{{\rjb_i}-1}}{({\rjb_i}-1)!}  \\
    \vdots  &  \vdots  &  \ddots  &  \vdots  &  \vdots   \\
   0 & 0 &  \ldots& e^{\hat \lambda_i t} & e^{\hat \lambda_it} t\\
   0 & 0 & \ldots& 0& e^{\hat \lambda_it}
   \end{array}\right] \in {\mathbb{C}}^{{\rjb_i}\times {\rjb_i}},\quad i=1,2,...,\njb \ .
\]

By substituting \eqref{pf:eq:vp}, \eqref{pf:eq:wp},
\eqref{app:eq:jordanexp} in
\eqref{pf:eq:solxi2}, one arrives at
\eqref{sssa:eq:sol3}.

%
\item
From 
\eqref{sssa:eq:sol3}, the evolution of $\x_k(t)$, i.e. the $k$-th element of  $\xs (t)$, is:
\begin{equation}
\x_k(t) = \sum_{i=1}^\njb e^{\hat \lambda_it}\sum_{j=1}^{\rjb_i}
\Big(
\sum_{\sigma=1}^j \, t^{\sigma-1}
\, \leigv_{i}^
{[j-\sigma+1]} \, \bfg E \, \xs(0) 
\Big) \,
\reigvel_{k,i}^{[j]} \, ,
\label{sssa:eq:solk}
\end{equation}
where ${ \reigvel_{k,i}}^{[j]} \in \reigv_i^{[j]}$.

Partial differentiation of this equation with respect to $e^{\hat \lambda_i t}$
leads to:
\begin{equation}
\frac
{\partial \x_k(t)}
{\partial e^{\hat \lambda_i t}}
=\sum_{j=1}^{\rjb_i}
\big(
\sum_{\sigma=1}^j t^{\sigma-1}
\, 
\leigv_{i}^
{[j-\sigma+1]} \,  \bfg E \, 
\xs(0) 
\big) \, 
\reigvel_{k,i}^{[j]} 
\, ,
\label{pf:eq:partialk}
\end{equation}
which is the \ac{pf} of $\hat \lambda_i$, $i=1,2,...,\njb$, in $\x_k(t)$, $k=1,2,...,\nd$. 
\hfill
\eop

\end{enumerate}

\section{Proof of Theorem \ref{foc:theorem:pencil}}

Let $\mathcal{L}\{ {\xs}(t)\}$ be the Laplace transform of $\xs(t)$. 
Using the Caputo fractional derivative, by applying the
Laplace transform $\mathcal{L}$ as defined in \eqref{foc:eq:caputol} for $\upmu=1$ into \eqref{foc:eq:final2t}, one gets \cite{Kac}:
\begin{equation}
\label{foc:eq:lap_sing}
\mathcal{L}\{ \Efr \, {\xs}^\Delta (t)\}=\mathcal{L}\{ \bfi { A} \, {\xs}(t)\}
\, .  
\end{equation}
Note that 
\[
\xs^\Delta (t)
=
\begin{bmatrix}
\frac{d^\gamma}{dt^\gamma}  \bfg I_{{\rho}} &  \bfg 0_{\rho,\rho} \\
\bfg 0_{{\rho,\rho}} & \frac{d^\beta}{dt^\beta}  \bfg I_{{\rho}}
\end{bmatrix}
\xs(t) \, , 
\]
and hence
\[
\Efr \xs^\Delta (t)=
\begin{bmatrix}
   \bfg I_{{\rho}} 
   &  \bfg 0_{\rho,\rho} 
    \\
      \bfg 0_{\rho,\rho} 
      &  \bfg M
\end{bmatrix}
\begin{bmatrix}
\frac{d^\gamma}{dt^\gamma}  \bfg I_{\rho} &  \bfg 0_{\rho,\rho} \\
 \bfg 0_{\rho,\rho} & \frac{d^\beta}{dt^\beta}  \bfg I_{\rho}
\end{bmatrix}
\xs(t) \, , 
\]
or, equivalently,
\[
\Efr \xs^\Delta (t)=
\begin{bmatrix}
  \frac{d^\gamma}{dt^\gamma} \bfg I_{\rho}
&  \bfg 0_{\rho,\rho} 
\\
      \bfg 0_{\rho,\rho} 
      &  \frac{d^\beta}{dt^\beta} 
      \bfg M
\end{bmatrix}
\begin{bmatrix}
\xs_1 \\
\xs_2 
\end{bmatrix}
=
\begin{bmatrix}
  \frac{d^\gamma \xs_1}{dt^\gamma}
&  \bfg 0_{\rho,\rho} \\
      \bfg 0_{\rho,\rho} 
&  \bfg M \frac{d^\beta \xs_2}{dt^\beta}
\end{bmatrix} \, . 
\]
Thus, $\mathcal{L}\{\Efr \xs^\Delta(t)\}=\Efr \mathcal{L}\{\xs^\Delta(t)\}$ and
\eqref{foc:eq:lap_sing} becomes:

\[
\Efr \, \mathcal{L}\{ {\xs}^\Delta (t)\}=\bfi { A}\mathcal{L}\{ {\xs}(t)\}
\, ,
\]
or, equivalently,
\[
\Efr\mathcal{L} 
\Big\{
\begin{bmatrix}
  \xs_1^{(\gamma)}(t) \\
  \xs_2^{(\beta)}(t)  \\
  \end{bmatrix}
\Big\}
=\bfi { A} \, \mathcal{L} 
\{ {\xs}(t)\} \, ,
\]
or, equivalently,
\[
\Efr
\begin{bmatrix}
  s^\gamma\mathcal{L}\{ \xs_1(t)\}-s^{\gamma-1} \xs_1(0) \\
  s^\beta \mathcal{L}\{\xs_2(t)\}-s^{\beta-1}\xs_2(0)
 \end{bmatrix}=\Afr \, \mathcal{L}\{\xs(t)\} \, ,
\]
or, equivalently,
\[
\Efr
\begin{bmatrix}
  s^\gamma\mathcal{L}\{ \xs_1(t)\} \\
  s^\beta \mathcal{L}\{\xs_2(t)\}
\end{bmatrix}
-\Efr
\begin{bmatrix}
  s^{\gamma-1} \xs_1(0) \\
 s^{\beta-1}   \xs_2(0)
\end{bmatrix}
=\bfi { A}\mathcal{L}\{ {\xs}(t)\} \, ,
\]
or, equivalently,
\[
\Efr
\begin{bmatrix}
  s^\gamma \bfg I_{\rho} & \bfg 0_{\rho,\rho}  \\
  \bfg 0_{\rho,\rho} & s^\beta \bfg I_{\rho}\\
\end{bmatrix} 
\begin{bmatrix}
  \mathcal{L}\{\xs_1(t)\} \\
  \mathcal{L}\{\xs_2(t)\}
\end{bmatrix}
-
\Efr
\begin{bmatrix}
  s^{\gamma-1} \bfg I_{\rho} & 
\bfg 0_{\rho,\rho}  \\
\bfg 0_{\rho,\rho} & s^{\beta-1}\bfg I_{\rho}\\
\end{bmatrix} 
\begin{bmatrix}
\xs_1(0) \\
\xs_2(0)
\end{bmatrix}
=  \Afr \,
\mathcal{L}\{\xs(t)\} \, ,
\]
or, equivalently,
\begin{equation}
\begin{aligned}
\label{foc:eq:pencilproof}
\Bigg(\Efr
\begin{bmatrix}
s^\gamma \bfg I_{\rho} & 
\bfg 0_{\rho,\rho}    \\
\bfg 0_{\rho,\rho} & 
s^\beta \bfg I_{\rho} \\
\end{bmatrix} 
-
\bfi {A} \Bigg) \, \mathcal{L}\{ {\xs}(t)\}
= 
\begin{bmatrix}
s^{\gamma-1} \bfg I_{\rho} & 
\bfg 0_{\rho,\rho}  \\
\bfg 0_{\rho,\rho} & 
s^{\beta-1} \bfg I_{\rho} \\
\end{bmatrix} 
{\xs}(0)  \, .
\end{aligned}
\end{equation}
\hfill
\eop

\newpage
\chapter{Map of the All-Island Irish Transmission System}
\label{app:data}

\begin{figure}[ht]
    \centering
    \resizebox{0.62\linewidth}{!}{\includegraphics{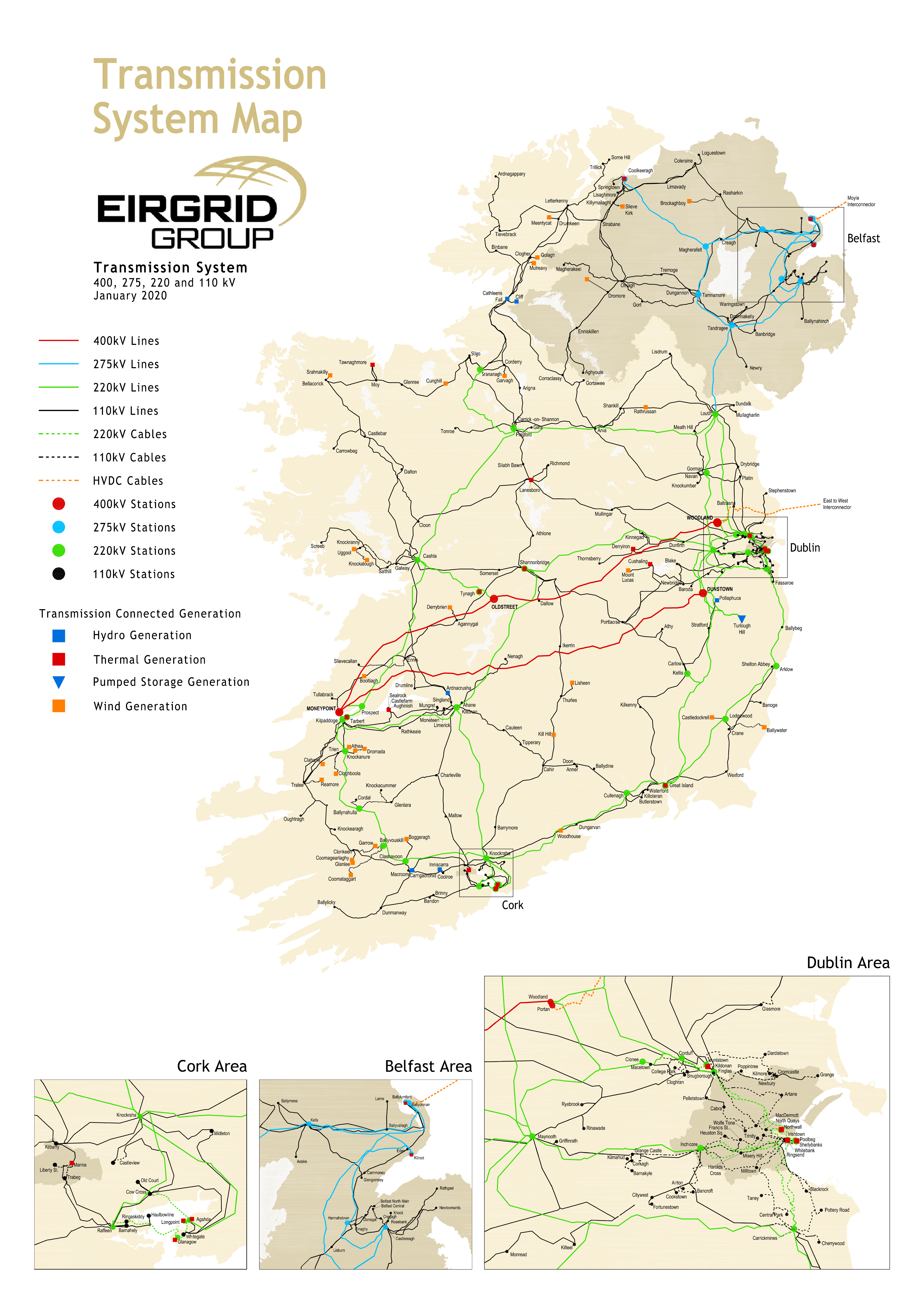}}
    \caption[AIITS: transmission system map]{AIITS: transmission system map, January 2020.}
    \label{appB:fig:aiits}
\end{figure}
\end{appendix}

\newpage

\cleardoublepage
\phantomsection
\addcontentsline{toc}{chapter}{Bibliography}
\bibliographystyle{references/IEEEtranS_custom}
\bibliography{references/thesis}

\begin{thebibliography}{100}
\providecommand{\url}[1]{#1}
\csname url@samestyle\endcsname
\providecommand{\newblock}{\relax}
\providecommand{\bibinfo}[2]{#2}
\providecommand{\BIBentrySTDinterwordspacing}{\spaceskip=0pt\relax}
\providecommand{\BIBentryALTinterwordstretchfactor}{4}
\providecommand{\BIBentryALTinterwordspacing}{\spaceskip=\fontdimen2\font plus
\BIBentryALTinterwordstretchfactor\fontdimen3\font minus
  \fontdimen4\font\relax}
\providecommand{\BIBforeignlanguage}[2]{{%
\expandafter\ifx\csname l@#1\endcsname\relax
\typeout{** WARNING: IEEEtranS.bst: No hyphenation pattern has been}%
\typeout{** loaded for the language `#1'. Using the pattern for}%
\typeout{** the default language instead.}%
\else
\language=\csname l@#1\endcsname
\fi
#2}}
\providecommand{\BIBdecl}{\relax}
\BIBdecl

\bibitem{abbsvc}
{ABB FACTS Division}, ``A matter of {FACTS} deliver more high quality power,''
  2015, product guide, available at library.e.abb.com.

\bibitem{Abdallah:1993}
C.~Abdallah, P.~Dorato, J.~Benites-Read, and R.~Byrne, ``Delayed positive
  feedback can sabilize oscillatory systems,'' in \emph{Proceedings of the
  American Control Conference}, 1993, pp. 3106--3107.

\bibitem{abed:00}
E.~H. Abed, D.~Lindsay, and W.~A. Hashlamoun, ``On participation factors for
  linear systems,'' \emph{Automatica}, vol.~36, no.~10, pp. 1489 -- 1496, 2000.

\bibitem{Abidi:2016-IJRNC}
K.~Abidi, Y.~Yildiz, and B.~E. Korpe, ``Explicit time-delay compensation in
  teleoperation: An adaptive control approach,'' \emph{International Journal of
  Robust and Nonlinear Control}, vol.~26, no.~15, pp. 3388--3403, 2016.

\bibitem{MUMPS}
P.~Amestoy, I.~S. Duff, J.~Koster, and J.-Y. L'Excellent, ``A fully
  asynchronous multifrontal solver using distributed dynamic scheduling,''
  \emph{SIAM Journal on Matrix Analysis and Applications}, vol.~23, no.~1, pp.
  15--41, 2001.

\bibitem{amirnaser}
Y.~Amirnaser and I.~Reza, \emph{{Voltage-Sourced Converters in Power Systems:
  Modeling, Control, and Applications}}.\hskip 1em plus 0.5em minus 0.4em\relax
  Wiley-IEEE Press, 2012.

\bibitem{anderson:2003}
P.~M. Anderson and A.~A. Fouad, \emph{\BIBforeignlanguage{English}{{Power
  System Control and Stability }}}, 2nd~ed.\hskip 1em plus 0.5em minus
  0.4em\relax IEEE Press: Wiley-Interscience, 2003.

\bibitem{lapack}
E.~Angerson, Z.~Bai, J.~Dongarra, A.~Greenbaum, A.~McKenney, J.~D. Croz,
  S.~Hammarling, J.~Demmel, C.~Bischof, and D.~Sorensen, ``{LAPACK}: A portable
  linear algebra library for high-performance computers,'' in \emph{Proceedings
  of the 1990 ACM/IEEE Conference on Supercomputing}, Nov. 1990, pp. 2--11.

\bibitem{petsc}
\BIBentryALTinterwordspacing
{Argonne National Laboratory}, ``{PETS}c users manual,'' 2020. [Online].
  Available: \url{https://www.mcs.anl.gov/petsc}
\BIBentrySTDinterwordspacing

\bibitem{Arnoldi}
W.~E. Arnoldi, ``The principle of minimized iterations in the solution of the
  matrix eigenvalue problem,'' \emph{Quarterly of Applied Mathematics}, vol.~9,
  no.~1, pp. 17--29, 1951.

\bibitem{asghari:2018}
R.~{Asghari}, B.~{Mozafari}, M.~{Salay Naderi}, T.~{Amraee}, V.~{Nurmanova},
  and M.~{Bagheri}, ``A novel method to design delay-scheduled controllers for
  damping inter-area oscillations,'' \emph{IEEE Access}, vol.~6, pp.
  71\,932--71\,946, 2018.

\bibitem{atangana:2016}
A.~Atangana and D.~Baleanu, ``New fractional derivatives with nonlocal and
  non-singular kernel: Theory and application to heat transfer model,''
  \emph{Thermal Science}, vol.~20, no.~2, pp. 763--769, Jan. 2016.

\bibitem{anasazi}
C.~Baker, U.~Hetmaniuk, R.~Lehoucq, and H.~Thornquist, ``Anasazi software for
  the numerical solution of large-scale eigenvalue problems,'' \emph{{ACM}
  Transactions on Mathematical Software}, vol.~36, no.~13, pp. 351--362, Jul.
  2009.

\bibitem{subspace}
K.~J. Bathe and E.~L. Wilson, ``Solution methods for large generalized
  eigenvalue problems in structural engineering,'' \emph{International Journal
  for Numerical Methods in Engineering}, vol.~6, pp. 213--226, 1973.

\bibitem{bellen:99}
A.~{Bellen}, N.~{Guglielmi}, and A.~E. {Ruehli}, ``Methods for linear systems
  of circuit delay differential equations of neutral type,'' \emph{IEEE
  Transactions on Circuits and Systems - I: Fundamental Theory and
  Applications}, vol.~46, no.~1, pp. 212--215, Jan. 1999.

\bibitem{bellen:2000}
A.~Bellen and S.~Maset, ``Numerical solution of constant coefficient linear
  delay differential equations as abstract {Cauchy} problems,''
  \emph{Numerische Mathematik}, vol.~84, no.~3, pp. 351--374, Jan. 2000.

\bibitem{Bellman:1963-BOOK}
R.~Bellman and K.~L. Cooke, \emph{{Differential-Difference Equations}}.\hskip
  1em plus 0.5em minus 0.4em\relax New York, USA: Academic Press, 1963.

\bibitem{scalapack}
L.~S. Blackford, J.~Choi, A.~Cleary, E.~D'Azevedo, J.~Demmel, I.~Dhillon,
  J.~Dongarra, S.~Hammarling, G.~Henry, A.~Petitet, K.~Stanley, D.~Walker, and
  R.~C. Whaley, \emph{{ScaLAPACK} Users' Guide}.\hskip 1em plus 0.5em minus
  0.4em\relax SIAM, 1997.

\bibitem{bode:1945}
H.~Bode, \emph{Network Analysis and Feedback Amplifier Design}.\hskip 1em plus
  0.5em minus 0.4em\relax Princeton, NJ: Van Nostrand, 1945.

\bibitem{Bo}
B.~Bonilla, M.~Rivero, and J.~Trujillo, ``On systems of linear fractional
  differential equations with constant coefficients,'' \emph{Applied
  Mathematics and Computation}, vol. 187, no.~1, pp. 68--78, 2007.

\bibitem{Breda:2015-BOOK}
D.~Breda, S.~Maset, and R.~Vermiglio, \emph{{Stability of Linear Delay
  Differential Equations: A Numerical Approach with MATLAB}}.\hskip 1em plus
  0.5em minus 0.4em\relax New York, USA: Springer, 2015.

\bibitem{camacho:07}
E.~F. Camacho and C.~Bordons, \emph{{Model Predictive Control}}.\hskip 1em plus
  0.5em minus 0.4em\relax Springer, 2007.

\bibitem{caputo:2015}
M.~Caputo and M.~Fabrizio, ``A new definition of fractional derivative without
  singular kernel,'' \emph{Progress in Fractional Differentiation and
  Applications}, vol.~1, no.~2, pp. 73--85, Apr. 2015.

\bibitem{chaib:2017}
L.~Chaib, A.~Choucha, and S.~Arif, ``{Optimal design and tuning of novel
  fractional order {PID} power system stabilizer using a new metaheuristic
  {Bat} algorithm},'' \emph{Ain Shams Engineering Journal}, vol.~8, no.~2, pp.
  113 -- 125, Jun. 2017.

\bibitem{chow:00}
J.~H. Chow, J.~J. Sanchez-Gasca, H.~Ren, and S.~Wang, ``Power system damping
  controller design-using multiple input signals,'' \emph{IEEE Control Systems
  Magazine}, vol.~20, no.~4, pp. 82--90, Aug. 2000.

\bibitem{book:chow:13}
J.~H. Chow, \emph{{Power System Coherency and Model Reduction}}, ser. Power
  Electronics and Power Systems 94.\hskip 1em plus 0.5em minus 0.4em\relax New
  York: Springer-Verlag, 2013.

\bibitem{Dab}
A.~Dabiri, B.~P. Moghaddam, and J.~A.~T. Machado, ``{Optimal variable-order
  fractional PID controllers for dynamical systems},'' \emph{Journal of
  Computational and Applied Mathematics}, vol. 339, pp. 40--48, 2018.

\bibitem{Dai}
L.~Dai, \emph{{Singular Control Systems}}.\hskip 1em plus 0.5em minus
  0.4em\relax M. Thoma, A. Wyner (Eds.), Lecture Notes in Control and
  information Sciences, 1988.

\bibitem{Das8}
I.~Dassios and D.~Baleanu, ``Caputo and related fractional derivatives in
  singular systems,'' \emph{Applied Mathematics and Computation}, vol. 337, pp.
  591--606, 2018.

\bibitem{moebius}
I.~Dassios, G.~Tzounas, and F.~Milano, ``{The M{\"o}bius transform effect in
  singular systems of differential equations},'' \emph{Applied Mathematics and
  Computation}, vol. 361, pp. 338--353, 2019.

\bibitem{foc2}
I.~{Dassios}, G.~{Tzounas}, and F.~{Milano}, ``Generalized fractional
  controller for singular systems of differential equations,'' \emph{Journal of
  Computational and Applied Mathematics}, vol. 378, p. 112919, 2020.

\bibitem{pfactors2}
I.~Dassios, G.~Tzounas, and F.~Milano, ``Participation factors for singular
  systems of differential equations,'' \emph{Circuits, Systems, and Signal
  Processing}, vol.~39, pp. 83--110, 2020.

\bibitem{dassios:robust}
I.~Dassios, G.~Tzounas, and F.~Milano, ``Robust stability criterion for
  perturbed singular systems of linearized differential equations,''
  \emph{Journal of Computational and Applied Mathematics}, vol. 381, p. 113032,
  2021.

\bibitem{Datko:1978-QAM}
R.~Datko, ``A procedure for determination of the exponential stability of
  certain differential-difference equations,'' \emph{Quarterly of Applied
  Mathematics}, vol.~36, no.~3, pp. 279--292, 1978.

\bibitem{Deb}
A.~Debbouche, ``Fractional evolution integro-differential systems with nonlocal
  conditions,'' \emph{Advances in Dynamical Systems and Applications}, vol.~5,
  no.~1, 2010.

\bibitem{delghavi:2016}
M.~B. {Delghavi}, S.~{Shoja-Majidabad}, and A.~{Yazdani}, ``Fractional-order
  sliding-mode control of islanded distributed energy resource systems,''
  \emph{IEEE Transactions on Sustainable Energy}, vol.~7, no.~4, pp.
  1482--1491, Oct. 2016.

\bibitem{Du}
G.~R. Duan, \emph{{The Analysis and Design of Descriptor Linear
  Systems}}.\hskip 1em plus 0.5em minus 0.4em\relax Springer, 2011.

\bibitem{FabozziLNewton}
D.~Fabozzi, A.~S. Chieh, B.~Haut, and T.~{Van Cutsem}, ``{Accelerated and
  localized Newton schemes for faster dynamic simulation of large power
  systems},'' \emph{IEEE Transactions on Power Systems}, vol.~28, no.~4, pp.
  4936--4947, Nov. 2013.

\bibitem{VCutsemCOI}
D.~Fabozzi and T.~{Van Cutsem}, ``On angle references in long-term time-domain
  simulations,'' \emph{IEEE Transactions on Power Systems}, vol.~26, no.~1, pp.
  483--484, Feb. 2011.

\bibitem{ajjarapu:11}
M.~{Fan}, V.~{Ajjarapu}, C.~{Wang}, D.~{Wang}, and C.~{Luo}, ``{RPM}-based
  approach to extract power system steady state and small signal stability
  information from the time-domain simulation,'' \emph{IEEE Transactions on
  Power Systems}, vol.~26, no.~1, pp. 261--269, 2011.

\bibitem{Fong1978}
J.~Fong and C.~Pottle, ``Parallel processing of power system analysis problems
  via simple parallel microcomputer structures,'' \emph{IEEE Transactions on
  Power Apparatus and Systems}, vol. PAS-97, no.~5, pp. 1834--1841, Sep. 1978.

\bibitem{Francis}
J.~G.~F. Francis, ``The {QR} transformation a unitary analogue to the {LR}
  transformation -- {Part} 1,'' \emph{The Computer Journal}, vol.~4, no.~3, pp.
  265--271, 1961.

\bibitem{Fridman2016}
E.~Fridman and L.~Shaikhet, ``Simple \textsc{LMI}s for stabilization by using
  delays,'' in \emph{Proceedings of the IEEE Conference on Decision and
  Control}, 2016, pp. 3240--3245.

\bibitem{zpares_guide}
Y.~Futamura and T.~Sakurai, \emph{{z-Pares} Users’ Guide Release
  0.9.5}.\hskip 1em plus 0.5em minus 0.4em\relax University of Tsukuba, 2014.

\bibitem{gantmacher:59}
R.~F. Gantmacher, \emph{{The Theory of Matrices I, II}}.\hskip 1em plus 0.5em
  minus 0.4em\relax New York: Chelsea, 1959.

\bibitem{garofal:O2}
F.~Garofalo, L.~Iannelli, and F.~Vasca, ``Participation factors and their
  connections to residues and relative gain array,'' \emph{IFAC Proceedings
  Volumes}, vol.~35, no.~1, pp. 125 -- 130, 2002, 15th IFAC World Congress.

\bibitem{Gopalsamy:1992-BOOK}
K.~Gopalsamy, \emph{{Stability and Oscillations in Delay Differential Equations
  of Population Dynamics}}.\hskip 1em plus 0.5em minus 0.4em\relax Norwell, MA:
  Kluwer, 1992.

\bibitem{grigsby:2007}
L.~L. Grigsby, \emph{{Power System Stability and Control}}.\hskip 1em plus
  0.5em minus 0.4em\relax Boca Raton, CA: CRC Press, 2007.

\bibitem{Guglielmi2001}
N.~Guglielmi and E.~Hairer, ``Implementing {Radau} {IIA} methods for stiff
  delay differential equations,'' \emph{Computing}, vol.~67, no.~1, pp. 1--12,
  Jul. 2001.

\bibitem{gurrala:10}
G.~{Gurrala} and I.~{Sen}, ``Power system stabilizers design for interconnected
  power systems,'' \emph{IEEE Transactions on Power Systems}, vol.~25, no.~2,
  pp. 1042--1051, 2010.

\bibitem{Hale:1993-BOOK}
J.~K. Hale and S.~M.~V. Lunel, \emph{{Introduction to Functional Differential
  Equations}}.\hskip 1em plus 0.5em minus 0.4em\relax New York, USA:
  Springer-Verlag, 1993.

\bibitem{hamdan:86}
A.~M.~A. Hamdan, ``Coupling measures between modes and state variables in
  power-system dynamics,'' \emph{International Journal of Control}, vol.~43,
  no.~3, pp. 1029--1041, 1986.

\bibitem{hamdan:87}
H.~Hamdan and A.~Hamdan, ``On the coupling measures between modes and state
  variables and subsynchronous resonance,'' \emph{Electric Power System
  Research}, vol.~13, no.~3, pp. 165 -- 171, 1987.

\bibitem{abed:09}
W.~A. Hashlamoun, M.~A. Hassouneh, and E.~H. Abed, ``New results on modal
  participation factors: Revealing a previously unknown dichotomy,'' \emph{IEEE
  Transactions on Automatic Control}, vol.~54, no.~7, pp. 1439--1449, Jul.
  2009.

\bibitem{Helbing:2004-PRE}
D.~Helbing, S.~L\"{a}mmer, T.~Seidel, P.~\u{S}eba, and T.~P{\l}atkowski,
  ``Physics, stability, and dynamics of supply networks,'' \emph{Physical
  Review E}, vol.~70, no.~6, 2004, art. no. 066116.

\bibitem{slepc}
V.~Hernandez, J.~E. Roman, and V.~Vidal, ``{SLEPc}: A scalable and flexible
  toolkit for the solution of eigenvalue problems,'' \emph{{ACM} Transactions
  on Mathematical Software}, vol.~31, no.~3, pp. 351--362, 2005.

\bibitem{Hi}
R.~Hilfe, \emph{Applications of Fractional Calculus in Physics, p. 463}.\hskip
  1em plus 0.5em minus 0.4em\relax World Scientific, River Edge, NJ, USA, 2000.

\bibitem{hsu:87}
Y.~Hsu and C.~Chen, ``Identification of optimum location for stabiliser
  applications using participation factors,'' \emph{IEE Proceedings C -
  Generation, Transmission and Distribution}, vol. 134, no.~3, pp. 238--244,
  May 1987.

\bibitem{stability:20}
{IEEE}, ``{IEEE Task Force on Stability definitions and characterization of
  dynamic behavior in systems with high penetration of power electronic
  interfaced technologies},'' \emph{Technical Report PES-TR77}, Apr. 2020.

\bibitem{ionescu:17}
C.~Ionescu, A.~Lopes, D.~Copot, J.~A.~T. Machado, and J.~H.~T. Bates, ``The
  role of fractional calculus in modeling biological phenomena: A review,''
  \emph{Communications in Nonlinear Science and Numerical Simulation}, vol.~51,
  pp. 141--159, 2017.

\bibitem{ishihara:02}
J.~Y. {Ishihara} and M.~H. {Terra}, ``On the {Lyapunov} theorem for singular
  systems,'' \emph{IEEE Transactions on Automatic Control}, vol.~47, no.~11,
  pp. 1926--1930, 2002.

\bibitem{jesus:2008}
I.~S. Jesus and J.~A.~T. Machado, ``Fractional control of heat diffusion
  systems,'' \emph{Nonlinear Dynamics}, vol.~54, no.~3, pp. 263--282, Nov.
  2008.

\bibitem{Kac}
T.~Kaczorek, \emph{{Selected Problems of Fractional Systems Theory: Fractional
  Continuous-Time Linear Systems, 27--52}}.\hskip 1em plus 0.5em minus
  0.4em\relax Springer Berlin Heidelberg, 2011.

\bibitem{Kallmann:1940}
H.~E. Kallmann, ``Transversal filters,'' \emph{Proceedings of the Institute of
  Radio Engineers}, vol.~28, pp. 302--10, 1940.

\bibitem{kamwa:2001}
I.~{Kamwa}, R.~{Grondin}, and Y.~{Hebert}, ``Wide-area measurement based
  stabilizing control of large power systems-a decentralized/hierarchical
  approach,'' \emph{IEEE Transactions on Power Systems}, vol.~16, no.~1, pp.
  136--153, Feb. 2001.

\bibitem{8274156}
U.~{Karaagac}, J.~{Mahseredjian}, I.~{Kocar}, G.~{Soykan}, and O.~{Saad},
  ``Partial refactorization based machine modeling techniques for
  electromagnetic transients,'' in \emph{Proceedings of the IEEE PES General
  Meeting}, Jul. 2017.

\bibitem{LOBPCG}
A.~V. Knyazev, ``Toward the optimal preconditioned eigensolver: Locally optimal
  block preconditioned conjugate gradient method,'' \emph{SIAM Journal on
  Scientific Computing}, vol.~23, no.~2, p. 517–541, 2001.

\bibitem{6624122}
I.~{Kocar}, J.~{Mahseredjian}, U.~{Karaagac}, G.~{Soykan}, and O.~{Saad},
  ``Multiphase load-flow solution for large-scale distribution systems using
  {MANA},'' \emph{IEEE Transactions on Power Delivery}, vol.~29, no.~2, pp.
  908--915, Apr. 2014.

\bibitem{Kokame:2001-TAC}
H.~Kokame, K.~Hirata, K.~Konishi, and T.~Mori, ``Difference feedback fan
  stabilize uncertain steady states,'' \emph{IEEE Transactions on Automatic
  Control}, vol.~46, no.~12, pp. 1908--1913, 2001.

\bibitem{kressner}
D.~Kressner, \emph{Numerical Methods for General and Structured Eigenvalue
  Problems}, 4th~ed.\hskip 1em plus 0.5em minus 0.4em\relax Springer, 2015.

\bibitem{kron:1963}
G.~Kron, \emph{{The Piecewise Solution of Large-Scale Systems}}.\hskip 1em plus
  0.5em minus 0.4em\relax London, UK: Macdonald, 1963.

\bibitem{stability:04}
P.~{Kundur}, J.~{Paserba}, V.~{Ajjarapu}, G.~{Andersson}, A.~{Bose},
  C.~{Ca{\~n}izares}, N.~{Hatziargyriou}, D.~{Hill}, A.~{Stankovic},
  C.~{Taylor}, T.~{Van Cutsem}, and V.~{Vittal}, ``Definition and
  classification of power system stability {IEEE/CIGRE} joint task force on
  stability terms and definitions,'' \emph{IEEE Transactions on Power Systems},
  vol.~19, no.~3, pp. 1387--1401, 2004.

\bibitem{kundur:1990}
P.~{Kundur}, G.~J. {Rogers}, D.~Y. {Wong}, L.~{Wang}, and M.~G. {Lauby}, ``A
  comprehensive computer program package for small signal stability analysis of
  power systems,'' \emph{IEEE Transactions on Power Systems}, vol.~5, no.~4,
  pp. 1076--1083, 1990.

\bibitem{kundur:94}
P.~Kundur, \emph{{Power System Stability and Control}}.\hskip 1em plus 0.5em
  minus 0.4em\relax New York: Mc-Grall Hill, 1994.

\bibitem{pal:16}
L.~P. {Kunjumuhammed}, B.~C. {Pal}, C.~{Oates}, and K.~J. {Dyke}, ``Electrical
  oscillations in wind farm systems: Analysis and insight based on detailed
  modeling,'' \emph{IEEE Transactions on Sustainable Energy}, vol.~7, no.~1,
  pp. 51--62, Jan. 2016.

\bibitem{317667}
R.~H. {Lasseter} and J.~{Zhou}, ``{TACS} enhancements for the electromagnetic
  transient program,'' \emph{IEEE Transactions on Power Systems}, vol.~9,
  no.~2, pp. 736--742, 1994.

\bibitem{blas}
C.~L. Lawson, R.~J. Hanson, D.~R. Kincaid, and F.~T. Krogh, ``Basic linear
  algebra subprograms for {FORTRAN} usage,'' University of Texas at Austin,
  USA, Tech. Rep., 1977.

\bibitem{lazarevic:2006}
M.~Lazarevi{\'c}, ``Finite time stability analysis of {$PD^\alpha$} fractional
  control of robotic time-delay systems,'' \emph{Mechanics Research
  Communications}, vol.~33, no.~2, pp. 269 -- 279, Mar. 2006.

\bibitem{Lehoucq}
R.~B. Lehoucq and D.~C. Sorensen, ``Deflation techniques for an implicitly
  restarted {Arnoldi} iteration,'' \emph{SIAM Journal on Matrix Analysis and
  Applications}, vol.~17, no.~4, 1996.

\bibitem{arpack}
R.~B. Lehoucq, D.~C. Sorensen, and C.~Yang, ``{ARPACK} users' guide: Solution
  of large-scale eigenvalue problems with implicitly restarted {A}rnoldi
  methods,'' in \emph{SIAM}, 1998.

\bibitem{leibniz:1695}
G.~W. Leibniz, ``{Letter from Hanover, Germany, September 30, 1695 to G. A.
  L'Hospital},'' \emph{Leibnizen Mathemutische Schriften}, vol.~2, no.~1, pp.
  301--302, 1962, {Olms Verlag., Hildesheim, Germany, Published in 1849}.

\bibitem{Lew}
F.~L. Lewis, ``A survey of linear singular systems,'' \emph{Circuits, Systems,
  and Signal Processing}, vol.~5, pp. 3--36, 1986.

\bibitem{li:2019}
C.~{Li}, Y.~{Chen}, T.~{Ding}, Z.~{Du}, and F.~{Li}, ``A sparse and low-order
  implementation for discretization-based eigen-analysis of power systems with
  time-delays,'' \emph{IEEE Transactions on Power Systems}, vol.~34, no.~6, pp.
  5091--5094, Nov. 2019.

\bibitem{li:2007}
C.~Li and W.~Deng, ``Remarks on fractional derivatives,'' \emph{Applied
  Mathematics and Computation}, vol. 187, no.~2, pp. 777 -- 784, Apr. 2007.

\bibitem{liouville:1832}
J.~Liouville, ``{M{\'e}moire sur quelques qu{\'e}stions de g{\'e}ometrie et de
  m{\'e}canique, et sur un nouveau genre de calcul pour r{\'e}soudre ces
  qu{\'e}stions},'' \emph{Journal Ecole Polytechnique}, vol.~13, pp. 1--69,
  1832, {Sect. 21}.

\bibitem{muyang:wams}
M.~{Liu}, I.~{Dassios}, G.~{Tzounas}, and F.~{Milano}, ``{Stability analysis of
  power systems with inclusion of realistic-modeling WAMS delays},'' \emph{IEEE
  Transactions on Power Systems}, vol.~34, no.~1, pp. 627--636, Jan. 2019.

\bibitem{muyang:compensation:mdpi}
M.~Liu, I.~Dassios, G.~Tzounas, and F.~Milano, ``Model-independent derivative
  control delay compensation methods for power systems,'' \emph{Energies},
  vol.~13, no.~2, p. 342, 2020.

\bibitem{muyang:powertechmilan}
M.~{Liu}, G.~{Tzounas}, and F.~{Milano}, ``A model-independent delay
  compensation method for power systems,'' in \emph{Proceedings of the IEEE
  PowerTech Conference}, 2019.

\bibitem{Liz1}
C.~Lizama, M.~Murillo‐Arcila, and C.~Leal, ``Lebesgue regularity for
  differential difference equations with fractional damping,''
  \emph{Mathematical Methods in the Applied Sciences}, vol.~41, no.~7, pp.
  2535--2545, 2018.

\bibitem{1583706}
J.~{Mahseredjian}, L.~{Dube}, {Ming Zou}, S.~{Dennetiere}, and G.~{Joos},
  ``Simultaneous solution of control system equations in {EMTP},'' \emph{IEEE
  Transactions on Power Systems}, vol.~21, no.~1, pp. 117--124, Feb 2006.

\bibitem{niculescu}
W.~Michiels and S.-I. Niculescu, \emph{{Stability and Stabilization of
  Time-Delay Systems: An Eigenvalue-Based Approach}}.\hskip 1em plus 0.5em
  minus 0.4em\relax Philadelphia, PA: SIAM, 2009.

\bibitem{milano:10}
F.~Milano, \emph{{Power System Modelling and Scripting}}.\hskip 1em plus 0.5em
  minus 0.4em\relax London: Springer, 2010.

\bibitem{vancouver}
F.~Milano, ``{A Python-based software tool for power system analysis},'' in
  \emph{Proceedings of the IEEE PES General Meeting}, Jul. 2013.

\bibitem{milano:psa:16}
F.~{Milano}, ``Delay-based numerical stability of the partitioned-solution
  approach,'' in \emph{Proceedings of the IEEE PES General Meeting}, 2016, pp.
  1--5.

\bibitem{semi:2016}
F.~{Milano}, ``Semi-implicit formulation of differential-algebraic equations
  for transient stability analysis,'' \emph{IEEE Transactions on Power
  Systems}, vol.~31, no.~6, pp. 4534--4543, Nov. 2016.

\bibitem{multiple:16}
F.~Milano, ``Small-signal stability analysis of large power systems with
  inclusion of multiple delays,'' \emph{IEEE Transactions on Power Systems},
  vol.~31, no.~4, pp. 3257--3266, Jul. 2016.

\bibitem{delay1}
F.~Milano and M.~Anghel, ``Impact of time delays on power system stability,''
  \emph{IEEE Transactions on Circuits and Systems - I: Fundamental Theory and
  Applications}, vol.~59, no.~4, pp. 889--900, Apr. 2012.

\bibitem{hessenberg:16}
F.~{Milano} and I.~{Dassios}, ``{Small-signal stability analysis for non-index
  1 Hessenberg form systems of delay differential-algebraic equations},''
  \emph{IEEE Transactions on Circuits and Systems - I: Regular Papers},
  vol.~63, no.~9, pp. 1521--1530, Sep. 2016.

\bibitem{fdf}
F.~Milano and {\'A}.~Ortega, ``Frequency divider,'' \emph{IEEE Transactions on
  Power Systems}, vol.~32, no.~2, pp. 1493--1501, Mar. 2017.

\bibitem{dual:17}
F.~Milano and I.~Dassios, ``Primal and dual generalized eigenvalue problems for
  power systems small-signal stability analysis,'' \emph{IEEE Transactions on
  Power Systems}, vol.~32, no.~6, pp. 4626 -- 4635, 2017.

\bibitem{book:eigenvalue}
F.~Milano, I.~Dassios, M.~Liu, and G.~Tzounas, \emph{Eigenvalue Problems in
  Power Systems}.\hskip 1em plus 0.5em minus 0.4em\relax CRC Press, Taylor \&
  Francis Group, 2020.

\bibitem{milano_ortega:2019}
F.~Milano and {\'A}.~Ortega~Manjavacas, \emph{Converter-Interfaced Energy
  Storage Systems: Context, Modelling and Dynamic Analysis}.\hskip 1em plus
  0.5em minus 0.4em\relax Cambridge University Press, 2019.

\bibitem{minasian:06}
R.~A. {Minasian}, ``Photonic signal processing of microwave signals,''
  \emph{IEEE Transactions on Microwave Theory and Techniques}, vol.~54, no.~2,
  pp. 832--846, Feb. 2006.

\bibitem{molerqz}
C.~B. Moler and G.~W. Stewart, ``An algorithm for generalized matrix eigenvalue
  problems,'' \emph{SIAM Journal on Numerical Analysis}, vol.~10, no.~2, pp.
  241--256, 1973.

\bibitem{monje:2010}
C.~A. Monje, Y.~Chen, B.~M. Vinagre, D.~Xue, and V.~Feliu,
  \emph{Fractional-order Systems and Controls, Fundamentals and
  Applications}.\hskip 1em plus 0.5em minus 0.4em\relax Springer, 2010.

\bibitem{ahsan:2019}
M.~A.~A. {Murad} and F.~{Milano}, ``{Modeling and simulation of PI-controllers
  limiters for the dynamic analysis of VSC-Based devices},'' \emph{IEEE
  Transactions on Power Systems}, vol.~34, no.~5, pp. 3921--3930, Sep. 2019.

\bibitem{ahsan:gm2019}
M.~A.~A. {Murad}, G.~{Tzounas}, M.~{Liu}, and F.~{Milano}, ``{Frequency control
  through voltage regulation of power system using SVC devices},'' in
  \emph{Proceedings of the IEEE PES General Meeting}, Aug. 2019.

\bibitem{foc:ifac}
M.~A.~A. Murad, G.~Tzounas, and F.~Milano, ``Modeling and simulation of
  fractional order {PI} control limiters for power systems,'' in
  \emph{Proceedings of the 21st IFAC World Congress}, 2020, pp. 1--6.

\bibitem{ahsan:2018}
M.~A.~A. Murad, {\'A}.~Ortega, and F.~Milano, ``{Impact on power system
  dynamics of PI control limiters of VSC-based devices},'' \emph{Proceedings of
  the Power Systems Computation Conference}, pp. 1--7, Jun. 2018.

\bibitem{mili:19}
M.~Netto, Y.~Susuki, and L.~Mili, ``Data-driven participation factors for
  nonlinear systems based on {K}oopman mode decomposition,'' \emph{IEEE Control
  Systems Letters}, vol.~3, no.~1, pp. 198--203, Jan. 2019.

\bibitem{Niculescu:2001}
S.~I. Niculescu, \emph{{Delay Effects on Stability: A Robust Control
  Approach}}.\hskip 1em plus 0.5em minus 0.4em\relax Springer-Verlag, 2001.

\bibitem{Olfati:2005-ACC}
R.~Olfati-Saber, ``Ultrafast consensus in small-world networks,'' in
  \emph{Proceedings of the American Control Conference}, 2005, pp. 2371--2378.

\bibitem{Olgac:2005}
N.~Olgac, A.~F. Ergenc, and R.~Sipahi, ``{``Delay scheduling'': A new concept
  for stabilization in multiple delay systems},'' \emph{Journal of Vibration
  and Control}, vol.~11, no.~9, pp. 1159--1172, 2005.

\bibitem{Orosz:2010-PTRSL}
G.~Orosz, R.~E. Wilson, and G.~St\'{e}p\'{a}n, ``Traffic jams: dynamics and
  control,'' \emph{Philosophical Transactions of the Royal Society of London
  A}, vol. 368, no. 1928, pp. 4455--4479, 2010.

\bibitem{pll:18}
{\'A}.~{Ortega} and F.~{Milano}, ``{Comparison of different PLL implementations
  for frequency estimation and control},'' in \emph{Proceedings of the
  International Conference on Harmonics and Quality of Power}, May 2018.

\bibitem{oustaloup:1991}
A.~Oustaloup, \emph{{La commande CRONE (commande robuste d’ordre non
  entier)}}.\hskip 1em plus 0.5em minus 0.4em\relax {Herm{\'e}s, Paris}, 1991.

\bibitem{oustaloup:2000}
A.~{Oustaloup}, F.~{Levron}, B.~{Mathieu}, and F.~M. {Nanot}, ``Frequency-band
  complex noninteger differentiator: characterization and synthesis,''
  \emph{IEEE Transactions on Circuits and Systems - I: Fundamental Theory and
  Applications}, vol.~47, no.~1, pp. 25--39, Jan. 2000.

\bibitem{cronetool}
A.~{Oustaloup}, P.~{Melchior}, P.~{Lanusse}, O.~{Cois}, and F.~{Dancla}, ``{The
  CRONE toolbox for Matlab},'' in \emph{Proceedings of the IEEE International
  Symposium on Computer-Aided Control System Design}, Sep. 2000, pp. 190--195.

\bibitem{arriaga:89}
F.~L. Pagola, I.~J. P{\'e}rez-Arriaga, and G.~C. Verghese, ``On sensitivities,
  residues and participations: applications to oscillatory stability analysis
  and control,'' \emph{IEEE Transactions on Power Systems}, vol.~4, no.~1, pp.
  278--285, Feb. 1989.

\bibitem{pan:2013}
I.~Pan and S.~Das, ``{Frequency domain design of fractional order PID
  controller for AVR system using chaotic multi-objective optimization},''
  \emph{International Journal of Electrical Power and Energy Systems}, vol.~51,
  pp. 106--118, Oct. 2013.

\bibitem{PANDEY:2017}
S.~Pandey, P.~Dwivedi, and A.~Junghare, ``A novel 2-{DOF} fractional-order
  {$PI^\lambda$-$D^\mu$} controller with inherent anti-windup capability for a
  magnetic levitation system,'' \emph{AEU - International Journal of
  Electronics and Communications}, vol.~79, pp. 158 -- 171, 2017.

\bibitem{graphtool:14}
\BIBentryALTinterwordspacing
T.~P. Peixoto, ``{The graph-tool Python library},'' 2014. [Online]. Available:
  \url{graph-tool.skewed.de}
\BIBentrySTDinterwordspacing

\bibitem{arriaga:82_1}
I.~J. P{\'e}rez-Arriaga, G.~C. Verghese, and F.~C. Schweppe, ``Selective modal
  analysis with applications to electric power systems, part i: heuristic
  introduction,'' \emph{IEEE Transactions on Power Apparatus and Systems}, vol.
  PAS-101, no.~9, pp. 3117--3125, Sep. 1982.

\bibitem{petras:2009}
I.~P{\'e}tras, ``Stability of fractional-order systems with rational orders: A
  survey,'' \emph{Fractional Calculus and Applied Analysis}, vol.~12, no.~3,
  2009.

\bibitem{podlubny:book99}
I.~{Podlubny}, \emph{Fractional Differential Equations, Volume 198: An
  Introduction to Fractional Derivatives, Fractional Differential equations, to
  Methods of Their Solution and Some of Their Applications}.\hskip 1em plus
  0.5em minus 0.4em\relax Academic Press, 1999.

\bibitem{podlubny:1999}
I.~{Podlubny}, ``{Fractional-order systems and {$PI^\lambda D^\mu
  $}-controllers},'' \emph{IEEE Transactions on Automatic Control}, vol.~44,
  no.~1, pp. 208--214, Jan. 1999.

\bibitem{polizzifeast}
E.~Polizzi, ``Density-matrix-based algorithm for solving eigenvalue problems,''
  \emph{Physical Review B, American Physical Society}, vol.~79, no.~11, 2009.

\bibitem{feast_guide}
E.~Polizzi, ``{FEAST} eigenvalue solver v4.0 user guide,'' 2020.

\bibitem{Pyragas:1992-PLA}
K.~Pyragas, ``Continuous control of chaos by self-controlling feedback,''
  \emph{Physical Letters A}, vol. 170, no.~6, pp. 421--428, 1992.

\bibitem{qiao2016consensus}
W.~Qiao and R.~Sipahi, ``Consensus control under communication delay in a
  three-robot system: Design and experiments,'' \emph{IEEE Transactions on
  Control Systems Technology}, vol.~24, no.~2, pp. 687--694, 2016.

\bibitem{Ramirez:2015-ISA}
A.~Ram\'{i}rez, R.~Garrido, and S.~Mondi\'{e}, ``Velocity control of servo
  systems using an integral retarded algorithm,'' \emph{ISA Transactions},
  vol.~58, pp. 357--366, 2015.

\bibitem{Ramirez:2016-TAC}
A.~Ram\'{i}rez, S.~Mondi\'{e}, R.~Garrido, and R.~Sipahi, ``Design of
  proportional-integral-retarded ({PIR}) controllers for second-order {LTI}
  systems,'' \emph{IEEE Transactions on Automatic Control}, vol.~61, no.~6, pp.
  1688--1693, 2016.

\bibitem{Ramirez:2018-CYBERNETICS}
A.~Ram\'{i}rez and R.~Sipahi, ``Multiple intentional delays can facilitate fast
  consensus and noise reduction in a multiagent system,'' \emph{IEEE
  Transactions on Cybernetics}, vol.~49, no.~4, pp. 1224--1235, 2019.

\bibitem{Ramirez:2017-SICON}
A.~Ram\'{i}rez, R.~Sipahi, S.~Mondi\'{e}, and R.~Garrido, ``An analytical
  approach to tuning of delay-based controllers for {LTI-SISO} systems,''
  \emph{SIAM Journal Control Optim.}, vol.~55, no.~1, pp. 397--412, 2017.

\bibitem{roy:2019}
S.~{Roy}, A.~{Patel}, and I.~N. {Kar}, ``Analysis and design of a wide-area
  damping controller for inter-area oscillation with artificially induced time
  delay,'' \emph{IEEE Transactions on Smart Grid}, vol.~10, no.~4, pp.
  3654--3663, Jul. 2019.

\bibitem{saad:2011}
Y.~Saad, \emph{Numerical Methods for Large Eigenvalue Problems}.\hskip 1em plus
  0.5em minus 0.4em\relax SIAM, 2011.

\bibitem{sakurai1}
T.~Sakurai and H.~Sugiura, ``A projection method for generalized eigenvalue
  problems using numerical integration,'' \emph{Journal of Comput. and Applied
  Mathematics}, vol. 159, no.~1, pp. 119--128, 2003.

\bibitem{sakurai2}
T.~Sakurai and H.~Tadano, ``{CIRR}: a {Rayleigh-Ritz} type method with contour
  integral for generalized eigenvalue problems,'' \emph{Hokkaido Mathematical
  Journal}, vol.~36, no.~4, pp. 745--757, 2007.

\bibitem{saxena:2019}
S.~Saxena, ``Load frequency control strategy via fractional-order controller
  and reduced-order modeling,'' \emph{International Journal of Electrical Power
  and Energy Systems}, vol. 104, pp. 603 -- 614, Jan. 2019.

\bibitem{pardiso}
O.~Schenk and K.~Gartner, ``{Solving unsymmetric sparse systems of linear
  equations with PARDISO},'' \emph{Journal of Future Generation Computer
  Systems}, vol.~20, no.~3, 2004.

\bibitem{Shahidehpour:2003}
M.~Shahidehpour and Y.~Wang, \emph{{Communication and Control in Electric Power
  Systems}}.\hskip 1em plus 0.5em minus 0.4em\relax Hoboken, NJ: John Wiley \&
  Sons, 2003.

\bibitem{Shiri}
B.~Shiri and D.~Baleanu, ``Numerical solution of some fractional dynamical
  systems in medicine involving non-singular kernel with vector order,''
  \emph{Results in Nonlinear Analysis}, vol.~2, no.~4, pp. 160--168, 2019.

\bibitem{Sipahi2011_CSM}
R.~Sipahi, S.-I. Niculescu, C.~T. Abdallah, W.~Michiels, and K.~Gu, ``Stability
  and stabilization of systems with time delay,'' \emph{IEEE Control Systems
  Magazine}, vol.~31, no.~1, pp. 38--65, 2011.

\bibitem{Sipahi:2005-AUTOMATICA}
R.~Sipahi and N.~Olgac, ``Complete stability robustness of third-order lti
  multiple time-delay systems,'' \emph{Automatica}, vol.~41, no.~8, pp.
  1413--1422, 2005.

\bibitem{Smith:1959}
O.~J. Smith, ``A controller to overcome dead time,'' \emph{ISA Journal},
  vol.~6, no.~2, pp. 28--33, 1959.

\bibitem{mpi}
M.~Snir, S.~Otto, S.~Huss-Lederman, D.~Walker, and J.~Dongarra, \emph{MPI-The
  Complete Reference, Volume 1: The MPI Core}.\hskip 1em plus 0.5em minus
  0.4em\relax MIT Press, 1998.

\bibitem{Sorensen}
D.~C. Sorensen, ``Implicit application of polynomial filters in a k-step
  {Arnoldi} method,'' \emph{SIAM Journal on Matrix Analysis and Applications},
  vol.~13, no.~1, pp. 357--385, 1992.

\bibitem{heydt_stochastic:08}
J.~W. Stahlhut, T.~J. Browne, G.~T. Heydt, and V.~Vittal, ``Latency viewed as a
  stochastic process and its impact on wide area power system control
  signals,'' \emph{IEEE Transactions on Power Systems}, vol.~23, no.~1, pp.
  84--91, Feb. 2008.

\bibitem{Stepan:1989-BOOK}
G.~St\'{e}p\'{a}n, \emph{{Retarded Dynamical Systems: Stability and
  Characteristic Function}}.\hskip 1em plus 0.5em minus 0.4em\relax New York,
  USA: Longman Scientific \& Technical, 1989.

\bibitem{KrylovSchur}
G.~W. Stewart, ``A {Krylov--Schur} algorithm for large eigenproblems,''
  \emph{SIAM Journal on Matrix Analysis and Applications}, vol.~23, no.~3, pp.
  601--614, 2002.

\bibitem{stott:1979}
B.~{Stott}, ``Power system dynamic response calculations,'' \emph{Proceedings
  of the IEEE}, vol.~67, no.~2, pp. 219--241, Feb. 1979.

\bibitem{Su}
G.~Su, L.~Lu, B.~Tang, and Z.~Liu, ``{Quasi-linearization technique for solving
  nonlinear Riemann-Liouville fractional-order problems},'' \emph{Applied
  Mathematics and Computation}, vol. 378, pp. 125--199, 2020.

\bibitem{sart:11}
G.~{Sulligoi}, M.~{Chiandone}, and V.~{Arcidiacono}, ``New {SART} automatic
  voltage and reactive power regulator for secondary voltage regulation: Design
  and application,'' in \emph{Proceedings of the IEEE PES General Meeting},
  Jul. 2011.

\bibitem{taher:2014}
S.~A. Taher, M.~H. Fini, and S.~F. Aliabadi, ``{Fractional order {PID}
  controller design for {LFC} in electric power systems using imperialist
  competitive algorithm},'' \emph{Ain Shams Engineering Journal}, vol.~5,
  no.~1, pp. 121 -- 135, Mar. 2014.

\bibitem{TAKABA199549}
K.~Takaba, N.~Morihira, and T.~Katayama, ``A generalized {Lyapunov} theorem for
  descriptor system,'' \emph{Systems {\&} Control Letters}, vol.~24, no.~1, pp.
  49 -- 51, 1995.

\bibitem{tang:2012}
Y.~Tang, M.~Cui, C.~Hua, L.~Li, and Y.~Yang, ``{Optimum design of fractional
  order $PI^\lambda D^\mu$ controller for {AVR} system using chaotic ant
  swarm},'' \emph{Expert Systems with Applications}, vol.~39, no.~8, pp.
  6887--6896, 2012.

\bibitem{1601717}
C.~W. {Taylor} and R.~L. {Cresap}, ``Real-time power system simulation for
  automatic generation control,'' \emph{IEEE Transactions on Power Apparatus
  and Systems}, vol.~95, no.~1, pp. 375--384, Jan 1976.

\bibitem{fomcon:2011}
A.~Tepljakov, E.~Petlenkov, and J.~Belikov, ``{FOMCON: Fractional-order
  modeling and control toolbox for Matlab},'' \emph{Proceedings of the 18th
  International Conference Mixed Design of Integrated Circuits and Systems},
  Jun. 2011.

\bibitem{tian:18}
T.~Tian, X.~Kestelyn, O.~Thomas, H.~Amano, and A.~R. Messina, ``An accurate
  third-order normal form approximation for power system nonlinear analysis,''
  \emph{IEEE Transactions on Power Systems}, vol.~33, no.~2, pp. 2128--2139,
  Mar. 2018.

\bibitem{trilinos}
T.~{T}rilinos~{P}roject {T}eam, \emph{The {T}rilinos {P}roject {W}ebsite}.

\bibitem{tzou:pfactors}
G.~{Tzounas}, I.~{Dassios}, and F.~{Milano}, ``Modal participation factors of
  algebraic variables,'' \emph{IEEE Transactions on Power Systems}, vol.~35,
  no.~1, pp. 742--750, 2020.

\bibitem{foc1}
G.~{Tzounas}, I.~{Dassios}, M.~A.~A. {Murad}, and F.~{Milano}, ``Theory and
  implementation of fractional order controllers for power system
  applications,'' \emph{IEEE Transactions on Power Systems}, vol.~35, no.~6,
  pp. 4622--4631, Nov. 2020.

\bibitem{freqpss:powertech}
G.~{Tzounas}, M.~{Liu}, M.~A.~A. {Murad}, and F.~{Milano}, ``Impact of
  realistic bus frequency measurements on wide-area power system stabilizers,''
  in \emph{Proceedings of the IEEE PowerTech Conference}, 2019.

\bibitem{freqpss:gm}
G.~{Tzounas} and F.~{Milano}, ``Impact of the estimation of synchronous machine
  rotor speeds on wide-area damping controllers,'' in \emph{Proceedings of the
  IEEE PES General Meeting}, 2019, pp. 1--5.

\bibitem{osda}
G.~{Tzounas} and F.~{Milano}, ``Delay-based decoupling of power system models
  for transient stability analysis,'' \emph{IEEE Transactions on Power
  Systems}, vol.~36, no.~1, pp. 464--473, Jan. 2021.

\bibitem{delaydamp}
G.~{Tzounas}, R.~Sipahi, and F.~{Milano}, ``Damping power system
  electromechanical oscillations using time delays,'' \emph{IEEE Transactions
  on Circuits and Systems - I: Regular Papers}, 2021, accepted in Feb.~2021, in
  press.

\bibitem{app10217592}
G.~Tzounas, I.~Dassios, M.~Liu, and F.~Milano, ``Comparison of numerical
  methods and open-source libraries for eigenvalue analysis of large-scale
  power systems,'' \emph{Applied Sciences}, vol.~10, no.~21, 2020.

\bibitem{tzounas2018}
G.~Tzounas, M.~Liu, M.~A.~A. Murad, and F.~Milano, ``Stability analysis of wide
  area damping controllers with multiple time delays,''
  \emph{IFAC-PapersOnLine}, vol.~51, no.~28, pp. 504 -- 509, 2018, 10th IFAC
  Symposium on Control of Power and Energy Systems.

\bibitem{Ulsoy:2015-JDSMC}
A.~G. Ulsoy, ``Time-delayed control of \textsc{SISO} systems for improved
  stability margins,'' \emph{Journal of Dynamic Systems, Measurement, and
  Control}, vol. 137, no.~4, pp. 558--563, 2015.

\bibitem{URIARTE2012146}
F.~Uriarte, ``On {Kron}'s diakoptics,'' \emph{Electric Power System Research},
  vol.~88, pp. 146 -- 150, 2012.

\bibitem{uriante:13}
F.~Uriarte, \emph{{Multicore Simulation of Power System Transients}}.\hskip 1em
  plus 0.5em minus 0.4em\relax London, UK: The IET, 2013.

\bibitem{ninteger:2004}
D.~Val{\'e}rio and J.~Costa, ``{Ninteger: a non-integer control toolbox for
  Matlab},'' \emph{Proceedings of the 1st IFAC Workshop on Fractional
  Differentiation and its Application}, Jan. 2004.

\bibitem{venk:94}
V.~{Venkatasubramanian}, H.~{Schattler}, and J.~{Zaborszky}, ``A time-delay
  differential-algebraic phasor formulation of the large power system
  dynamics,'' in \emph{Proceedings of IEEE International Symposium on Circuits
  and Systems}, vol.~6, May 1994, pp. 49--52.

\bibitem{arriaga:82_2}
G.~C. Verghese, I.~J. P{\'e}rez-Arriaga, and F.~C. Schweppe, ``Selective modal
  analysis with applications to electric power systems, part ii: the dynamic
  stability problem,'' \emph{IEEE Transactions on Power Apparatus and Systems},
  vol. PAS-101, no.~9, pp. 3126--3134, Sep. 1982.

\bibitem{vinagre:2000}
B.~M. Vinagre, I.~Podlubny, and V.~Feliu, ``Some approximations of fractional
  order operators used in control theory and applications,'' \emph{Journal of
  Fractional Calculus and Applied Analysis}, pp. 231--248, Jan. 2000.

\bibitem{wang:2012}
S.~{Wang}, X.~{Meng}, and T.~{Chen}, ``Wide-area control of power systems
  through delayed network communication,'' \emph{IEEE Transactions on Control
  Systems Technology}, vol.~20, no.~2, pp. 495--503, Mar. 2012.

\bibitem{WatsonEMTP}
N.~Watson and J.~Arrillaga, \emph{{Power Systems Electromagnetic Transients
  Simulation}}.\hskip 1em plus 0.5em minus 0.4em\relax London, UK: The IET,
  2003.

\bibitem{Wei}
Y.~Wei, W.~T. Peter, Z.~Yao, and Y.~Wang, ``The output feedback control
  synthesis for a class of singular fractional order systems,'' \emph{ISA
  transactions}, vol.~69, pp. 1--9, 2017.

\bibitem{heydt:2002}
H.~Wu, H.~Ni, and G.~T. {Heydt}, ``The impact of time delay on robust control
  design in power systems,'' in \emph{Proceedings of the IEEE PES Winter
  Meeting}, 2002.

\bibitem{heydt:04}
H.~Wu, K.~S. Tsakalis, and G.~T. Heydt, ``Evaluation of time delay effects to
  wide-area power system stabilizer design,'' \emph{IEEE Transactions on Power
  Systems}, vol.~19, no.~4, pp. 1935--1941, Nov. 2004.

\bibitem{chen:06}
D.~{Xue}, C.~{Zhao}, and Y.~{Chen}, ``A modified approximation method of
  fractional order system,'' in \emph{International Conference on Mechatronics
  and Automation}, 2006, pp. 1043--1048.

\bibitem{ajjarapu:06}
D.~{Yang} and V.~{Ajjarapu}, ``A decoupled time-domain simulation method via
  invariant subspace partition for power system analysis,'' \emph{IEEE
  Transactions on Power Systems}, vol.~21, no.~1, pp. 11--18, 2006.

\bibitem{zamani:2009}
M.~Zamani, M.~Karimi-Ghartemani, N.~Sadati, and M.~Parniani, ``{Design of a
  fractional order {PID} controller for an {AVR} using particle swarm
  optimization},'' \emph{Control Engineering Practice}, vol.~17, no.~12, pp.
  1380 -- 1387, Dec. 2009.

\bibitem{zhou:99}
K.~Zhou and J.~C. Doyle, \emph{{Essentials of Robust Control}}.\hskip 1em plus
  0.5em minus 0.4em\relax Prentice Hall, 1998.

\end{thebibliography}
\end{spacing}

\newpage

\end{document}